\pdfoutput=1
\documentclass[AER,a4paper, draftmode]{style/umatter_AEA}
\AtBeginDocument{\setlength{\overfullrule}{0pt}} % keep draftmode's wide spacing but hide overfull-rule black bars

\usepackage{tocloft}
\usepackage{xcolor}
 
\usepackage[margin=1.5in]{geometry}

\usepackage[authoryear]{natbib}
\usepackage{authblk}
\usepackage{amssymb}
\usepackage{setspace}
\usepackage{graphicx}
\usepackage{float}
\usepackage{booktabs}
\usepackage{lscape}
\usepackage{ragged2e}
\usepackage{threeparttable}
\usepackage{longtable}

\usepackage{afterpage}

\usepackage{titletoc}

\usepackage{scrwfile}
\TOCclone[\textbf{}]{toc}{atoc}

\AfterTOCHead[toc]{
  
}
\AfterTOCHead[atoc]{
  \edef\maintocdepth{\the\value{tocdepth}}
  \value{tocdepth}=-10000\relax
  
}

\usepackage{mathptmx}
\PassOptionsToPackage{hyphens}{url}
\usepackage{url}
\usepackage{hyperref}
\hypersetup{
    colorlinks=true,
    linkcolor=black,
    filecolor=black,      
    urlcolor=black,
    citecolor=teal,
}
\usepackage[utf8]{inputenc}

\usepackage[toc,page,header]{appendix}
\usepackage{minitoc}

\renewcommand\thepart{}
\def\partname{}

\renewcommand*{\theparagraph}{\theparagraph . ---}

\draftSpacing{1.5}

\begin{document}

\title{Who Owns the Online Media?}

% General points:

% Timeline/observation period
% We are somewhat unclear on the timeline/observation period(s):
% Around line 109: States ownership data is "as of 2020 and as of 2023"
% Around line 109 footnote: Says "different data sources... ranging from 2018 to 2020"
% around line 193 says: Traffic data is "average... from 2018 through 2020"
% around line 296 says: Content analysis covers "2019 to 2023"
% around line 303 refers to periods as "2020/21 and 2022/23" but earlier says "2019/2020"
% We should have a clear timeline table in the Data section or in the data appendix showing exactly when each data source was collected and what it covers.

% Potential sample size discrepancies and/or lack of clarity
% - Initial sample: 11,967 domains (line 149)
% - Main analysis: 4,578 "hard news" websites (line 111, 151)
% - But I had the feeling that various tables and analyses seem to use different subsamples without always clarifying which

\author[1,2]{\authorsize\textsc{Ulrich Matter}}
\author[3,4]{\authorsize\textsc{Philine Widmer}}

\affil[1]{\Large Bern University of Applied Sciences}
\affil[2]{\Large University of St.Gallen}
\affil[3]{\Large UniDistance Suisse}
\affil[4]{\Large Paris School of Economics}

\date{\today}

% \version{Preliminary version. Please do not cite or distribute.}

\begin{abstract}
% Previous: We examine the ownership structure, reach, and news content of thousands of news websites in the U.S., Canada, and Europe. Documenting the ownership network of each website, we find that more than half of news websites have a single ultimate owner. Otherwise, they tend to have highly diversified ownership structures, making it unclear who is ultimately responsible for the content. We link the ownership networks to a large database of online news content. News websites (partly) owned by the same entities exhibit systematically more similar news content, an association that holds even when exploiting within-outlet-pair variation over time.
% Suggested by Philine:
Ownership matters for the media's watchdog role. We map the ownership networks behind thousands of online news outlets in the U.S., Canada, and Europe. The networks reveal who is ultimately responsible for the news: for over half of the outlets, a single entity. The rest sit behind multi-layered structures, making responsibility hard to trace. Market concentration, measured comparably across countries, is largely low to moderate. Looking at content, we find that co-owned outlets report more similarly, even within fixed outlet pairs, as ownership changes -- not least in the U.S., where reader demand is often thought dominant.
\end{abstract}

\footnotetext[1]{
%We acknowledge seed grants from the University of St.Gallen Basic Research Fund (GFF) and the Wachter Stiftung. We thank three anonymous reviewers, Elliott Ash, Stefan Bühler, Julia Cagé, Marcel Garz, Malka Guillot, Roland Hodler, Ekaterina Zhuravskaya, participants at the European Public Choice Society Annual Meeting, the Annual Meeting of the Verein fuer Socialpolitik, the Annual Congress of the Swiss Society of Economics and Statistics, the Public Governance Group online seminar at Paris-Dauphine PSL Institute, the Workshop of the Swiss Network on Public Economics, the Workshop on the Political Economy of Attention and Electoral Accountability, and the ZEW-IFO Young Scholar Political Economy Workshop, as well as seminar participants at ETH Zurich and at the Universities of Basel, Bern, Konstanz, St.Gallen, and Zurich for helpful comments and suggestions. Noel Ackermann, Hewson Duffy, Anton Heimerdinger, Nobin Kachirayil, Johannes Ladwig, Andrei Rafikov, Elena Trevisani, and Dzhon Vi provided excellent research assistance. \protect\\ \indent 
\textit{Email addresses:} \href{mailto:ulrich.matter@unisg.ch}{ulrich.matter@unisg.ch} (Ulrich Matter) \href{mailto:philine.widmer@psemail.eu}{philine.widmer@psemail.eu} (Philine Widmer)}

\maketitle

\doparttoc
\faketableofcontents

\newgeometry{margin=1.1in}

%%%%%%%%%%%%%%%%%%%%% INTRODUCTION %%%%%%%%%%%%%%%%%%%%%%%%%%%%%%%%%%%%%%%%%%%

\section{Introduction} \label{sec:introduction}

News media are a key resource for citizens seeking to hold their governments accountable,\footnote{See \cite{eisensee2007news}, \cite{ferraz2008exposing}, \cite{gerber2009does}, \cite{Snyder_Stroemberg_2010}, \cite{kendall2015how}, \cite{arias2018does}, \cite{arias2018priors}, \cite{cruz2021buying}, \cite{chen2019impact}, and \cite{knight2019opposition} for the disciplining effect of mass media.} making the ownership and concentration of news markets longstanding matters of public concern. Questions about who owns the media, and how concentrated that ownership is, are therefore closely tied to the effectiveness of the media as the ``fourth estate'' \citep{noam_2016}. Diversity in media ownership is widely considered crucial for political accountability \citep{McMillan_Zoido_2004,Besley_Prat_2006}. In line with this view, particular owners have been shown to bend reporting toward their special interests \citep{Gilens_2000,Szeidl_Szucs_2021}. This evidence is largely case-based, while systematic evidence on the link between ownership and content (both which events outlets cover and the tone in which they report them) across many outlets and countries remains surprisingly limited. The broadest existing evidence comes from a single setting, U.S.\ newspapers, where slant has been linked largely to reader demand \citep{Gentzkow_Shapiro_2010}. In this paper, we propose a parsimonious way to describe media ownership and to compute comparable market-concentration figures across countries, and we apply it to the near-universe of online news outlets in the largest Western democracies.\footnote{We cover North America (the U.S.\ and Canada) and the seven most populous (former) European Union members (the U.K., France, Germany, Italy, Poland, Romania, and Spain). Our sample also includes Switzerland, which we used as a small pilot sample to design the data-collection procedures, since both authors were based there at the time.} We then also combine these ownership data with news-content data to study whether outlets under shared ownership cover more similar sets of stories, both cross-sectionally and within fixed outlet pairs over time, exploiting only \textit{changes} in shared ownership.

A key characteristic of the existing evidence on media ownership and concentration is that studies differ widely in their samples, units, and methods, making comparisons across countries difficult. \cite{Djankov_et_al_2003} examine the \textit{top five} television stations and \textit{top five} newspapers in 97 countries, focusing on the distinction between public and private ownership. \cite{noam_2016} documents market concentration in thirty countries from the market shares of leading firms (covering traditional industries such as television and newspapers as well as Internet-related industries such as Internet service providers and search engines) with country-level data assembled by different national experts. \cite{cageetal2017} studies ownership structures in France and Spain, highlighting a lack of transparency, and \cite{durante2020media} compiles ownership data for 20 outlets across four European countries. Each contribution is informative, but the heterogeneity in coverage and methods makes it hard to place countries on a common scale, or to look beyond the leading firms to the full structure of ownership. Scale also matters for the link between ownership and content: testing whether co-owned outlets produce similar news requires observing many outlets, and thus many co-owned pairs, in comparable data across many markets.

Our cross-country, near-universe approach is possible because we focus on online news outlets and assemble the sample from openly available, cross-nationally comparable directories of news websites rather than from country-specific sources: information on web domain registrants. Online news outlets have become an increasingly important source of (political) information (the majority of U.S.\ citizens now consume news on digital devices, primarily via news websites or apps; \citealp[see][with data as of 2020]{Shearer_2021}; the pattern persists in more recent data, with news websites or apps still the leading digital pathway, \citealp{Pew_2025}). Yet, little is known about who owns them. Near-universal coverage is built into our sampling: we start from ABYZ News Links, a directory of news websites by country, and cross-check this list with Media Cloud and with domains labeled as news or media by Amazon's Alexa Web Information Services (AWIS), arriving at 11,967 news domains across the ten countries. Given our political-economy focus, our main analyses then concentrate on the 4,578 ``hard news'' outlets among them, those where at least one in four stories covers politics, so that we retain, for example, nytimes.com but drop foxsports.com. The outlets we cover indeed account for a substantial share of online attention: for every million unique visitors present on any website worldwide, roughly 110,000 cumulatively went to the news websites in our sample (roughly 85,000 for hard news outlets); for reference, Google's reach is roughly 575,000 \citep{AWIS}. The market we characterize thus carries a highly relevant share of the political information consumed in these democracies.

When collecting ownership data (which we do in two waves, as of July 2020 and July 2023), we start from a domain (e.g., nytimes.com) and identify its immediate owner by matching domain registration records to Bureau van Dijk's Orbis database (now Moody's Orbis database). We then trace the entire ownership structure behind that initial owner until we reach entities that are not themselves owned by any other entity (i.e., the endpoints of the control chains), which we call the outlet's \emph{ultimate owners}. Ultimate owners need not be natural persons: they comprise individuals, trusts, foundations, and state bodies, as well as firms for which no further owners are recorded (notably, widely held corporations). The two vantage points capture different aspects of control: the immediate (domain) owner is closest to an outlet's day-to-day editorial and operational decisions, whereas the ultimate owners sit at the top of the chain of voting rights and thus hold the residual control rights over the outlet. In principle, ultimate owners determine who gets to make those day-to-day decisions, but whether they exercise this control in practice (particularly when ownership is dispersed across many small or passive investors) is an empirical question to which our content analysis returns. For each ownership link, we record voting (control) rights rather than cash-flow rights wherever Orbis distinguishes the two, following \cite{aminadav2020corporate} -- in isolated cases, Orbis reports financial ownership instead (see the New York Times example in Section~\ref{sec:data:collection}); our premise is that control over votes, not cash flow, positions an owner to shape editorial decisions. This procedure yields 23,229 unique owning firms (i.e., firms that appear somewhere in an ownership structure) and 20,048 unique ultimate owners. In contrast to most of the existing literature, our procedure allows us to detail the individual ownership structure behind each outlet. Work in media economics has typically assumed, at least implicitly, that a single entity owns and controls an outlet \citep[e.g.,][]{Prat_2018}. Our approach lets us document whether that is the case or whether highly diversified ownership structures are also present.

Our results show that for 57\% of outlets, ownership is tractable: they have a single ultimate owner, either directly (for 39\% of outlets, the registering entity is itself the ultimate owner) or through a relatively simple chain. A substantial share, however, has what we call \emph{hard-to-track} ownership. By this, we do not mean that ownership is (necessarily) deliberately concealed, but that it passes through many intermediaries and many layers, so that, in consequence, it is \emph{complex}. Concretely, 45\% of outlets have ownership networks with five or more nodes (among them, a mean of 1,689 and a median of 511 nodes), and outlets with more than one ultimate owner have a mean (median) of 98 (71) ultimate owners, reached through up to 22 ownership levels. Importantly, the two facets of complexity go together: node counts and ultimate-owner counts are correlated at 91\%, and outlets very rarely combine many nodes with a single ultimate owner. Hard-to-track ownership in our data thus reflects genuinely dispersed ownership (many small shareholders) rather than a single controlling owner hiding behind layers of intermediary companies.

Such ownership structures are not unusual: studies of corporate ownership across industries document similarly complex and dispersed structures \citep{Vitali_et_al_2011,aminadav2020corporate,Backus2020,dall2022using}. If anything, ownership concentration is higher in the media: whereas \cite{aminadav2020corporate} find that, as of 2012, slightly more than half of the listed firms they study worldwide are widely held (no owner holding more than 20\% of voting rights), 87\% of the outlets in our data have at least one controlling owner by the same 20\% standard, and 71\% even have a majority owner. Our ownership-concentration figures exceed their cross-sector figures in all ten countries. This is in line with the long-standing argument that controlling a media outlet carries benefits beyond profits, its ``amenity potential'', so that private media ownership is expected to be concentrated \citep{demsetz_lehn1985,grossman_hart1988,Djankov_et_al_2003}.

Hard-to-track ownership structures do raise, however, the question of who is ultimately responsible for news content, and thereby speak to a second, complementary concern: ownership transparency. Only if news consumers can tell who stands behind an outlet can they rationally discount potential owner-driven reporting biases \citep{cageetal2017,Prat_2018}. When ownership passes through many entities and many layers, it becomes prohibitively costly for ordinary readers, or even regulators, to establish who ultimately stands behind an outlet, which makes it hard to anticipate, and thus to discount, potential bias. Tracing ownership to its endpoints, we find that a large share of the online news market is ultimately owned not by media companies but by the financial sector, which is the most or second-most important sector among ultimate owners in every country in our sample except Switzerland (see Section~\ref{sec:results:sectors}). While much of this may reflect passive investment rather than strategic control, ownership still matters for content even at the ultimate-owner level, as we show below (Section~\ref{sec:coverage}).

Turning from individual outlets to markets, we measure market shares based on web traffic (our main measure is unique visitors, averaged over 2018--2020, given by data availability). Relying on web traffic data lets us measure readership consistently across countries without requiring country-specific revenue or circulation data.\footnote{Traditionally, media economists used newspaper circulation to quantify an outlet's audience; reach (unique visitors) is its closest online counterpart. While reach is our primary measure, our results are robust to using page views instead.} One consequence is that we also capture part of the attention an outlet receives indirectly, for example through social media: if someone ends up clicking on a link they saw in a social-media post and lands on the outlet's site, we record that reader just as if they had navigated directly to the website, say, to nytimes.com. What we would not capture is a reader who only sees a headline embedded in a social-media post without visiting the site.

We find market concentration to be mostly low to moderate. At the news-site level, the Herfindahl-Hirschman Index (HHI, in terms of reach) lies between 0.02 and 0.14 for most countries, and is moderate to high only in Poland and the U.K.\ (0.26 and 0.20, respectively). Concentration increases somewhat when shares are measured at the level of the sites' immediate owners (Poland, Switzerland, and the U.K.\ then show moderate to high values) and declines again when measured at the level of \emph{ultimate} owners (the endpoints of the ownership chains), where eight of the ten countries exhibit low concentration (0.10 or below) and only the U.K.\ (0.18) and Poland (0.16) remain in the moderate band. These levels echo prior cross-country evidence: in \cite{noam_2016}, Western democracies sit toward the lower end of the international concentration distribution. Also in line with previous research (the U.S.\ is the least concentrated of the thirty countries covered by \citealp{noam_2016}), we find concentration to be particularly low in the United States, the least concentrated market in our sample at the news-site and domain-owner levels.

Finally, we turn to perhaps the most important question: whether ownership matters for content at large. Given our consistent-across-countries, near-universe approach, we cannot exploit quasi-experimental variation in ownership, since such variation typically applies only to specific outlets, events, or media markets. What we can offer is evidence on whether outlets owned by the same entity feature systematically more similar content, at both the level of immediate (domain) owners and ultimate owners. Linking our ownership networks to a large database of online news coverage (the Global Database of Events, Language, and Tone, or GDELT), we find that two outlets sharing a domain owner cover substantially more similar sets of events and report in a more similar tone. The association is large: same-owner pairs have a mean coverage similarity of 0.42, on a zero-to-one scale, against 0.04 for pairs with different owners. It also survives outlet-pair fixed effects: exploiting only \emph{changes} in shared ownership between our 2020 and 2023 waves, coverage similarity is 0.33 higher when a pair shares a domain owner (relative to when the same pair does not). The estimates are essentially unchanged when we restrict the sample to within-country pairs, and pairs of outlets that later come under common ownership show no pre-existing convergence in coverage; if anything, the opposite. Smaller but precisely estimated associations hold at the ultimate-owner level; if these associations reflect owner influence, they imply that the ease with which readers can establish who is ultimately responsible for their news matters in practice. Because these patterns emerge across thousands of news websites in ten countries and under various demanding specifications, we conclude that owner influence on content is a likely feature of the online news market in Western democracies.

Regarding the mechanisms that drive this alignment, co-owned outlets may share editorial inputs, a syndication or efficiency channel that should operate regardless of topic. Alternatively, common owners may impose a shared political line, which should matter mainly for political content. We find evidence for both channels (Section~\ref{sec:mechanisms}). In choosing which events to cover, the same-owner association is large for both political and non-political events, consistent with content sharing. For the tone of reporting, the association is roughly twice as large for political as for non-political events, consistent with owners specifically aligning the tone of political coverage. Our results are thus consistent with the interpretation that syndication shapes \emph{what} co-owned outlets cover, while ideology shapes \emph{how} they slant politics.

Our work relates to three literatures. First, it speaks to research on media market structure, ownership, and their political consequences \citep{Djankov_et_al_2003,Stroemberg_2004_Mass,OberholzerGee_Waldfogel_2009,Gentzkow_et_al_2011,rolnik2019protecting,cage2020media}. Cross-country evidence in this literature has characterized the leading firms in a few, mostly traditional, media industries \citep{Djankov_et_al_2003,noam_2016}; as noted above, our concentration results align with its findings for Western democracies. Our contribution is to map the near-universe of online news outlets rather than the leading firms. We trace ownership through the corporate chain to the ultimate owners and focus on voting (control) rights. Our harmonized pipeline makes the resulting figures comparable across countries. This yields individual ownership \emph{structures}, not only concentration ratios, including the prevalence of hard-to-track ownership, which prior cross-country work could not observe. Relatedly, recent work argues that media power is best assessed through the attention outlets command rather than through traditional market shares \citep{Prat_2018,Kennedy_Prat_2019}. Our market shares are traffic-based rather than revenue-based, and Section~\ref{sec:extensions} shows that they correlate strongly with attention shares constructed in the spirit of \cite{Prat_2018}.

% Second, our content results connect to work on how owners shape news. Much of this evidence comes from particular cases: individual owners imprinting their interests on coverage \citep{Gilens_2000,Szeidl_Szucs_2021}, station groups reshaping the content of the local TV outlets they acquire \citep{martin_mccrain_2019,Martin_et_al_2024}, or newspaper mergers homogenizing content \citep{Garz_Ots_2025}. Beyond particular cases, comparative work links broad ownership forms (market, private, civil society, and public) to systematically different modes of news production \citep{Benson_et_al_2024}. Relative to the demand-side benchmark of \citet[who find that the slant of U.S.\ newspapers is driven mainly by reader demand, with common ownership explaining little]{Gentzkow_Shapiro_2010}, our online evidence points to a detectable supply side even in the U.S.\ setting: restricting the analysis to U.S.\ outlet pairs yields same-owner associations comparable to those in the rest of our sample, for coverage and tone alike (Section~\ref{sec:mechanisms}).

Second, our content results connect to work on how owners shape news. Much of this evidence comes from particular cases: individual owners imprinting their interests on coverage \citep{Gilens_2000,Szeidl_Szucs_2021}, station groups reshaping the content of the local TV outlets they acquire \citep{martin_mccrain_2019,Martin_et_al_2024}, or newspaper mergers homogenizing content \citep{Garz_Ots_2025}. Beyond particular cases, comparative work links broad ownership forms (market, private, civil society, and public) to systematically different modes of news production \citep{Benson_et_al_2024}. Relative to the demand-side benchmark of \citet[who find that the slant of U.S.\ newspapers is driven mainly by reader demand, with common ownership explaining little]{Gentzkow_Shapiro_2010}, our online evidence points to a detectable supply side even in the U.S.\ setting: restricting the analysis to U.S.\ outlet pairs yields same-owner associations comparable to those in the rest of our sample, for coverage and tone alike (Section~\ref{sec:mechanisms}). A plausible reconciliation is technological: online production has sharply lowered the marginal cost of reusing content across co-owned outlets (\citet{cage_herve_viaud_2019} document pervasive and near-instantaneous copying of online news content) so that supply-side ownership effects that were negligible in the print era become detectable online. In addition, we study a different margin (i.e., the pairwise similarity of event coverage and tone across co-owned outlets, rather than the slant of a given outlet relative to reader preferences). Accordingly, our findings should be seen as complementing rather than overturning the demand-side evidence. We return to this in Section~\ref{sec:mechanisms}.

Third, we contribute to the literature in economics and finance that characterizes corporate ownership structures \citep{Vitali_et_al_2011,aminadav2020corporate,Backus2020,dall2022using} by providing ownership-structure facts specific to the media sector, where, as discussed above, controlled ownership is even more prevalent than across sectors generally.

Our paper is organized as follows: Section~\ref{sec:data} describes our data collection procedure, the construction of ownership networks, and the measurement of market concentration. Section~\ref{sec:results} presents our main results regarding media ownership structure and market concentration. Section \ref{sec:coverage} presents our results on the association between ownership and news content, and Section~\ref{sec:mechanisms} separates the channels behind this association. Section~\ref{sec:extensions} shows robustness checks and extensions. Section~\ref{sec:discussion} puts our findings into perspective and concludes.

%%%%%%%%%%%%%%%%%%%%% DATA AND MEASUREMENT %%%%%%%%%%%%%%%%%%%%%%%%%%%%%%%%%%%%%%%%%%%

\section{Data and Measurement} \label{sec:data}

Our study employs domain registry data, data on firm ownership, content data, and domain-level web traffic data. We describe our data collection and measurement approach step-by-step.

\subsection{Data Collection, Matching, and Coding} \label{sec:data:collection}

We collect data on all major news websites (i.e., domains) associated with a given country. Specifically, we collect the domain registry information of these websites and match the registrant information with a state-of-the-art firm database. We also collect traffic and content data for the corresponding news websites. In the following, we provide details on our data collection procedure.

\paragraph{Initial sample of ``news'' websites. }  We use three sources for our initial set of news websites: ABYZ News Links \citep{ABYZ_2021}, Media Cloud \citep{mediacloud_2021}, and domains broadly labeled with ``news'' or ``media'' by Amazon's Alexa Web Information Services (AWIS; \citealt{AWIS}). Our observation units are canonical domains: for example, we treat nytimes.com as a single website and do not treat open.nytimes.com and krugman.blogs.nytimes.com separately. First, we compile a list of domains from ABYZ that uniquely identify news websites by country. We then cross-check whether these websites are also covered in \cite{AWIS} and Media Cloud, and extend the list if necessary. We also drop some domains from this list, such as news aggregators (see Online Appendix \ref{sec:app:data} for details). Our approach identifies an initial set of 11,967 news domains/websites across 10 countries: the U.S., Canada, Germany, France, the U.K., Italy, Spain, Poland, Romania, and Switzerland. We focus on North America and Europe because of their economic and technological development (a well-established online media industry), political institutions (free media and internet access, but variation in the role of public broadcasters), and the availability of standardized firm-level data. Within Europe, we have chosen the most populous (former) members of the European Union (Germany, France, the U.K., Italy, Spain, Poland, Romania), and Switzerland. The latter served in this project's pilot study, since both authors were based in Switzerland at that time.

\paragraph{Identifying hard news websites. } ABYZ, AWIS, and Media Cloud, define ``news'' broadly. Some listed websites focus primarily on sports (e.g., espn.com). We therefore categorize news websites as ``soft news'' (sports, entertainment, etc.) and ``hard news'' providers (outlets that engage in journalism that fosters political accountability). As a proxy for hard-news-providing websites, we filter for news sites where at least one in four articles covers politics: we collect the content of the news websites (from 2018 through 2020) via the GDELT database\footnote{The Global Database on Events, Language, and Tone (GDELT) monitors worldwide news by scraping content from online newspapers, TV station websites, and other online news sources.}. Online Appendix \ref{sec:app:political_tags} describes the procedure to filter for political content from automated annotations performed by GDELT and provides examples of how articles and outlets were classified, including examples. Our main analyses focus on 4,578 hard news websites.
%  It also provides examples of outlets that were categorized as (not) covering the political process.

\paragraph{Domain registrant and direct ownership information. }
To collect ownership information for each news website, we exploit the fact that web domains must be registered with the Internet Corporation for Assigned Names and Numbers (ICANN).\footnote{The Internet Assigned Numbers Authority (IANA, overseen by the ICANN) manages the registration of domains. IANA delegates most of its domain name authority to other domain name registries, such as registries for country-code top-level domains (e.g., DENIC in Germany for the top-level domain .de). Many such registries can be queried via the WHOIS protocol. A WHOIS lookup provides information about who (an individual or organization) is officially registered as the domain name holder. The \textit{registrant of a domain} is the entity that controls the content hosted at the domain (e.g., for nytimes.com, the registrant is The New York Times Company).} For each domain on our list, we programmatically query data on the registrant. We collect the data via a commercial provider (WHOIS API LLC) and by directly querying the WHOIS service. Next, we match the registrant information to firms in Bureau van Dijk's Orbis database.\footnote{Orbis is a global database for standardized information on 280 million firms.}  Identifying owners involves a lot of manual data handling due to unsuitable formatting of the raw registry data and a lack of exact matches with Orbis. That is, when linking registrants to the Orbis database, there are instances in which a direct match can be found via an exact-string search, such as when both the registrant and the Orbis entry share the same name, such as ``The New York Times Company.'' However, it is essential to note that string matching often yields imperfect results, as in ``The New York Times Company LTD'' versus ``The New York Times Company.'' Consequently, manual verification is necessary when searching for the registrant's string in Orbis, where a human evaluator must confirm a close match or select from a list of potential matches. Two independent coders validated every manual match to ensure accuracy.\footnote{Registrants can in principle be privacy or proxy services, legal holding entities, or IT service providers rather than the operating media company. In our sample, this is rare: registrants flagged as privacy services or IT service providers account for less than 2\% of the outlets we match to Orbis, and the matched entity is itself a privacy company in three cases across the two waves. Where such registrants prevent a match, the outlet enters our data as missing ownership information, whose consequences we assess in the simulations of Online Appendix~\ref{sec:robustness_missing}.}

Importantly, our approach to working with \emph{Orbis} parallels the approach of \citeauthor{aminadav2020corporate} in their seminal work. Like in \cite{aminadav2020corporate}, we particularly focus on voting rights, while acknowledging the inherent limitations of BvD's data (as also outlined in \cite{aminadav2020corporate}). BvD meticulously aggregates ownership data from numerous sources, including corporate reports, stock exchange disclosures, official company websites, media publications, private correspondence, and specialized agencies that monitor corporate performance and ownership structures. Thus, BvD's data collection primarily focuses on voting rights associated with ownership, rather than on the full range of share types. As elucidated by \cite[p.~1196]{aminadav2020corporate}, the BvD's user manual explicitly states its intent to ``track control relationships rather than patrimonial relationships,'' emphasizing that in instances of dual share classifications (Voting/Nonvoting), the database predominantly records percentages linked to voting shares.

Nevertheless, in certain instances, coding errors in Orbis may result in the inclusion of financial ownership rights rather than voting rights. For example, the New York Times Company has a dual-class share structure: a family trust holds approximately 90\% of the Class B common stock, granting the Ochs-Sulzberger family the ability to elect 70\% of the board of directors \citep{NYT10K2015}. However, Orbis reports financial ownership stakes (primarily Class A shares) rather than voting control for this company. % For example, this seems to be the case for nytimes.com (and the corresponding domain owner, the New York Times Company. The class B shares held by the Sulzberger family should be in Orbis rather than those held by Slim Helú. 

Furthermore, academic discourse frequently distinguishes between entities categorized as ``controlled'' and those exhibiting widely distributed ownership. This distinction is central to the analysis conducted by \citeauthor{aminadav2020corporate} (\citeyear{aminadav2020corporate}), who specifically identify shareholders possessing more than 20\% of voting rights as proxies for ``controlled'' firms. We consider all owners in our baseline results. Robustness checks consider only the largest owners who collectively hold 20\% or 50\% of an outlet. Specifically, we aggregate ownership percentages from the largest shareholder, successively including subsequent major shareholders until at least 20\% or 50\% of ownership is achieved.

Ownership data coverage varies across countries, ranging from almost 40\% of domains in the U.S. to 90\% in Germany (see Online Appendix Table~\ref{tab:covstat} for details). When weighted by traffic, coverage is substantially higher, at least 83\% across all countries in the 2020 wave, on which our baseline ownership analyses rest (in the 2023 wave, the minimum is 66\%, for France), as ownership information is more often unavailable for smaller outlets. Section~\ref{sec:robustness_missing} addresses potential biases from missing data via simulation and confirms that our concentration estimates are robust up to relatively large error margins.

\paragraph{Data on events covered by news websites}.  
We draw on the Global Database of Events, Language, and Tone (GDELT) to identify which news outlets covered which events.\footnote{For other use cases of GDELT data in economics research, see, for example \cite{besley2020terror}.} In line with our two waves of ownership data collection, we concentrate on two distinct time frames, an earlier period from January 2019 to June 2020 and a later period from January 2022 to June 2023. We merge these event mentions with our previously curated list of outlets, which we have matched to firm-level ownership data from Orbis. To ensure the resulting dataset is both substantively meaningful and systematically comparable, we implement a series of increasingly stringent filters. We first focus on events with high reporting intensity, retaining only those covered by more than 50 distinct outlets in our matched hard-news sample. As in our main analysis of ownership structures, we further filter for news websites that cover hard news to a substantial degree (more than 25\% of the content is related to politics and related topics). Given the remaining news sites and the reported events they cover, we focus on major, widely reported stories, those in the top two quintiles of the distinct-outlet coverage count (quintiles computed on the pooled data across both periods). This refined dataset, organized at the news site-event level, later enables us to compute pairwise similarity measures of coverage sets.\footnote{GDELT visibility varies across countries (Online Appendix Table~\ref{tab:covstat}): traffic-weighted content coverage is lowest for Poland (62\%) and Italy (77\%). Our content results are robust to excluding these two countries; the within-dyad coverage estimate on the remaining within-country pairs is 0.330, identical to the baseline. The excluded subsample contains only 14 switching pairs, too few for informative separate estimates.} Online Appendix Table~\ref{tab:sample_overview} traces the path from the initial outlet universe to each estimation sample used below.

In a second step, we complement the coverage data with information on reporting tone. Drawing again on GDELT and its integrated Global Knowledge Graph (GKG) database, we compile measures of article-level sentiment (tone) that GKG provides based on the full text of the news articles. We merge these indicators back into our event-outlet pairs, retaining only the reporting tone for events that are also in our event-coverage set.\footnote{GDELT GKG does not provide reporting tone data for all events/mentions in GDELT.} We apply the same selection filters as before and maintain the temporal delineation between the earlier and later periods. When an outlet mentions an event in several articles, we assign the outlet--event tone as the mean article tone. This produces a second news site-event-level dataset for more nuanced comparisons of editorial convergence or divergence (not only in the choice of events/stories covered but also in the tone of that coverage).

\subsection{Drawing Ownership Networks} \label{sec:drawing_networks}

Once we have identified a registrant in Orbis, we recursively extract the entire ownership structure from the database (the entities that own the registrant, their owners, the owners of these owners, etc.). We can thus capture the ownership relationships of a news website as an \emph{ownership network}. To draw ownership networks, we devise the following graph traversal algorithm (starting at the domain registrant entity): $(1)$ find all entities that have a stake in the current entity, $(2)$ draw a directed edge (weighted with the corresponding ownership share) from the current (owned) entity to each of the owning entities, $(3)$ follow each of these links and continue with step $(1)$.\footnote{With circular ownership, the entry point into the circle is coded as the owner. For example, let UBS own X\% of an outlet, and let Y\% of these X\% be owned by Julius Baer, which is, in turn, owned by Z\% by UBS. Then, the $X\times Y\times Z$\% are assigned to UBS.} Based on recursion, the algorithm continues traversing and drawing the ownership network behind a given domain until there are no more entities to visit that own more than 1\% of the current entity. We consider the leaf nodes in this network, that is, the last firms that held a share of 0.01 or more, as the \emph{ultimate owners} of the corresponding registrant and thus the registered website.\footnote{The halting criterion (1\% of shares) is, of course, somewhat arbitrary. It is a pragmatic choice as smaller thresholds would be computationally demanding, yet not necessarily informative.} To calculate the share of an ultimate owner, we compute the product of the ownership percentages from each level in a branch leading to that owner. If several branches link to the same ultimate owner, these products are aggregated under this owner's total share. Ultimate owners are essential for considerations of transparency, ownership concentration, and potential influence on news content. Suppose a news domain is registered by a media corporation that is ultimately owned by one person (either directly or through a network of firms and trusts). We would consider the ownership of this news website to be highly concentrated, and this owner's influence on the content as \textit{potentially} big. Alternatively, suppose that the registrant of a news domain is a media corporation owned by many other firms, which are owned by a myriad of retail investors. In that case, one might argue that the potential influence of a single ultimate owner on the content is relatively small, whereas ownership (and ultimate responsibility for the content) is considered rather opaque.

\begin{figure}[htbp]
\centering
\caption{News Website Ownership Networks}
\label{fig:trees}
\includegraphics[width=0.80\textwidth]{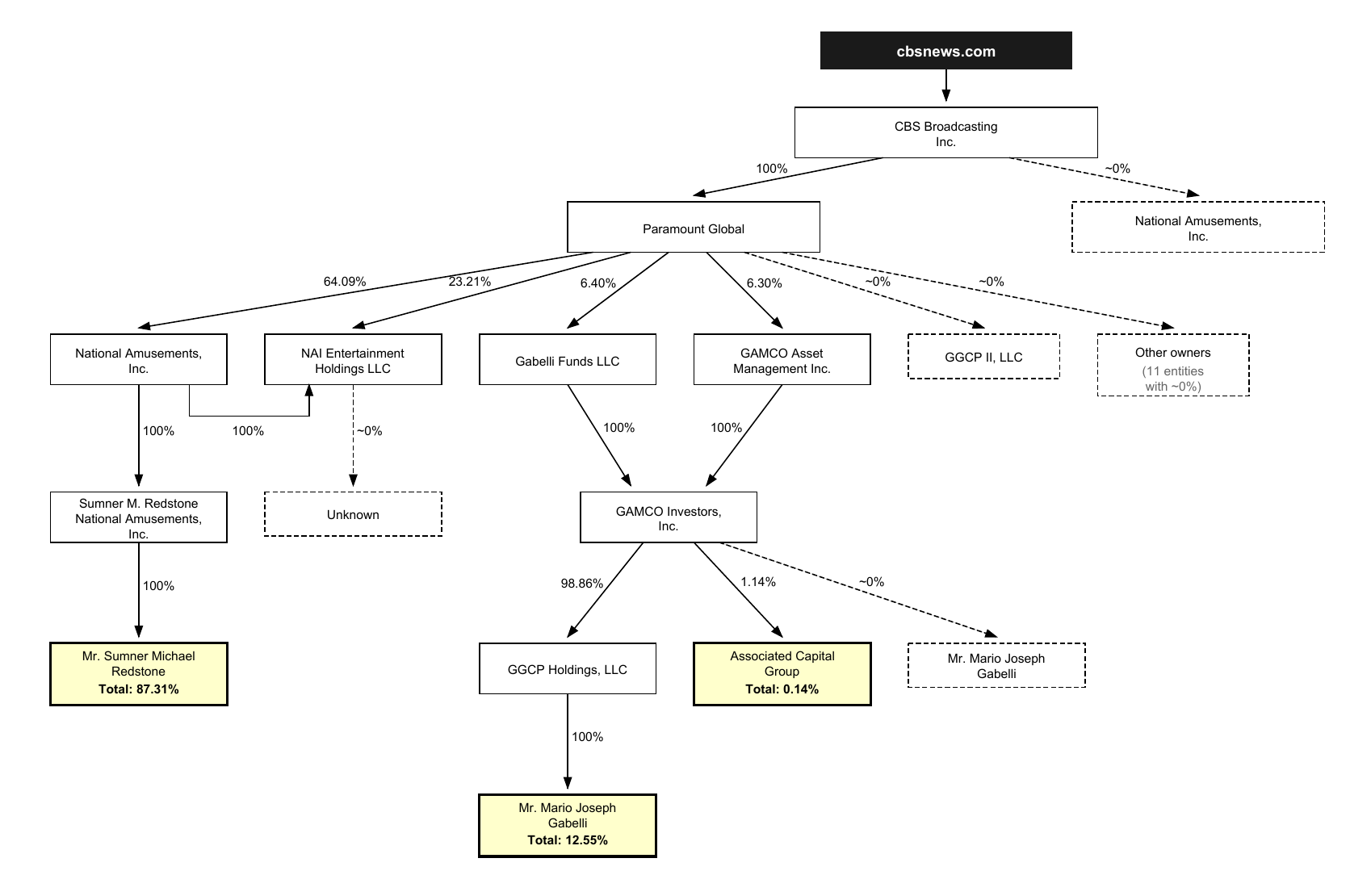}
\vskip 0.8cm
\includegraphics[width=0.99\textwidth]{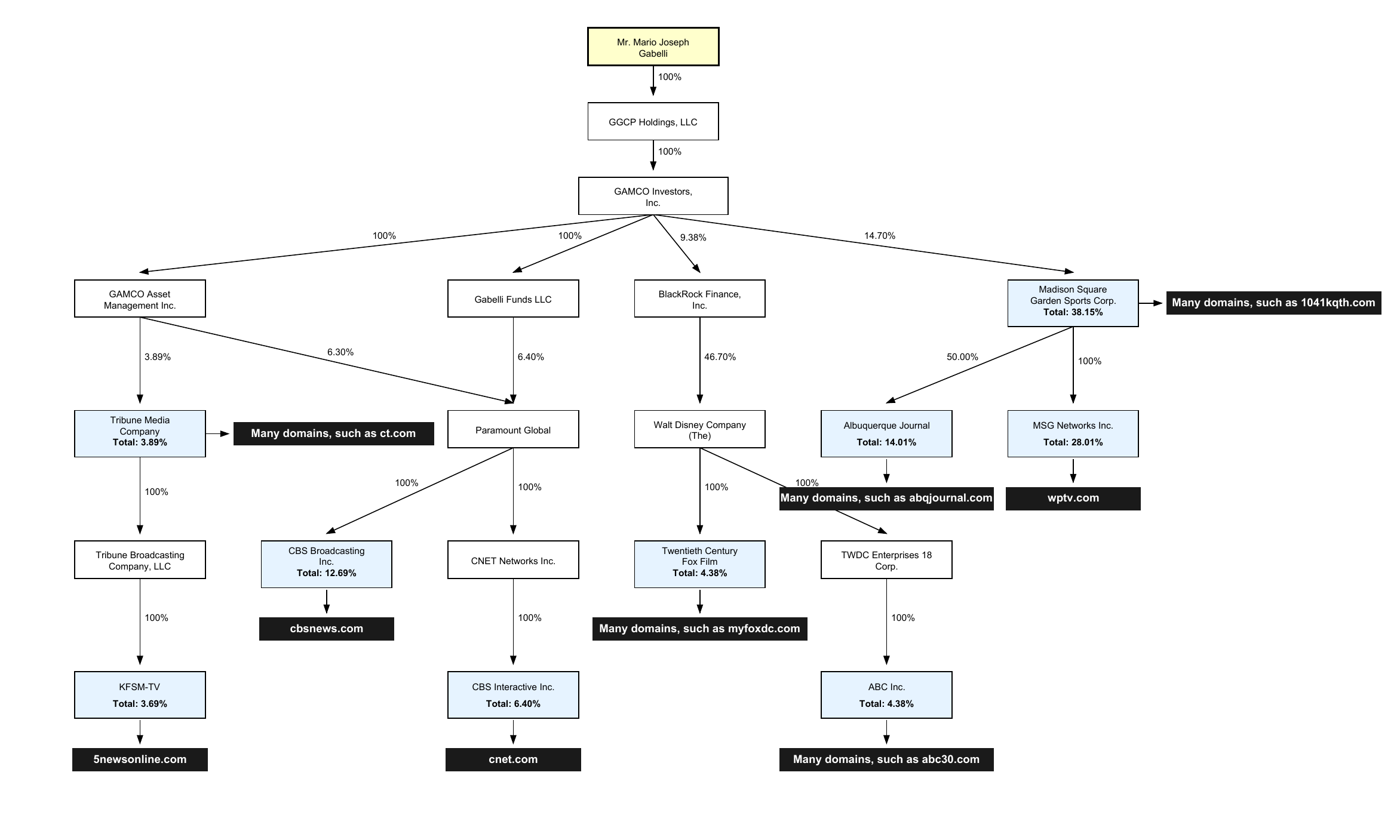}
\vskip 0.8cm
\begin{figurenotes}
The upper panel shows the ownership network for cbsnews.com as of July 2020. CBS Broadcasting Inc. is the registrant. The network traces ownership to three ultimate owners (yellow nodes): Sumner Michael Redstone (87.31\%), Mario Joseph Gabelli, and Associated Capital Group. The Gabelli investment firms channel the remaining 12.69\%, of which 12.55\% accrues to Mario Joseph Gabelli personally and 0.14\% to Associated Capital Group.
 The lower panel shows the portfolio of news domains partially owned by Mario Joseph Gabelli as of July 2020. In the upper panel, edges (arrows) point from an entity to its owners. In the lower panel, the arrows indicate what an entity owns.
\end{figurenotes}
\end{figure}

The ownership networks allow us to examine media ownership from two complementary perspectives. First, starting from a news website, we can trace ownership to identify its ultimate owners. Second, starting from an individual or firm, we can trace ownership to identify all news websites in their portfolio. Figure~\ref{fig:trees} illustrates both perspectives.

The upper panel shows the ownership network behind cbsnews.com as of July 2020. The domain is registered to CBS Broadcasting Inc., which, at the time, was wholly owned by Paramount Global. Tracing ownership further reveals three ultimate owners (highlighted in yellow): Sumner Michael Redstone holds 87.31\% through National Amusements, Inc.; the remaining 12.69\% is channeled through Mario Joseph Gabelli's investment firms, Gabelli Funds LLC and GAMCO Asset Management, with 12.55\% accruing to Gabelli personally and a negligible 0.14\% to Associated Capital Group, which enters the network through a 1.14\% stake in those firms.\footnote{Associated Capital Group's sub-1\% total does not contradict the 1\% halting criterion: the criterion governs the traversal, whereas an ultimate owner's reported share is the product of the stakes along the chain -- here, $1.14\% \times 12.69\% \approx 0.14\%$.}

This domain-rooted view reveals that cbsnews.com has concentrated ownership, with one individual controlling the vast majority.\footnote{Data used in earlier versions of the paper, based on older Orbis versions, also included fractional investors (e.g., institutional investors with very small shares). In the older data, owner counts were typically higher, and networks were more complex due to these fractional investors. As of 2023, when the data for this paper version were collected, those investors had been removed from Orbis. However, none of the substantial findings change as a result of this change in the data (e.g., regarding the ranges of market concentration). Further note that in 2025, Paramount (and thus cbsnews.com) was taken over by Skydance \citep{koblin2025skydance} in a widely observed takeover-battle that, according to some sources, was at least in part motivated by controlling the broadcasting/publishing content of CBS news \citep{malone2026weiss}.}

The lower panel illustrates the \textit{opposite perspective}, showing the portfolio of news domains partially owned by a single individual, Mario Joseph Gabelli. Through his holding company GGCP Holdings and investment firm GAMCO Investors, Gabelli holds stakes in multiple media companies, such as 12.69\% of CBS Broadcasting (cbsnews.com). This owner-rooted view complements the outlet-rooted view and reveals that a single investor can hold partial stakes across a diverse portfolio of news outlets.

\subsection{Measuring Ownership Concentration, Transparency, and Market Shares}\label{sec:measuring_market_concentration}

Based on the drawing of ownership networks, we identify 23,229 unique firms with ownership stakes in news websites, appearing across 953,089 nodes in the ownership networks. Tracing these networks to their endpoints, we find 20,048 unique ultimate owners across 257,929 outlet-owner pairs. We analyze these ownership structures, assessing ownership concentration, transparency, and market shares (of registrants, ultimate owners, and industry sectors). We also compute market concentration at the country level.
To assess ownership concentration, we count the number of ultimate owners for each ownership network and identify ultimate owners with majority shares in an outlet. To measure ownership transparency, we follow \cite{cageetal2017} and count the number of nodes (owning entities) as well as the number of ownership levels per network (i.e., the ``depth'' of the network). We extend our database with domain-level web traffic data to understand which news websites, domain registrants, ultimate owners, and industry sectors reach how many internet users. We acquire traffic information (estimates of website reach and page views) from AWIS.\footnote{These metrics are used in digital marketing and are often the basis for advertisement placement decisions. These metrics are, therefore, closely monitored by an outlet's management (an indicator of success vis-à-vis competitors). We use web traffic (e.g., as opposed to revenues) because online news is often freely accessible, and inferring reach from advertising spending would not be straightforward.} As we are primarily interested in level differences between individual outlets, we compute the average reach and page views from 2018 through 2020 per domain. Web traffic data is normalized by one million active website visitors or one million page views (on all domains tracked by AWIS) during the measurement period. For a reading example, consider nytimes.com and its average reach (page view) figure of 6,818 (396): for every million unique visitors present (every million page views) on any website between 2018 and 2020, an average of 6,818 (396) went to nytimes.com. For comparison, the reach (page view) figures for google.com are 574,893 (195,269). For country-level analyses, we scale all traffic data by visitors that originate from that country. For instance, when calculating the market share of nytimes.com, we re-scale the reach (page views) by the share of visitors located in the U.S. We then infer market shares based on the number of unique domain visitors relative to the total number of visitors of all news domains in a country. At the owner levels, we assign each outlet's traffic to its owner; at the ultimate-owner level, an outlet's traffic is split across its ultimate owners in proportion to their chain-product ownership shares (renormalized to sum to one) and summed by owner. That is, we compute market shares for each news website $i$ in country $c$ based on the reach of the website:\footnote{Our choice of using reach as the basis for the market share measure is motivated as follows. Traditionally, media economists used newspaper circulation to quantify an outlet's audience. Reach is the closest proxy of circulation, as it refers to individual users. Importantly, page views and reach are \textit{strongly} correlated. Our results are robust to using page views (see Online Appendixes~\ref{sec:app:data} and~\ref{sec:app:addl_fig_tab}).}

\begin{equation}
marketShare_{ci} = \frac{reach_{ci}}{\sum_{i=1}^N{reach_{ci}}}.
\end{equation} 

We measure market concentration using the Herfindahl-Hirschman Index (HHI):
\begin{equation}
HHI_c = \sum_{i=1}^{N} marketShare_{ci}^2.
\end{equation}
HHI values range from 0 to 1, with values below 0.15 typically indicating low concentration, 0.15--0.25 moderate concentration, and above 0.25 high concentration. As a robustness check, we also compute the four-firm concentration ratio ($C_4$), which captures the cumulative market share of the top four entities.

For each country, we compute these measures at three levels of the ownership networks: $(i)$ news domains, $(ii)$ domain registrants (i.e., several domains can belong to the same registrant), and $(iii)$ ultimate owners (considering the entire ownership tree).

%%%%%%%%%%%%%%%%%%%%% RESULTS %%%%%%%%%%%%%%%%%%%%%%%%%%%%%%%%%%%%%%%%%%%

\section{Results: Media Ownership}\label{sec:results}

Equipped with the ownership data, we now analyze ownership concentration, ownership network tractability, owners' industry-sector affiliation, and market concentration. Section~\ref{sec:coverage} then asks whether these ownership patterns matter for news content.

\subsection{Ownership Structures and Their Tractability} \label{sec:results:structures}

\begin{figure}[htbp]
\centering
\caption{Ownership Structures of Online News Outlets, by Country}
\label{fig:ownership_anatomy}
\includegraphics[width=0.99\textwidth]{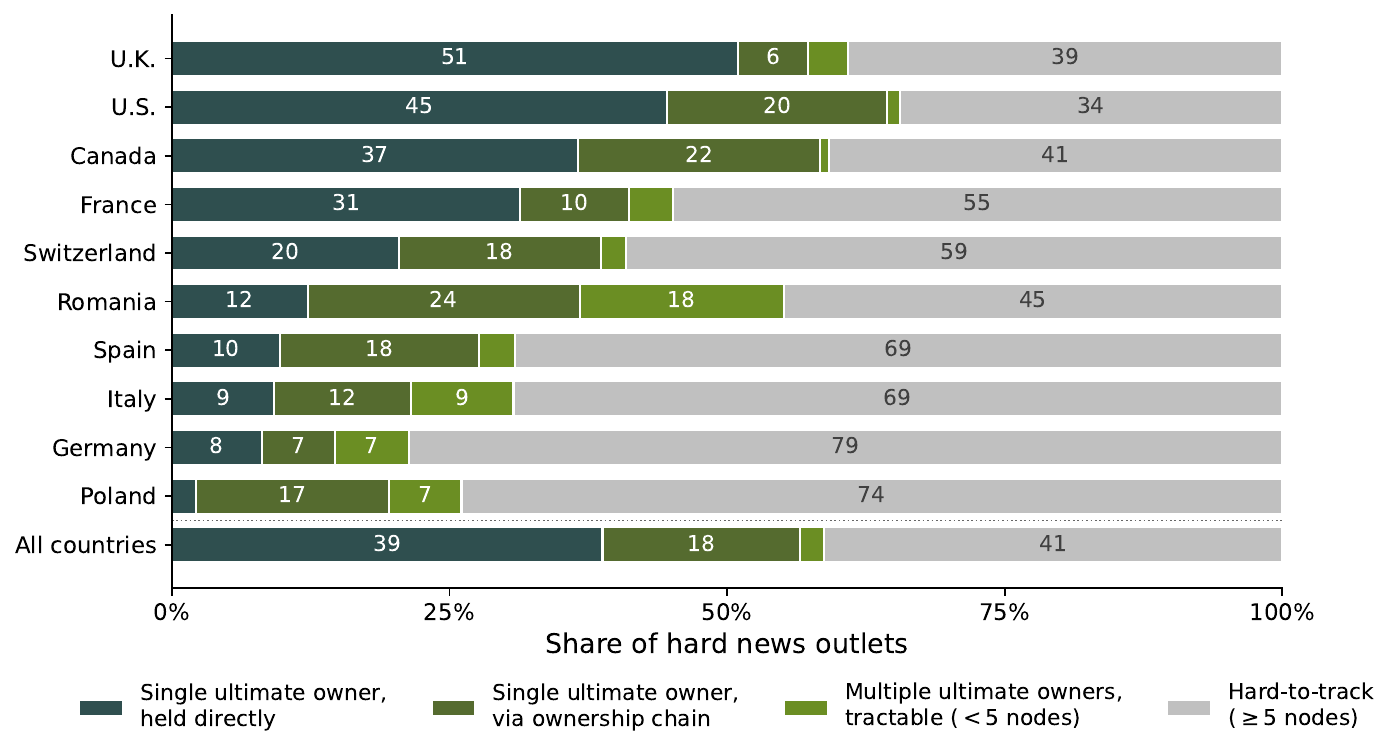}
\begin{figurenotes}
Each bar partitions a country's hard news outlets into four categories: outlets whose registering entity is itself the single ultimate owner (``held directly''); outlets with a single ultimate owner reached through a multi-node ownership network (``via ownership chain''); outlets with multiple ultimate owners whose network nonetheless has fewer than five nodes; and outlets with multiple ultimate owners and a network of five or more nodes (``hard to track''). Because single-owner outlets are classified by owner count regardless of node count, this last share (41\% pooled) is below the overall share of outlets with five or more nodes (45\%, Online Appendix Table~\ref{tab:ownership_transparency}). Countries are sorted by the share of directly held outlets; the bottom bar pools all ten countries. The sample comprises hard news outlets with both node-count and ultimate-owner data. Exact values: Online Appendix Table~\ref{tab:anatomy_by_country}.
\end{figurenotes}
\end{figure}

From a corporate control perspective, ultimate owners are particularly relevant because they finance a news website and, arguably, have the (as the expression implies) ultimate responsibility for the site's content. Figure~\ref{fig:ownership_anatomy} summarizes the ownership structures behind the outlets in our sample, by country.\footnote{Online Appendix Table~\ref{tab:anatomy_by_country} reports the exact values shown in the figure. Online Appendix Table~\ref{tab:ownership_transparency} provides complementary ownership concentration and network complexity metrics, including traffic-weighted versions.} Across countries, for 39\% of the news websites examined, the registering entity is directly the ultimate owner (often a media company or an individual). An additional 18\% of outlets have one ultimate owner reached through an ownership network. In total, 57\% of outlets have a single ultimate owner. These single-owner outlets, however, account for only 45\% of traffic: larger outlets tend to have more dispersed ownership.

Following \citet{aminadav2020corporate}, who define ``controlled'' firms as those in which a single entity holds more than 20\% of ownership, we find that 87\% of outlets have at least one controlling owner (76\% have exactly one). If we apply a stricter threshold of 50\% (majority ownership), 71\% of outlets still have a majority owner. For outlets with more than one ultimate owner, the mean (median) number of owners is 98 (71).\footnote{Our data also allow to replicate Aminadav's and Papaioannou's findings for the online media market: As Figure~\ref{fig:corr_ap_mw} shows, ownership concentration in the online media sector (our data) and ownership concentration across sectors (\citeauthor{aminadav2020corporate}, \citeyear{aminadav2020corporate}) correlates positively across the ten countries (49\% correlation coefficient). Our ownership concentration figures for the online media are higher compared to their cross-sector figures, in all ten countries.}

From a political economy viewpoint, high ownership concentration presents a trade-off. On the one hand, concentrated ownership means that a single entity can potentially influence editorial decisions, a concern for media pluralism. On the other hand, concentrated ownership is more tractable: when a news consumer wants to understand who is responsible for a website's content, a single majority owner is easier to identify than many dispersed shareholders. In the latter case, it is more challenging for readers to understand who is responsible for the content and to discount potential bias. Apart from the sheer \textit{number} of ultimate owners, how tractable an ownership network is also depends on the number of nodes and levels through which ownership passes before reaching the ultimate owner.

Can readers actually trace who stands behind an outlet? We assess the tractability (or ``complexity'') of ownership relationships with the breadth (number of nodes) and depth (number of ownership levels) of the networks (see Panel B of Online Appendix Table~\ref{tab:ownership_transparency}).  While 55\% of outlets have fewer than five nodes in their ownership network, the remaining 45\% have a mean (median) of 1,689 (511) nodes. This implies that for nearly half of the hard news domains in our sample, the underlying ownership structure is hard to trace. Minority shareholders account for only a modest share of this complexity: when counting only nodes that lead to the majority owner (50\% ownership or more), almost two-thirds of outlets (64\%) have fewer than five nodes, and among outlets with at least one controlling owner (20\% threshold), the path to one can be identified for 77\% by traversing fewer than five nodes.

Hard-to-track ownership is more prevalent among large outlets and varies substantially across countries. Weighted by traffic, only 45\% of the market has networks of fewer than five nodes, and the share of outlets with fewer than five nodes ranges from 64\% in the U.S. and 58\% in the U.K. to between 21\% and 26\% in Germany, Poland, Italy, and Spain. Online Appendix~\ref{app:sec:topology_detail} provides the full country-level and traffic-weighted detail, including ownership levels as an additional tractability measure; Online Appendix Table~\ref{tab:ownership_transparency_all} shows that hard news outlets and all outlets display similar patterns.

Figure~\ref{fig:ownership_anatomy} also shows that there is little in between the two main configurations: outlets either have a single ultimate owner (which is usually tractable) or many ultimate owners behind a complex network; outlets with multiple ultimate owners but a simple network are rare (2\% of outlets pooled across the ten countries). Does complexity, then, conceal a single owner? In principle, an outlet could have countless nodes that all lead to the same (or a few) ultimate owner(s), the configuration in which a controlling owner would be hidden behind layers of intermediaries. This does not seem to be a dominant pattern: the number of nodes and the number of ultimate owners are strongly correlated (91\%), and outlets scarcely have one ultimate owner but more than five nodes (see Online Appendix Figure~\ref{fig:node_counts_all}). Complex ownership networks thus overwhelmingly reflect genuinely dispersed ownership (many small shareholders) rather than concealment of a single controller. The concern they raise is, accordingly, one of transparency: with dispersed, multi-layered ownership, it is costly for readers to establish who is responsible for the content and to discount potential bias.

We conclude that, on the one hand, between one-half and two-thirds of news domains are (mainly) owned by one ultimate owner, and this tends to be tractable. On the other hand, we find highly diverse ownership structures for most of the remaining news domains. Section \ref{sec:discussion} discusses potential implications.

\subsection{Market Shares by Sector} \label{sec:results:sectors}

Identifying ultimate owners allows us to assess the involvement of industry sectors in the online media market. Consistent with the concern that majority owners might imprint their views in their outlet's reporting, it is questionable whether news outlets can act as the fourth estate if their owners have significant economic interests outside of news reporting.\footnote{Historically, it was considered essential that shareholders of media outlets did not have substantial financial interests in other economic sectors (see \citeauthor{cageetal2017}, 2017).} A crucial question is thus: to which sectors do ultimate owners belong?

Across all ten countries, at the level of direct domain owners (registrants), media companies account for more than three-quarters of the market.\footnote{The Orbis categories ``Printing \& Publishing'' and ``Media \& Broadcasting'' are grouped into ``Media Sector (broad)'' (manual inspection reveals that the two categories describe outlets interchangeably, e.g., they do \textit{not} distinguish native print or TV).} In some countries, this figure amounts to more than 95\% (U.K., Poland, Italy, and France). Additionally, manual checks reveal that cases in which domain owners are not in the media sector sometimes reflect sector-coding idiosyncraticities (e.g., Orbis codes some newspapers as ``Travel, Personal, and Leisure'', such as the ``Journal Gazette'' from Indiana, U.S.). In conclusion, the vast majority of domain owners are from the media sector.

\begin{figure}[htbp]
\centering
\caption{Sector Composition of the Online News Market: Domain Owners versus Ultimate Owners}
\label{fig:sector_shift}
\includegraphics[width=0.99\textwidth]{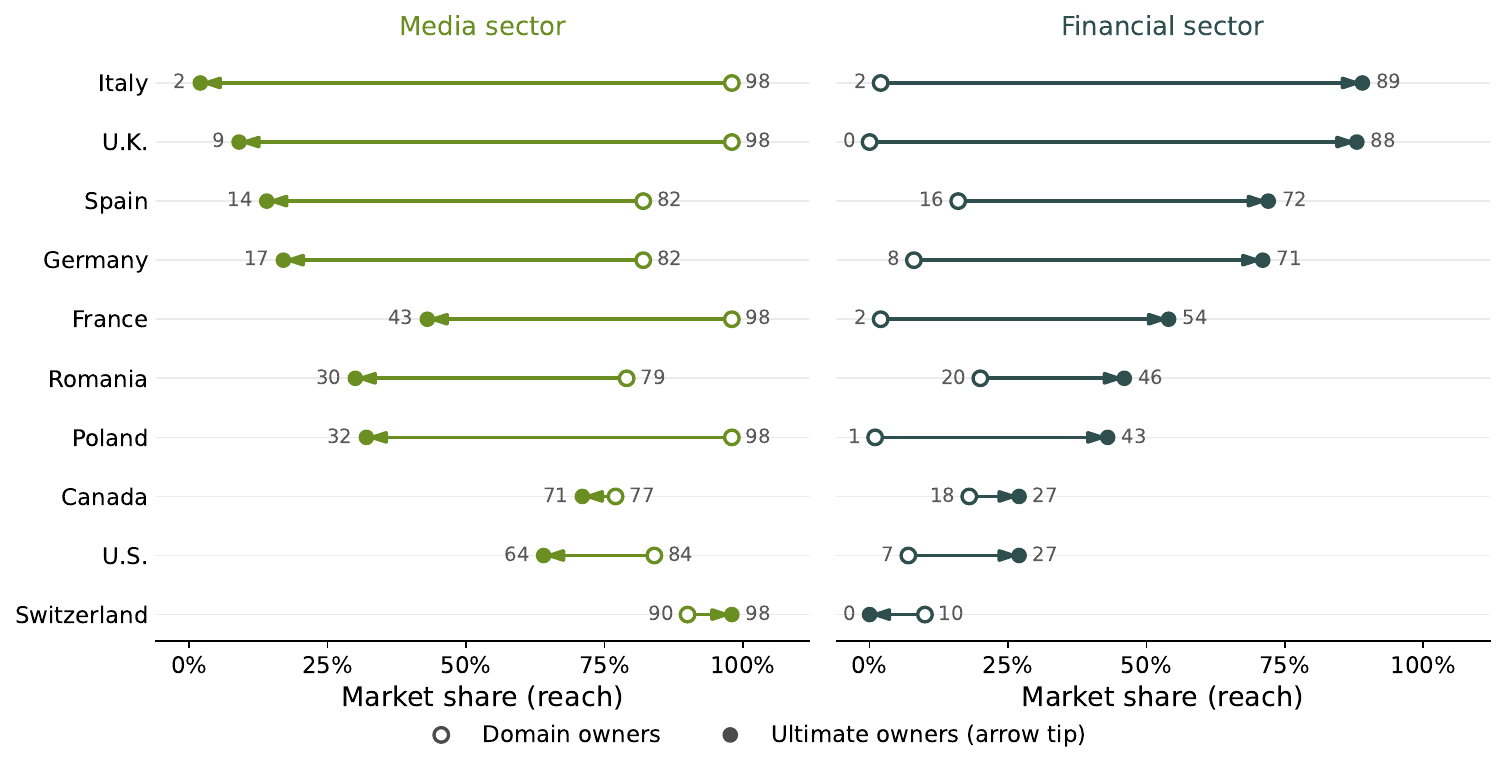}
\begin{figurenotes}
For each country, the arrow runs from the sector's market share at the domain-owner level (open circle) to its share at the ultimate-owner level (filled circle); market shares are based on reach. The left panel shows the media sector (broad), the right panel the financial sector (broad); the sector groupings are defined in the text. Countries are sorted by the financial sector's ultimate-owner share. Exact values: Online Appendix Table~\ref{tab:shares_sectors_political_grouped}.
\end{figurenotes}
\end{figure}

However, when moving to the ultimate owner level, the media sector share drops in all countries except Switzerland, see Figure~\ref{fig:sector_shift}. Media companies retain around 15\% in the U.K. and Germany, and between one-third and one-half of the market in Spain, Romania, and Poland. In the U.S., Canada, France, and Switzerland, they still own 60\% or more of the market. Across all ten countries (except Switzerland), the financial industry ultimately holds a significant share of the online media market.\footnote{We group ``Banking, Insurance \& Financial Services'' and ``Business Services'' into ``Financial Sector (broad)'' (manual inspection reveals that business services mostly refer to holding or asset management companies). We find similar patterns when only considering ``Banking, Insurance \& Financial Services'', see Online Appendix Table \ref{tab:shares_sectors_political}.}
In all countries (except for Switzerland), the financial industry is the most or second-most dominant sector for ultimate owners. It is the most dominant sector in the U.K., Poland, Italy, and Germany. It is the second-most dominant sector after media companies in the U.S., Romania, France, and Canada.\footnote{Online Appendix Table \ref{tab:shares_sectors_political_grouped} reports the exact shares shown in Figure~\ref{fig:sector_shift}; Online Appendix Tables \ref{tab:shares_sectors_political} and \ref{tab:shares_sectors} provide more disaggregated sector shares, for outlets specialized in ``hard'' news and for all outlets (similar patterns emerge for the former and the latter).}

In sum, examining domain owners is not enough to uncover whether those ultimately financing the news have economic interests outside of the news sector. Given the often hard-to-track ownership networks, such interests remain hard to trace for consumers. Our results would suggest that the answer to ``who owns the online media?'' is, to a significant degree, the financial industry. Yet, this alone, of course, does not mean that an ``influence motive'' is driving this pattern. It may well be that many firms in the financial sector view shares in news sites (or in the corporations behind them) as lucrative investments, independent of any influence on news content through corporate control. 

\subsection{Market Concentration} \label{sec:results:concentration}

How concentrated are national online news markets, and does the answer change as we move up the ownership chain? We now turn to country-level market concentration, measured by the Herfindahl-Hirschman Index (HHI). Following standard thresholds used by competition authorities, we classify markets as having low concentration for HHI values between 0 and 0.15, moderate concentration for values between 0.15 and 0.25, and high concentration for values of 0.25 and above \citep{doj2010horizontal}.

\begin{figure}[htbp]
\centering
\caption{Market Concentration (HHI) Across Ownership Levels, by Country}
\label{fig:hhi_slope}
\includegraphics[width=0.85\textwidth]{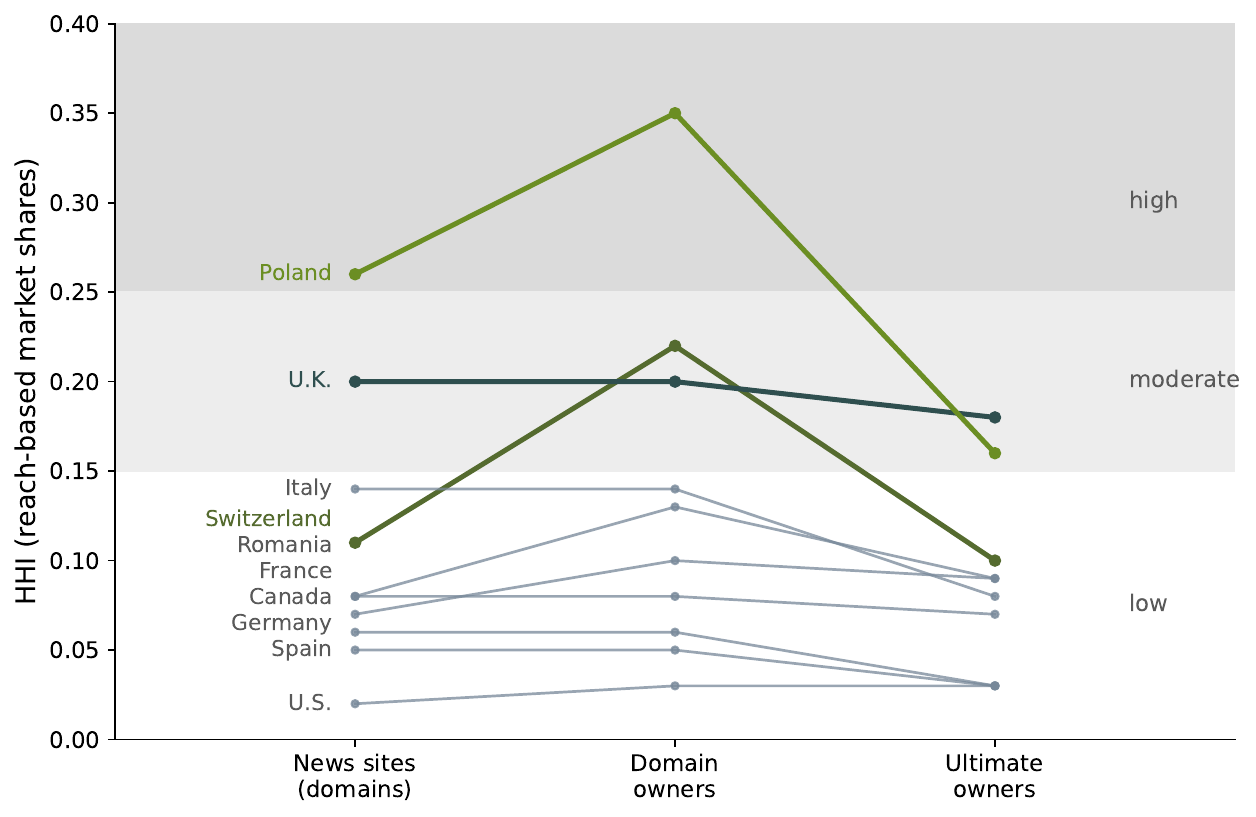}
\begin{figurenotes}
Each line traces a country's HHI from the news-site (domain) level to the domain-owner and ultimate-owner levels; market shares are based on reach. Shaded bands mark the standard thresholds for moderate (0.15--0.25) and high ($>$0.25) concentration \citep{doj2010horizontal}. Poland, the U.K., and Switzerland (the only countries that leave the low-concentration band at any level) are highlighted. The figure includes all hard news outlets (more than 25\% political reporting). Exact values: Online Appendix Table~\ref{tab:concentration_all}.
\end{figurenotes}
\end{figure}

Figure~\ref{fig:hhi_slope} traces the HHI (in terms of reach) across the three levels at which we measure market shares: news sites (domains), their immediate domain owners, and their ultimate owners.\footnote{Online Appendix Table~\ref{tab:concentration_all} provides a comprehensive tabular summary of market concentration values across all countries and ownership levels.} Consider the U.S., for example: the domain-level HHI is 0.02, the domain-owner-level HHI is 0.03, and the ultimate-owner-level HHI is 0.03, all indicating low concentration.

At the domain level, concentration is highest in Poland (HHI of 0.26, indicating high concentration) and the U.K. (HHI of 0.20, moderate concentration). In Italy (0.14) and Switzerland (0.11), concentration is low but relatively elevated compared to other countries. Domain-level concentration is lowest in the U.S. (0.02). The other countries lie between these poles, all with low concentration (HHI below 0.10).

In all ten countries, market concentration increases or remains stable when moving from domains to domain owners, by construction, since each domain owner holds one or more domains. At the domain-owner level, concentration is highest in Poland (HHI of 0.35, high), followed by Switzerland (0.22, moderate) and the U.K. (0.20, moderate). Again, concentration is lowest in the U.S. (0.03), with the remaining countries exhibiting low concentration.

Finally, when considering ultimate owners, concentration in the most concentrated markets falls relative to the domain-owner level, reflecting the dispersed ownership networks of many outlets. Eight of the ten countries fall into the low-concentration category, with HHI values at or below 0.10; the U.K. (0.18) and Poland (0.16) are in the moderate band. For robustness checks, see Section~\ref{sec:extensions}.

In sum, market concentration at the level of news sites is moderate to high only in Poland and the U.K., and low in all other countries. At the domain-owner level, Poland, Switzerland, and the U.K. exhibit moderate to high concentration, while all other countries, including the U.S., show low concentration. The ultimate owner analysis reveals low market concentration in most countries, with only moderate concentration in the U.K. and Poland. Adding to the evidence from our ownership network analysis, our findings indicate that the online media markets in the U.S., Canada, and Europe are not dominated by a few influential ultimate owners.

A natural question is which ownership level matters most for media pluralism. Our content analysis below reveals that ownership at the domain level has the strongest association with editorial alignment, suggesting that (in the countries we observe, with their dispersed ultimate ownership structures) domain-owner concentration is immediately policy-relevant. At the same time, we still find a significant content-ownership association at the ultimate-owner level. Ultimate ownership is much harder for readers to track, especially when dispersed across many small shareholders. Hence, while the mostly low ultimate-owner concentration we document is reassuring from a traditional market-power perspective, it raises concerns about accountability. We revisit these questions in Section~\ref{sec:discussion}.

%%%%%%%%%%%%%%%%%%%%%%%%%%%%%%%%%%%%%%%%%%%%%%%%%%%%%%%%%%%%%%%%%%5
\section{Ownership and News Content Result}
\label{sec:coverage}

To assess the similarity or distinctiveness of news content between news sites under different ownership, we again leverage the GDELT database. Given our data collection timeframe, our focus is on news articles from 2019 to 2023. Specifically, for each news site with complete ownership information for both waves of our ownership survey in Orbis, we create a vector containing all event IDs covered by the site across the two distinct periods. The first period covers all of 2019 and the first half of 2020 (matching the first wave of ownership data collection); the second period covers all of 2022 and the first half of 2023 (matching the second wave of ownership data collection). We then compute the Dice Similarity Coefficient (DSC; \citealt{dice_1945,sorensen_1948}) between the sets of events/stories covered for each news site pair in either time period.\footnote{The DSC (or Sorensen-Dice Index) is a similarity measure closely related to the Jaccard Index \citep{Jaccard1901}. Note that using the Jaccard Index instead does not qualitatively change any of our main results.} For example, over the given time frame, a hypothetical news website $X$ has covered events A, B, and C, while news website $Y$ has covered events A and D. The resulting DSC in news coverage for the news websites $X$ and $Y$ would then be given as:

\[
\text{DSC}(X,Y) \;=\; \frac{2 \cdot | \{A,B,C\} \cap \{A,D\} |}{|\{A,B,C\}| \;+\; |\{A,D\}|}.
\]

In this example, the intersection of the sets of covered events is $\{A\}$, hence $|\{A,B,C\} \cap \{A,D\}| = 1$. The total number of events covered by website $X$ is 3 (i.e., $|\{A,B,C\}| = 3$), and the total number of events covered by website $Y$ is 2 (i.e., $|\{A,D\}| = 2$). Thus , the DSC in news coverage between $X$ and $Y$ would amount to $\frac{2 \cdot 1}{3 + 2} =  0.4$. In other words, the two hypothetical news websites share approximately 40\% similarity in their coverage of events, scaled by their overall coverage.
The resulting measure of \emph{news coverage similarity} for each news website pair and either period (2019/20 and 2022/23) serves as the dependent variable in a first set of panel regressions with news site pairs as the unit of observation, taking the following form (and variants thereof):

\begin{equation*}
\begin{split}
NewsCoverageSimilarity_{it}  = & \beta_{1}Ownership_{it} + \\
%& + \beta_{2}SameRegion_{i} +\beta_{3}SameMediaType_{i} \\
&  NewsSitePairFE_{i} + TimeFE_{t} +\epsilon_{it}.\\
\end{split}
\end{equation*}

$NewsCoverageSimilarity_{it}$ is the DSC by news site pair $i$ of events covered by either news site in the pair in period $t$. In the simplest specifications, $Ownership_{it}$ is a dummy variable equal to 1 if both news sites in pair $i$ have the same domain owner and equal to 0 otherwise. Thus, we can exploit this within-news-site-pair variation to identify how ownership relates to news coverage when accounting for news site pair fixed effects ($NewsSitePairFE_{i}$).\footnote{Our balanced panel contains 3434 pairs of news sites that were owned by the same domain owner in the first period, but not in the second period as well as 853 news site pairs that were not owned by the same entity in the first period but are owned by the same entity in the second period. These switching pairs trace back to 92 distinct owner-level ownership events involving 326 outlets. No single event drives our results: the largest (that is, the restructuring of a portfolio of 102 U.S.\ local news outlets, accounting for 43\% of all switching pairs) amplifies but does not create the association; excluding it, the within-dyad estimates remain 0.18 for coverage and 0.22 for tone, both highly significant (Online Appendix Table~\ref{tab:switcher_anatomy}). Standard errors clustered at the level of the ownership events leave the estimates significant (coverage 0.330, SE 0.092; tone 0.342, SE 0.106). Weighting each ownership event equally instead of each pair yields smaller estimates (0.061 and 0.072), because content alignment is concentrated in large portfolio events; the two weightings answer different questions---market-level content homogenization versus the effect of a typical ownership transaction (Online Appendix Table~\ref{tab:switcher_events}).} Our pair fixed effects absorb time-invariant characteristics of each news site pair, including any baseline similarity that might attract common ownership. For reverse causality to bias our estimates, acquirers in the later period would need to systematically target outlets that are increasingly similar, and this targeting behavior would need to differ from whatever drove ownership patterns in the earlier period. While we cannot rule out such time-varying selection, we note that this is a considerably more demanding assumption than the simple story that acquirers buy similar outlets. Directly probing this, we find no evidence of pre-acquisition convergence: pairs that become co-owned show, if anything, a \emph{declining} differential coverage-similarity trend in the pre-period ($-0.015$ per quarter, $p<0.01$), the opposite of what reverse causality would predict (Online Appendix Figure~\ref{fig:pretrend} and Table~\ref{tab:pretrend_trend}).

In a second set of specifications, we focus on the ultimate owner level and define $Ownership_{it}$ as either the shared percentage in ultimate owner shares or the cosine similarity between the ownership vectors of the two news sites in pair $i$. The shared percentage is computed as follows: for each ultimate owner holding stakes in both outlets, we take the smaller of the two stakes and sum these minima. For the cosine similarity, we represent each outlet's ownership as a vector in which each element is the stake held by one ultimate owner (with zeros for owners holding no stake in that outlet) and compute the cosine of the angle between the two outlets' vectors. The measure is increasing in similarity: it equals 1 when the two outlets have proportionally identical ownership compositions and 0 when they share no ultimate owners. In specifications of this type, we only consider news site pairs that are not directly owned by the same domain owner. That is, we examine whether news sites within the same corporate construct (though not directly owned by the same entity) tend to cover news more similarly than those outside the same construct. 

The two ownership margins are conceptually distinct. The domain owner is the operational-control margin: it is closest to hiring, editorial routines, and day-to-day production decisions. Shared ultimate ownership, by contrast, captures overlapping financial exposure at the top of the control chain, typically without operational involvement, and our shared-percentage measure aggregates proportional stakes rather than effective voting control. We therefore expect content alignment to be strongest at the domain-owner margin, with the ultimate-owner margin providing a complementary, lower-powered test of whether ownership overlap matters even absent direct operational control; the mechanism evidence in Section~\ref{sec:mechanisms} sharpens this distinction further.

Finally, for the subset of news site pairs and events with available sentiment score data, we compile the tone of the corresponding news article (as assigned by GDELT's sentiment analysis of the article's text). Following the same logic of our news coverage approach, we then generate vectors of sentiment scores, one for each news website and period. By tracking which sentiment score was assigned to each event ID, we conduct two types of analyses. First, we compute similarity in news-reporting tone across all news-site pairs and events, enabling us to assess whether news sites owned by the same entity tend to report in a similar tone. Second, we compute the similarity in news reporting as the cosine similarity between the two outlets' event-level tone vectors. Only events covered by both outlets contribute to the numerator of this measure, while each outlet's norm is computed over the tone scores of all events it covers; the measure is therefore highest when two outlets cover largely overlapping event sets and report on the jointly covered events in a similar tone, and pairs without jointly covered events are treated as missing. We then estimate regressions of the form

\begin{equation*}
\begin{split}
ReportingToneSimilarity_{it}  = & \beta_{1}Ownership_{it} + \\
%& + \beta_{2}SameRegion_{i} +\beta_{3}SameMediaType_{i} \\
&  WebsitePairFE_{i} + TimeFE_{t} +\epsilon_{it}.\\
\end{split}
\end{equation*}

, whereby $ReportingToneSimilarity_{it}$ measures the similarity between the tone in reporting of the two news websites in pair $i$ as the cosine similarity between their corresponding sets of sentiment scores. As in the news coverage similarity regressions, we employ different definitions of $Ownership_{it}$ and corresponding sample restrictions across specifications. We discuss this additional set of results in Subsection~\ref{sec:results_tone}. Because every news website appears in many pairs, the regression residuals are not independent across pairs that share an outlet. We therefore use dyadic clustering: standard errors are clustered two-way on the identities of the two outlets in each pair, allowing for arbitrary correlation between any two pairs that have an outlet in common \citep{cameron_miller_2015}.

\subsection{News coverage results}
\label{sec:results_coverage}

The main results from our news coverage regression analyses are presented in Table~\ref{tab:coverage_pairs}; Online Appendix Table~\ref{tab:coverage} provides descriptive statistics for the sample. Overall, we see that news websites (partially) owned by the same entity tend to be much more aligned in what they cover than news sites owned by different entities.\footnote{This pattern is consistent with recent evidence from print media. \cite{Garz_Ots_2025} analyze over two million articles from 108 Swedish newspapers during a period of extensive ownership changes (2014--2022) and find that mergers lead to content homogenization: acquired newspapers decrease their provision of local news while relying more on content shared with co-owned titles.} Columns 1 and 2 present specifications with the ``same domain owner'' indicator as explanatory variable. Specification 1 shows the baseline model, which also exploits between-dyad variation in news coverage (while accounting only for period fixed effects). Specification 2 accounts for period and dyad fixed effects, exploiting the fact that our panel contains news site pairs that are once observed as belonging to the same owner, and once as belonging to different owners. The within-dyad estimate (0.33, specification 2) is somewhat smaller than the cross-sectional estimate (0.38), consistent with modest positive selection, but the effect remains economically and statistically significant. The coefficient estimate in specification 2 suggests that news outlets are systematically more frequently observed to cover the same events when directly owned by the same entity (same domain owner). Holding all other factors constant, a given news site pair's news coverage exhibits a 0.33 higher similarity on the DSC scale when owned by the same domain owner than when not owned by the same domain owner. To contextualize the magnitude of our estimates, we compare the within-dyad coefficient to the cross-sectional difference in coverage similarity between same-owner and different-owner pairs. Outlet pairs with the same domain owner have a mean DSC of 0.416, compared to 0.035 for pairs with different owners, a gap of 0.381 (specification 1; Online Appendix Table~\ref{tab:dsc_by_ownership} reports these descriptive statistics). Our within-dyad estimate of 0.33 amounts to approximately 87\% of this cross-sectional gap (for scale, the within-dyad standard deviation of the DSC is 0.043 overall and 0.245 among switcher pairs; Online Appendix Table~\ref{tab:dsc_by_ownership}). Because the within-dyad estimate is identified from ownership switchers while the gap is computed across all pairs, we read this comparison as a benchmark of magnitudes rather than a formal decomposition; it nevertheless suggests that the ownership-similarity association is not primarily driven by time-invariant selection. For comparison, outlet pairs in the same country but with different owners have a mean DSC of 0.055, higher than the overall different-owner mean (0.035) but still an order of magnitude below that of same-owner pairs (0.416). This suggests that shared ownership is associated with substantially greater content alignment than geographic proximity alone.

Moving to specifications 3 and 4, we observe a consistent pattern in similarity among ultimate owners (restricting the sample to domain pairs that do not share the same domain owner). In these linear specifications, a one-standard-deviation increase in shared ownership is associated with a 0.0133 increase in DSC. A one standard deviation increase in the cosine similarity of ownership structures is related to an increase in DSC of 0.0126 (thus, in either case a roughly 35\% to 40\% increase relative to the dependent variable's mean). Finally, for specifications 5 to 7, we restrict the sample to pairs in which the same domain owner does not own both and that have at least some shared ultimate owners. The results in specification 6 suggest that news site pairs that exceed a 50\% shared ownership threshold have a DSC approximately 0.191 higher, compared with pairs with lower levels of common ownership. Based on a more granular classification by quartiles of shared ownership percentage in specification 7, higher quartiles consistently correspond to higher DSC values. In particular, pairs in the highest shared-ownership quartile (Q3+) are associated with the largest increase, with a DSC approximately 0.258 higher. These results underscore a monotonic relationship, suggesting that greater ownership overlaps are associated with incrementally greater alignments in content coverage decisions. 

% , and specification 3 additionally for covariates capturing whether the two news sites in the pair are to be located in the same media market (an indicator that is equal to 1 if both news sites in the pair are located in the same state and 0 otherwise, an indicator that is equal to 1 if both news sites are of the same media type\footnote{News sites are categorized into three possible media types: broadcast (news sites of radio or TV stations/networks, such as cnn.com; print (news sites of print media, such as nytimes.com); ``digital native'' news sites without an analog equivalent or history, such as buzzfeed.com }.

\begin{center}
    
\begin{table}[htbp]\centering
\begin{threeparttable}
\caption{Similarity in News Coverage} \label{tab:coverage_pairs}
\small

% Table created by stargazer v.5.2.3 by Marek Hlavac, Social Policy Institute. E-mail: marek.hlavac at gmail.com
% Date and time: Fri, May 08, 2026 - 11:17:14 AM
\begin{tabular}{@{\extracolsep{0.0pt}}lccccccc} 
\\[-1.8ex]\hline 
\hline \\[-1.8ex] 
 & \multicolumn{7}{c}{Dependent variable:} \\ 
\cline{2-8} 
\\[-1.8ex] & \multicolumn{7}{c}{DSC(events covered by news site a, events covered by news site b)} \\ 
\\[-1.8ex] & (1) & (2) & (3) & (4) & (5) & (6) & (7)\\ 
\hline \\[-1.8ex] 
 Same domain owner & 0.381$^{***}$ & 0.330$^{***}$ &  &  &  &  &  \\ 
  & (0.013) & (0.031) &  &  &  &  &  \\ 
  & & & & & & & \\ 
 Ownership shared perc. &  &  & 0.001$^{***}$ &  & 0.003$^{***}$ &  &  \\ 
  &  &  & (0.0001) &  & (0.0003) &  &  \\ 
  & & & & & & & \\ 
 Ownership cosine simil. &  &  &  & 0.131$^{***}$ &  &  &  \\ 
  &  &  &  & (0.012) &  &  &  \\ 
  & & & & & & & \\ 
 Above 50 shared perc. &  &  &  &  &  & 0.191$^{***}$ &  \\ 
  &  &  &  &  &  & (0.035) &  \\ 
  & & & & & & & \\ 
 Own. sh. perc. Q1-2 &  &  &  &  &  &  & 0.041$^{***}$ \\ 
  &  &  &  &  &  &  & (0.012) \\ 
  & & & & & & & \\ 
 Own. sh. perc. Q2-3 &  &  &  &  &  &  & 0.119$^{***}$ \\ 
  &  &  &  &  &  &  & (0.013) \\ 
  & & & & & & & \\ 
 Own. sh. perc. Q3+ &  &  &  &  &  &  & 0.258$^{***}$ \\ 
  &  &  &  &  &  &  & (0.029) \\ 
  & & & & & & & \\ 
\hline \\[-1.8ex] 
Period FE & Yes & Yes & Yes & Yes & Yes & Yes & Yes \\ 
Dyad FE & No & Yes & Yes & Yes & Yes & Yes & Yes \\ 
Sample & All & All & Not same d. o. & Not same d. o. & Some sh. o. & Some sh. o. & Some sh. o. \\ 
Mean dep. var. & 0.037 & 0.037 & 0.035 & 0.035 & 0.132 & 0.132 & 0.132 \\ 
No. news sites & 2,019 & 2,019 & 2,019 & 2,019 & 721 & 721 & 721 \\ 
No. observations & 4,074,342 & 4,074,342 & 4,045,660 & 4,045,660 & 63,118 & 63,118 & 63,118 \\ 
Within R$^2$ & 0.111 & 0.03 & 0.034 & 0.031 & 0.067 & 0.029 & 0.094 \\ 
R$^{2}$ & 0.111 & 0.765 & 0.759 & 0.759 & 0.732 & 0.721 & 0.739 \\ 
\hline 
\hline \\[-1.8ex] 
\end{tabular} 

\begin{tablenotes}[flushleft] \footnotesize
\item This table reports estimates from panel regressions of news website pairs' similarity in news reporting on variables indicating the degree to which the news websites are owned by the same entity. The sample includes pairs of hard news outlets with ownership and content data for both periods (2019/2020 and 2022/2023). The dependent variable is the DSC of event IDs covered by the two news websites in the pair. \emph{Same domain owner} is an indicator variable equal to 1 if the same domain owner owns the news websites in the pair and zero otherwise. \emph{Ownership shared perc.} is the sum of the minimum ownership stakes across all shared ultimate owners: for each ultimate owner holding stakes in both outlets, we take the smaller of the two stakes and sum these minima. \emph{Ownership shared perc.} is expressed in percentage points (0--100). \emph{Ownership cosine similarity} is the cosine similarity between the ownership vectors of the two news websites, where each vector contains the ownership shares of all ultimate owners. Columns 3 and 5 show specifications where we impose linearity in the relation between \emph{Ownership shared percentage} and the dependent variable. Columns 6 and 7 present results from more flexible specifications, in which we regress the dependent variable on dummy variables constructed from \emph{Ownership shared percentage}. In specification 6, we regress the DSC on an indicator equal to 1 if \emph{ownership shared percentage} exceeds 50. In specification 7, we use dummies indicating whether an observation falls between the first and the second quartile (Q1-2), between the second and the third (Q2-3), or above the third quartile (Q3+), only considering observations that have  \emph{Ownership shared perc.} values higher than 0 (observations with values lower than or equal to the first quartile are in the reference category). The results in the first two columns are based on the full news coverage sample, the results in columns 3 and 4 are based on the subsample of cases that the same domain owner does not own, and the results in columns 5 to 7 are based on pairs that are not owned by the same domain owner, but have at least some shared ultimate owners. Fixed effects are accounted for as indicated. Standard errors are shown in parentheses below the respective coefficient estimates and are dyadic two-way clustered by domain (\texttt{domain\_a} and \texttt{domain\_b}), following \citet{cameron_miller_2015}. The statistical significance of the coefficient estimates is indicated as follows: $^{*}$p$<$0.1; $^{**}$p$<$0.05; $^{***}$p$<$0.01.
\end{tablenotes}
\end{threeparttable}
\end{table}
\end{center}

\subsection{Results on reporting tone}
\label{sec:results_tone}
% \\ 

We now look at whether news websites report in a more similar tone when (partially) being owned by the same entity, using the tone-similarity measure defined above. The results are presented in Table~\ref{tab:reporting_tone}.

\begin{center}
    
\begin{table}[htbp]\centering
\begin{threeparttable}
\caption{Similarity in News Tone} 
\label{tab:reporting_tone}
\small

% Table created by stargazer v.5.2.3 by Marek Hlavac, Social Policy Institute. E-mail: marek.hlavac at gmail.com
% Date and time: Fri, May 08, 2026 - 11:18:18 AM
\begin{tabular}{@{\extracolsep{0.0pt}}lccccccc} 
\\[-1.8ex]\hline 
\hline \\[-1.8ex] 
 & \multicolumn{7}{c}{Dependent variable:} \\ 
\cline{2-8} 
\\[-1.8ex] & \multicolumn{7}{c}{Cosine‐Similarity(tone news site a, tone news site b)} \\ 
\\[-1.8ex] & (1) & (2) & (3) & (4) & (5) & (6) & (7)\\ 
\hline \\[-1.8ex] 
 Same domain owner & 0.440$^{***}$ & 0.342$^{***}$ &  &  &  &  &  \\ 
  & (0.015) & (0.043) &  &  &  &  &  \\ 
  & & & & & & & \\ 
 Ownership shared perc. &  &  & 0.002$^{***}$ &  & 0.002$^{***}$ &  &  \\ 
  &  &  & (0.0002) &  & (0.001) &  &  \\ 
  & & & & & & & \\ 
 Ownership cosine simil. &  &  &  & 0.184$^{***}$ &  &  &  \\ 
  &  &  &  & (0.019) &  &  &  \\ 
  & & & & & & & \\ 
 Above 50 shared perc. &  &  &  &  &  & 0.033$^{**}$ &  \\ 
  &  &  &  &  &  & (0.016) &  \\ 
  & & & & & & & \\ 
 Own. sh. perc. Q1-2 &  &  &  &  &  &  & 0.036$^{*}$ \\ 
  &  &  &  &  &  &  & (0.020) \\ 
  & & & & & & & \\ 
 Own. sh. perc. Q2-3 &  &  &  &  &  &  & 0.135$^{***}$ \\ 
  &  &  &  &  &  &  & (0.020) \\ 
  & & & & & & & \\ 
 Own. sh. perc. Q3+ &  &  &  &  &  &  & 0.115$^{***}$ \\ 
  &  &  &  &  &  &  & (0.024) \\ 
  & & & & & & & \\ 
\hline \\[-1.8ex] 
Period FE & Yes & Yes & Yes & Yes & Yes & Yes & Yes \\ 
Dyad FE & No & Yes & Yes & Yes & Yes & Yes & Yes \\ 
Sample & All & All & Not same d. o. & Not same d. o. & Some sh. o. & Some sh. o. & Some sh. o. \\ 
Mean dep. var. & 0.077 & 0.077 & 0.072 & 0.072 & 0.163 & 0.163 & 0.163 \\ 
No. news sites & 1,346 & 1,346 & 1,346 & 1,346 & 471 & 471 & 471 \\ 
No. observations & 986,512 & 986,512 & 976,206 & 976,206 & 23,280 & 23,280 & 23,280 \\ 
Within R$^2$ & 0.138 & 0.018 & 0.059 & 0.053 & 0.021 & 0.001 & 0.067 \\ 
R$^{2}$ & 0.138 & 0.746 & 0.730 & 0.729 & 0.640 & 0.633 & 0.657 \\ 
\hline 
\hline \\[-1.8ex] 
\end{tabular} 

\begin{tablenotes}[flushleft] \footnotesize
\item This table reports estimates from panel regressions of news website pairs' similarity in the tone of news reporting on variables indicating to what degree the news websites are owned by the same entity. The sample includes pairs of hard news outlets with ownership and content data for both periods (2019/2020 and 2022/2023) and, for the tone analysis, with tone data and at least one jointly covered tone-scored event. The dependent variable is the cosine similarity between the two news websites' event-level tone vectors (tone scores provided by GDELT), as described in the text. \emph{Same domain owner} is an indicator variable equal to 1 if the same domain owner owns the news websites in the pair and zero otherwise. \emph{Ownership shared perc.} is the sum of the minimum ownership stakes across all shared ultimate owners: for each ultimate owner holding stakes in both outlets, we take the smaller of the two stakes and sum these minima. \emph{Ownership shared perc.} is expressed in percentage points (0--100). \emph{Ownership cosine similarity} is the cosine similarity between the ownership vectors of the two news websites, where each vector contains the ownership shares of all ultimate owners. Columns 3 and 5 show specifications where we impose linearity in the relation between \emph{Ownership shared perc.} and the dependent variable. Columns 6 and 7 present results from more flexible specifications, in which we regress the dependent variable on dummy variables constructed from \emph{Ownership shared perc.} In specification 6, we regress the dependent variable on an indicator equal to 1 if \emph{Ownership shared perc.} exceeds 50. In specification 7, we use dummies indicating whether an observation falls between the first and the second quartile (Q1-2), between the second and the third (Q2-3), or above the third quartile (Q3+), only considering observations that have \emph{Ownership shared perc.} values higher than 0 (observations with values lower than or equal to the first quartile are in the reference category). The results in the first two columns are based on the full sample, the results in columns 3 and 4 are based on the subsample of pairs that are not owned by the same domain owner, and the results in columns 5 to 7 are based on pairs that are not owned by the same domain owner but have at least some shared ultimate owners. Fixed effects are accounted for as indicated. Standard errors are shown in parentheses below the respective coefficient estimates and are dyadic, two-way clustered on the identities of the two outlets in the pair, following \citet{cameron_miller_2015}. The statistical significance of the coefficient estimates is indicated as follows: $^{*}$p$<$0.1; $^{**}$p$<$0.05; $^{***}$p$<$0.01.
\end{tablenotes}
\end{threeparttable}
\end{table}
\end{center}

The results on reporting tone are consistent with the news coverage results, both in statistical significance and robustness across specifications, as well as in terms of economic significance. Common domain ownership is associated with a higher cosine similarity of tone of about 0.44 in the baseline model, and the association remains at 0.342 after controlling for dyad fixed effects in specification 2. Moving beyond the binary indicator of same-domain ownership, the continuous Ownership Shared Percentage (Column 3) shows that increases in shared stakes are associated with greater tone similarity. The alternative specification (Column 4), which uses the cosine similarity of ownership structures as the explanatory variable, confirms a strong positive effect. Finally, splitting the sample into segments based on thresholds for shared ownership (Columns 6 and 7) confirms the overall pattern. However, unlike in the coverage similarity analysis, very high ownership similarity does not appear to be more strongly associated with similarity in news content tone than moderately high ownership similarity. 

The tone-similarity measure combines two margins: which events both outlets cover (the extensive margin documented in Table~\ref{tab:coverage_pairs}) and how similarly they report the events both of them cover (the intensive margin). To isolate the intensive margin, Online Appendix Table~\ref{tab:tone_intersection} re-estimates all specifications of Table~\ref{tab:reporting_tone} with an intersection-normalized tone similarity, in which both outlets' tone vectors are restricted to the jointly covered events before normalizing. The same-owner association remains positive and precisely estimated ($0.089$, dyadic standard error $0.027$, in the fixed-effects specification 2), though smaller than the baseline estimate (as expected, since the baseline measure also reflects the event-selection margin). The intersection-normalized estimate is, in turn, conservative: the jointly covered set is itself an outcome of ownership (becoming co-owned expands a pair's jointly covered set roughly six-fold and shifts its composition toward less widely covered events) and, for pairs with different owners, the jointly covered set is selected toward the most widely covered stories, whose tone is strongly anchored by the event itself (Online Appendix Tables~\ref{tab:intersection_descriptives} and~\ref{tab:intersection_ownership}). The baseline and intersection-normalized estimates therefore serve as complementary upper- and lower-magnitude benchmarks (not formal bounds) for the tone effect. The event-level regressions below, which include event fixed effects and condition only on the focal outlet's own coverage, provide the cleanest evidence on the intensive margin. Sample selection into the tone panel does not drive comparisons between the coverage and tone results either: re-estimating the coverage specification (Table~\ref{tab:coverage_pairs}, column 2) on the tone estimation sample yields 0.321 (dyadic standard error 0.041), close to the baseline 0.330 (Online Appendix Table~\ref{tab:sample_overview}).

Taken together, these results suggest that both the choice of what to report and the tone of reporting are much more closely aligned between news websites that are owned by the same domain owner than when not. Although of smaller magnitude, we also document a consistent and highly statistically significant positive relationship between similarity in ownership at the level of ultimate owners and the similarity of news site pairs' reporting, in event coverage and in tone alike. These findings collectively suggest that greater overlap in ownership structures is associated with significantly more consonant reporting styles.

The pair-level results leave two questions that the dyadic design cannot answer: how large is ownership-linked tone alignment relative to the other forces that align reporting (such as shared geography, market, and medium) and does it hold within individual events, once the tone that a story imposes on every outlet covering it is absorbed? To answer both, we estimate a complementary set of regressions at the level of individual outlet--event observations, which also deliver the intensive-margin evidence anticipated above. The estimation sample builds on the event universe described in Section~\ref{sec:data}, excluding the quarter of events with the lowest coverage (threshold computed on the pooled data across both periods). For each event, we relate the tone with which a given news site reports it (we refer to this site as the \emph{focal} outlet, the outlet whose reporting tone the regression explains) to the ``upstream'' tone of the event, that is, the average tone with which the other outlets sharing the focal outlet's domain-level or ultimate owner report the same event. We replace continuous measures of upstream tone with quartile indicators and include website-specific controls and fixed effects. Formally, let $i$ index news websites and $e$ index events. The model is:

% \begin{equation*}
% \begin{split}
% Tone_{i,e} = 
% & \sum_{q=2}^4 \beta_{q}^{DO}\,\text{I}\bigl(ToneSameDO_{i,e} \in Q_q\bigr) 
% + \sum_{q=2}^4 \beta_{q}^{UO}\,\text{I}\bigl(ToneSameUO_{i,e} \in Q_q\bigr) \\
% & + \gamma_1 ToneSameState_{i,e} 
% + \gamma_2 ToneSameCountry_{i,e} 
% + \gamma_3 ToneSameMediaType_{i,e} \\
% & + WebSiteFE_i + EventFE_e + \varepsilon_{i,e}.
% \end{split}
% \end{equation*}

% Here, $Tone_{i,e}$ is the focal outlet \(i\)’s tone in covering event \(e\). The terms $ToneSameDO_{i,e}$ and $ToneSameUO_{i,e}$ represent the upstream tone from other outlets under the same domain-level or ultimate owner, respectively, divided into quartiles \(Q_q\). The indicator functions \(\text{I}(\cdot)\) equal one if the upstream tone falls into the given quartile. The variables $ToneSameState_{i,e}$, $ToneSameCountry_{i,e}$, and $ToneSameMediaType_{i,e}$ control for the average tone of other outlets in the same state, country, and media type. Beyond their use as controls, the coefficients of these variables also serve as a useful point of reference. That is, we can assess the extent to which the tone of news sites serving the same market (same country, region, media type) predicts the focal news site's tone, ceteris paribus, as opposed to the tone of news sites with the same/similar owners. We also include website (domain) fixed effects and event fixed effects, allowing us to compare coverage of the same event across different ownership structures. The standard errors are clustered at the domain and event levels.

\begin{equation*}
\begin{split}
Tone_{ie} = 
& \sum_{q=2}^4 \beta_{q}^{DO}\,\text{I}\bigl(ToneSameDO_{ie} \in Q_q\bigr) 
+ \sum_{q=2}^4 \beta_{q}^{UO}\,\text{I}\bigl(ToneSameUO_{ie} \in Q_q\bigr) \\
& + \gamma_1 ToneSameState_{ie} 
+ \gamma_2 ToneSameCountry_{ie} 
+ \gamma_3 ToneSameMediaType_{ie} \\
& + WebSiteFE_i + EventFE_e + \varepsilon_{ie}.
\end{split}
\end{equation*}

Here, $Tone_{ie}$ is the focal outlet \(i\)’s tone in covering event \(e\). The terms $ToneSameDO_{ie}$ and $ToneSameUO_{ie}$ represent the upstream tone from other outlets under the same domain-level or ultimate owner, respectively, divided into quartiles \(Q_q\). The indicator functions \(\text{I}(\cdot)\) equal one if the upstream tone falls into the given quartile. The variables $ToneSameState_{ie}$, $ToneSameCountry_{ie}$, and $ToneSameMediaType_{ie}$ control for the average tone of other outlets in the same state, country, and media type. Beyond their use as controls, the coefficients of these variables also serve as a useful point of reference. That is, we can assess the extent to which the tone of news sites serving the same market (same country, region, media type) predicts the focal news site's tone, ceteris paribus, as opposed to the tone of news sites with the same/similar owners. We also include website (domain) fixed effects and event fixed effects, allowing us to compare coverage of the same event across different ownership structures. The standard errors are clustered at the domain and event levels.

\begin{figure}[htbp]
\centering
\caption{Common Ownership and Reporting Tone: Quartile-Bin Regressions}
\label{fig:events_tone_bin}
\includegraphics[width=0.99\textwidth]{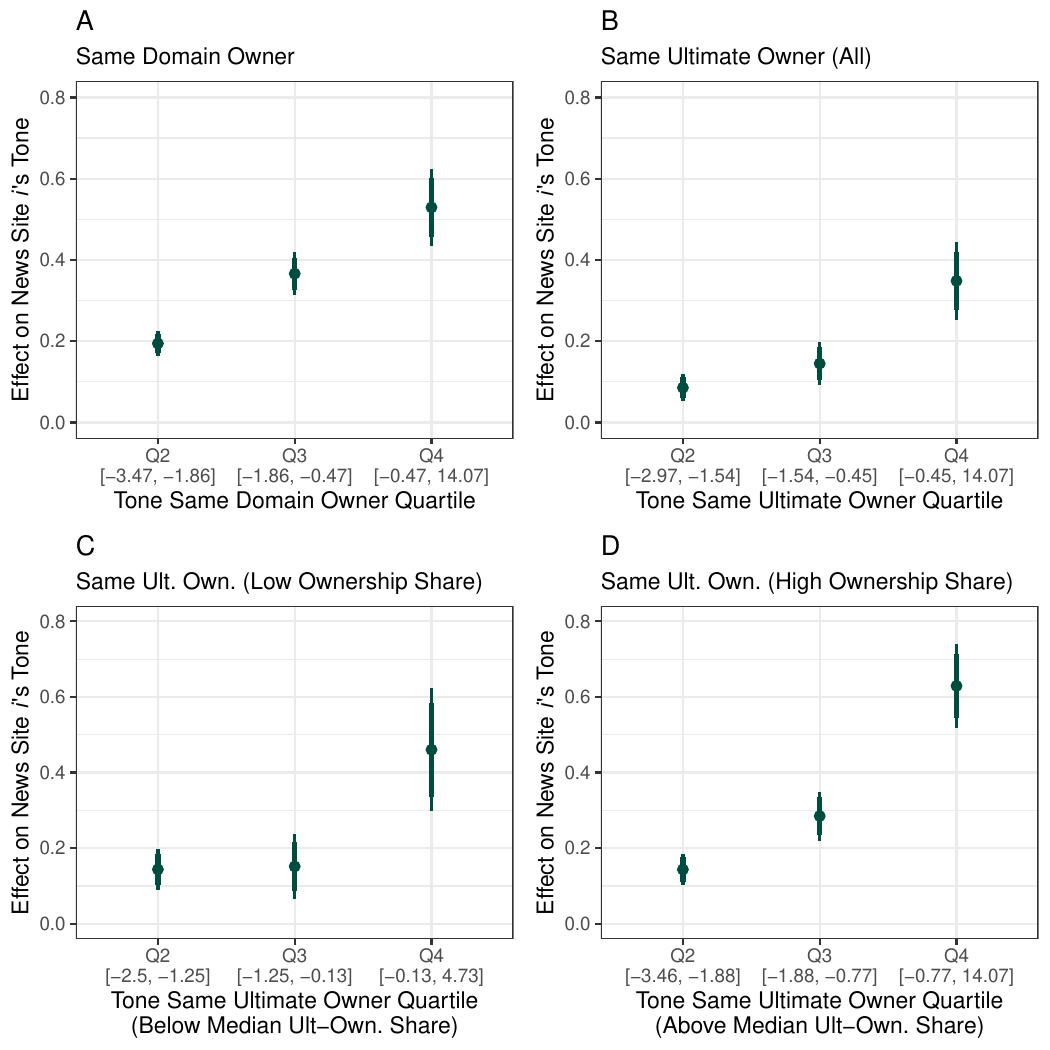}
\begin{figurenotes}
This figure reports estimates from regressions relating the focal outlet’s tone to quartile indicators of upstream tone under common ownership. Panel A uses domain-level owners and Panel B uses ultimate owners. Panels C and D split the sample according to whether the largest ultimate owner's share is above or below the median. Each dot represents a coefficient for the 2nd, 3rd, or 4th quartile of upstream tone relative to the 1st quartile; vertical lines show 95\% (thick) and 99\% (thin) confidence intervals. The numbers in square brackets below each quartile label report the endpoints of the corresponding quartile's range of the upstream-tone variable in the respective estimation sample; in Panels C and D, quartiles are formed within the below- and above-median subsamples, so these ranges differ across panels. All specifications include controls for average tone in the same state, country, and media type, as well as domain- and event-level fixed effects. Standard errors are clustered by domain and event. The sample includes hard news outlets ($>$25\% political content) with ownership and content data for both waves.
\end{figurenotes}
\end{figure}

Figure \ref{fig:events_tone_bin} presents the main results, plotting the coefficients for the second, third, and fourth quartiles of upstream tone relative to the first quartile.\footnote{Table~\ref{tab:events_tone_bin} in the Online Appendix shows the corresponding regression output in a regression table, including alternative specifications as robustness checks.} Panels (A) and (B) show that, as upstream tone becomes more positive, the focal outlet's tone also increases (reports have a more positive sentiment), again suggesting stronger editorial alignment under common ownership. Panels (C) and (D) split the sample at the median share of the largest ultimate owner, suggesting that this alignment effect tends to be stronger when ownership is more concentrated.

 Moreover, additional results based on the continuous measures of average tone under the same domain and ultimate owners remain significant and economically sizable. This is evident when we compare their coefficients to those on the average tone of news sites serving the same market (i.e., the same region, same country, and same media type; see Table~\ref{tab:events_tone_bin}). For example, a 1-point increase in the average tone of news sites in the same country as the focal news site increases the focal news site's tone by approximately 0.245. In contrast, an increase of the same magnitude in the average tone of news sites under the same domain owner increases the focal news site's tone by approximately 0.145.

\section{Mechanisms: Syndication versus Ideology} \label{sec:mechanisms}

A natural question is whether common ownership aligns content because co-owned outlets
share editorial inputs (a syndication / efficiency channel that operates regardless of
topic) or because common owners impose a shared political line (an ideology channel that
should bite mainly on political content). We separate the two by classifying each event as
\emph{political} if a majority ($>50\%$) of its GDELT mentions carry the POLITICS theme
tag, and re-estimating the headline within-dyad specification within each theme's event
universe.

The two channels leave distinct signatures. For \emph{coverage}, the same-domain-owner effect is large in both themes and only modestly larger for political events ($0.262$ vs.\ $0.327$): co-owned outlets cover the same events whatever the topic, consistent with content sharing. For \emph{tone}, the pair-level effect is concentrated in political content, where it is roughly twice the non-political effect ($0.265$ vs.\ $0.127$). A stacked specification estimating orthogonal theme-specific slopes with dyad-by-theme and period-by-theme fixed effects (Table~\ref{tab:bytheme_interaction}) confirms this formally: the political$-$non-political difference is $0.065$ for coverage ($p=0.025$) but $0.138$ for tone ($p<0.0001$). Part of the pair-level tone contrast, however, reflects the event-selection margin discussed in Section~\ref{sec:results_tone}: with the intersection-normalized tone measure, the theme difference is small and imprecise ($0.017$, dyadic SE $0.030$), although this test conditions on jointly covered event sets that are themselves outcomes of ownership and is therefore likely conservative. The sharpest evidence comes from the event level: re-estimating the event-level tone regressions of Figure~\ref{fig:events_tone_bin} separately by theme (where event fixed effects absorb event-specific tone and conditioning is on the focal outlet's own coverage only) the within-event co-movement of tone among outlets under the same domain owner is significantly stronger for political than for non-political events ($0.156$ vs.\ $0.131$; difference $0.026$, $p<0.001$), while the corresponding ultimate-owner co-movement shows no political premium (Online Appendix Table~\ref{tab:events_tone_bytheme}). Because these specifications relate an outlet's tone to the average tone of its co-owned peers on the same event, the coefficients measure equilibrium tone alignment within owner groups rather than a directional pass-through---the reflection problem of \citet{manski1993}; the political$-$non-political \emph{difference} remains interpretable under the assumption that the joint determination of co-owned outlets' tone operates symmetrically across themes.\footnote{Additional checks are consistent with a genuine political premium rather than a reflection artifact, though not uniformly conclusive. The theme premium is robust to requiring at least two reporting co-owned peers per event (difference $0.028$, $p=0.01$). A specification that replaces the same-event peer mean with the peers' average tone on \emph{other} events---which is immune to the reflection concern---yields a larger but imprecisely estimated premium ($0.122$, SE $0.075$). The subsample of outlet-event cells based on two or more articles (13\% of cells) is uninformative: the premium there is $0.010$ (SE $0.012$), consistent with both zero and the baseline estimate.} Together this points to \emph{both} channels operating on different margins: syndication shapes \emph{what} co-owned outlets cover and how they report it across topics, while direct owners align the tone of \emph{political} coverage somewhat more than that of other content---an ideology channel operating through operational control, more modest than the raw pair-level contrast suggests. Two caveats continue to apply: GDELT tone is a general sentiment measure rather than a direct measure of ideological position, and shared editorial inputs used more intensively for political content could generate a similar signature. Full per-theme estimates are in Online Appendix Tables~\ref{tab:cov_bytheme} and~\ref{tab:tone_bytheme}.

This supply-side reading contrasts with \citet{Gentzkow_Shapiro_2010}, who found U.S.\
newspaper slant to be largely demand-driven and ownership-invariant. Restricting our
within-dyad specification to U.S.\ dyads, the same geographic setting, still yields a precise same-owner effect ($0.325$ for coverage and $0.267$ for tone), comparable to the pooled non-U.S.\ estimate (Online Appendix Table~\ref{tab:gs_contrast}). One interpretation is that the shift from print to online has lowered the marginal cost of producing common content across commonly-owned outlets, making supply-side ownership effects more detectable than in the print era. Our data also offer a within-sample benchmark for the demand-side view: in a cross-sectional specification, the same-owner association is roughly 10 to 50 times the size of the observable shared-market proxies for common reader demand (same country, state, or media type; Online Appendix Table~\ref{tab:demand_supply}). The comparison is descriptive, as these proxies capture demand overlap only coarsely, but it points the same way: observable market commonality explains far less of content alignment than common ownership does.

\begin{table}[htbp]
\centering
\begin{threeparttable}
\caption{Content Similarity by Event Theme: Stacked Interaction Test}
\label{tab:bytheme_interaction}
\footnotesize
\singlespacing

% Table created by stargazer v.5.2.3 by Marek Hlavac, Social Policy Institute. E-mail: marek.hlavac at gmail.com
% Date and time: Thu, May 28, 2026 - 02:17:04 PM
\begin{tabular}{@{\extracolsep{0.0pt}}lcc} 
\\[-1.8ex]\hline 
\hline \\[-1.8ex] 
 & \multicolumn{2}{c}{Dependent variable: content similarity (theme-stacked)} \\ 
\cline{2-3} 
 & Coverage (DSC) & Tone (cosine) \\ 
\\[-1.8ex] & (1) & (2)\\ 
\hline \\[-1.8ex] 
 Same domain owner $\times$ political & 0.327$^{***}$ & 0.265$^{***}$ \\ 
  & (0.034) & (0.036) \\ 
  & & \\ 
 Same domain owner $\times$ non-political & 0.262$^{***}$ & 0.127$^{***}$ \\ 
  & (0.037) & (0.032) \\ 
  & & \\ 
\hline \\[-1.8ex] 
Dyad $\times$ theme FE & Yes & Yes \\ 
Period $\times$ theme FE & Yes & Yes \\ 
Diff. (political $-$ non-pol.) & 0.065$^{**}$ & 0.138$^{***}$ \\ 
\quad (dyadic SE) & (0.029) & (0.034) \\ 
No. observations & 6,894,752 & 2,921,006 \\ 
\hline 
\hline \\[-1.8ex] 
\end{tabular} 

\begin{tablenotes}[flushleft]
\footnotesize{
\item \textit{Notes}: The political and non-political theme observations are stacked into a
single regression with orthogonal theme-specific same-domain-owner indicators and
dyad-by-theme and period-by-theme fixed effects. Column 1 is the Dice coverage similarity; column 2 the cosine tone similarity. Each theme panel is balanced within theme (dyad-theme cells observed in both periods): the coverage stack comprises 1{,}775{,}670 political and 1{,}671{,}706 non-political pairs (3{,}452 and 3{,}316 of which switch domain-ownership status), the tone stack 817{,}091 and 643{,}412 pairs (1{,}822 and 1{,}241 switchers). The \emph{Diff.} row reports the tested political$-$non-political difference, whose dyadic standard error uses the joint covariance $\mathrm{Var}(\hat b_{\text{pol}}-\hat b_{\text{npol}}) = V_{11}+V_{22}-2V_{12}$. Standard errors are dyadic two-way clustered by domain (\texttt{domain\_a} and \texttt{domain\_b}), following \citet{cameron_miller_2015}. $^{*}$p$<$0.1; $^{**}$p$<$0.05; $^{***}$p$<$0.01.
}
\end{tablenotes}
\end{threeparttable}
\end{table}

\section{Extensions}
\label{sec:extensions}

\subsection{Sector Shares and Market Concentration: Robustness to Other Measures}

To verify the robustness of our main results, we examine alternative measures of market shares and concentration.

First, we recompute market shares using page views rather than reach. As reach and page views are generally highly correlated (see Online Appendix Figure \ref{fig:corr_reach_views}), we find similar sector shares (see Online Appendix Tables \ref{tab:shares_sectors_political} and \ref{tab:shares_sectors}).

Regarding market concentration, we also find qualitatively similar patterns when recalculating our HHI measures with page views instead of reach (see Online Appendix Figure~\ref{fig:hhi_political_views}). The key finding that concentration increases at the domain-owner level and then declines at the ultimate-owner level remains robust. The United Kingdom is the country where page views and reach yield the most different HHI values: at the domain level, the HHI is 0.35 with page views versus 0.20 with reach; at the domain-owner level, 0.35 versus 0.20; and at the ultimate-owner level, 0.33 versus 0.18. Italy also shifts from 0.14 (reach) to 0.17 (page views) at both the domain and domain-owner levels, moving from the low into the moderate band. For the remaining countries, the differences are minor and leave the concentration classifications unchanged. From a political economy perspective, reach may be the more relevant measure, as it captures the number of individual readers, and hence potential voters, exposed to an outlet's content, rather than total page consumption.

Next, we verify that our findings are robust to using 2023 ownership data instead of 2020 ownership data (see Online Appendix Figure~\ref{fig:hhi_2023_robustness}). The patterns remain qualitatively unchanged. The largest shift occurs in Poland, where the domain-owner-level HHI decreases from 0.35 to 0.28, a move from high to moderate concentration. All other countries exhibit small changes, and the qualitative classification of concentration levels remains stable. Note that we cannot directly interpret any changes in market concentration between 2020 and 2023 as actual changes since the underlying traffic data is the same (recall that we only have one cross-section of traffic, specifically averages for 2018-2020). Hence, we think of market concentration calculations based on 2023 data as a robustness check.

We also verify that our results hold when considering all media outlets rather than only hard news outlets (see Online Appendix Figure~\ref{fig:hhi_all_outlets}). The qualitative patterns are similar: concentration increases when moving from domains to domain owners, and decreases when moving to ultimate owners. Concentration levels are generally lower when including non-political outlets, as the sample size increases substantially. The largest difference appears in Poland, where the domain-owner-level HHI falls from 0.35 (hard news only) to 0.20 (all outlets), shifting from high to moderate concentration. For all other countries, the differences are smaller (at most 0.05). Beyond robustness, this comparison is informative in itself: ownership is systematically more concentrated among hard news outlets than in the broader outlet population. If concentrated ownership (and the content alignment it enables) were partly driven by ideological motives to control and slant political coverage, this is the pattern one would expect, and it complements the evidence in Section~\ref{sec:mechanisms}, where the tone alignment among co-owned outlets is concentrated in political content. Because part of the difference also reflects the mechanically larger market when non-political outlets are included, we regard this pattern as suggestive rather than conclusive.

As an alternative to the HHI, we examine concentration using the $M$-firm concentration ratio ($C_M$), defined as the cumulative market share of the top $M$ firms in a market. We focus on the four-firm concentration ratio ($C_4$), a widely used measure in industrial organization. Following standard thresholds, a $C_4$ below 40\% indicates low concentration (competitive market), between 40\% and 70\% indicates medium concentration (loose oligopoly), and above 70\% indicates high concentration (tight oligopoly) \citep{legalclarity2024cr4}.

Figure~\ref{fig:c4_domain_to_ultimate} shows the $C_4$ measure across countries for domains, domain owners, and ultimate owners. The $C_4$ findings are consistent with the HHI results in terms of which countries exhibit the highest concentration and the overall pattern across ownership levels. At the domain level, Poland (70\%) and the United Kingdom (68\%) show the highest concentration, the same two countries flagged by the HHI. At the domain-owner level, Poland (87\%) and Switzerland (83\%) exhibit the highest concentration, both falling into the tight oligopoly category; these are also the countries with the highest HHI values at this level. The pattern of concentration increasing from domains to domain owners and then declining at the ultimate-owner level is evident with both measures.

However, the $C_4$ measure flags more countries as having elevated concentration than the HHI. The $C_4$ measure places seven countries in the medium-concentration category (loose oligopoly): the U.K., Italy, Switzerland, Romania, France, and Canada all have $C_4$ values between 40\% and 70\%, in addition to Poland at the threshold. Similarly, at the domain-owner level, while HHI flags only Poland, Switzerland, and the U.K. as non-low, the $C_4$ measure classifies seven countries as medium or high concentration. This discrepancy arises because the HHI captures additional information beyond the $C_4$.\footnote{This is also why it is more commonly used in antitrust practice. Since 1982, the U.S. Department of Justice has relied on HHI rather than concentration ratios in its merger guidelines \citep{doj2010horizontal}.} While $C_4$ measures only the combined share of the top four firms, HHI, by summing squared market shares across all firms, is sensitive to two additional features of the market structure. First, HHI accounts for how evenly the top firms share the market: four firms each holding 15\% yield a lower HHI than one firm holding 45\% and three firms holding 5\% each, even though both configurations produce a $C_4$ of 60\%. Second, HHI captures information about the tail: markets where outlets beyond the top four retain meaningful shares will have lower HHI values than markets where the top four dominate entirely.

Our data illustrate both mechanisms. Romania at the domain-owner level exemplifies the first mechanism: the top four owners hold nearly identical shares (18.6\%, 18.4\%, 15.3\%, and 14.5\%), yielding a high $C_4$ of 67\% but a moderate HHI of 0.13. Canada at the domain level exemplifies the second mechanism: the fourth-ranked outlet holds 5.4\% of the market, yet the outlets beyond the top four -- 375 in total, led by the fifth and sixth with comparable shares of 4.6\% and 4.2\% -- collectively account for 56\% of the market. This long tail produces a $C_4$ of 44\% but an HHI of only 0.073.\footnote{Of the 461 Canadian hard news outlets, 379 have traffic data and enter the market-share computation.}

Crucially, the core finding largely holds, though less starkly than the HHI alone suggests: at the ultimate-owner level, the HHI places eight of the ten countries in the low-concentration band. The $C_4$ paints a more nuanced picture: three countries fall below the 40\% threshold (the U.S. at 25\%, Spain at 28\%, and Germany at 30\%), while the remaining seven lie in the 40--70\% band (medium concentration), led by the U.K. and Poland at 57\%. This qualifies our main conclusion: dominance by a single owner or a very small group remains absent -- no country reaches the high-concentration band on reach -- but a handful of large ultimate owners do account for a substantial share of the market in most countries, consistent with the long-tail mechanism above.

Finally, Online Appendix Figures \ref{fig:cum_shares} and \ref{fig:cum_shares_nonpolitical} show the $C_M$ measures for $M = 1, \ldots, 20$.

\subsection{Market Concentration: Robustness to Missing Domains} \label{sec:robustness_missing}

Missing news domains in our seed sample or missing data (on traffic or ownership) could produce imprecise market concentration estimates. Given the way in which we compile our sample, we expect (if at all) hypothetically missing outlets to attract low to moderate traffic. To estimate the impact of missing data in a ``worst case scenario'', we run a simulation. Suppose several hypothetical outlets are missing in our data. We assume that the missing websites would receive traffic amounting to the median of all outlets for which we have complete data. This is a generous assumption since missed outlets are likely small (see above). Then, we ask how the market concentration results would change if $Z$ such outlets were missing. The simulation confirms the robustness of our estimates against such missing data (see Online Appendix \ref{app:sec:simulation}).

\subsection{Market Concentration Based on Attention Shares}

\cite{Prat_2018} emphasizes the importance of voters' attention dedicated to specific outlets when assessing the potential political influence (``power'') of media outlets. Suppose outlet $A$ is read by 10 individuals, who also dedicate equal amounts of attention to 9 other outlets ($A$ gets 10\% of each individual's attention). Then, $A$ influences 1 voter (10 voters $\times$ 10\%). In contrast, only two individuals read outlet $B$, but they do not consume any other news. $B$ then influences 2 voters (2 $\times$ 100\%).

In this framework, to interpret our findings on market shares as a proxy for the potential influence of a news website, we must assume that news websites' reach is correlated with voter attention. Because we can only measure web traffic at the domain level (not at the individual-user level), we do not directly observe attention shares. We, therefore, proxy for whether an individual visits a news domain by whether they follow the same outlet on Twitter. We justify this proxy by noting that the number of followers of a news site and our measure of reach correlate almost perfectly (91\%).\footnote{We collect data on the total number of Twitter followers of each outlet in our sample by manually matching news domains and their Twitter accounts. Then, we scrape the follower number for each account.} We draw a random sample of 6,475 U.S. Twitter users, extract all the accounts that they follow, identify accounts of news sites, and match these accounts to our news domains.\footnote{To obtain a seed sample of U.S. Twitter users, we scrape \textit{all} English-language Tweets on three days (1 October, 1 November, and 1 December 2020). We randomly sampled 100,000 users active on any of these days. For each such user, we query their location via the Twitter API. We send the location information to the Google Location API to determine if the user is based in the U.S. We match 11,230 individuals (out of the 100,000) to the U.S. Thus, we drop English-tweeting users from other countries or without location information. Out of the 11,230 U.S. users, we focus on the 6,475 users that use Twitter for news consumption--defined as following $\geq1$ outlet(s) from our sample.} Hence, for each random Twitter user, we know the combination of outlets they follow. We then calculate the attention shares and find that their correlation with our market share figures is 69\%, see left panel of Online Appendix Figure~\ref{fig:attention_reach}.  The relationship is robust to excluding the five strongest outliers (correlation of 64\%), such as npr.org (National Public Radio) and oann.com (One America News Network), which receive disproportionately high Twitter attention relative to their web traffic (see also left panel). The correlation increases to 81\% when we exclude very small outlets (reach $<$ 1 per million), as shown in the right panel.

\subsection{Robustness of Ownership and News Content Results}

We perform a series of robustness checks to verify our baseline results on the relationship between media ownership and news content. First, we vary the measure to compute the events set similarity (e.g., Jaccard Index instead of DSC) and re-run our baseline models with the resulting alternative dependent variables. Tables~\ref{tab:coverage_alt_similarity_pairs} and~\ref{tab:coverage_alt_similarity_pairs_II} in the Online Appendix show that our main results do not change qualitatively in these alternative specifications. Second, we verify the robustness of our main results across different subsamples (partially constraining the origin of variation in change of ownership) and contrast them with a simpler model that also exploits between-website-dyad variation. Tables~\ref{tab:coverage_robustness} and \ref{tab:tone_robustness} show that our conclusions from our baseline analyses are also robust to these alternative specifications. Finally, we show that our baseline results on news coverage are not driven by any single type of news website. Table~\ref{tab:coverage_similarity_media_types} shows the regression results of our baseline models when splitting the sample according to media type: Either ``Digital'' (no historical offline equivalent), ``Broadcast'' (e.g., cnn.com or npr.com), or ``Print'' (e.g., nytimes.com or lemonde.fr). In both regression estimates, we observe a precisely estimated positive partial correlation between common ownership and alignment in news coverage. Our estimates are likewise robust to restricting the sample to within-country news-site pairs, which removes cross-country dyads that mechanically contribute little same-owner variation: the headline within-dyad coefficients are essentially unchanged (0.328 for coverage and 0.341 for tone; Online Appendix Tables~\ref{tab:cov_withincountry} and~\ref{tab:tone_withincountry}). Finally, because ownership is measured as of July 2020 and July 2023 while the content windows end in June of those years, treatment status could be misclassified for pairs whose ownership changed close to the snapshot dates. Re-estimating the co-ownership effect on the quarterly coverage panel underlying Online Appendix Figure~\ref{fig:pretrend}, excluding the quarters adjacent to each snapshot, leaves the estimate essentially unchanged (0.23 with all quarters, 0.23 when excluding two quarters per window, and 0.25 when excluding three).

\section{Discussion and Conclusion} \label{sec:discussion}

We present first evidence on the current composition of the online news media landscape in the United States, Canada, and eight European countries. Our results have several politico-economic implications.

Regarding concerns of media capture by owners, our results highlight two crucial aspects. On the one hand, we find that more than two-thirds of outlets have a single ultimate owner or at least one majority owner. However, many of those have a relatively small reach. On the other hand, many online media outlets with large market shares tend to have diverse ultimate owners, with no single majority owner. Therefore, nearly half of the online news market has ownership networks that are difficult to trace. While seminal contributions on the persuasive power of outlets typically assume that a single owner controls the content of a news outlet \citep[e.g.,][]{Prat_2018}, this may be a rather strong abstraction, especially for many large outlets. Thus, diverse ownership may reduce the risk that any single owner captures editorial policy, but it also obscures accountability and may not preclude influence through other channels (e.g., advertiser pressure, coordinated minority shareholders).

Regarding transparency and the involvement of non-media industries, our findings are consistent with prior findings on general cross-sector firm ownership.\footnote{\cite{aminadav2020corporate} find that, as of 2012, slightly more than half of the listed firms they study worldwide are widely held (their work does not inform on media sector ownership).} Accordingly, while the hard-to-track ownership networks in the media sector may not stand out as special relative to other sectors, they still raise political concerns. With such ownership structures, it becomes prohibitively costly for ordinary consumers or regulators to understand who is (ultimately) responsible for the content. Accordingly, it is difficult to anticipate potential bias. The extent to which consumers can (and want) to correct for possible biases in reporting (by switching away from a particular outlet, for instance) has been proposed as one key determinant of the persuasive power of the media \citep[see, e.g., ][]{Prat_2018}. % Even if readers were highly motivated to correct for biases they are aware of, it is prohibitively costly to do so with the prevailing lack of transparency. 
Our work highlights a trade-off in media ownership: a single owner may exert substantial political influence. However, with the lack of transparency for diversely held outlets, correcting for biases is prohibitively costly (even if readers were highly motivated to do so).  %This holds unless we assume that shareholders beyond immediate domain owners are not involved in editorial decisions and only invest in news as a financial asset.

Finally, with respect to market concentration, at the news-site level, concentration is moderate to high in Poland and the U.K., and low elsewhere. At the domain-owner level, Poland, Switzerland, and the U.K. exhibit moderate to high concentration, while all other countries show low concentration. When analyzing ultimate owners, concentration falls in most countries, and most strongly where domain-owner concentration is highest (in Poland from 0.35 to 0.16 and in Switzerland from 0.22 to 0.10). Consequently, the online media markets in the U.S., Canada, and Europe do not seem to be dominated by just a few influential ultimate owners. %Currently, the primary concern regarding ownership in the online media market is, therefore, arguably lacking transparency.

Which ownership level matters most for media pluralism? Our content analysis shows that ownership at the domain level has the strongest association with editorial alignment. Outlets sharing the same domain owner exhibit substantially more similar coverage and tone. This suggests that domain-owner concentration is the most immediately policy-relevant metric. Ownership at the ultimate-owner level is also associated with content, though the association is smaller. The mostly low concentration at this level is reassuring from a traditional market-power perspective. However, the fact that content alignment extends to the ultimate-owner level reinforces the transparency concerns discussed above: dispersed ultimate ownership mitigates concentration risks but still presents accountability challenges. The mechanism evidence in Section~\ref{sec:mechanisms} refines this picture: shared ownership aligns which events outlets cover for political and non-political content alike, consistent with content sharing across co-owned outlets, whereas the alignment of reporting tone is concentrated in political content. Owner influence on online news thus appears to operate both through content sharing and through the tone of political coverage.

\clearpage

\bibliographystyle{style/aea}
\bibliography{ref.bib}

\clearpage

\pagestyle{plain}
\setcounter{footnote}{0}

%%%%%%%%%%%%%%%%%%%%%%%% APPENDIX %%%%%%%%%%%%%%%%%%%%%%%%%%%%%%%%%%%%%%%%%%%%%%%%%

\setcounter{table}{0}
\setcounter{figure}{0}
\setcounter{page}{1}
\renewcommand{\thetable}{\arabic{table}}
\renewcommand{\thefigure}{\arabic{figure}}
\renewcommand{\thesubsection}{\arabic{subsection}}
\renewcommand{\thesubsubsection}{\arabic{subsubsection}}
\renewcommand{\thesection}{\arabic{section}}

\clearpage
\appendix
% Note: \part{} removed to avoid empty page; parttoc provides TOC directly
\parttoc{}

\section{Details on Data Collection and Database}
\label{sec:app:data}

This appendix section documents the data collection in detail. It reports the timeline of all data collections, coverage statistics for ownership, traffic, and content data, summary statistics of the resulting sample, and the list of domains excluded from the sample together with the reasons for their exclusion.

\paragraph{Data collection timeline } Our analyses draw on multiple data sources collected at different points in time. Table \ref{tab:data_timeline} summarizes the timing of each data source.

\begin{table}[htbp]\centering
\caption{Data Collection Timeline}
\label{tab:data_timeline}
\begin{threeparttable}
\small
\begin{tabular}{@{}llll@{}}
\toprule
Data Source & Collection & Coverage Period & Used In \\
\midrule
Ownership (Orbis) & July 2020 & Cross-section, July 2020 & Main descriptives, Panel wave 1 \\
Ownership (Orbis) & July 2023 & Cross-section, July 2023 & Panel wave 2 \\
Traffic (AWIS) & 2020 & 2018--2020 (averaged) & All analyses \\
Political content (GDELT) & 2020 & 2018--2020 (averaged) & Political filtering ($>$25\%) \\
Event coverage (GDELT) & 2025 & Jan 2019 -- June 2020 & Panel wave 1 \\
Event coverage (GDELT) & 2025 & Jan 2022 -- June 2023 & Panel wave 2 \\
\bottomrule
\end{tabular}
\begin{tablenotes}[flushleft]
\footnotesize{\item Notes: This table summarizes the collection dates and coverage periods for each data source.}
\end{tablenotes}
\end{threeparttable}
\end{table}

Two details deserve note. First, the traffic weights (2018--2020 averages of weekly AWIS measurements) are used throughout, including in analyses involving the July 2023 ownership wave; traffic patterns for established news outlets are relatively stable, and updated traffic data were not available. Second, the hard news classification (political-content ratio above 25\%, based on 2018--2020 GDELT annotations) is held time-invariant across waves.

\paragraph{Data coverage } Table \ref{tab:covstat} shows the data coverage for the news outlet characteristics. Specifically, how many canonical domains are assigned to a given country (\textit{No.}), for how many of these domains we can identify an owner in the Orbis database (\textit{Owners}), for how many domains we obtain traffic information from AWIS (\textit{Traffic}), for how many domains weighted by traffic we identify an owner (\textit{Traffic} $\times$ \textit{owners}), for how many domains we can collect information on the reporting from GDELT (\textit{\%Content}), and for how many domains weighted by traffic we obtain reporting information (\textit{\%(Traffic $\times$ content)}). Ownership data availability ranges from almost 40\% (U.S.) to 90\% (Germany). Ownership information is often unavailable for small outlets. This is reflected in the fifth column (\textit{Traffic} $\times$ \textit{owners}). This column zooms in on outlets for which traffic data is available and then calculates traffic-weighted ownership coverage. It is high for all countries, for some even (close to) 100\% (Germany, Spain, Poland, and Switzerland).

Traffic data coverage (see fourth columns) ranges from 85\% in Canada to 96\% in Poland.
The stories featured by an outlet are covered in GDELT for 40\% (Canada) to 69\% (U.K.) of outlets. Again, content coverage appears to correlate with outlet traffic size: when we weight GDELT coverage by traffic, we obtain values ranging from 62\% (Poland) to 95\% (UK). The positive relationship between traffic and coverage implies that errors in our market concentration estimates arising from limited coverage are likely minor.

Finally, a note on the coverage of sector information. Unfortunately, sector information is not available for all Orbis entries. For domains with an Orbis match, but without sector information, we manually collect sector information at the domain level. Accordingly, we use data from Bloomberg and, at times, other online sources. Finally, sector information for domain owners is available for $>$90\% of domains in seven countries (U.S., Germany, France, Spain, Poland, Romania, and Switzerland). For the U.K., Italy, and Canada, it is available for $>$75\%. At the level of ultimate owners, sector information is available for a minority of the ultimate owners that have any stakes in outlets specializing in politics, and its availability varies across countries (from around 40\% of ultimate owners in the U.S., Canada, France, Spain, and Poland to 16\% in Italy and 3\% in Romania). Sector shares are computed among owners with sector information, so that unclassified owners enter neither the numerator nor the denominator; at the ultimate-owner level, classified owners account for roughly a fifth to a third of ownership-weighted traffic in most countries. For the ultimate owners, the sector information comes only from Orbis. We do not engage in a manual search for ultimate owners, given their number (20,048 unique ultimate owners).

\begin{table}[htbp]
\centering
\caption{Data Coverage}
\label{tab:covstat}
\resizebox{\textwidth}{!}{
\begin{tabular}{lrrrrrrrr}
\hline
\toprule
\textit{Country} & \textit{No.} & \textit{\%Owners} & \textit{\%Owners} & \textit{\%Traffic} & \textit{\%(Traffic $\times$} & \textit{\%(Traffic $\times$} & \textit{\%Content} & \textit{\%(Traffic $\times$} \\
 &  & \textit{2020} & \textit{2023} &  & \textit{owners)} & \textit{owners)} & & \textit{content)} \\
 &  & & &  & \textit{2020} & \textit{2023} & & \\
\midrule
U.S. & 8803 & 0.39 & 0.38 & 0.86 & 0.83 & 0.77 & 0.41 & 0.85 \\
CANADA & 1283 & 0.61 & 0.56 & 0.85 & 0.85 & 0.83 & 0.40 & 0.86 \\
U.K. & 598 & 0.79 & 0.77 & 0.91 & 0.99 & 0.98 & 0.69 & 0.95 \\
SPAIN & 350 & 0.65 & 0.61 & 0.89 & 0.97 & 0.96 & 0.39 & 0.93 \\
GERMANY & 341 & 0.89 & 0.83 & 0.87 & 0.98 & 0.93 & 0.48 & 0.94 \\
ITALY & 144 & 0.75 & 0.74 & 0.88 & 0.92 & 0.92 & 0.60 & 0.77 \\
ROMANIA & 142 & 0.58 & 0.52 & 0.89 & 0.85 & 0.82 & 0.46 & 0.85 \\
FRANCE & 114 & 0.79 & 0.69 & 0.90 & 0.87 & 0.66 & 0.55 & 0.88 \\
SWITZERLAND & 114 & 0.76 & 0.72 & 0.86 & 0.98 & 0.97 & 0.48 & 0.93 \\
POLAND & 78 & 0.83 & 0.83 & 0.96 & 1.00 & 1.00 & 0.63 & 0.62 \\
\bottomrule
\hline\hline
\end{tabular}
}
\justify \scriptsize {\emph{Notes:} The following data coverage statistics are shown: how many domains are assigned to a given country (\textit{No.}), for how many of these domains we can identify an owner in the Orbis database for the initial data collection in 2020 (\textit{\%Owners 2020}) and later for the update in 2023 (\textit{\%Owners 2023}), for how many domains we obtain traffic information (\textit{\%Traffic}), for how many domains weighted by traffic we identify an owner in 2020 (\textit{\%(Traffic} $\times$ \textit{owners}) 2020) and 2023 (\textit{\%(Traffic} $\times$ \textit{owners}) 2023), for how many domains we can collect information on the reporting (\textit{\%Content}), and for how many domains weighted by traffic we obtain reporting information (\textit{\%(Traffic $\times$ content)}).}
\end{table}

\begin{figure}[htbp]
\centering
\caption{Histograms of Reach per Million, by Country}
\label{fig:app:reach_hist}
\includegraphics[width=0.45\linewidth]{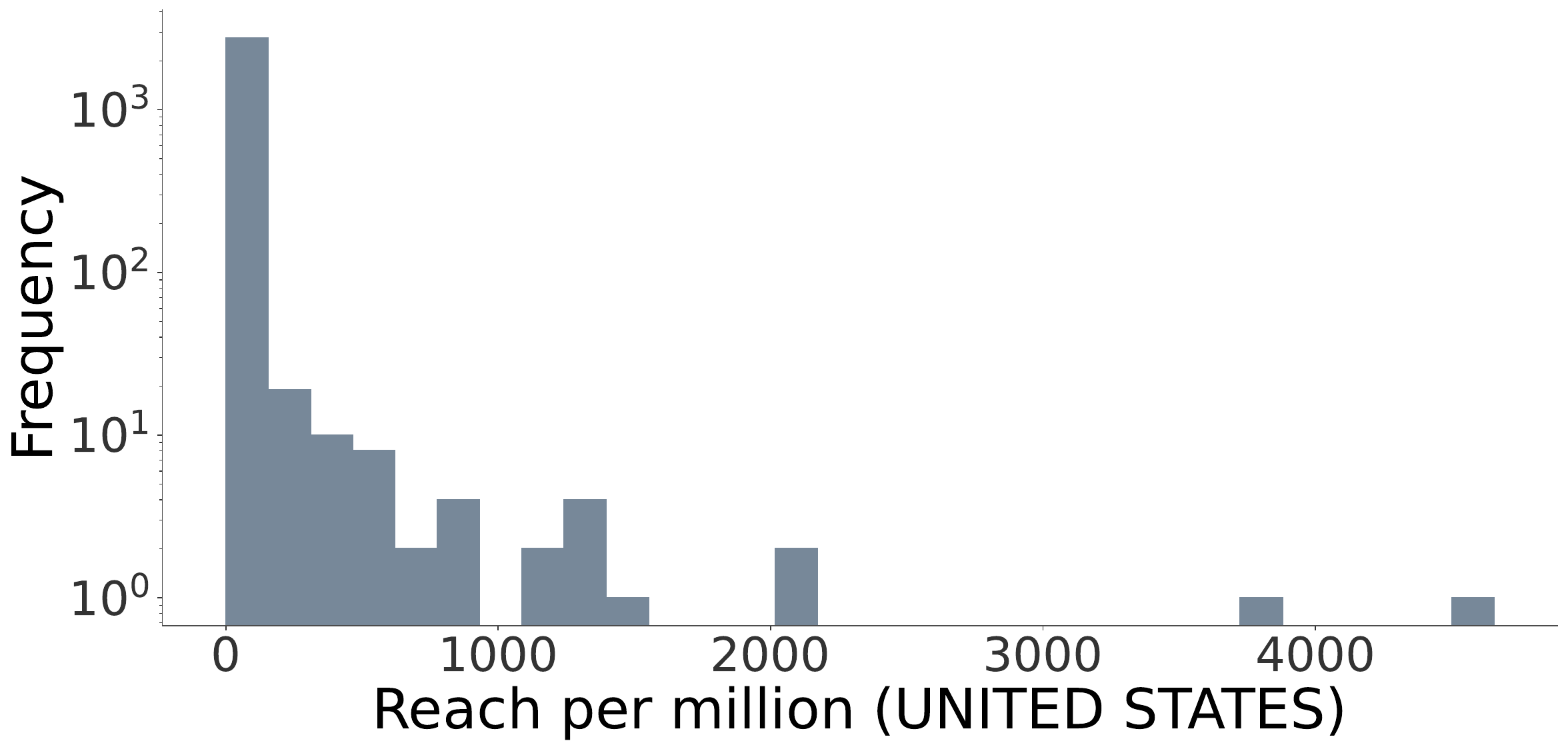}
\includegraphics[width=0.45\linewidth]{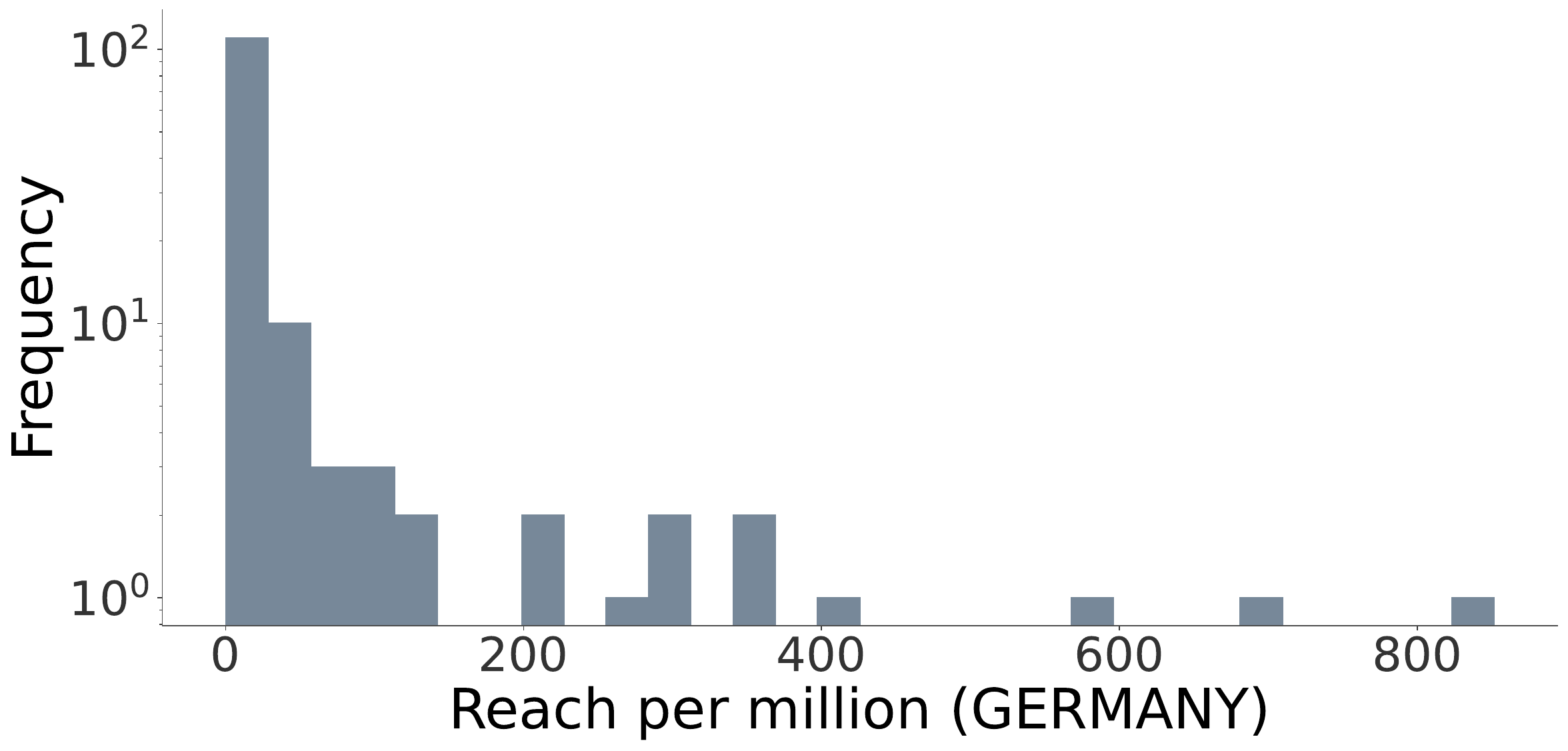}
\includegraphics[width=0.45\linewidth]{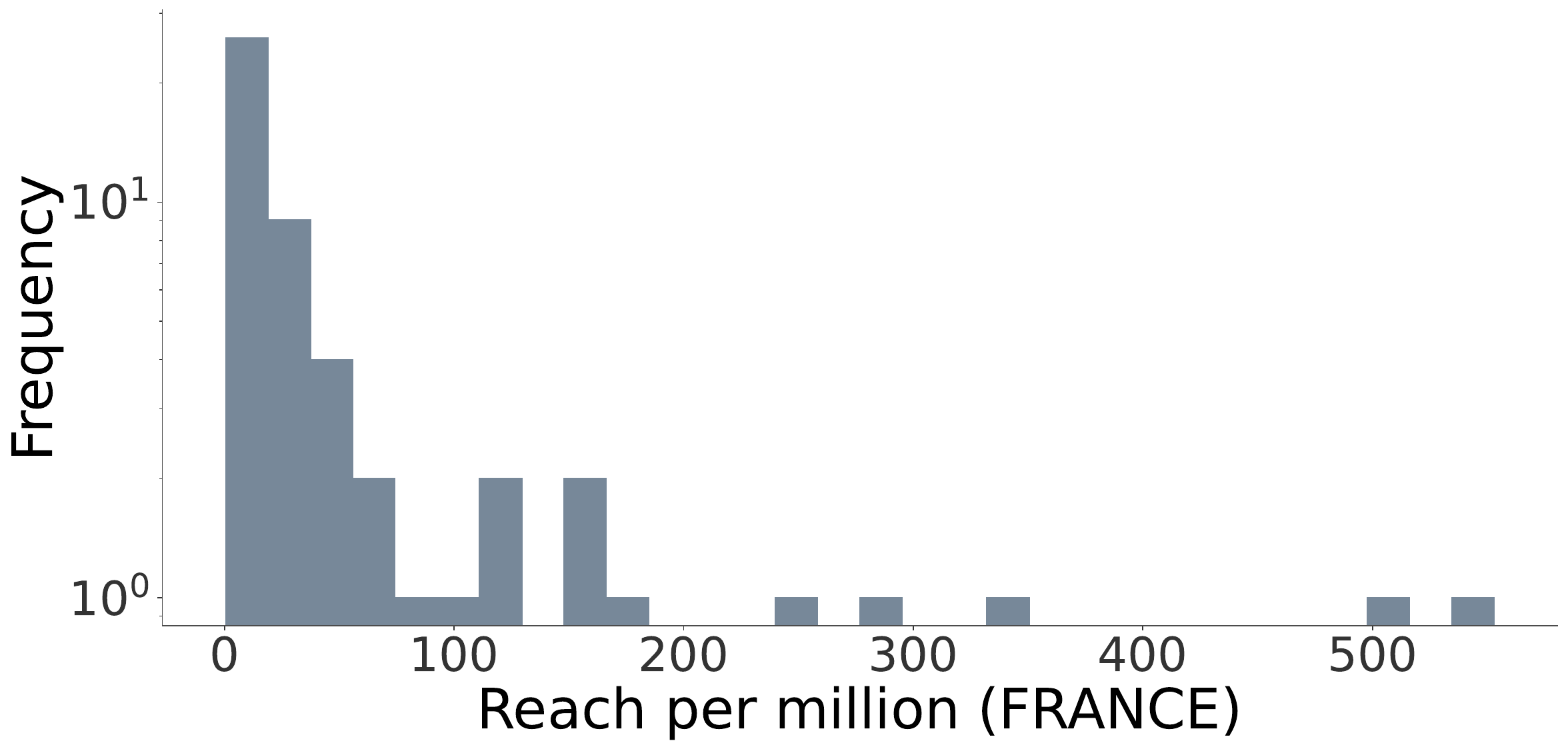}
\includegraphics[width=0.45\linewidth]{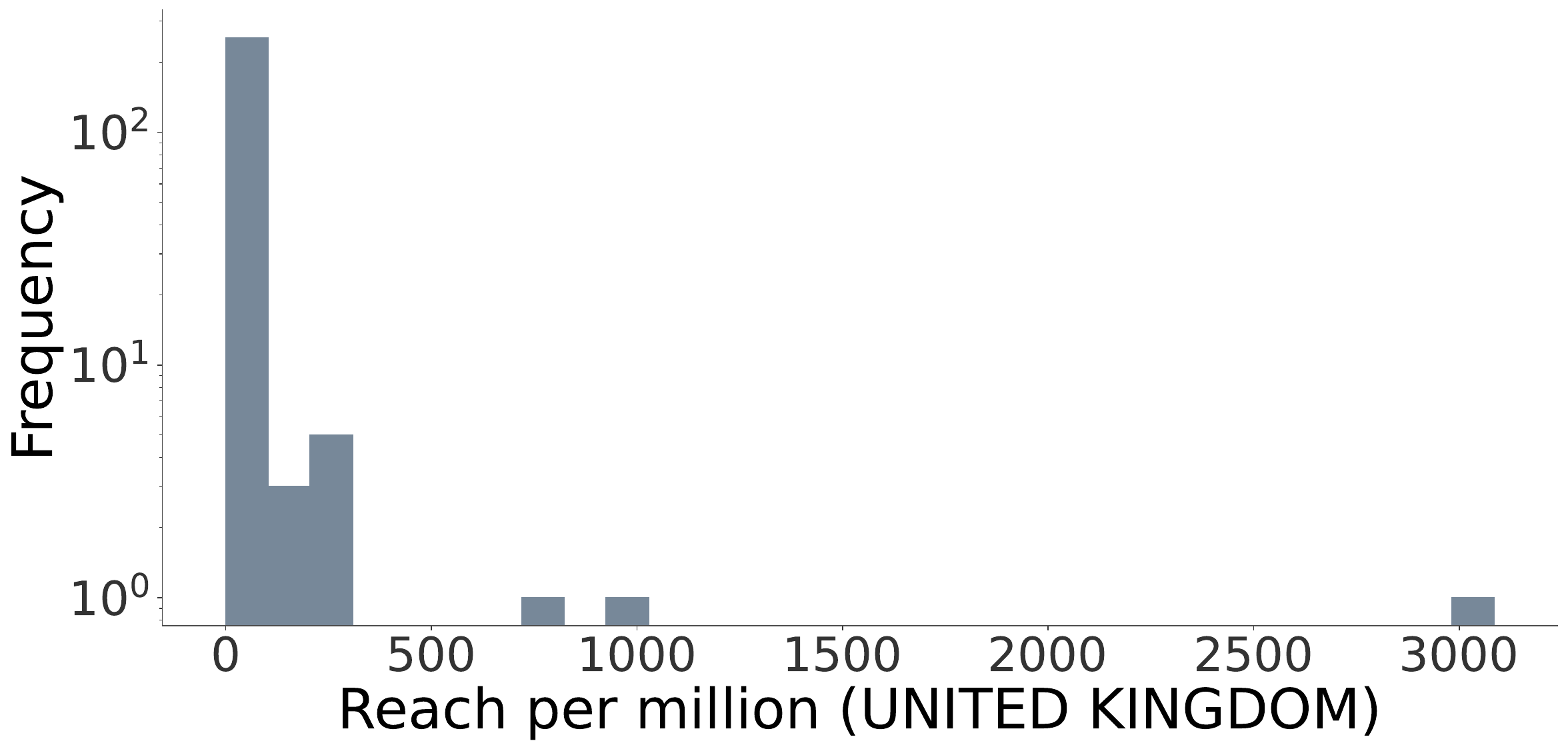}
\includegraphics[width=0.45\linewidth]{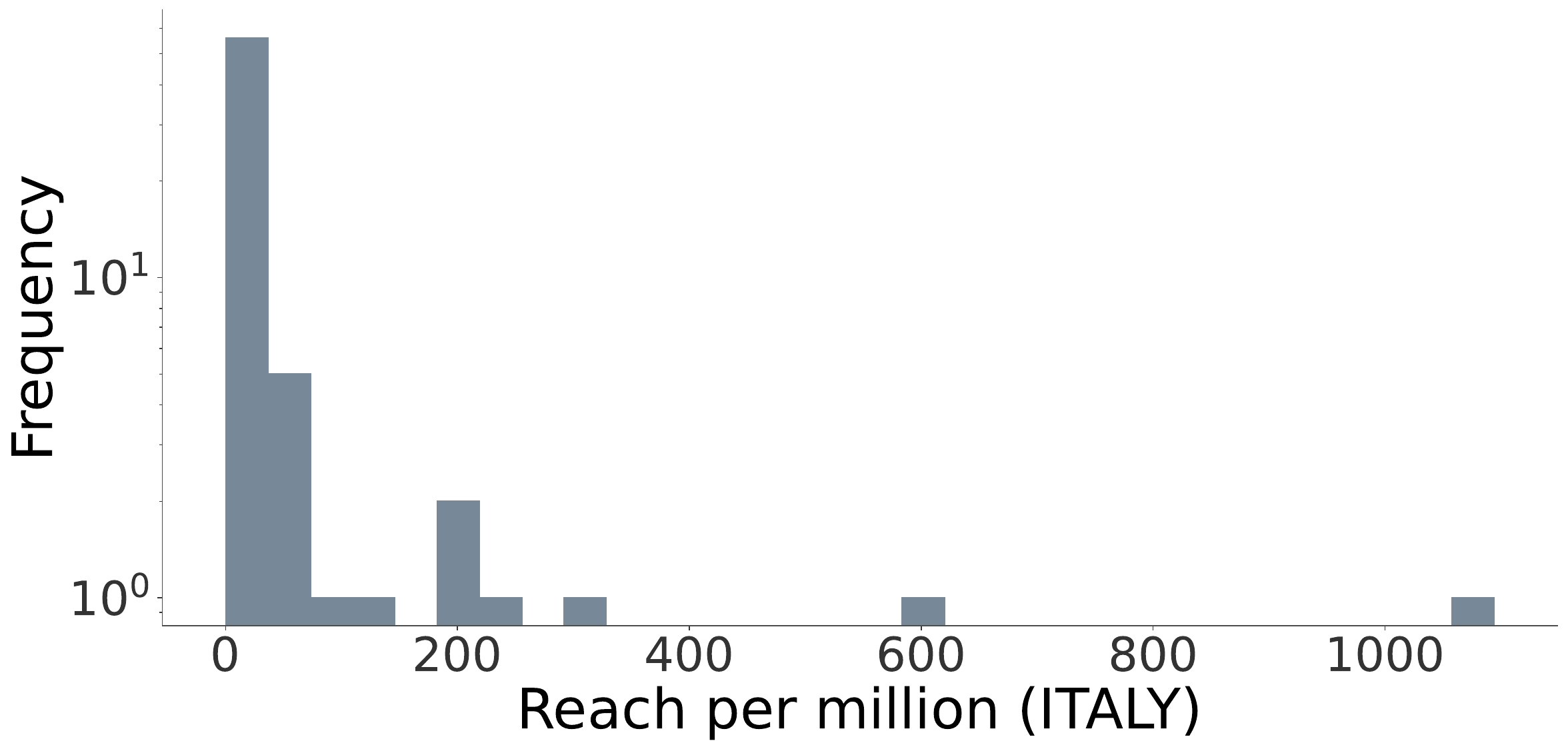}
\includegraphics[width=0.45\linewidth]{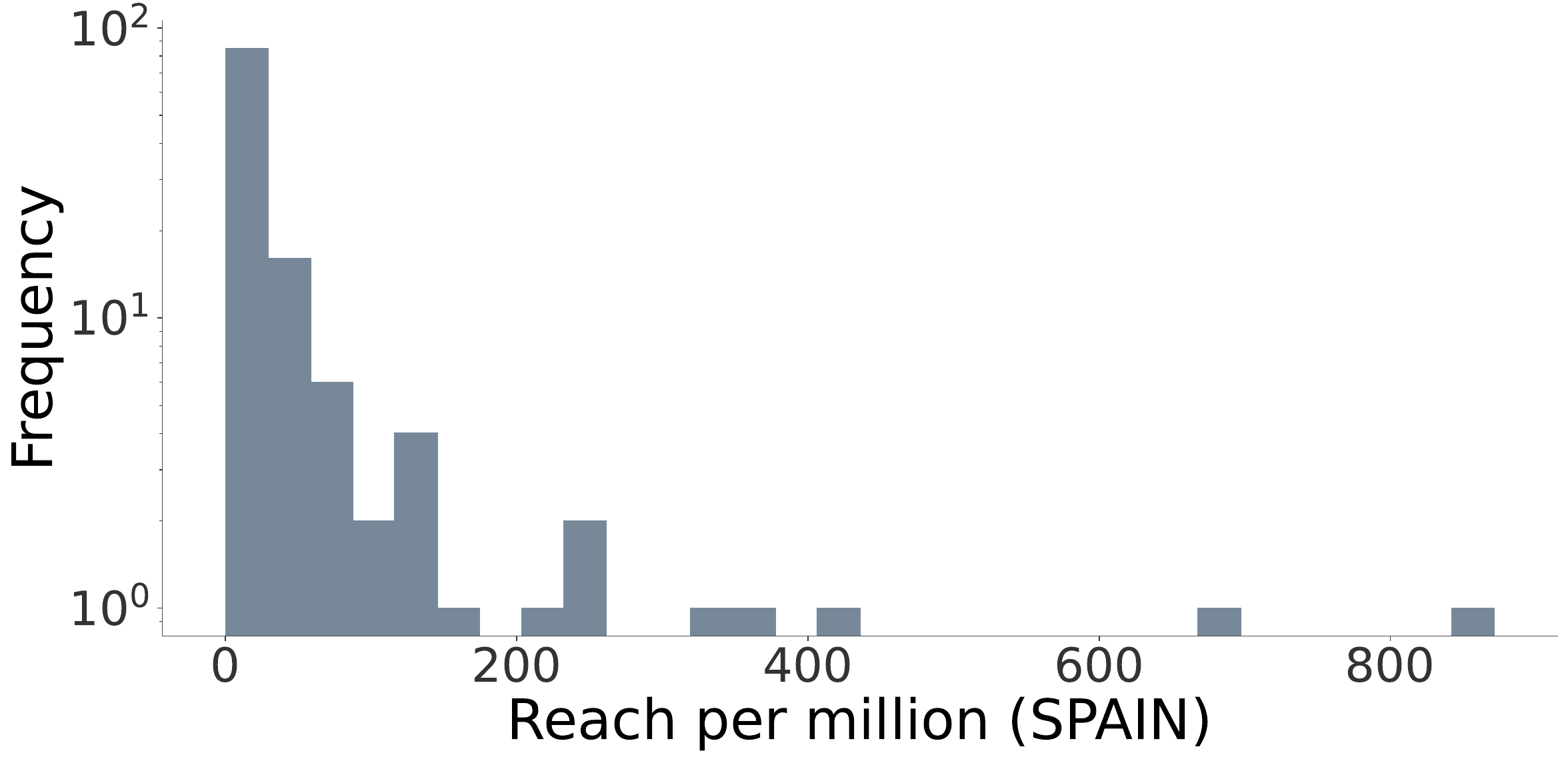}
\includegraphics[width=0.45\linewidth]{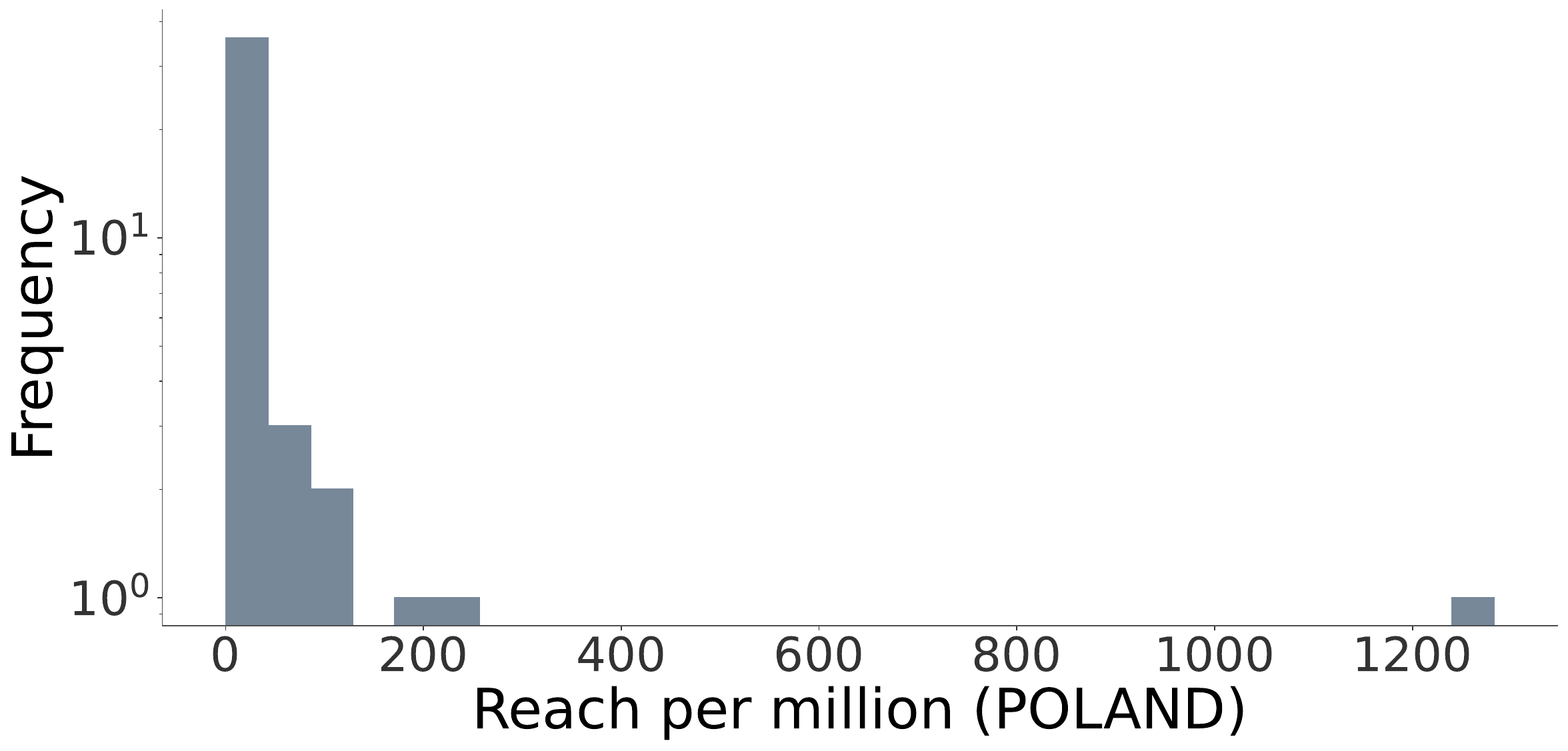}
\includegraphics[width=0.45\linewidth]{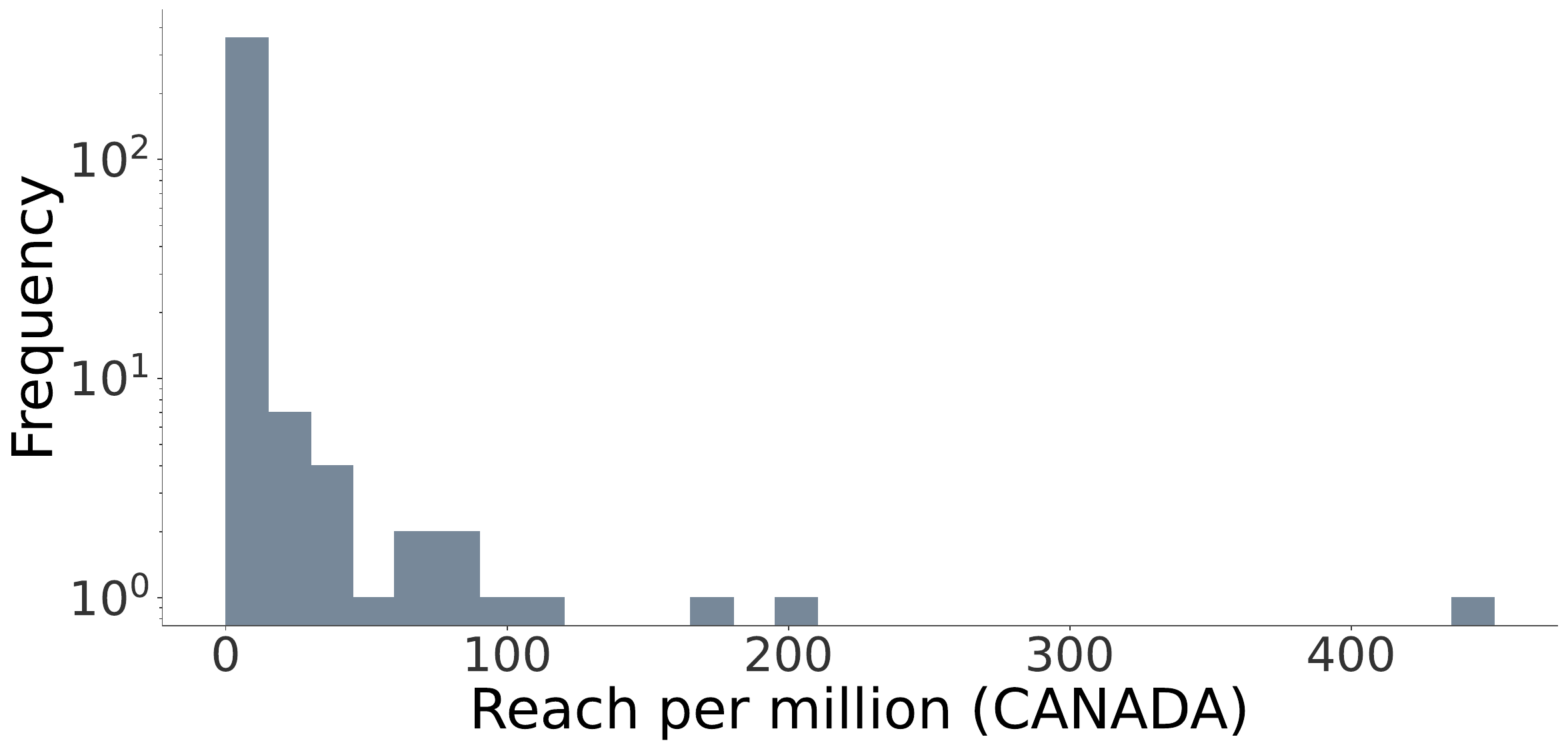}
\includegraphics[width=0.45\linewidth]{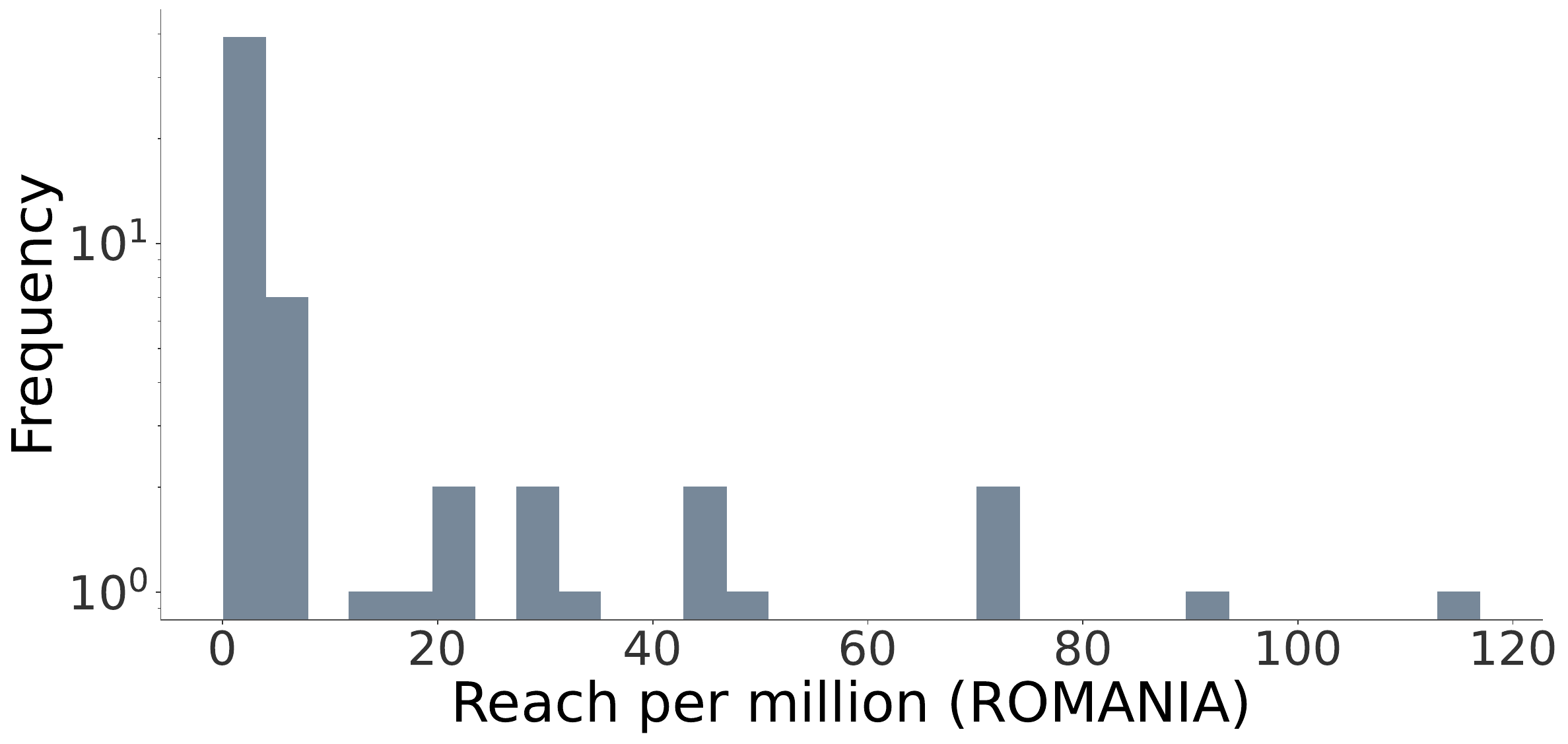}
\includegraphics[width=0.45\linewidth]{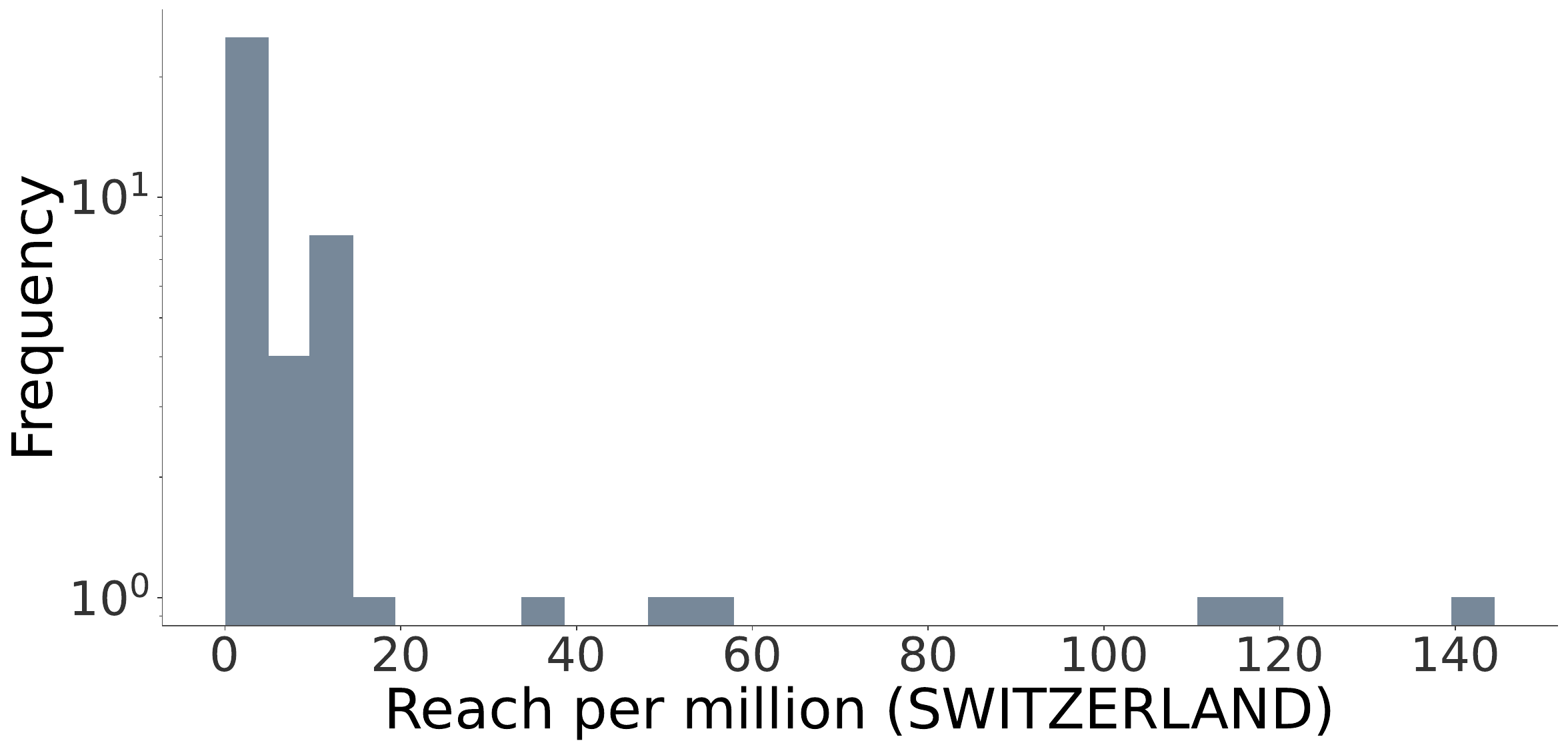}
\begin{figurenotes}
The figure shows the distribution of the reach figures by country, for all hard news outlets (more than 25\% political reporting). See Section~\ref{sec:app:data} for discussion of summary statistics.
\end{figurenotes}
\end{figure}

\begin{figure}[htbp]
\centering
\caption{Histograms of Page Views per Million, by Country}
\label{fig:app:pv_hist}
\includegraphics[width=0.45\linewidth]{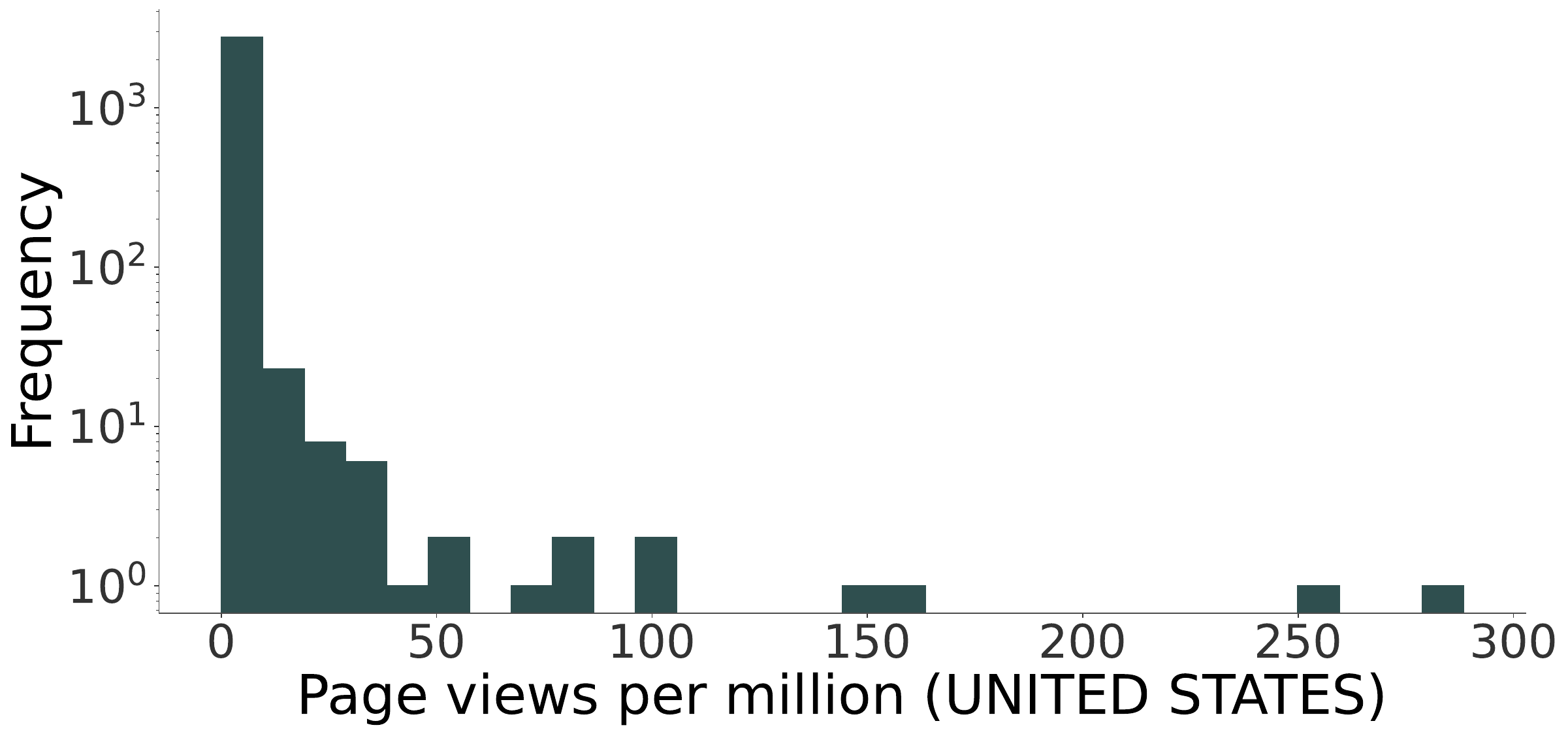}
\includegraphics[width=0.45\linewidth]{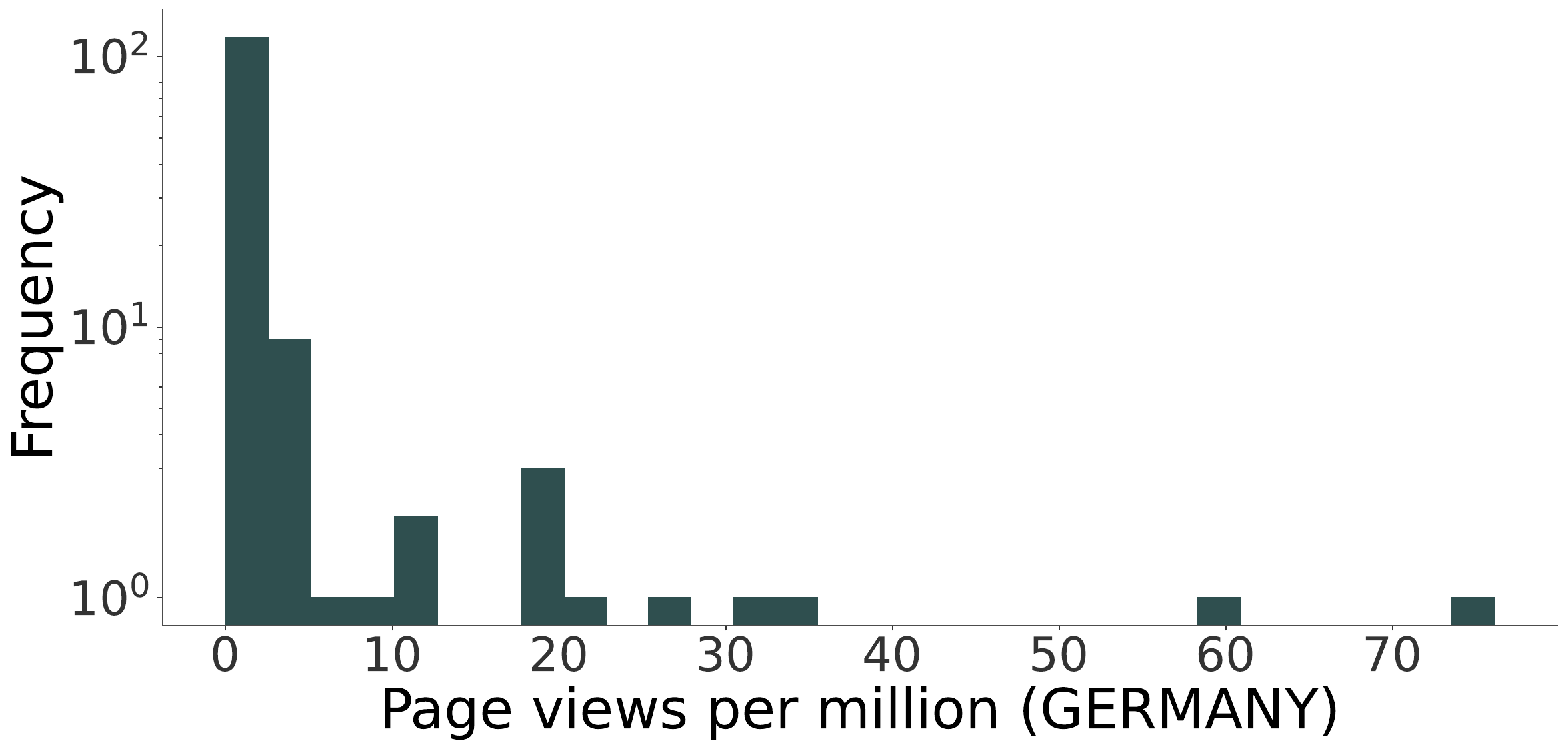}
\includegraphics[width=0.45\linewidth]{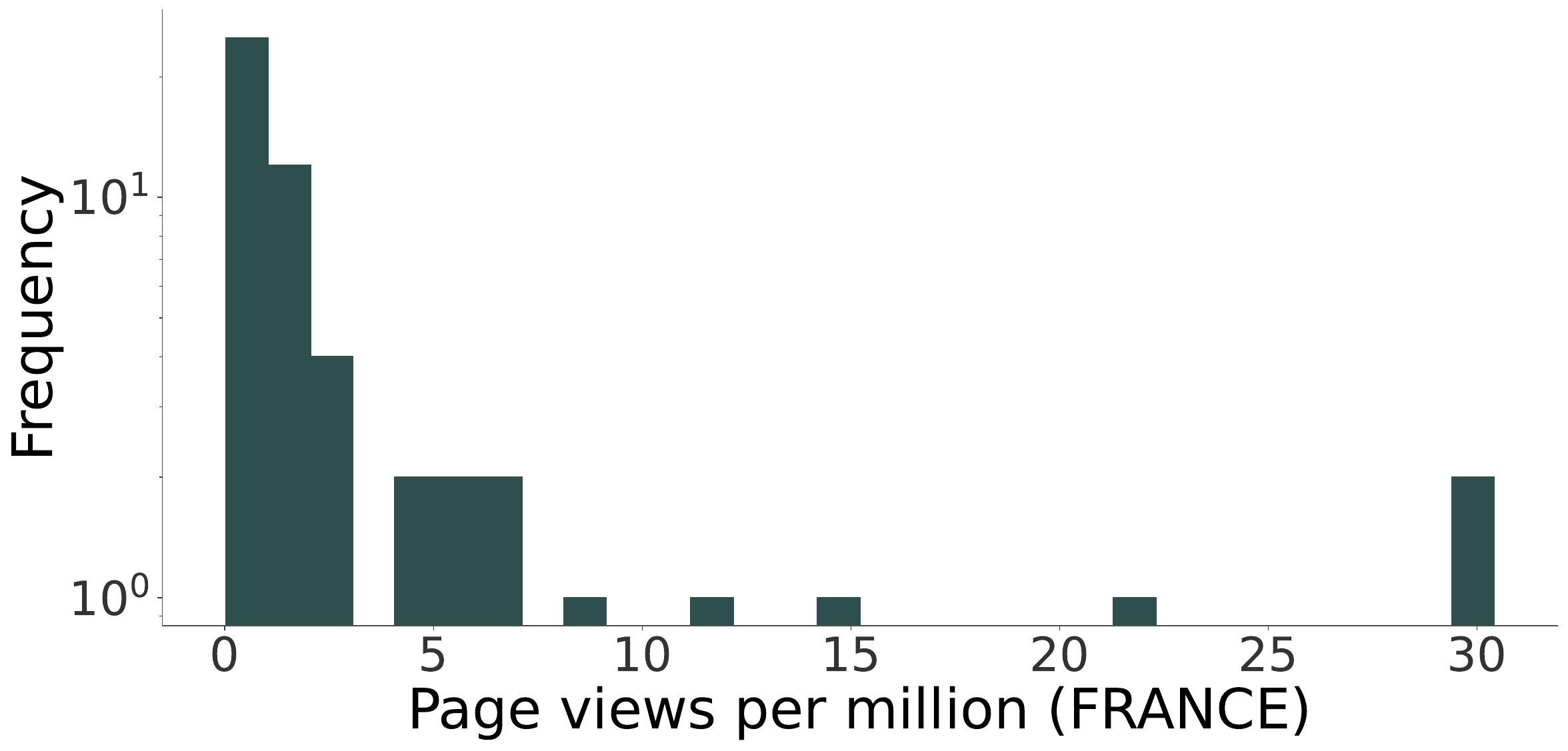}
\includegraphics[width=0.45\linewidth]{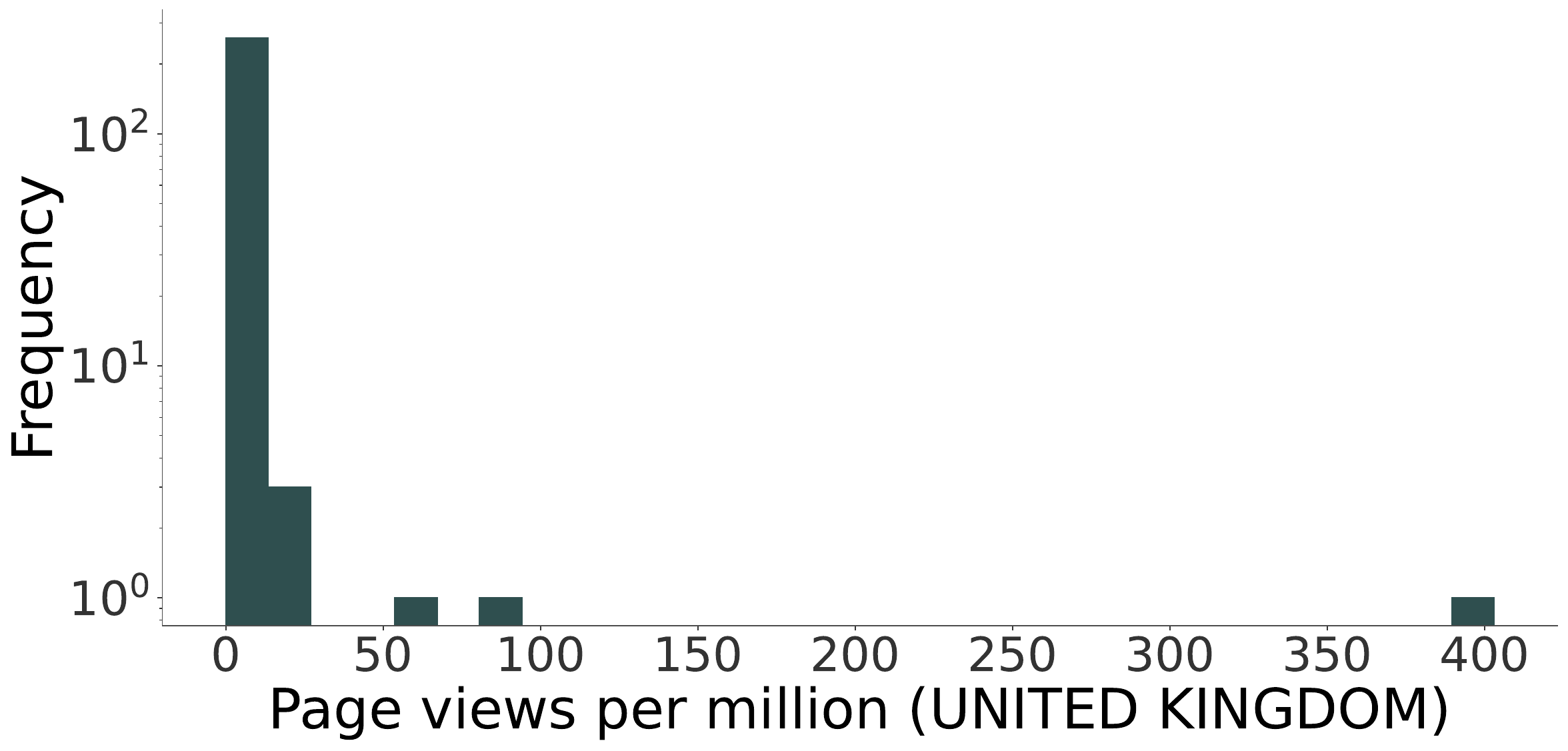}
\includegraphics[width=0.45\linewidth]{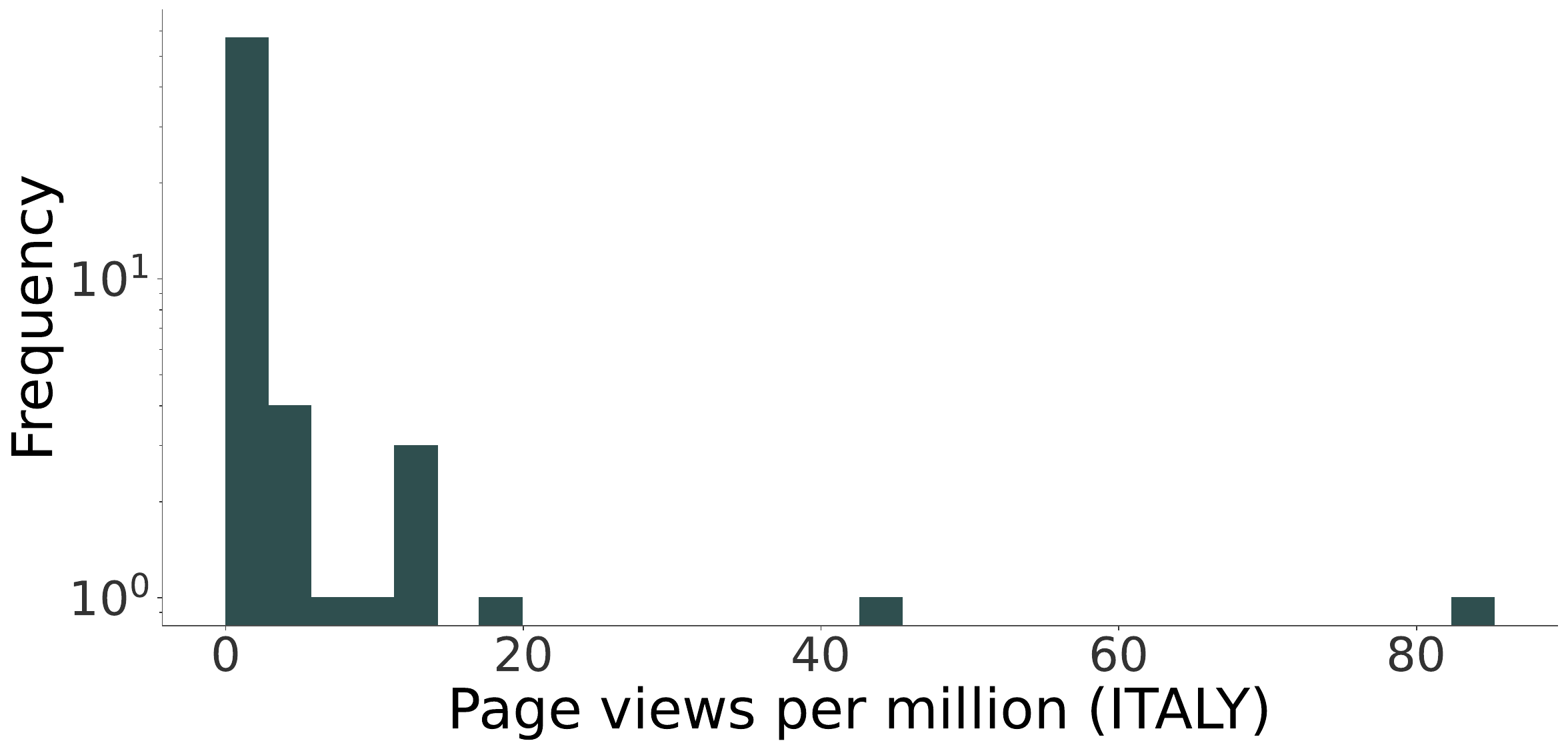}
\includegraphics[width=0.45\linewidth]{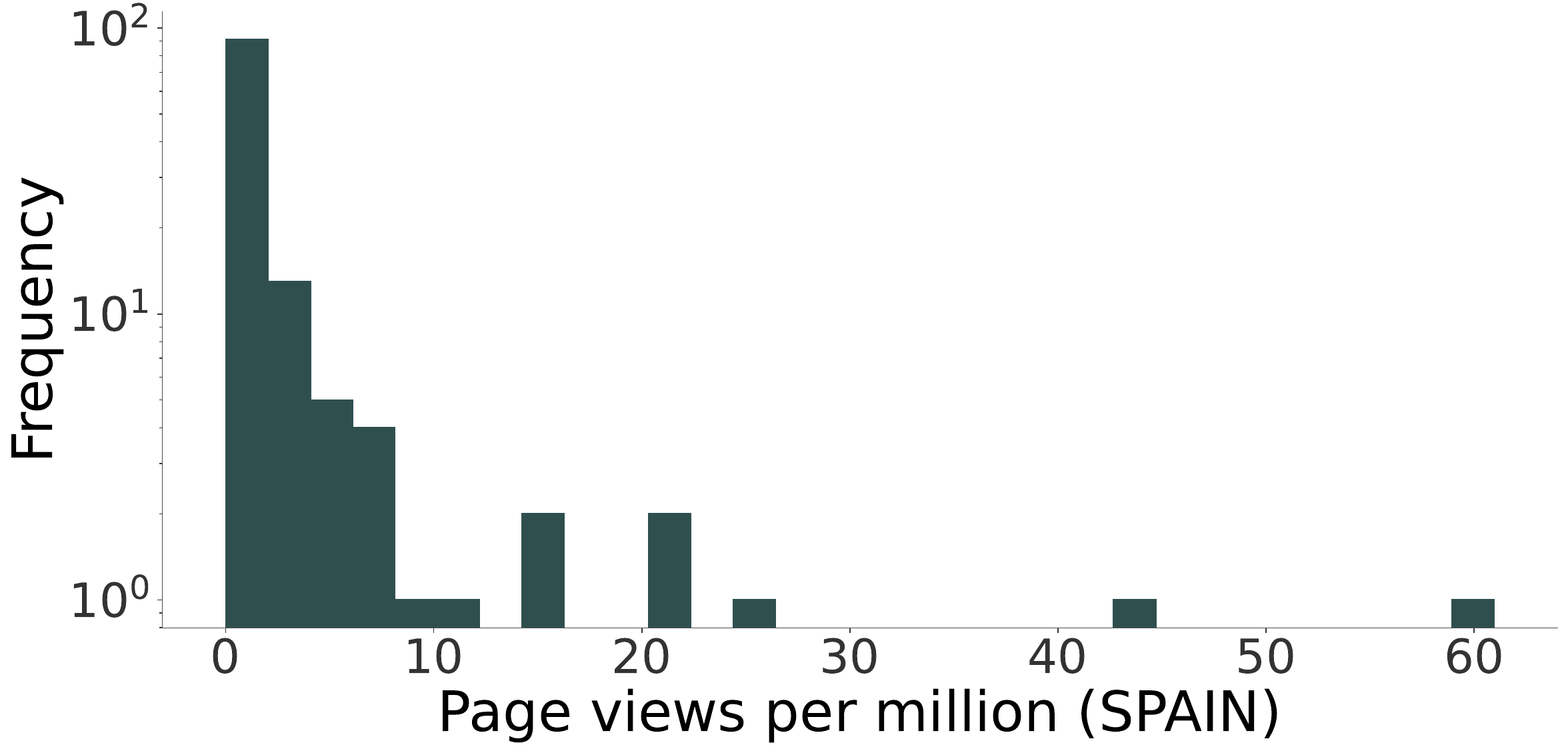}
\includegraphics[width=0.45\linewidth]{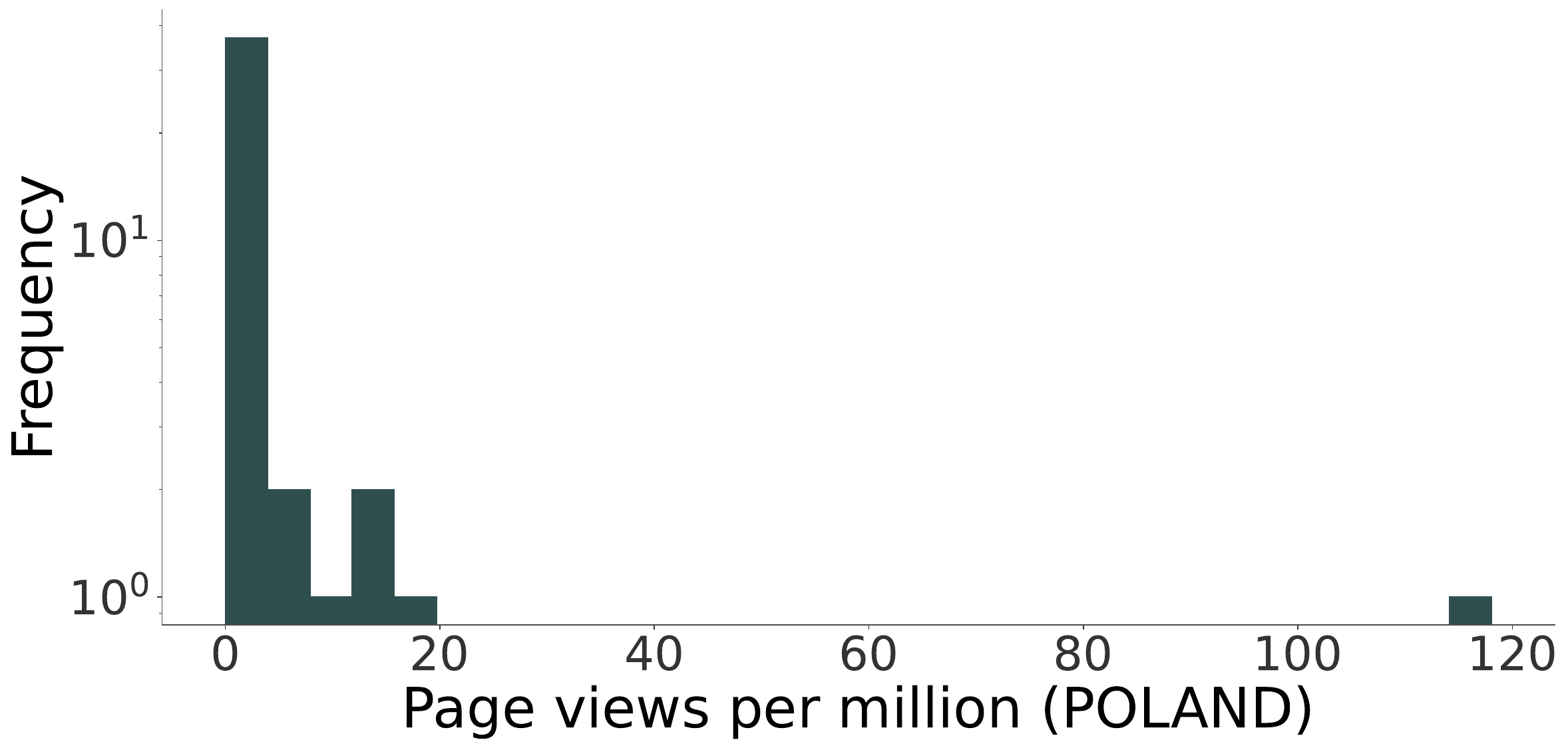}
\includegraphics[width=0.45\linewidth]{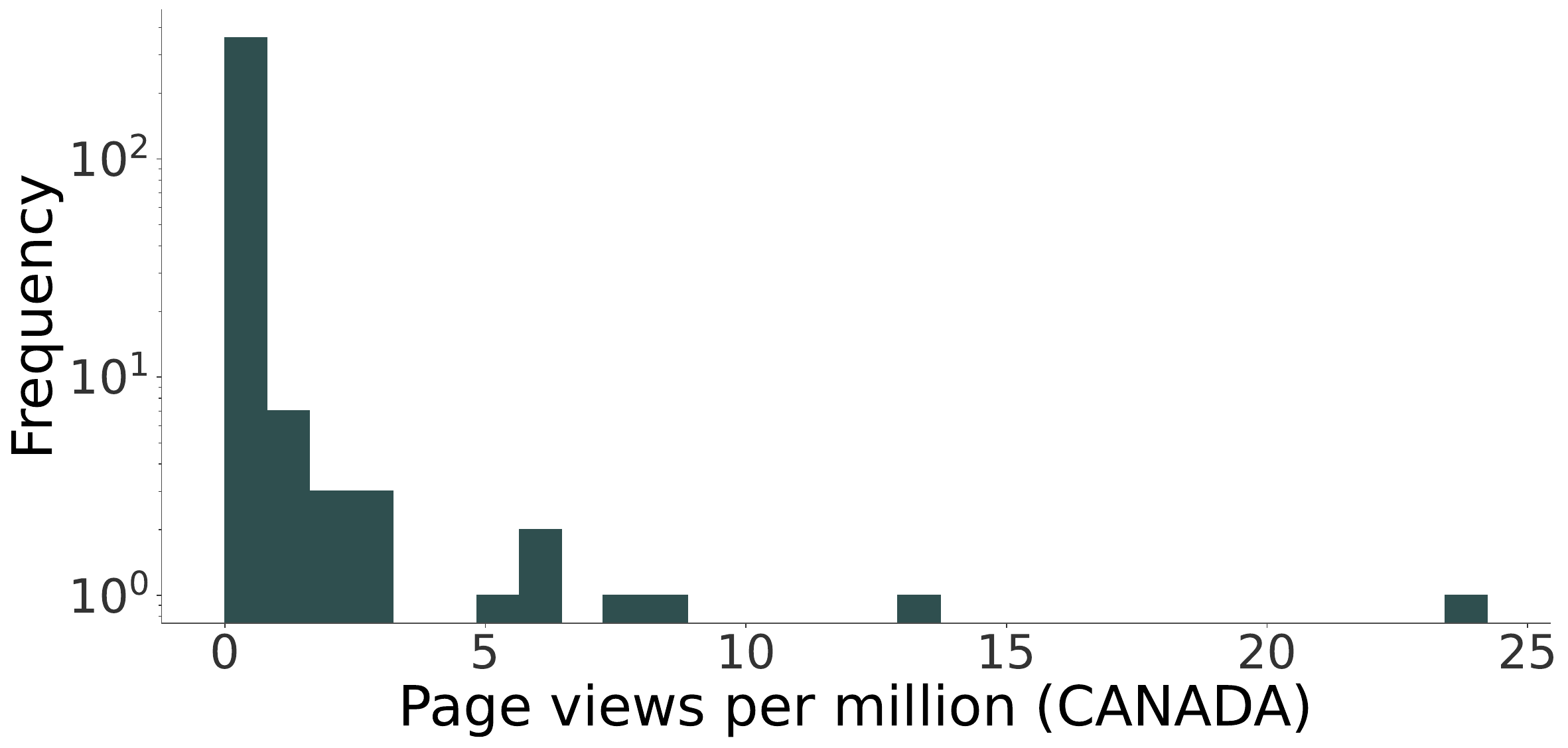}
\includegraphics[width=0.45\linewidth]{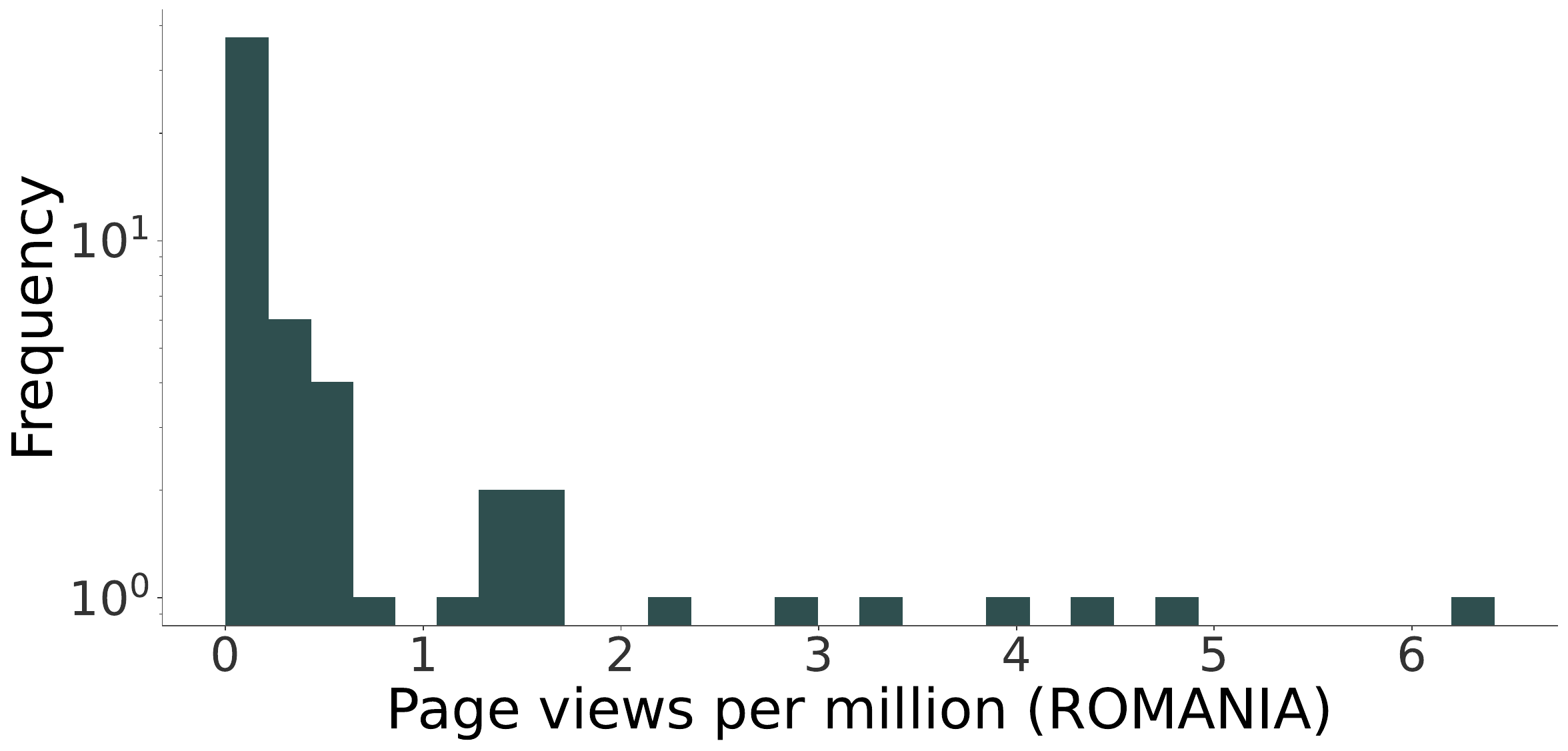}
\includegraphics[width=0.45\linewidth]{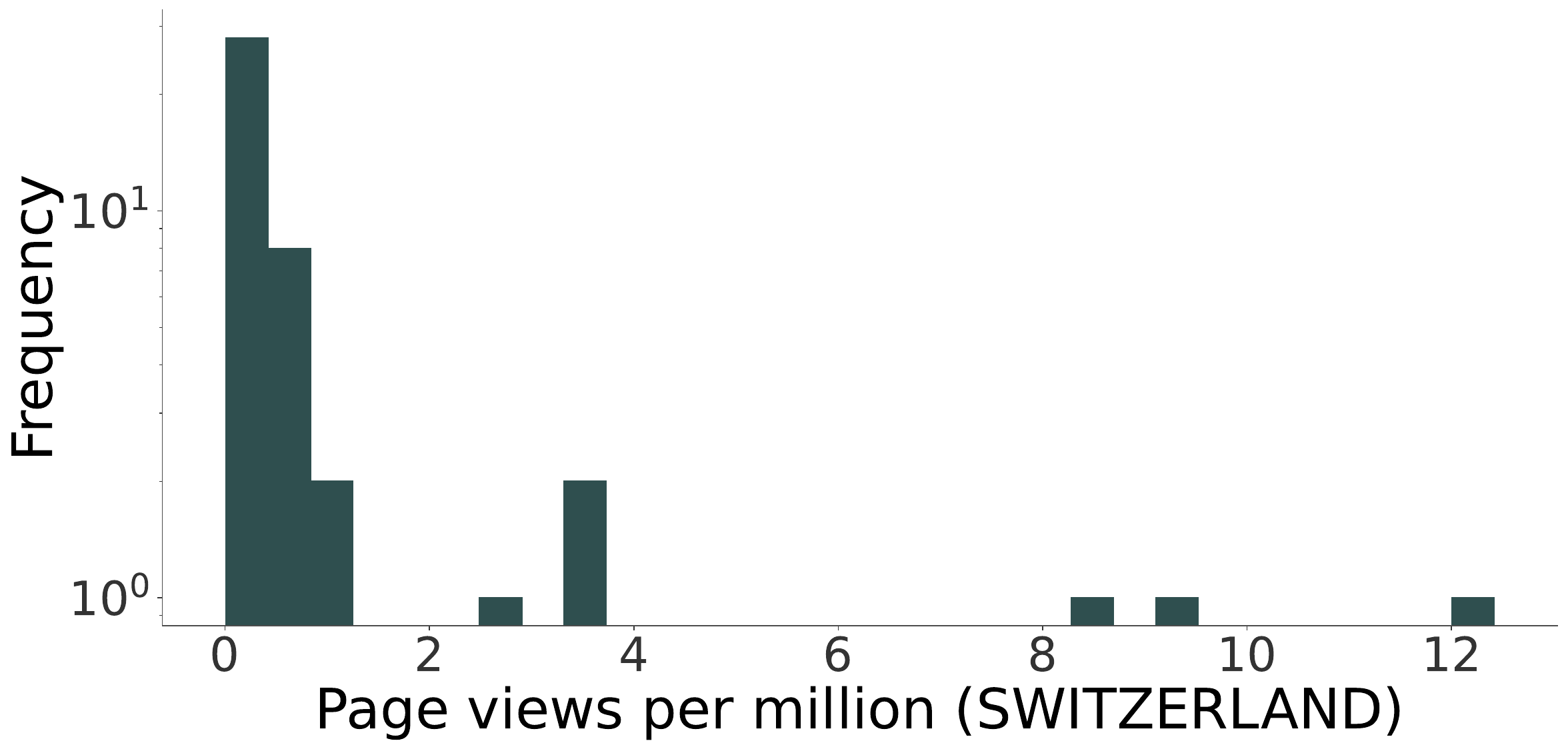}
\begin{figurenotes}
The figure shows the distribution of page views by country, for all hard news outlets (more than 25\% political reporting). See Section~\ref{sec:app:data} for discussion of summary statistics.
\end{figurenotes}
\end{figure}

\paragraph{Summary statistics } Table \ref{tab:sumstat} shows summary statistics for the number of domains, their traffic (in terms of page views and unique visitors), the share of traffic from domestic users, the stories published, and the share of political content. Traffic figures (page views, reach, and the domestic shares) are averages based on weekly measurements from 2018 through 2020; domain counts describe the sample, ultimate-owner counts are based on the July 2020 ownership data, and story figures are yearly averages from GDELT.
The number of domains ranges from 78 in Poland to 8,803 in the United States. \textit{Avg. page views} shows the number of page views for each country according to AWIS (normalized by one million page views across the internet worldwide, to account for seasonal fluctuations, changes in the number of internet users, etc.). Average page views vary between 0.23 (Canada) and 4.79 (Poland). For each of the 10 countries, page views are skewed; most domains get a low number of page views, and only a few get relatively many views (see Figures \ref{fig:app:reach_hist} and \ref{fig:app:pv_hist}). The average unique visitors per domain (again normalized by one million) ranges from around 4.3 in Canada to 59.3 in France (see \textit{Avg. reach}). In the next two columns, we present the share of page views from domestic users and the share of domestic unique visitors (out of all unique visitors). Both domestic page views and domestic reach are at three-quarters or higher for all countries (see \textit{Domestic page views} and \textit{Domestic reach}). Domestic shares are highest in Germany and Poland (approximately 90\% for both page views and reach). \textit{Ultimate owners} lists the average number of ultimate owners for each outlet. The minimum is 4 in Canada, and the maximum is 104 in Romania. \textit{No. stories} shows the average yearly number of stories published, ranging from 5,263 in Canada to 23,386 in Spain. In the last column (see \textit{Political stories}), we calculate the share of stories that cover political content. The shares are surprisingly stable across countries, ranging from roughly 40\% to one-half. In the main part, we focus on outlets that specialize in political reporting (defined as outlets with $>$ 25\% coverage of political content).

\begin{table}[htbp]
\centering
\caption{Summary Statistics}
\label{tab:sumstat}
\resizebox{\textwidth}{!}{
\begin{tabular}{lrrrrrrrrr}
\hline
\toprule
\textit{Country} & \textit{No.} & \textit{Avg. page} & \textit{Avg. reach} & \textit{Domestic} & \textit{Domestic} & \textit{Ultimate} & \textit{Ultimate} & \textit{No.} & \textit{Political} \\
& & \textit{views} & & \textit{page views} & \textit{reach} & \textit{owners} & \textit{owners} & \textit{stories} & \textit{stories} \\
& &  & & & & \textit{2020} & \textit{2023} & & \\
\midrule
U.S. & 8803 & 0.71 & 12.43 & 0.89 & 0.88 & 48 & 71 & 6236 & 0.42 \\
CANADA & 1283 & 0.23 & 4.27 & 0.81 & 0.79 & 4 & 61 & 5263 & 0.45 \\
UNITED KINGDOM & 598 & 2.67 & 38.72 & 0.75 & 0.74 & 56 & 97 & 9482 & 0.38 \\
SPAIN & 350 & 2.29 & 37.98 & 0.83 & 0.81 & 48 & 55 & 23386 & 0.48 \\
GERMANY & 341 & 1.83 & 28.38 & 0.90 & 0.89 & 29 & 27 & 21345 & 0.38 \\
ITALY & 144 & 2.99 & 45.15 & 0.89 & 0.89 & 81 & 76 & 12451 & 0.41 \\
ROMANIA & 142 & 0.61 & 10.00 & 0.83 & 0.83 & 104 & 3 & 11305 & 0.48 \\
FRANCE & 114 & 3.04 & 59.32 & 0.80 & 0.80 & 20 & 25 & 19371 & 0.39 \\
SWITZERLAND & 114 & 0.70 & 11.19 & 0.78 & 0.74 & 12 & 14 & 12392 & 0.42 \\
POLAND & 78 & 4.79 & 56.73 & 0.90 & 0.91 & 32 & 25 & 9735 & 0.46 \\
\bottomrule
\hline\hline
\end{tabular}
}
\justify \scriptsize {\emph{Notes:} The summary statistics are shown: how many domains are assigned to a given country (\textit{No.}), the average page views per domain per one million page views on the internet (\textit{Avg. page views}), the average reach (i.e., unique users) per one million unique users (\textit{Avg. reach}), the share of domestic page views (\textit{Domestic page views}), the share of domestic reach (\textit{Domestic reach}), the average number of ultimate owners as calculated with data from 2020 (\textit{Ultimate owners 2020}) and (\textit{Ultimate owners 2023}), the average yearly total of stories published by the outlet (\textit{No. stories}), and the share of these stories that cover political content (\textit{Political stories}). The traffic- and content-related figures reflect averages for the time period 2018 through 2020. The ownership-related figures are based on cross-sections in 2020 and 2023, respectively.}
\end{table}

\begin{table}[htbp]
\centering
\begin{threeparttable}
\caption{Descriptives: Ownership and Content, Sample Restrictions} \label{tab:coverage}
\footnotesize
\singlespacing

% Table created by stargazer v.5.2.3 by Marek Hlavac, Social Policy Institute. E-mail: marek.hlavac at gmail.com
% Date and time: Mon, Jun 02, 2025 - 02:34:56 PM
\begin{tabular}{@{\extracolsep{5pt}} lccccccc} 
\\[-1.8ex]\hline 
\hline \\[-1.8ex] 
  & Period & Full sample & (1) & (2) & (3) & (4) & (5) \\ 
\hline \\[-1.8ex] 
Average of Tone difference & 1 & 1.189 & 0.214 & 0.183 & 0.350 & 0.343 & 1.198 \\ 
Average of Tone difference & 2 & 1.500 & 0.735 & 0.923 & 0.341 & 0.966 & 1.510 \\ 
 &  &  &  &  &  &  &  \\ 
SD of Tone difference & 1 & 1.559 & 0.740 & 0.756 & 0.647 & 0.802 & 1.562 \\ 
SD of Tone difference & 2 & 1.779 & 1.272 & 1.419 & 0.746 & 1.241 & 1.782 \\ 
 &  &  &  &  &  &  &  \\ 
Minimum of Tone difference & 1 & 0.000 & 0.000 & 0.000 & 0.000 & 0.000 & 0.000 \\ 
Minimum of Tone difference & 2 & 0.000 & 0.000 & 0.000 & 0.000 & 0.000 & 0.000 \\ 
 &  &  &  &  &  &  &  \\ 
Maximum of Tone difference & 1 & 22.559 & 12.466 & 12.466 & 5.764 & 11.614 & 22.559 \\ 
Maximum of Tone difference & 2 & 25.115 & 12.025 & 12.025 & 8.908 & 13.256 & 25.115 \\ 
\hline &  &  &  &  &  &  &  \\ 
Average of Dice-similarity & 1 & 0.040 & 0.411 & 0.458 & 0.221 & 0.214 & 0.036 \\ 
Average of Dice-similarity & 2 & 0.035 & 0.136 & 0.079 & 0.364 & 0.176 & 0.034 \\ 
 &  &  &  &  &  &  &  \\ 
SD of Dice-similarity & 1 & 0.088 & 0.342 & 0.347 & 0.244 & 0.274 & 0.075 \\ 
SD of Dice-similarity & 2 & 0.088 & 0.221 & 0.166 & 0.263 & 0.255 & 0.083 \\ 
 &  &  &  &  &  &  &  \\ 
Minimum of Dice-similarity & 1 & 0.000 & 0.000 & 0.000 & 0.000 & 0.000 & 0.000 \\ 
Minimum of Dice-similarity & 2 & 0.000 & 0.000 & 0.000 & 0.000 & 0.000 & 0.000 \\ 
 &  &  &  &  &  &  &  \\ 
Maximum of Dice-similarity & 1 & 0.952 & 0.943 & 0.943 & 0.868 & 0.952 & 0.937 \\ 
Maximum of Dice-similarity & 2 & 1.000 & 1.000 & 1.000 & 0.937 & 1.000 & 1.000 \\ 
\hline &  &  &  &  &  &  &  \\ 
Average of "Same domain owner" & 1 & 0.007 & 0.801 & 1.000 & 0.000 & 0.216 & 0.000 \\ 
Average of "Same domain owner" & 2 & 0.005 & 0.199 & 0.000 & 1.000 & 0.200 & 0.000 \\ 
 &  &  &  &  &  &  &  \\ 
SD of "Same domain owner" & 1 & 0.081 & 0.399 & 0.000 & 0.000 & 0.411 & 0.000 \\ 
SD of "Same domain owner" & 2 & 0.073 & 0.399 & 0.000 & 0.000 & 0.400 & 0.000 \\ 
 &  &  &  &  &  &  &  \\ 
Minimum of "Same domain owner" & 1 & 0.000 & 0.000 & 1.000 & 0.000 & 0.000 & 0.000 \\ 
Minimum of "Same domain owner" & 2 & 0.000 & 0.000 & 0.000 & 1.000 & 0.000 & 0.000 \\ 
 &  &  &  &  &  &  &  \\ 
Maximum of "Same domain owner" & 1 & 1.000 & 1.000 & 1.000 & 0.000 & 1.000 & 0.000 \\ 
Maximum of "Same domain owner" & 2 & 1.000 & 1.000 & 0.000 & 1.000 & 1.000 & 0.000 \\ 
\hline &  &  &  &  &  &  &  \\ 
Average of shared ult. ownership & 1 & 0.016 & 0.850 & 1.000 & 0.247 & 0.452 & 0.009 \\ 
Average of shared ult. ownership & 2 & 0.017 & 0.386 & 0.247 & 0.945 & 0.446 & 0.012 \\ 
 &  &  &  &  &  &  &  \\ 
SD of shared ult. ownership & 1 & 0.115 & 0.355 & 0.000 & 0.424 & 0.456 & 0.082 \\ 
SD of shared ult. ownership & 2 & 0.126 & 0.486 & 0.431 & 0.225 & 0.481 & 0.107 \\ 
 &  &  &  &  &  &  &  \\ 
Minimum of shared ult. ownership & 1 & 0.000 & 0.000 & 1.000 & 0.000 & 0.000 & 0.000 \\ 
Minimum of shared ult. ownership & 2 & 0.000 & 0.000 & 0.000 & 0.000 & 0.000 & 0.000 \\ 
 &  &  &  &  &  &  &  \\ 
Maximum of shared ult. ownership & 1 & 1.000 & 1.000 & 1.000 & 1.000 & 1.000 & 1.000 \\ 
Maximum of shared ult. ownership & 2 & 1.000 & 1.000 & 1.000 & 1.000 & 1.000 & 1.000 \\ 
\hline &  &  &  &  &  &  &  \\ 
Number of obs. per period (tone) &  & 857,906 & 2,092 & 1,707 & 385 & 22,274 & 849,968 \\ 
Number of obs. per period (coverage) &  & 2,037,171 & 4,287 & 3,434 & 853 & 40,508 & 2,022,830 \\ 
\hline \\[-1.8ex] 
\end{tabular} 

\begin{tablenotes}[flushleft]
\footnotesize{
\item \textit{Notes}: This table presents descriptive statistics for the focal measures across the full sample and several restricted sub-samples. Each column corresponds to a different subset of observations: column (1) comprises switcher pairs, i.e., pairs whose domain-ownership status changes between the two periods; column (2) the switchers moving from same to different domain owner; column (3) the switchers moving from different to same domain owner; column (4) pairs with a positive share of ultimate ownership held in common (restricted to pairs observed in both periods); and column (5) pairs that are never owned by the same domain owner (restricted to pairs observed in both periods). For each subset, statistics are reported separately by period. ``Tone difference'' is the difference in tone between the focal outlet and its matched outlet. ``Dice-similarity'' measures the overlap in coverage between the two outlets. ``Same domain owner'' is an indicator for whether both outlets share the same domain-level owner. ``Shared ult. ownership'' measures the share of ultimate ownership that is held in common. Each panel shows the mean, standard deviation, minimum, and maximum of the variables for periods 1 and 2. The final rows report the number of observations per period. All variables are constructed using the data described in the main text.
}
\end{tablenotes}
\end{threeparttable}
\end{table}

\clearpage
\subsection*{Excluded Domains}

From the list of outlets provided by ABYZ, we drop the domains listed below. Note that the traffic data is available at the canonical domain level (e.g., google.com instead of news.google.com). Thus, we exclude the following domains because the majority of their traffic does not originate from their function as news providers. That is, the traffic figures for news.google.com or the blog tech.mit.edu also include all page views generated by Google as a search engine or by MIT as a university, respectively. Additionally, most of the domains we drop do not produce their own news content but instead aggregate it (e.g., Yahoo.com). Hence, some of their traffic is indirectly captured through the page views of the news outlets to which the aggregators link.

\begin{itemize}
\item All domains ending in reuters.com, such as de.reuters.com or es.reuters.com
\item All domains ending in yahoo.com, such as ca.news.yahoo.com or fr.news.yahoo.com
\item Blogs on university websites, identified as ending in .edu, such as tech.mit.edu
\item All domains ending in msn.com, such as uk.msn.com
\item All domains ending in google.com, such as news.google.com or sites.google.com
\item facebook.com
\item issuu.com
\item nasa.gov
\item expatica.com
\end{itemize}

\clearpage

\section{Tagging Political Content} \label{sec:app:political_tags}

This appendix section describes how we identify ``hard news'' outlets. It lists the GDELT theme tags used to classify articles as political, and provides examples of articles tagged as political and non-political as well as examples of outlets specializing in political coverage.
To identify content that covers the political process, we rely on the collection of news outlet links and their automated tagging by the Global Database on Events, Language, and Tone (\citeauthor{GDELT}). From all the available tags by GDELT, we select all that contain any of the following key words (the available tags are listed \href{http://data.gdeltproject.org/api/v2/guides/LOOKUP-GKGTHEMES.TXT}{here}).

\begin{itemize}
\item POLITIC
\item GOVERNMENT
\item POLICY
\item ELECTION\footnote{To avoid false positives, we delete all occurrences of ``SELECTION'' before filtering for ``ELECTION'', all occurrences of ``USPEC\_POLITICS\_GENERAL1'' before filtering for ``POLITIC'', and all occurrences of ``WB\_678\_DIGITAL\_GOVERNMENT'' before filtering for ``GOVERNMENT''.}
\end{itemize}

An article counts as political if at least one of its GDELT theme tags contains any of these keywords. A news site's political-content ratio is the share of its articles tagged as political, and we classify news sites with a ratio above 25\% as ``hard news'' outlets.

\subsection{Examples of Political/Non-political Articles}

The following table shows examples of news stories classified as political or non-political (based on the chosen tags).

\begin{center}
\begin{longtable}{p{.48\textwidth}p{.48\textwidth}}
\toprule

%%%%% nytimes

        \multicolumn{2}{c}{\textbf{\Large{nytimes.com}}}
    \\
          \emph{Political Articles} &  \emph{Non-Political Articles} \\
          \midrule
          \footnotesize{
          \href{https://www.nytimes.com/live/2021/01/12/us/impeachment-trump-25th-amendment}{\textbf{House Votes 223-205 to Call on Pence to Strip Trump of Power}}

Lawmakers adopted a resolution that would compel Vice President Mike Pence to invoke the 25th Amendment after President Trump incited a mob attack on the Capitol last week. In a letter to Speaker Nancy Pelosi earlier in the evening, Mr. Pence rejected the effort.\medskip

\noindent\href{https://www.nytimes.com/2020/10/13/world/pelosi-defends-demand-for-broad-stimulus-package-as-senate-prepares-scaled-down-bill.html}{\textbf{Pelosi defends demand for broad stimulus package as Senate prepares scaled-down bill}}

Speaker Nancy Pelosi of California defended her unwillingness to accept anything less than a broad coronavirus stimulus package in negotiations with the administration, on the same day that President Trump called on Twitter for negotiators to ``go big or go home!!!'' }

&
\footnotesize{
\href{https://www.nytimes.com/2020/12/06/science/space/jupiter-saturn-align-christmas-star.html}{\textbf{Jupiter and Saturn Head for Closest Visible Alignment in 800 Years}}

On Dec. 21, Jupiter and Saturn will appear to be no more than a dime's width apart in the night sky. The last time that could be seen was in 1226.\medskip

\noindent\href{https://www.nytimes.com/2020/12/05/us/winter-storm-noreaster.html}{\textbf{More Than 90,000 Without Power After New England Winter Storm}}

Around 85,000 customers in Maine alone still don't have electricity after a nor'easter brought high winds and heavy snow.} \\

\midrule

%%% CNN

\multicolumn{2}{c}{\textbf{\Large{cnn.com}}}
    \\
\emph{Political Articles} &  \emph{Non-Political Articles} \\
\midrule
\footnotesize{

\href{https://edition.cnn.com/2020/11/18/asia/thailand-protest-constitution-vote-intl-hnk/}{\textbf{Thousands protest in Bangkok after Thai parliament votes on constitutional reform}}

Thailand's parliament on Wednesday voted to move forward with two proposals on amending the constitution but stopped short of backing a motion that included monarchical reform, amid intensifying protests against the country's military-backed government.\medskip

\noindent\href{https://edition.cnn.com/2020/12/06/politics/geoff-duncan-trump-georgia-rally-cnntv/}{\textbf{Trump's ``mountains of misinformation'' at rally not helping Republican Senate chances, Georgia's GOP lieutenant governor says}}

Georgia Republican Lt. Gov. Geoff Duncan said Sunday that the ``mountains of misinformation'' about the election that President Donald Trump is spreading, most recently in his remarks at a rally in the state Saturday night, could hurt GOP chances in upcoming Senate runoff races.

}

&

\footnotesize{

\href{https://edition.cnn.com/2020/11/29/us/endangered-turtles-brought-to-florida-to-warm-up-trnd/}{\textbf{40 endangered sea turtles were brought to Florida to warm up after suffering from 'cold stunning'}}

Forty critically endangered sea turtles were brought to Florida after suffering from ``cold stunning'' off the coast of Massachusetts.\medskip

\noindent\href{https://edition.cnn.com/2020/12/05/us/fantasy-football-coronavirus-challenges-trnd/}{\textbf{Fantasy football is a billion-dollar pastime. Covid-19 is wreaking havoc with it}}

This NFL season has been a chaotic mix of coronavirus precautions, postponed games and infected players, all unfolding under a looming concern for the health of players, staff and spectators.

} \\

\midrule
%%% foxnews

\multicolumn{2}{c}{\textbf{\Large{foxnews.com}}}
    \\
\emph{Political Articles} &  \emph{Non-Political Articles} \\
\midrule
\footnotesize{

\href{https://www.foxnews.com/opinion/trump-voting-election-investigations-ronna-mcdaniel}{\textbf{RNC Chair McDaniel: Democrats wrong to attack Trump supporters who question election irregularities}}

States and the courts have a duty to ensure the election was conducted properly and that similar irregularities don't arise in future elections\medskip

\noindent\href{https://www.foxnews.com/politics/trump-surprised-congressional-republicans-biden-won-election-12-6-2020}{\textbf{Trump ``surprised'' some congressional Republicans think Biden won election}}

Last week, Senate Majority Leader Mitch McConnell, R-Ky., dodged questions about Trump's claims of fraud\medskip

}

&

\footnotesize{

\href{https://www.foxnews.com/sports/baker-mayfield-first-browns-qb-1951}{\textbf{Baker Mayfield becomes first Browns QB to do this since 1951}}

Mayfield followed in the footsteps of Hall of Fame signal-caller Otto Graham\medskip

\noindent\href{https://www.foxnews.com/us/california-man-punches-350-pound-bear-save-dog}{\textbf{California man punches 350-pound bear in face to save beloved dog ``Buddy''}}

The bear has returned to Kaleb Benham's home several times since the attack

} \\

\midrule
%%% Sun

\multicolumn{2}{c}{\textbf{\Large{thesun.co.uk}}}
    \\
\emph{Political Articles} &  \emph{Non-Political Articles} \\
\midrule
\footnotesize{

\href{https://www.thesun.co.uk/news/13395397/boris-johnson-urged-stand-firm-france-brexit/}{\textbf{Boris Johnson urged to hold nerve as France hijack Brexit talks at the eleventh hour}}

BORIS Johnson was urged to stand firm last night after Brexit talks stretched to the eleventh hour.\medskip

\noindent\href{https://www.thesun.co.uk/news/13396593/tory-mps-demand-keir-starmer-supports-deportation-flights/}{\textbf{Tory MPs demand Labour leader Sir Keir Starmer backs deportation flights for convicted foreign offenders}}

TORY MPs have written to Sir Keir Starmer demanding he backs the removal of convicted foreign offenders from the U.K.

}

&

\footnotesize{

\href{https://www.thesun.co.uk/tech/13386153/how-to-protect-iphone-security/}{\textbf{Five iPhone security settings you should check NOW to protect yourself from snoopers}}

APPLE is known for its top security - but there are ways you can secure your devices further.\medskip

\noindent\href{https://www.thesun.co.uk/money/13392217/ps5-litch-argos-app-restocked/}{\textbf{PS5 ``glitch'' spotted on Argos app lets shoppers buy consoles as soon as they are restocked}}

A SHOPPER has spotted a ``glitch'' on the Argos app that lets customers add a PlayStation 5 to their basket ready for when they're back in stock.

} \\

\midrule
%%% Guardian

\multicolumn{2}{c}{\textbf{\Large{theguardian.com}}}
    \\
\emph{Political Articles} &  \emph{Non-Political Articles} \\
\midrule
\footnotesize{

\href{https://www.theguardian.com/commentisfree/2020/dec/06/as-we-close-in-on-a-brexit-that-pleases-next-to-nobody-how-did-we-end-up-here}{\textbf{As we close in on a Brexit that pleases next to nobody, how did we end up here?}}

Blame staunch Leavers or purist Remainers, but the real fault might lie in our wantonly destructive politics\medskip

\noindent\href{https://www.theguardian.com/world/2020/dec/06/spain-marks-42-years-since-return-of-democracy-as-retired-officers-dissent}{\textbf{Spain marks 42 years since return of democracy as retired officers dissent}}

PM lauds 1978 constitution after some former armed forces members rue demise of Franco dictatorship

}

&

\footnotesize{

\href{https://www.theguardian.com/society/2020/dec/06/its-all-about-the-cracking-noise-the-unlikely-cult-of-the-online-chiropractor}{\textbf{`It's all about the cracking noise': the unlikely cult of the online chiropractor}}

Back pain has become rife in lockdown. But that's not the only reason chiropractors are being hailed as YouTube influencers – there's a horror movie thrill, too\medskip

\noindent\href{https://www.theguardian.com/tv-and-radio/2020/dec/06/netflix-has-no-plans-to-add-disclaimer-that-the-crown-is-work-of-fiction}{\textbf{Netflix has 'no plans' to add disclaimer that The Crown is work of fiction}}

Statement comes after the U.K. culture secretary said he was going to write to company to request caveat\medskip

} \\

\midrule
%%% CBC

\multicolumn{2}{c}{\textbf{\Large{cbc.ca}}}
    \\
\emph{Political Articles} &  \emph{Non-Political Articles} \\
\midrule
\footnotesize{

\href{https://www.cbc.ca/news/politics/us-diplomats-likely-targeted-in-energy-attack-1.5830651}{\textbf{Report says U.S. diplomats likely targeted in energy attack, unsure about Canadians}}

Stops short of saying the same about Canadian diplomats and families in Havana\medskip

\noindent\href{https://www.cbc.ca/news/world/venezuela-national-assembly-election-1.5830361}{\textbf{Venezuelans vote amid allegations of fraud against Nicolas Maduro}}

Opposition urges citizens to take part in referendum to push demands for new vote

}

&

\footnotesize{

\href{https://www.cbc.ca/news/canada/manitoba/simkin-centre-recovery-parades-covid-19-1.5830335}{\textbf{Winnipeg care home recovery parades bring hope — 'something that we all need right now,' says CEO}}

'We wanted to be the bright light during a very dark time,' says Laurie Cerqueti, Simkin Centre CEO\medskip

\noindent\href{https://www.cbc.ca/news/canada/nova-scotia/high-winds-power-outage-nova-scotia-1.5830366}{\textbf{High winds knock out power to homes, businesses}}

At one point overnight, about 2,200 customers were affected

} \\

\midrule
%%% Bild

\multicolumn{2}{c}{\textbf{\Large{bild.de}}}
    \\
\emph{Political Articles} &  \emph{Non-Political Articles} \\
\midrule
\footnotesize{

\href{https://www.bild.de/politik/ausland/politik-ausland/merkel-verbietet-firmenkauf-wollten-chinesen-an-bundeswehr-geheimnisse-74278398.bild.html}{\textbf{Wollten Chinesen Bundeswehr-Geheimnisse abfischen?}}

Bundeskanzlerin Angela Merkel und Bundeswirtschaftsminister Peter Altmaier (beide CDU) haben die Uebernahme eines deutschen Hightech-Unternehmens (u.a. Satelliten) durch eine chinesische Schachtel-Struktur verboten.\medskip

\noindent\href{https://www.bild.de/regional/hamburg/hamburg-aktuell/proteste-in-hamburg-2000-linke-gegen-g20-prozess-74329114.bild.html}{\textbf{2000 Demonstranten gegen G20-Prozess}}

Seit Donnerstag müssen sich in Hamburg fünf junge Leute im Alter von 19 bis 21 Jahren vor Gericht verantworten, weil sie sich 2017 an den Krawallen während des G20-Gipfels beteiligt haben sollen. Samstagnachmittag gingen in Hamburg zahlreiche Demonstranten auf die Straße, um gegen die Verhandlung zu protestieren.
}

&

\footnotesize{

\href{https://www.bild.de/geld/wirtschaft/wirtschaft/kein-corona-shopping-innenstaedte-machen-den-deutschen-angst-74327722.bild.html}{\textbf{Innenstädte machen den Deutschen Angst}}

Aus Angst vor Infektion gehen die Deutschen immer weniger in die Innenstädte.\medskip

\noindent\href{https://www.bild.de/regional/sachsen-anhalt/sachsen-anhalt-news/advent-in-corona-zeiten-erster-drive-in-weihnachtsmarkt-74337168.bild.html}{\textbf{Erster Drive-in-Weihnachtsmarkt}}

Sonst verkaufte Sandra Behrens (38) um diese Zeit auf dem Weihnachtsmarkt in Merseburg oder dem Christkind'l-Markt in Bad Lauchstädt Süßigkeiten oder Herzhaftes. So viel Zulauf wie an diesem Wochenende im Gewerbegebiet Stöbnitz, einem Ortsteil von Mücheln (Saalekreis), hatte die Unternehmerin dort nie.

} \\

\midrule
%%% Le Monde

\multicolumn{2}{c}{\textbf{\Large{lemonde.fr}}}
    \\
\emph{Political Articles} &  \emph{Non-Political Articles} \\
\midrule
\footnotesize{

\href{https://www.lemonde.fr/idees/article/2020/12/06/dans-les-pays-musulmans-la-trilogie-politique-religion-developpement-ne-peut-pas-fonctionner-harmonieusement_6062364_3232.html}{\textbf{Dans les pays musulmans, la trilogie politique-religion-développement ne peut pas fonctionner harmonieusement}}

L'économiste Taha Oudghiri analyse dans une tribune au Monde ce que sont les principes d'une gouvernance économique islamique, et ce qui pourrait en advenir\medskip

\noindent\href{https://www.lemonde.fr/international/article/2020/12/06/president-trump-an-iv-le-poison-du-desinteret_6062359_3210.html}{\textbf{Président Trump, an IV : le poison du désintérêt}}

Même s'il en a profité pour parler principalement de lui, et de nier à nouveau sa défaite à l'élection, le président américain est allé en Géorgie pour essayer de sauver la majorité républicaine au Sénat.
}

&

\footnotesize{

\href{https://www.lemonde.fr/m-styles/article/2020/12/06/trois-heures-et-20-km-cinq-balades-autour-de-bordeaux-lyon-nantes-strasbourg-et-paris_6062344_4497319.html}{\textbf{Trois heures et 20 km ? Cinq balades autour de Bordeaux, Lyon, Nantes, Strasbourg et Paris}}

La Matinale vous invite au voyage. Cette semaine, notre sélection d'itinéraires périurbains pour prendre l'air sur l'île Clémentine, dans la forêt de la Robertsau ou le long de l'Eau Bourde.\medskip

\noindent\href{https://www.lemonde.fr/idees/article/2020/12/05/roberto-saviano-maradona-etait-a-la-fois-le-meilleur-et-le-pire-de-tout-ce-que-ma-terre-a-genere_6062262_3232.html}{\textbf{Roberto Saviano : Maradona était à la fois le meilleur et le pire de tout ce que ma terre a généré}}

Les Napolitains se sont intensément identifiés au gosse misérable de Buenos Aires devenu un dieu du football. Ses vices et ses erreurs n'étaient que des ombres qui le rendaient plus lumineux encore, note, dans une tribune au Monde, l'écrivain et journaliste napolitain, menacé de mort par la Camorra.

} \\

\bottomrule
% \caption{Snippets of political and non-political news stories}
\label{tab:app:examples}
\end{longtable}
\end{center}

\clearpage

\subsection{Examples of Outlets Specializing in Political Articles}

The following table shows, the top five outlets that report on politics in \textit{more} than 25\% of their articles (to the left), and those that report on politics in less than 25\% of their articles (to the right).

{\centering
\begin{longtable}{p{.4\textwidth}p{.4\textwidth}}
\toprule
\emph{Specialized in Political Articles} &  \emph{Not Specialized in Political Articles} \\
\midrule
\multicolumn{2}{c}{\textbf{United States}} \\
cnn.com & buzzfeed.com \\
nytimes.com & bleacherreport.com \\
washingtonpost.com & cbssports.com \\
foxnews.com & people.com \\
forbes.com & tmz.com \\
\midrule
\multicolumn{2}{c}{\textbf{United Kingdom}} \\
bbc.co.uk & ft.com \\
theguardian.com & dailystar.co.uk \\
dailymail.co.uk & liverpoolecho.co.uk \\
telegraph.co.uk & espn.co.uk \\
sky.com & londonist.com \\
\midrule
\multicolumn{2}{c}{\textbf{Switzerland}} \\
20min.ch & bote.ch \\
blick.ch & shn.ch \\
srf.ch & jungfrauzeitung.ch \\
nzz.ch & rro.ch \\
tagesanzeiger.ch & news.ch \\
\midrule
\multicolumn{2}{c}{\textbf{Spain}} \\
elpais.com & marca.com \\
elmundo.es & as.com \\
abc.es & mundodeportivo.com \\
lavanguardia.com & hola.com \\
elconfidencial.com & cuatro.com \\
\midrule
\multicolumn{2}{c}{\textbf{Romania}} \\
adevarul.ro & gsp.ro \\
stirileprotv.ro & agerpres.ro \\
libertatea.ro & primatv.ro \\
hotnews.ro & (not more outlets in our sample) \\
realitatea.net & \\
\midrule
\multicolumn{2}{c}{\textbf{Poland}} \\
onet.pl & noizz.pl \\
tvn24.pl & (not more outlets in our sample) \\
wyborcza.pl & \\
fakt.pl & \\
naszemiasto.pl & \\
\midrule
\multicolumn{2}{c}{\textbf{Italy}} \\
repubblica.it & gazzetta.it \\
corriere.it & lanazione.it \\
ansa.it & napolitoday.it \\
ilsole24ore.com & corriereadriatico.it \\
lastampa.it & padovaoggi.it \\
\midrule
\multicolumn{2}{c}{\textbf{Germany}} \\
spiegel.de & wdr.de \\
bild.de & express.de \\
focus.de & rtl.de \\
welt.de & derwesten.de \\
heise.de & bz-berlin.de \\
\midrule
\multicolumn{2}{c}{\textbf{France}} \\
lefigaro.fr & lequipe.fr \\
lemonde.fr & tf1.fr \\
20minutes.fr & ledauphine.com \\
leparisien.fr & lavoixdunord.fr \\
ouest-france.fr & dna.fr \\
\midrule
\multicolumn{2}{c}{\textbf{Canada}} \\
cbc.ca & tsn.ca \\
globalnews.ca & rds.ca \\
ctvnews.ca & thescore.com \\
theglobeandmail.com & hellomagazine.com \\
thestar.com & chch.com \\
\bottomrule
\label{tab:app:top5_outlets_reach_political}
\end{longtable}
\par}

\clearpage

\section{Additional Figures and Tables: Traffic and Ownership}
\label{sec:app:addl_fig_tab}

This appendix section collects additional figures and tables on traffic and ownership. The first subsection shows the distributions of our traffic measures and the correlation between reach and page views. The second subsection contains the exact values and complementary metrics behind the ownership-structure results, as well as country-level topology detail. The third subsection reports sector shares, the fourth market concentration figures (the scatter version of the main-text HHI results and robustness to alternative traffic measures, data vintages, samples, and concentration measures), and the fifth a summary table of all concentration measures.

\subsection{Traffic}

\begin{figure}
\centering
\includegraphics[width=0.75\textwidth]{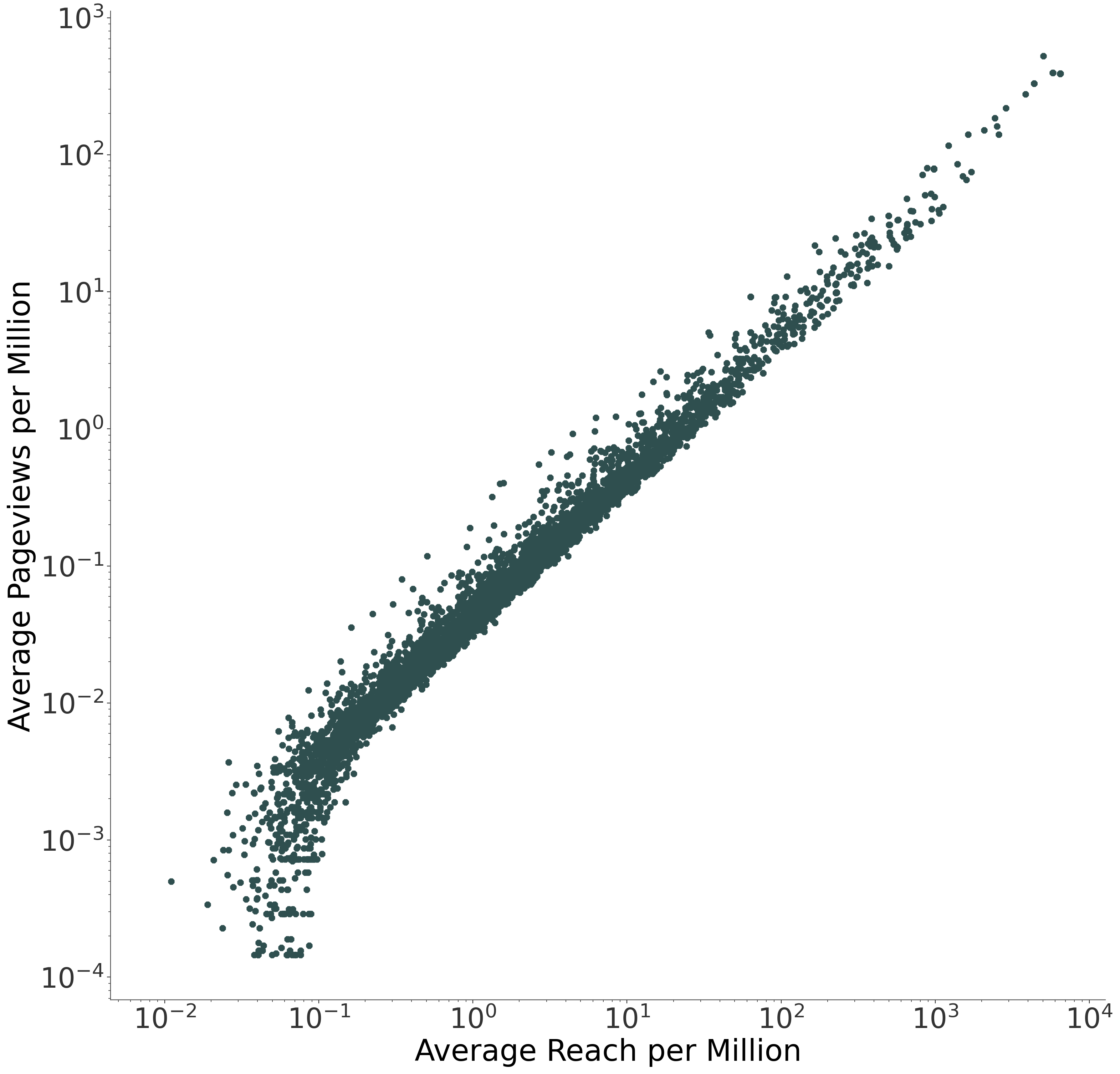}
\caption{Correlation Between Reach and Page Views}
\label{fig:corr_reach_views}
\begin{figurenotes}
This figure shows a scatterplot of news website reach (no. of unique visitors per million website visitors) on the x-axis vs. page views (per one million page views on all tracked websites) on the y-axis. Each dot represents one news website in our database. The figure shows that the extensive margin (reach) and the intensive margin (page views) correlate almost perfectly.no. of visits per one million website visits in the entire world wide web on the y-axis. Each dot represents one news website in our database. The figure shows that the extensive margin (reach) vs. intensive margin (no. visits) correlate almost perfectly.
\end{figurenotes}
\end{figure}

\clearpage

\subsection{Ownership Network Structure} \label{app:sec:topology_detail}

\begin{figure}[htbp]
\centering
\caption{Ownership Networks: Nodes and Ultimate Owners}
\label{fig:node_counts_all}
\includegraphics[width=0.79\textwidth]{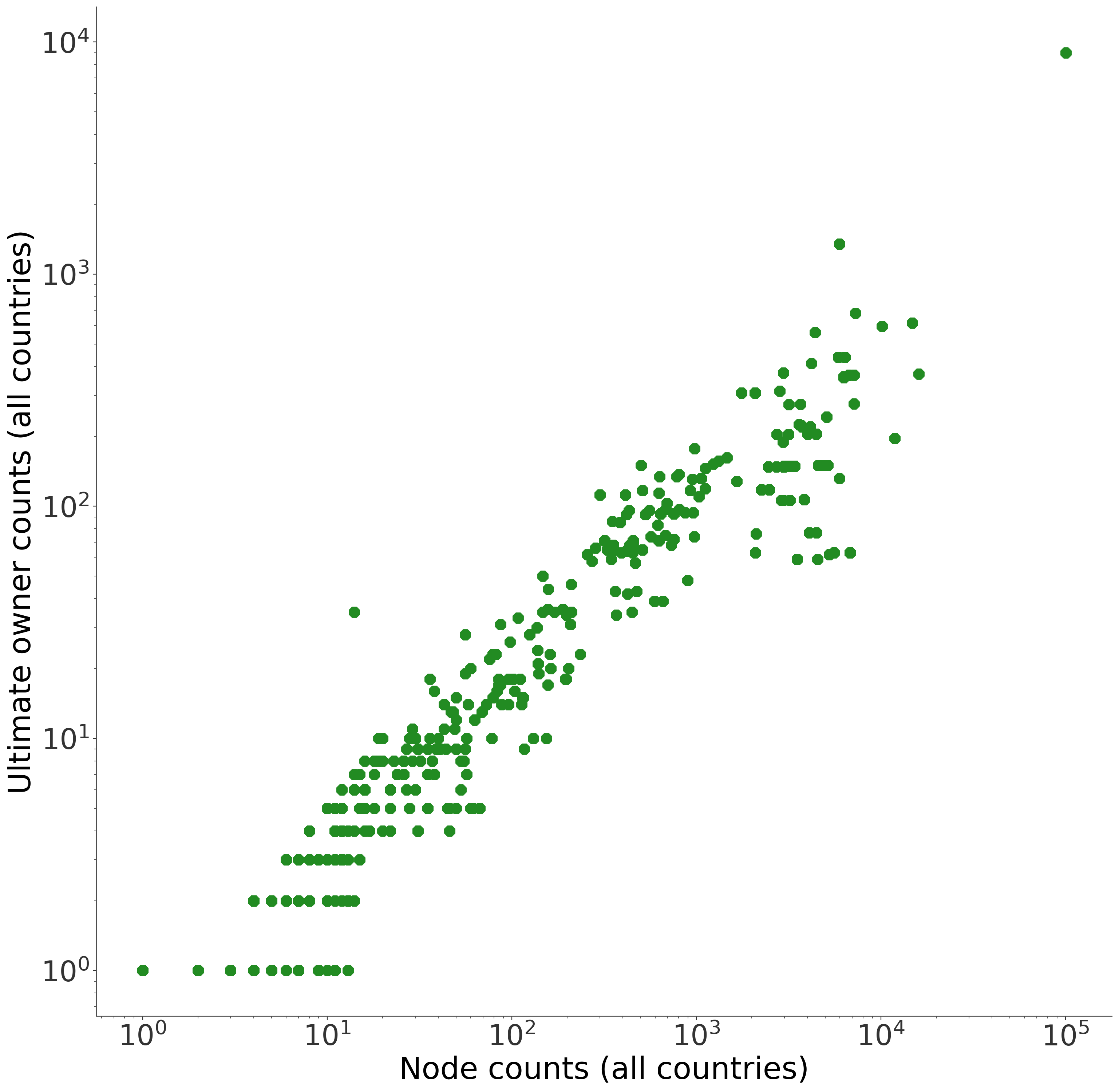}
\begin{figurenotes}
The figure plots the node count (the total number of owning entities in an outlet's ownership network, which is traversed link by link over stakes $>1\%$ in the entity directly below, so that an owner's multiplied-through stake in the outlet itself can be smaller; horizontal axis) against the ultimate owner count (the number of final nodes, or leaves, in the network; vertical axis). The two counts are strongly correlated (91\%). Figure~\ref{fig:app:node_level_distributions} shows the underlying distributions of node and level counts.
\end{figurenotes}
\end{figure}

This subsection collects the tables and figures behind the ownership-structure results of Section~\ref{sec:results}. Table~\ref{tab:anatomy_by_country} reports the exact values shown in main-text Figure~\ref{fig:ownership_anatomy}. Table~\ref{tab:ownership_transparency} reports the pooled ownership concentration metrics (Panel A) and network complexity metrics (Panel B), including traffic-weighted versions and node counts restricted to majority (50\%) and controlling (20\%) owners. Figure~\ref{fig:node_counts_all} plots node counts against ultimate-owner counts; the strong positive association underlies the conclusion that complex networks reflect dispersed ownership rather than concealment of a single controller. The remaining material provides country-level and traffic-weighted detail on network topology, the comparison with cross-sector ownership concentration, and country-specific node and level distributions.

\begin{table}[htbp]\centering
\caption{Ownership Structures by Country: Exact Shares} \label{tab:anatomy_by_country}
\begin{threeparttable}
\small
\begin{tabular}{lccccc}
\toprule
Country & Outlets & Single owner, & Single owner, & Multiple owners, & Hard-to-track \\
 & & held directly & via chain & tractable ($<$5 nodes) & ($\geq$5 nodes) \\
\midrule
U.K. & 302 & 51.0\% & 6.3\% & 3.6\% & 39.1\% \\
U.S. & 1,825 & 44.6\% & 19.8\% & 1.1\% & 34.5\% \\
Canada & 358 & 36.6\% & 21.8\% & 0.8\% & 40.8\% \\
France & 51 & 31.4\% & 9.8\% & 3.9\% & 54.9\% \\
Switzerland & 44 & 20.5\% & 18.2\% & 2.3\% & 59.1\% \\
Romania & 49 & 12.2\% & 24.5\% & 18.4\% & 44.9\% \\
Spain & 123 & 9.8\% & 17.9\% & 3.3\% & 69.1\% \\
Italy & 65 & 9.2\% & 12.3\% & 9.2\% & 69.2\% \\
Germany & 136 & 8.1\% & 6.6\% & 6.6\% & 78.7\% \\
Poland & 46 & 2.2\% & 17.4\% & 6.5\% & 73.9\% \\
\midrule
All countries & 2,991 & 38.8\% & 17.8\% & 2.2\% & 41.3\% \\
\bottomrule
\end{tabular}
\begin{tablenotes}[flushleft]
\footnotesize{
\item \textit{Notes}: Exact values underlying Figure~\ref{fig:ownership_anatomy} in the main text. The sample comprises hard news outlets with both node-count and ultimate-owner data. ``Single owner, held directly'' refers to outlets whose ownership network consists of a single node (the registering entity is the ultimate owner); ``via chain'' refers to outlets with one ultimate owner reached through a multi-node network; the remaining outlets have multiple ultimate owners and are split by whether their network has fewer than five nodes. Shares may not sum to 100\% due to rounding.
}
\end{tablenotes}
\end{threeparttable}
\end{table}

\begin{table}[htbp]
\centering
\caption{Ownership Concentration and Network Tractability}
\label{tab:ownership_transparency}
\begin{threeparttable}
\begin{tabular}{lcc}
\toprule
 & Share of & Share of \\
 & Outlets & Traffic \\
\midrule
\multicolumn{3}{l}{\textbf{Panel A: Ultimate Owners}} \\[0.5ex]
\quad With 1 ultimate owner & 57\% & 45\% \\
\quad \quad Direct ownership & 39\% & 32\% \\
\quad \quad Through network & 18\% & 12\% \\
\quad With 1 majority owner ($\geq$50\%) & 71\% & 78\% \\
\quad With 1 controlling owner ($\geq$20\%) & 76\% & 77\% \\[0.5ex]
\quad \textit{For outlets with $>$1 owner:} & & \\
\quad \quad Mean (median) no.\ of owners & 98 (71) & 102 (18) \\
\midrule
\multicolumn{3}{l}{\textbf{Panel B: Network Complexity}} \\[0.5ex]
\quad With $<$5 nodes (all owners) & 55\% & 45\% \\
\quad With $<$5 nodes (majority, $\geq$50\%) & 64\% & 61\% \\
\quad With $<$5 nodes (controlling, $\geq$20\%) & 77\% & 58\% \\[0.5ex]
\quad \textit{For outlets with $\geq$5 nodes:} & & \\
\quad \quad Mean (median) no.\ of nodes & 1,689 (511) & 1,470 (111) \\
\bottomrule
\end{tabular}
\begin{tablenotes}[flushleft]
\footnotesize
\item \textit{Notes}: This table reports ownership concentration and network tractability metrics for hard news outlets (more than 25\% political reporting) across 10 countries. Panel A focuses on ultimate owners: ``Direct ownership'' means the domain owner is itself the ultimate owner (single-node network); ``Through network'' means a single ultimate owner is reached through intermediary entities. ``Majority owner'' refers to shareholders with $\geq$50\% stake; ``Controlling owner'' follows \citet{aminadav2020corporate} and refers to shareholders with $\geq$20\% stake. Panel B focuses on network complexity: node counts are computed for all ownership paths, for paths leading to majority owners only, and for paths leading to controlling owners only. The first column reports unweighted shares; the second column weights outlets by web traffic (reach). See Online Appendix Table~\ref{tab:ownership_transparency_by_country} for country-level statistics.
\end{tablenotes}
\end{threeparttable}
\end{table}

\begin{figure}[htbp]
\centering
\caption{Ownership Networks: Distributions of Node and Level Counts}
\label{fig:app:node_level_distributions}
\includegraphics[width=0.49\textwidth]{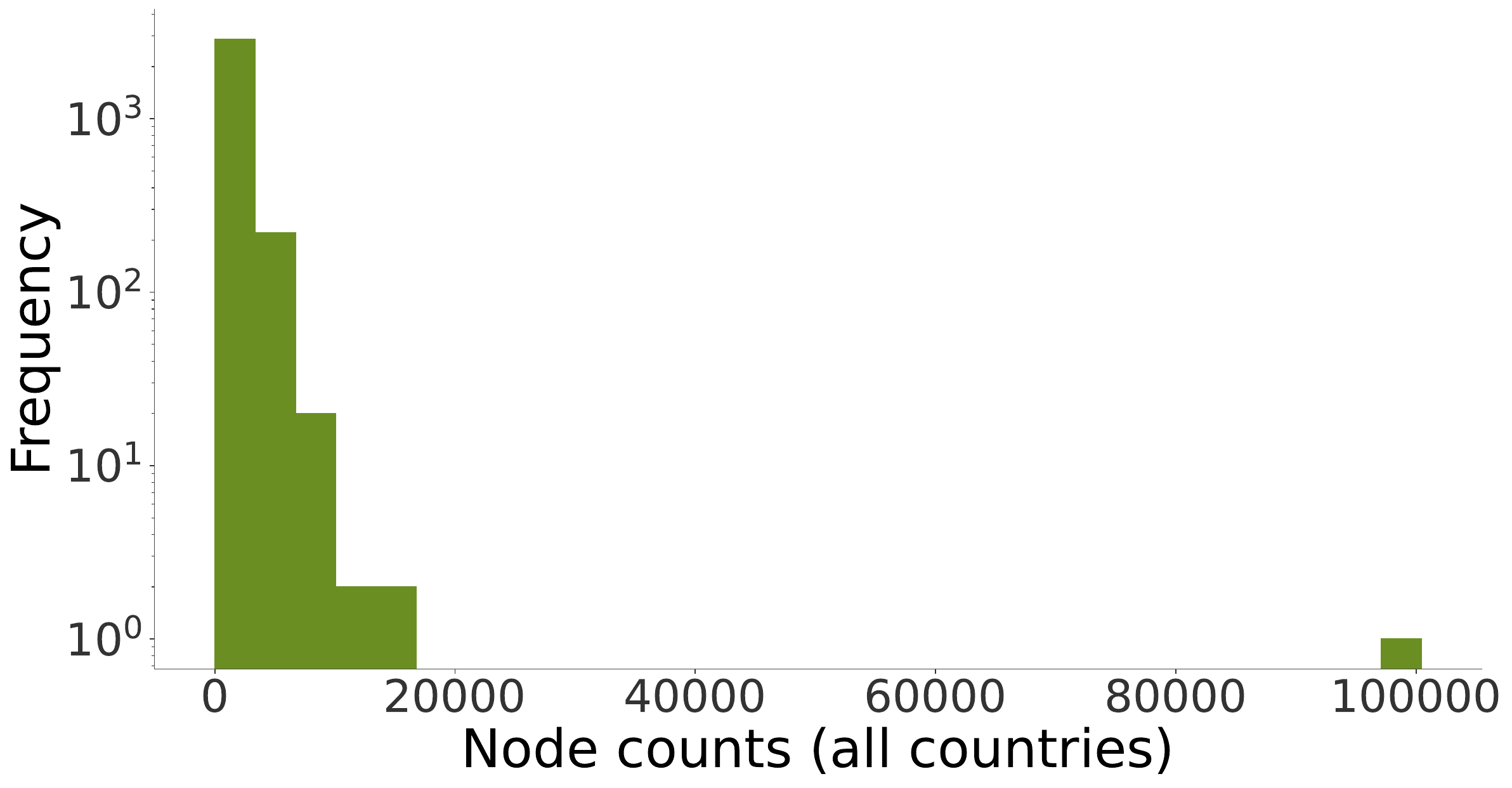}
\includegraphics[width=0.49\textwidth]{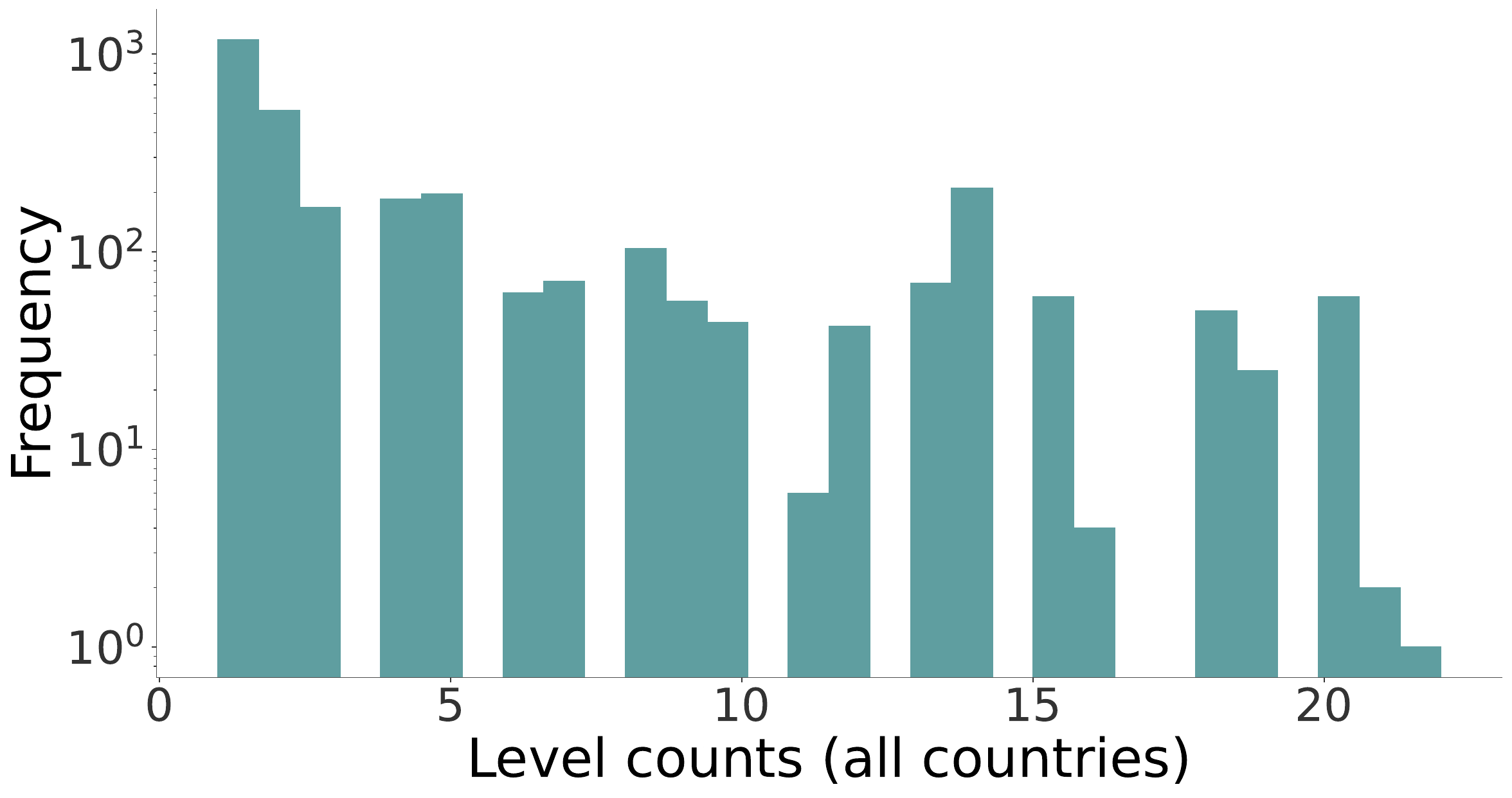}
\begin{figurenotes}
The left panel shows the distribution of node counts; the right panel shows the distribution of level counts. The node counts show the total number of owning entities in an outlet's ownership network; the network is traversed link by link, including every owner holding $>1\%$ of the entity directly below it, so that an owner's multiplied-through stake in the outlet itself can be below 1\%. The level counts show the number of levels in the same network. For example, if an immediate domain owner is owned by another entity with no further owners, the network has two levels.
\end{figurenotes}
\end{figure}

\begin{figure}[htbp]
\centering
\caption{Node Count Distribution, by Country}
\label{fig:app:node_counts_by_country}
\includegraphics[width=0.45\linewidth]{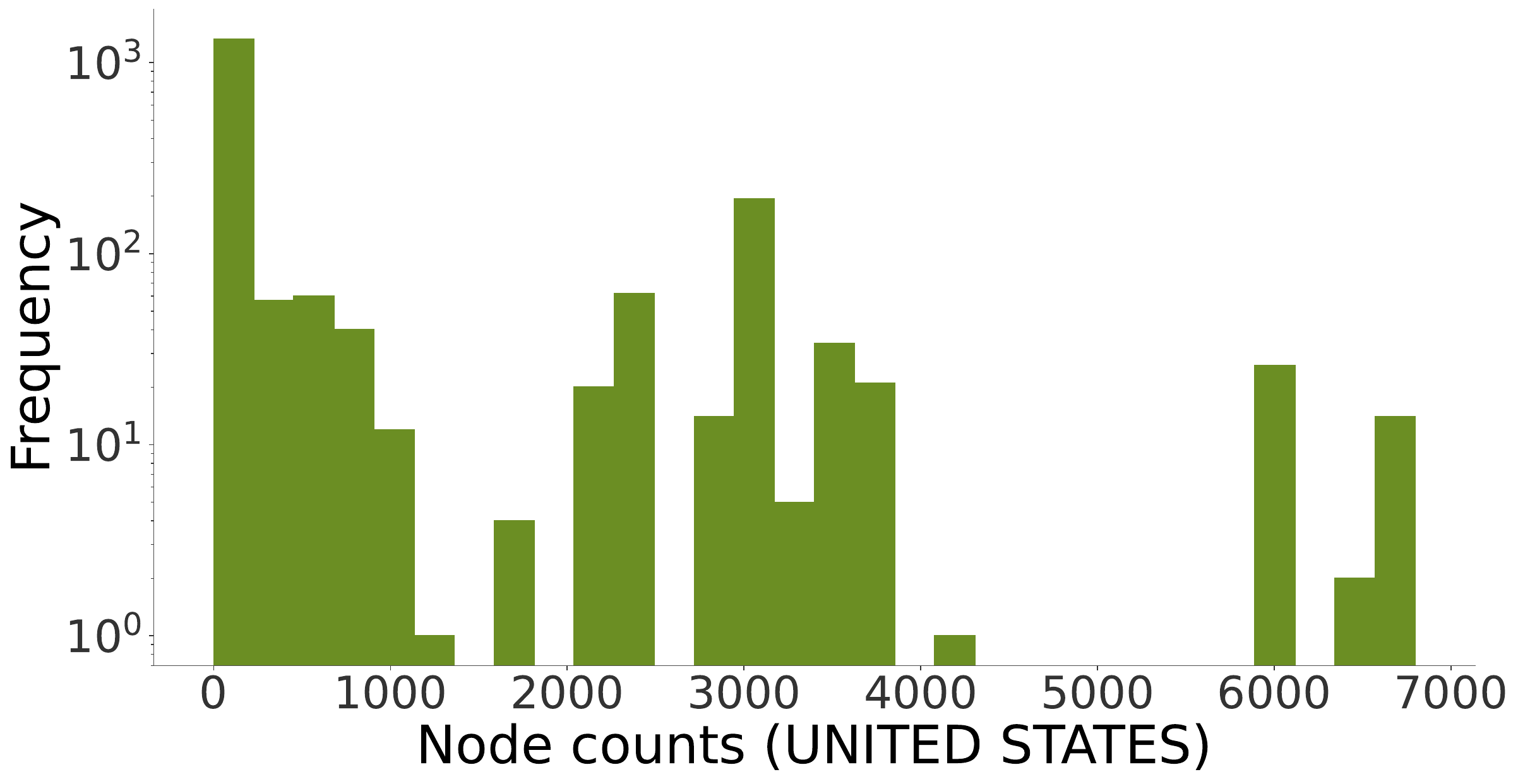}
\includegraphics[width=0.45\linewidth]{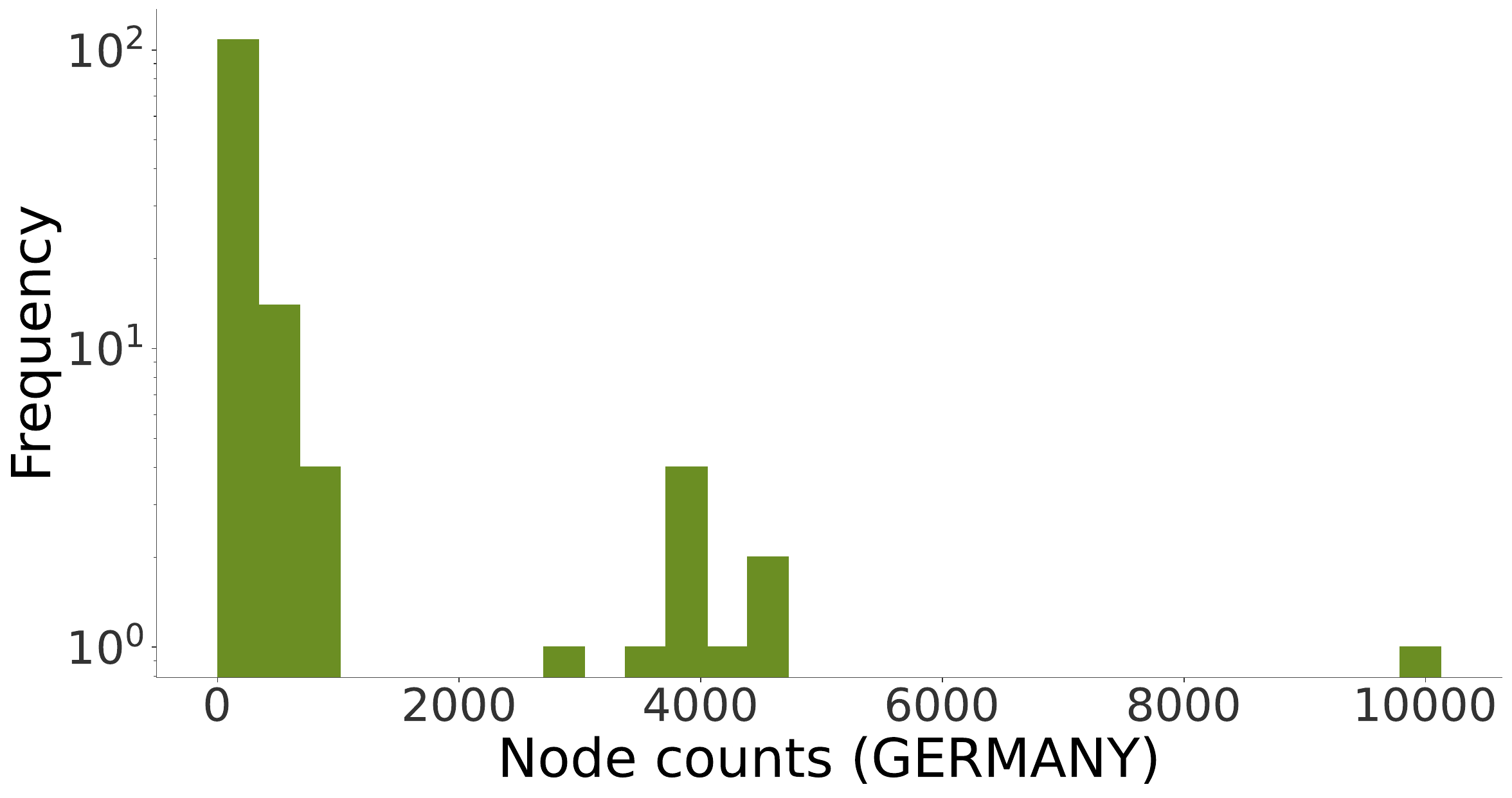}
\includegraphics[width=0.45\linewidth]{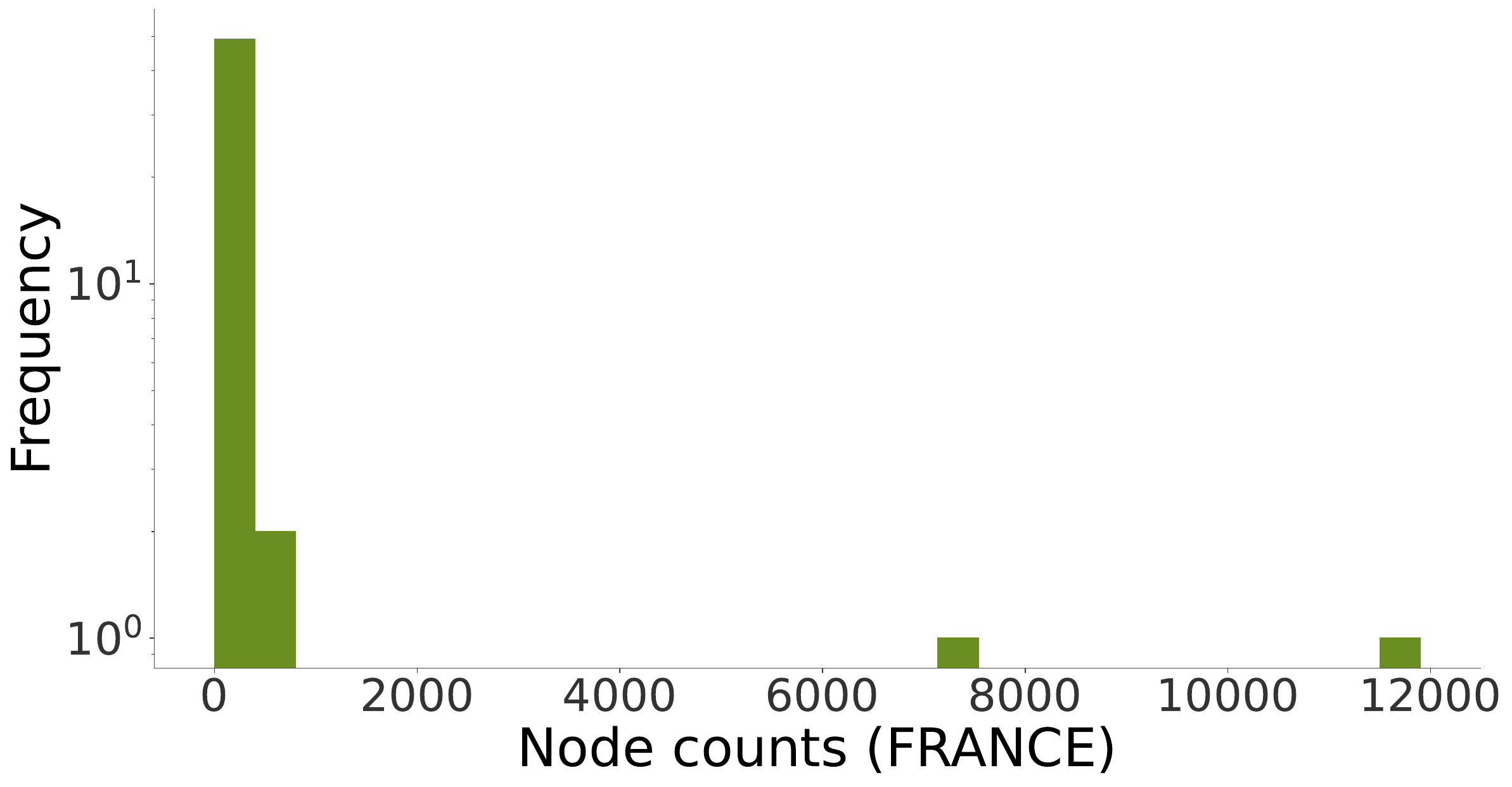}
\includegraphics[width=0.45\linewidth]{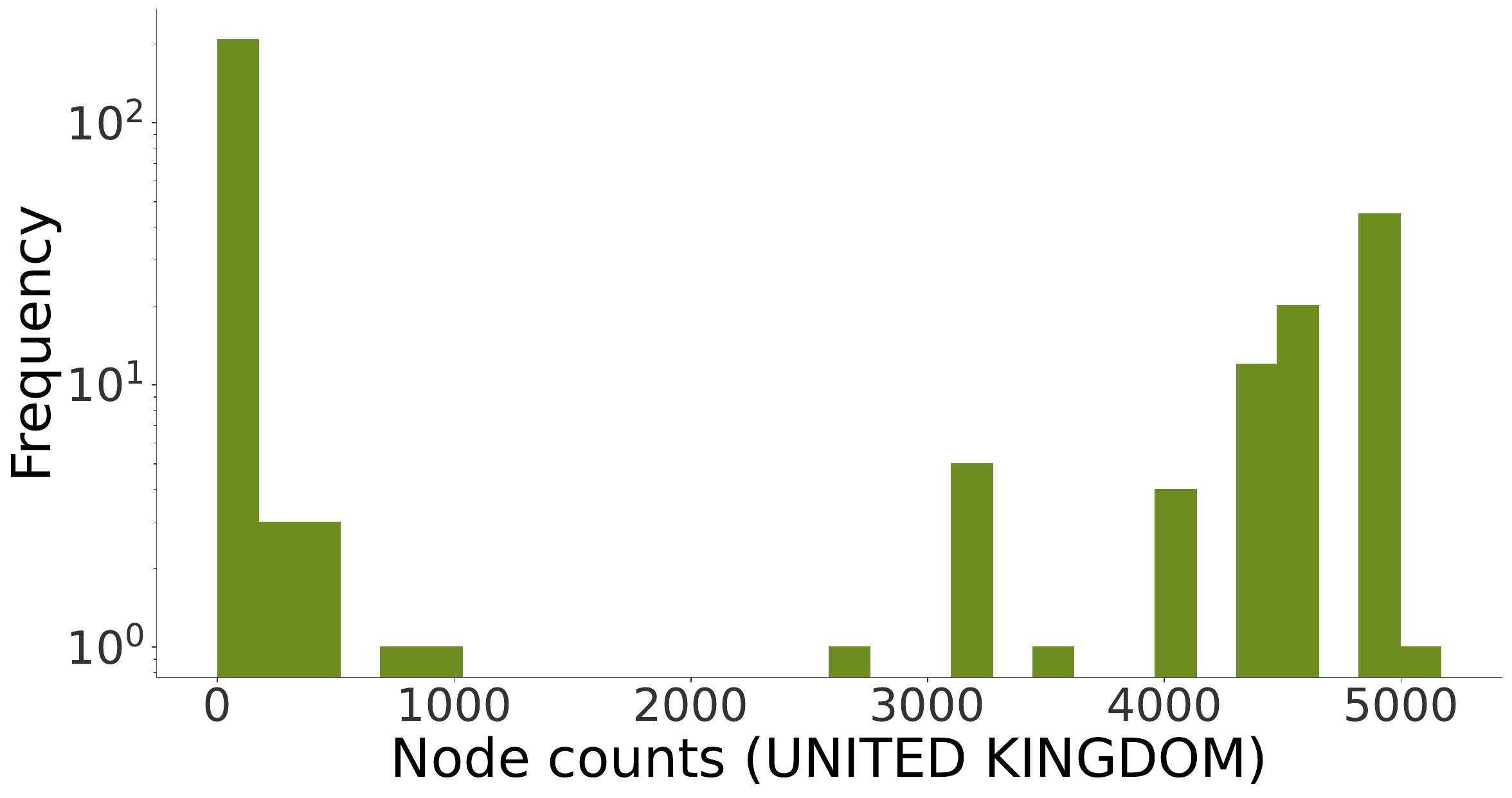}
\includegraphics[width=0.45\linewidth]{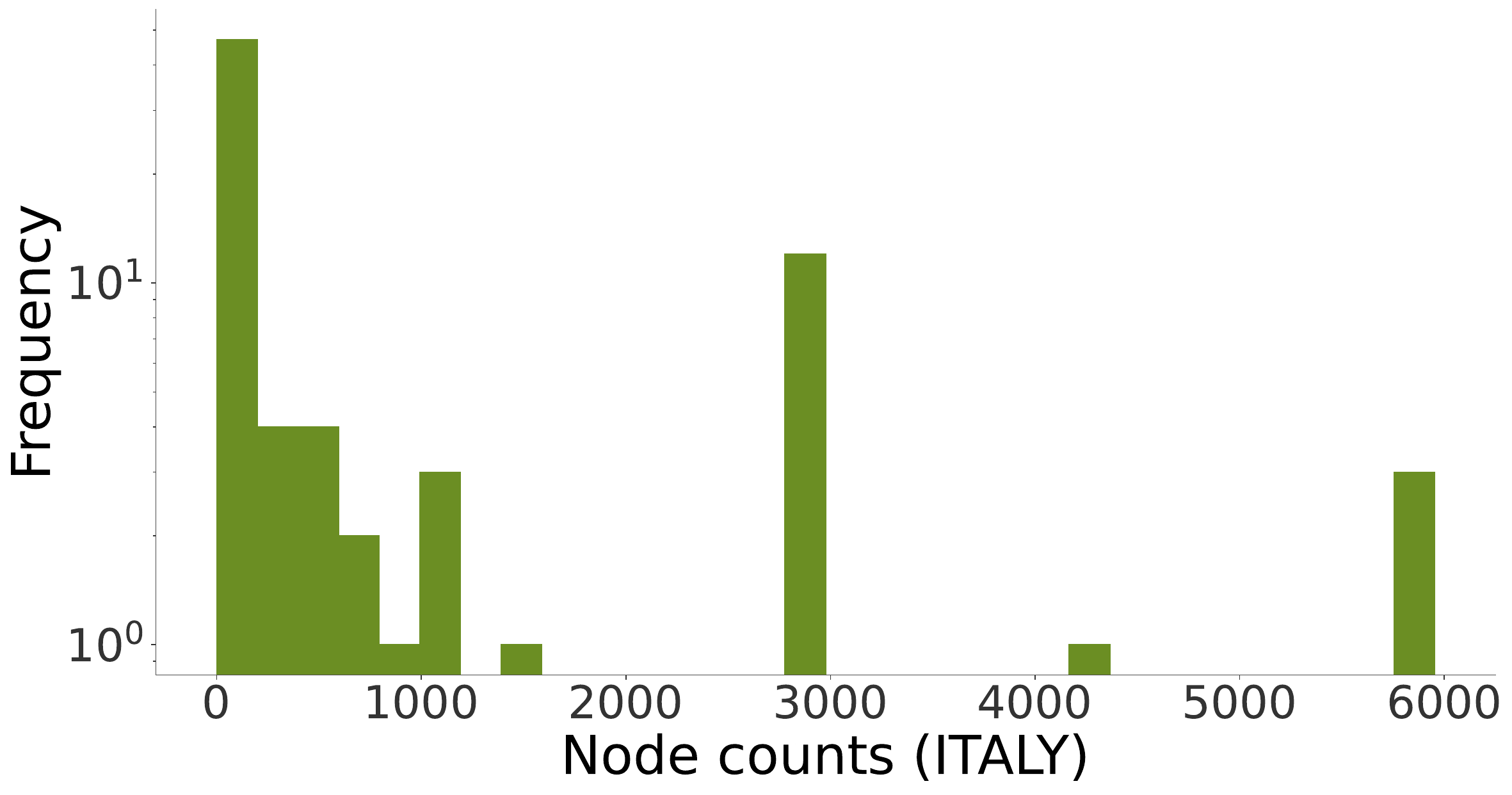}
\includegraphics[width=0.45\linewidth]{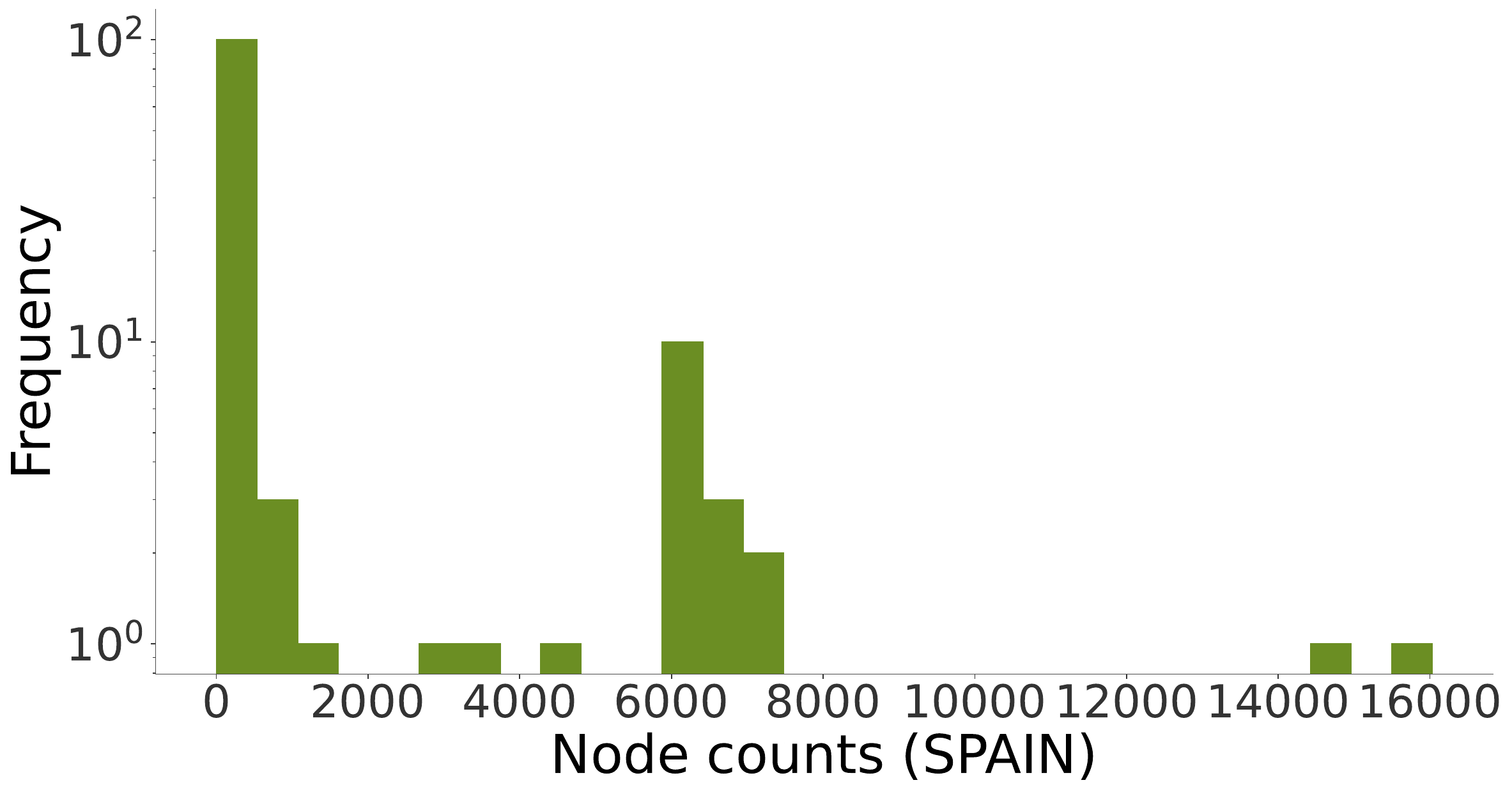}
\includegraphics[width=0.45\linewidth]{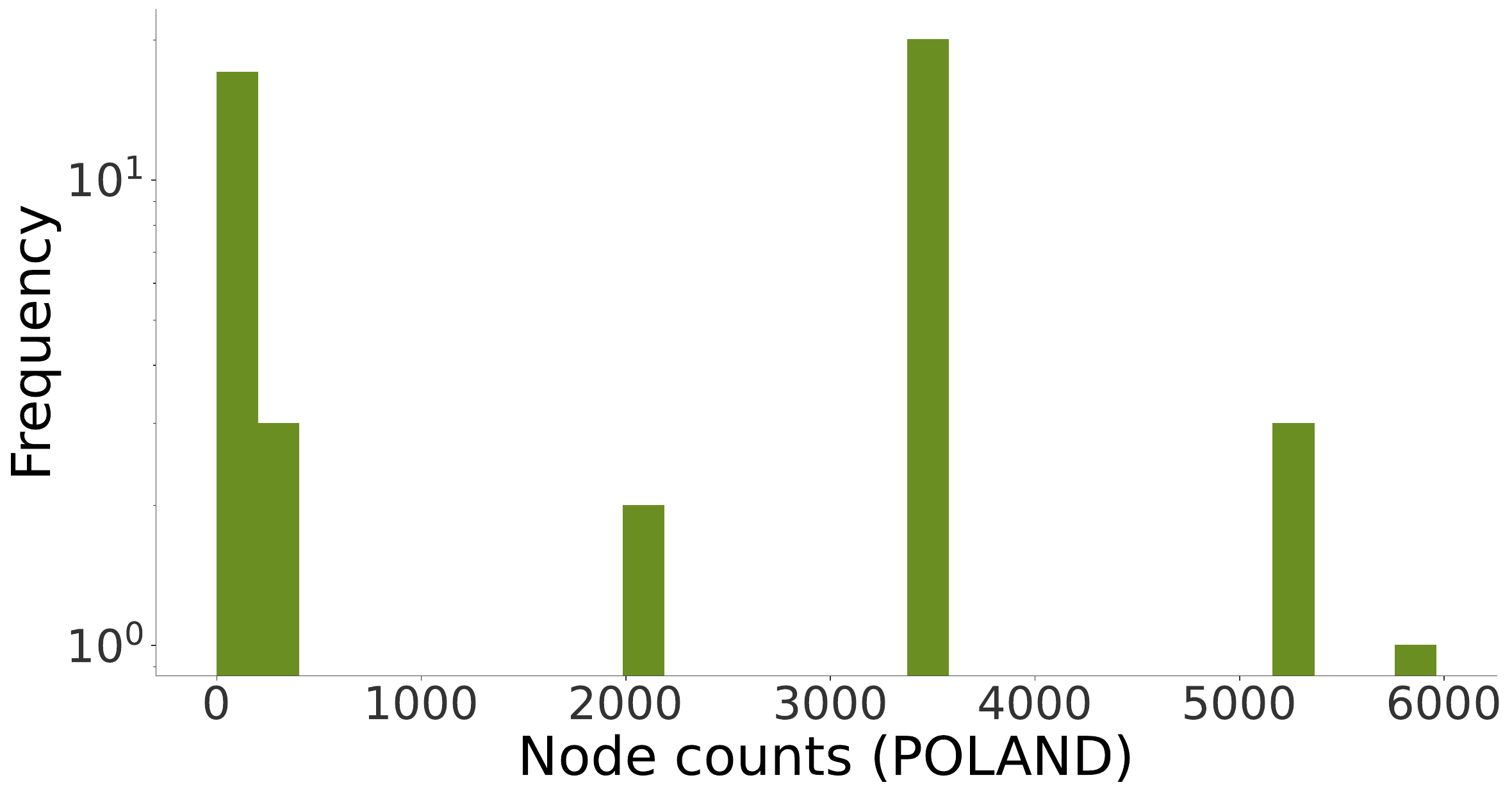}
\includegraphics[width=0.45\linewidth]{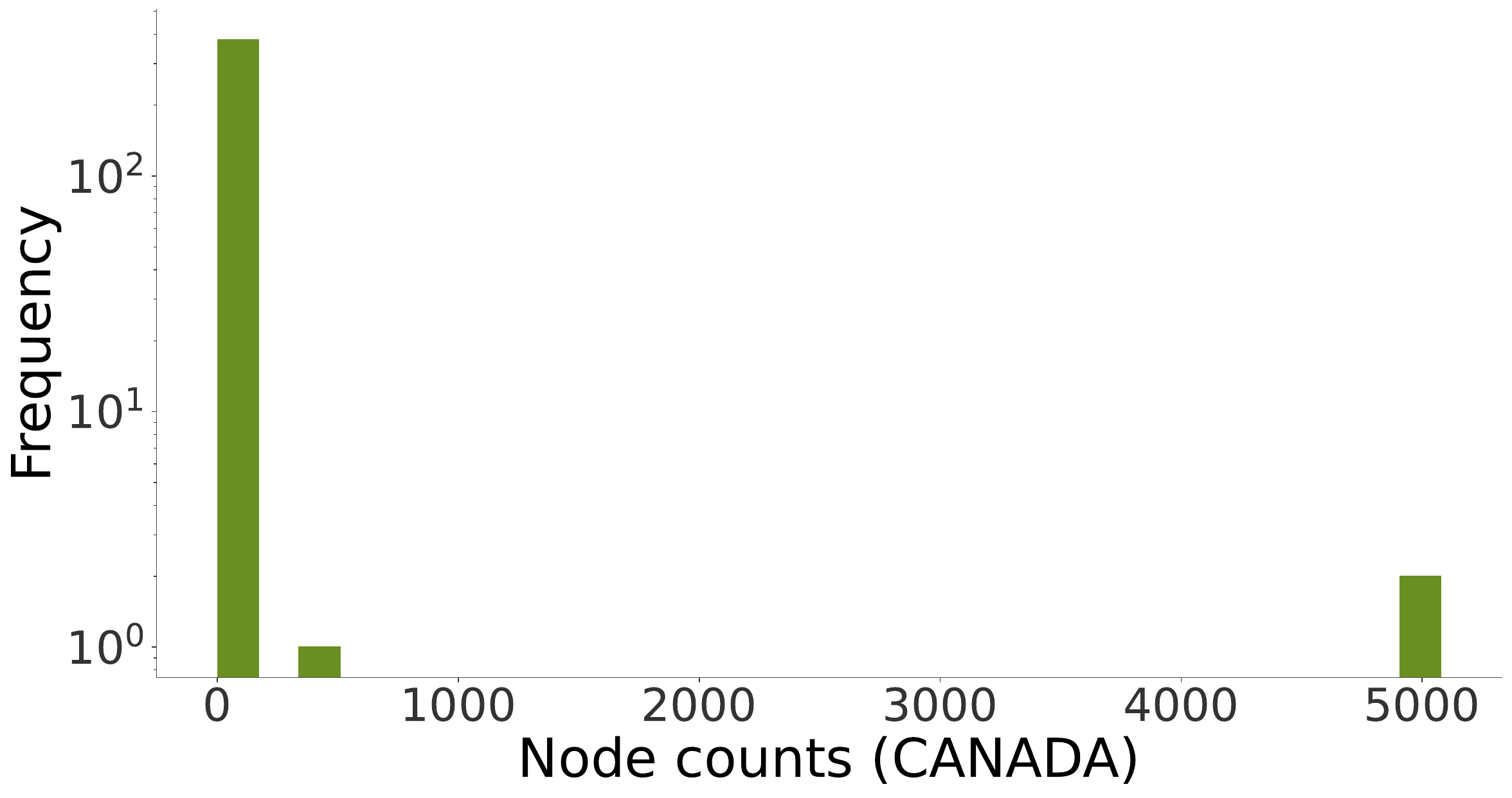}
\includegraphics[width=0.45\linewidth]{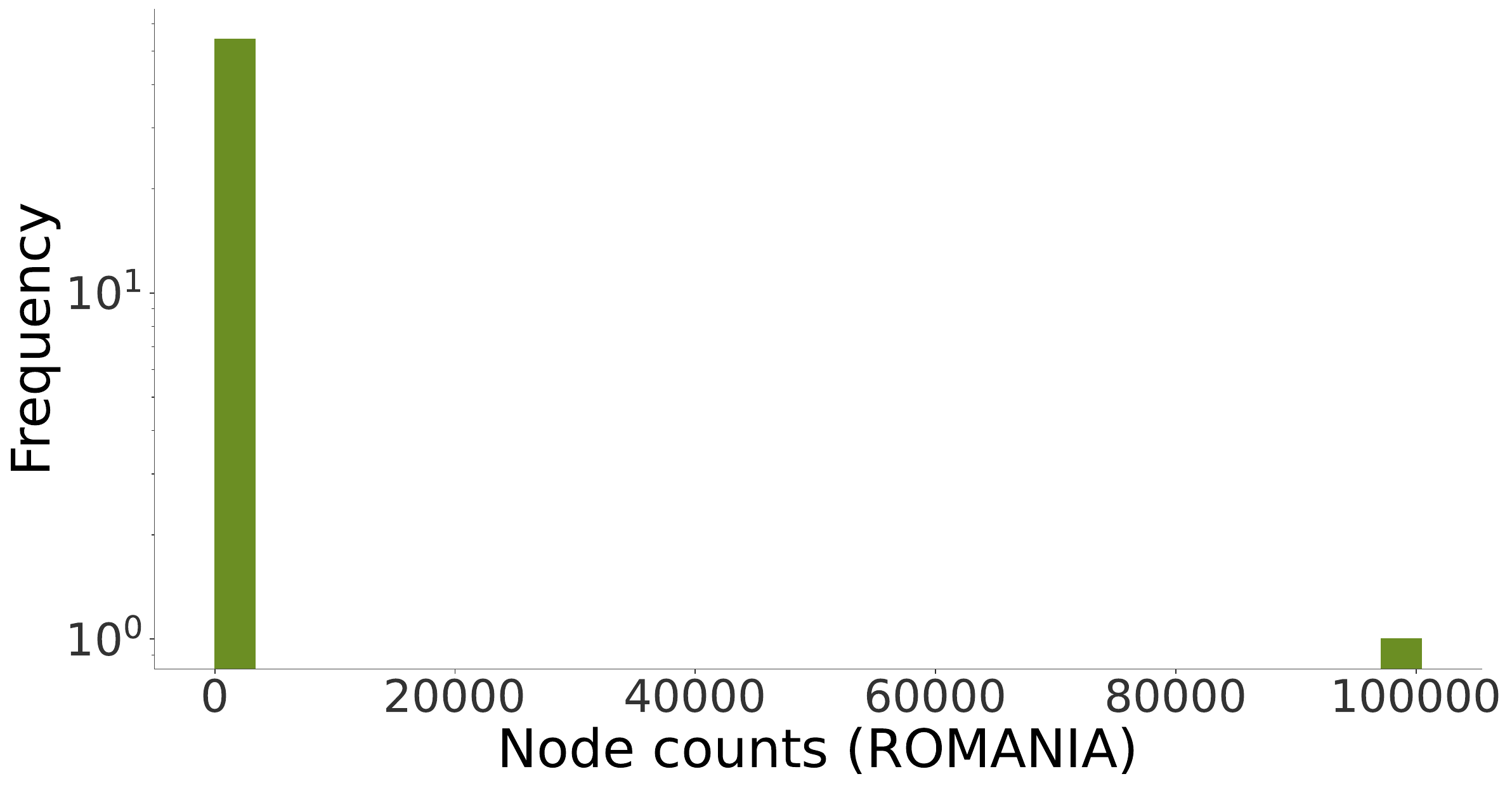}
\includegraphics[width=0.45\linewidth]{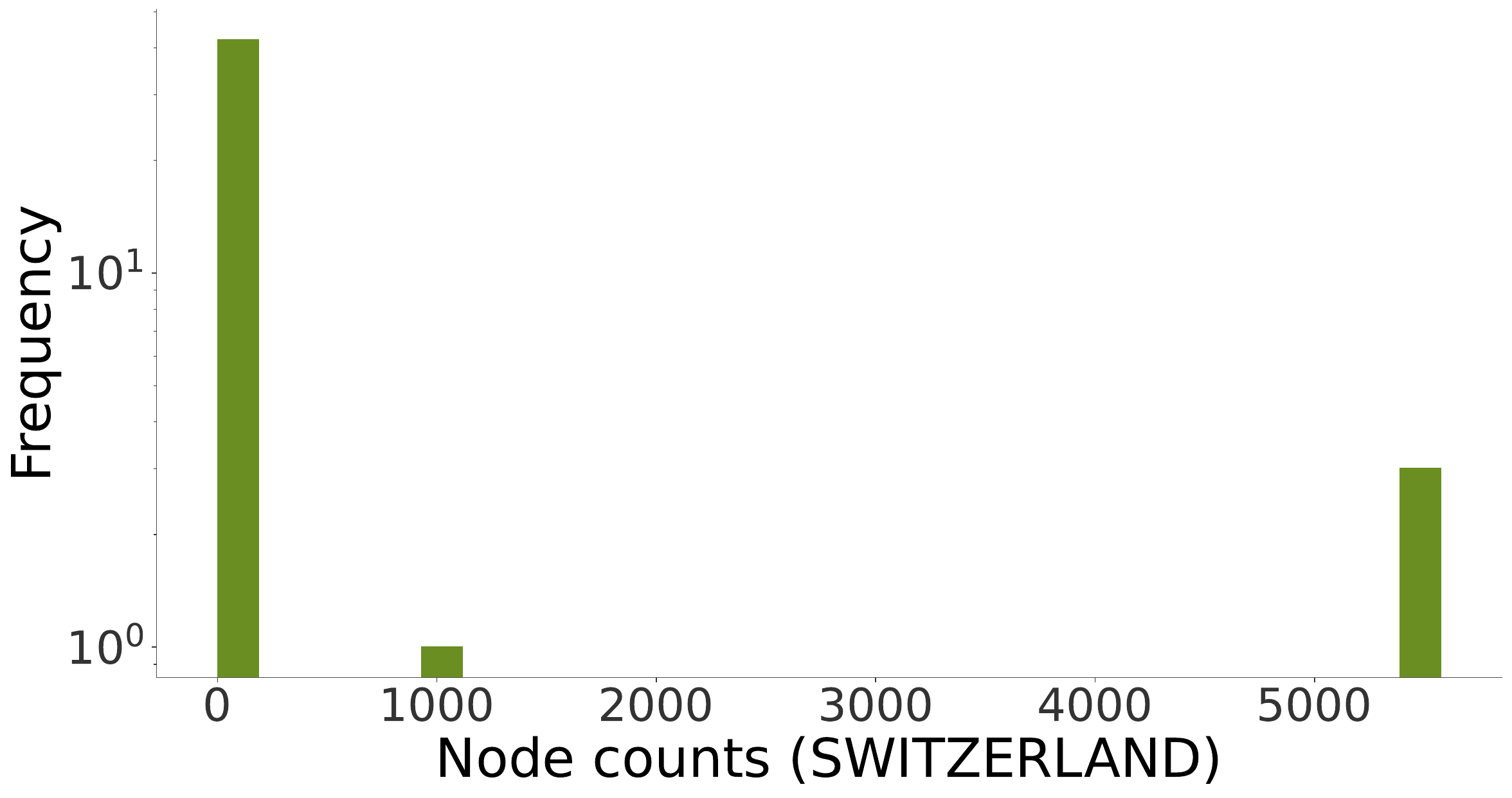}
\begin{figurenotes}
The figure shows the distribution of node counts by country, for all hard news outlets (more than 25\% political reporting). The node counts show the total number of owning entities in an outlet's ownership network, traversed link by link over stakes $>1\%$ in the entity directly below, so that multiplied-through stakes in the outlet itself can be smaller. This figure provides country-level detail for the aggregate distributions shown in Figure~\ref{fig:app:node_level_distributions}; see Section~\ref{sec:results} for discussion.
\end{figurenotes}
\end{figure}

\begin{figure}[htbp]
\centering
\caption{Node Count Distribution, by Country (All Outlets, Incl. Non-Political)}
\label{fig:app:node_counts_by_country_nonpol}
\includegraphics[width=0.45\linewidth]{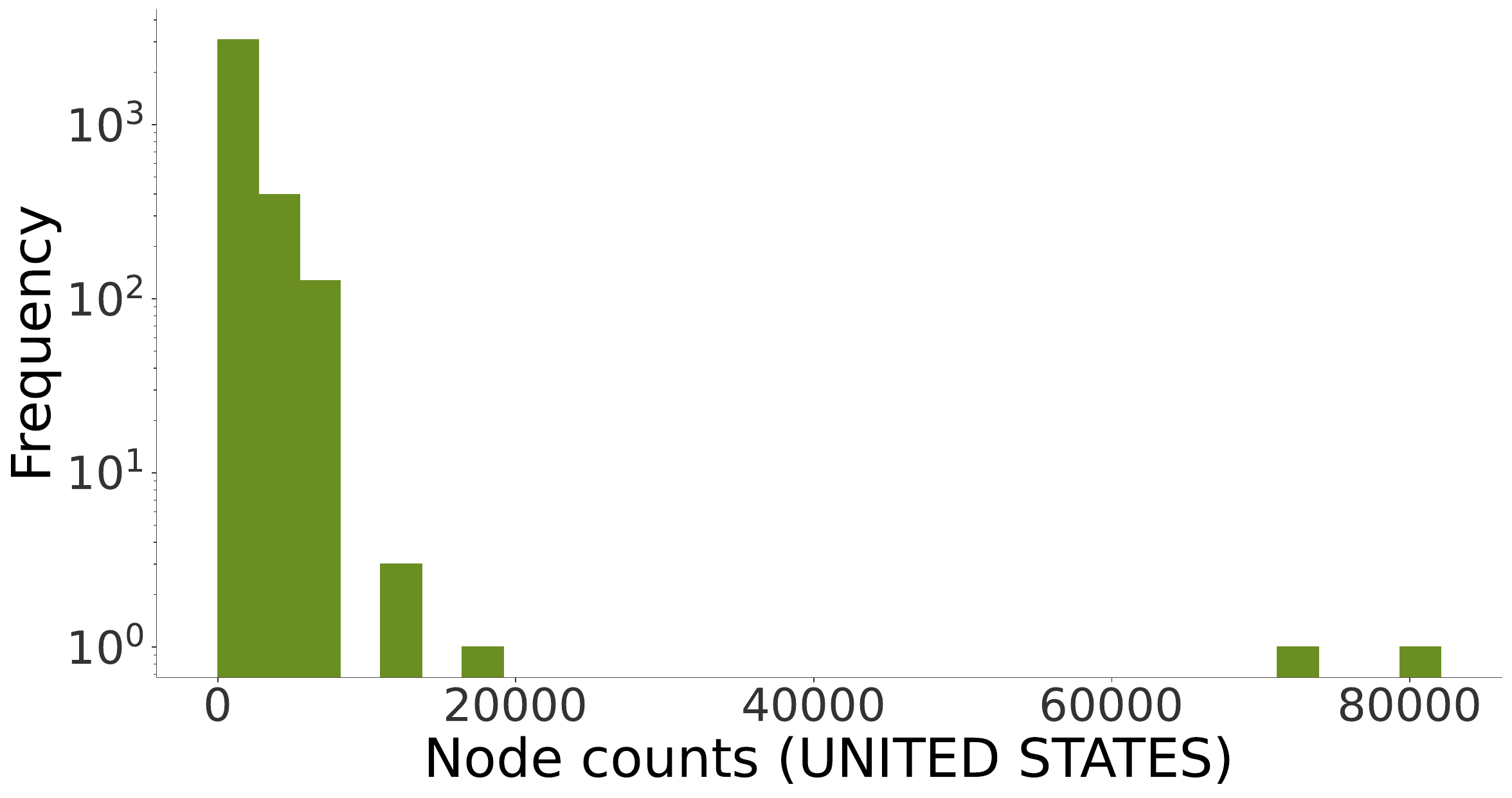}
\includegraphics[width=0.45\linewidth]{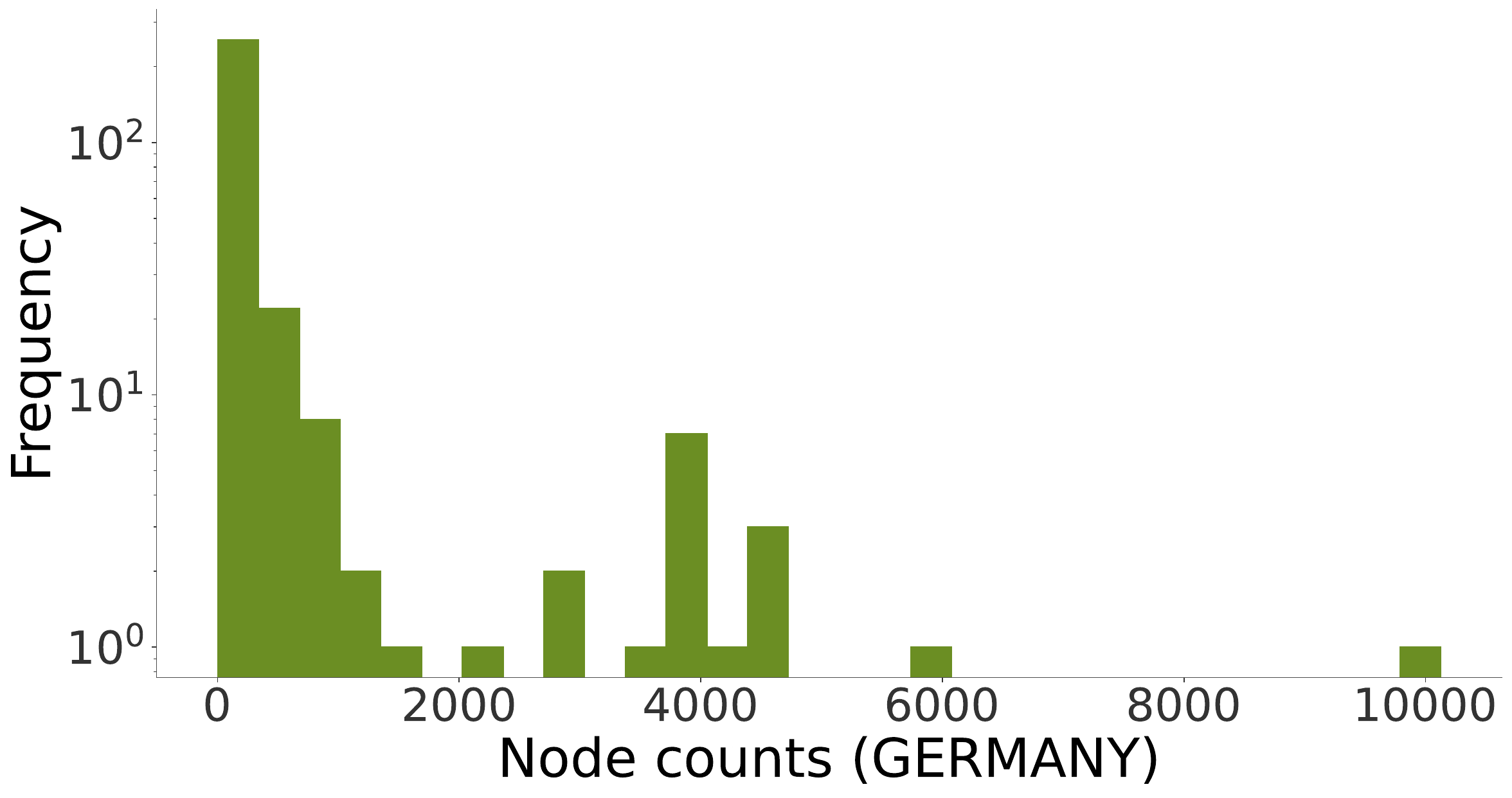}
\includegraphics[width=0.45\linewidth]{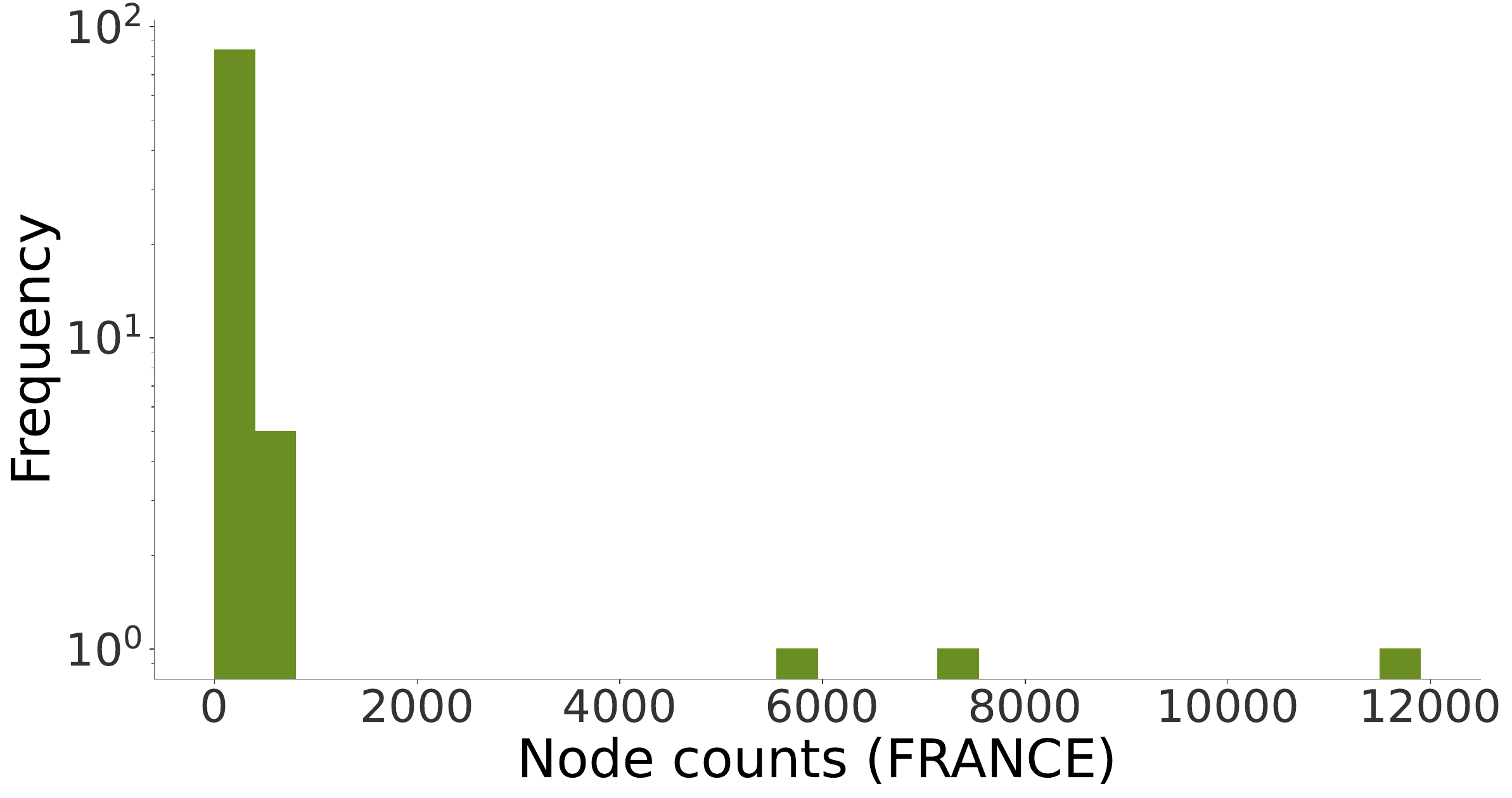}
\includegraphics[width=0.45\linewidth]{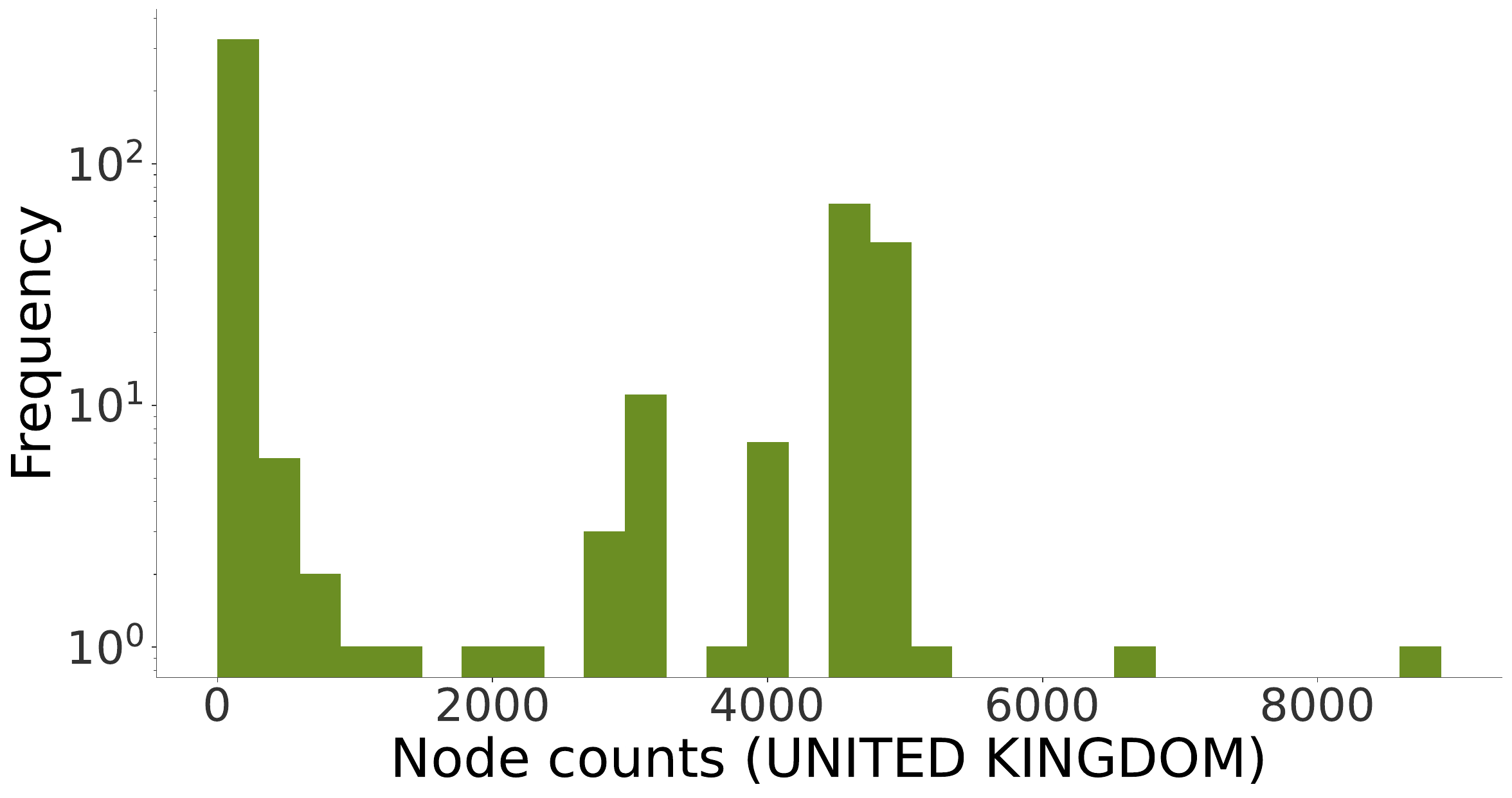}
\includegraphics[width=0.45\linewidth]{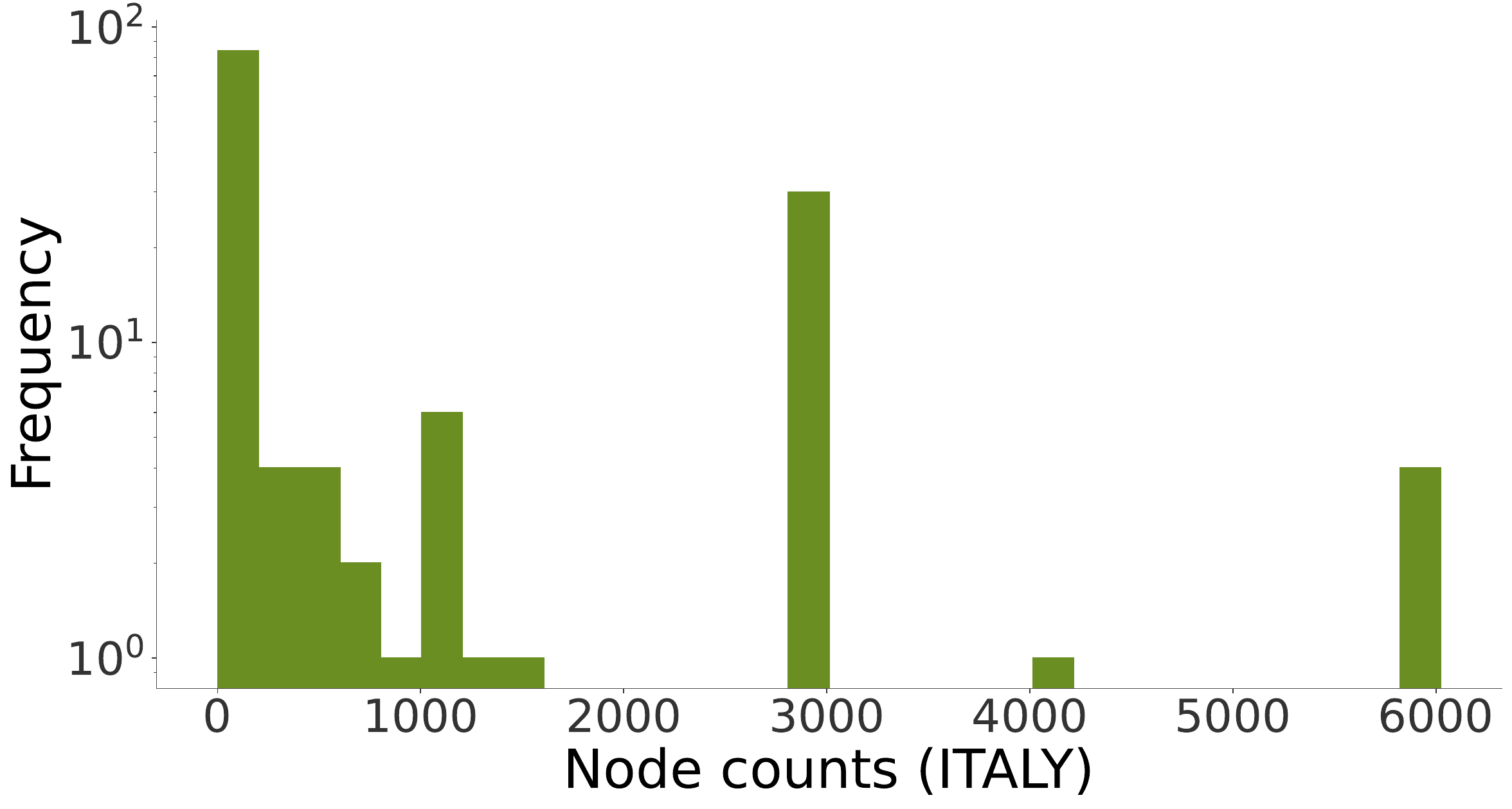}
\includegraphics[width=0.45\linewidth]{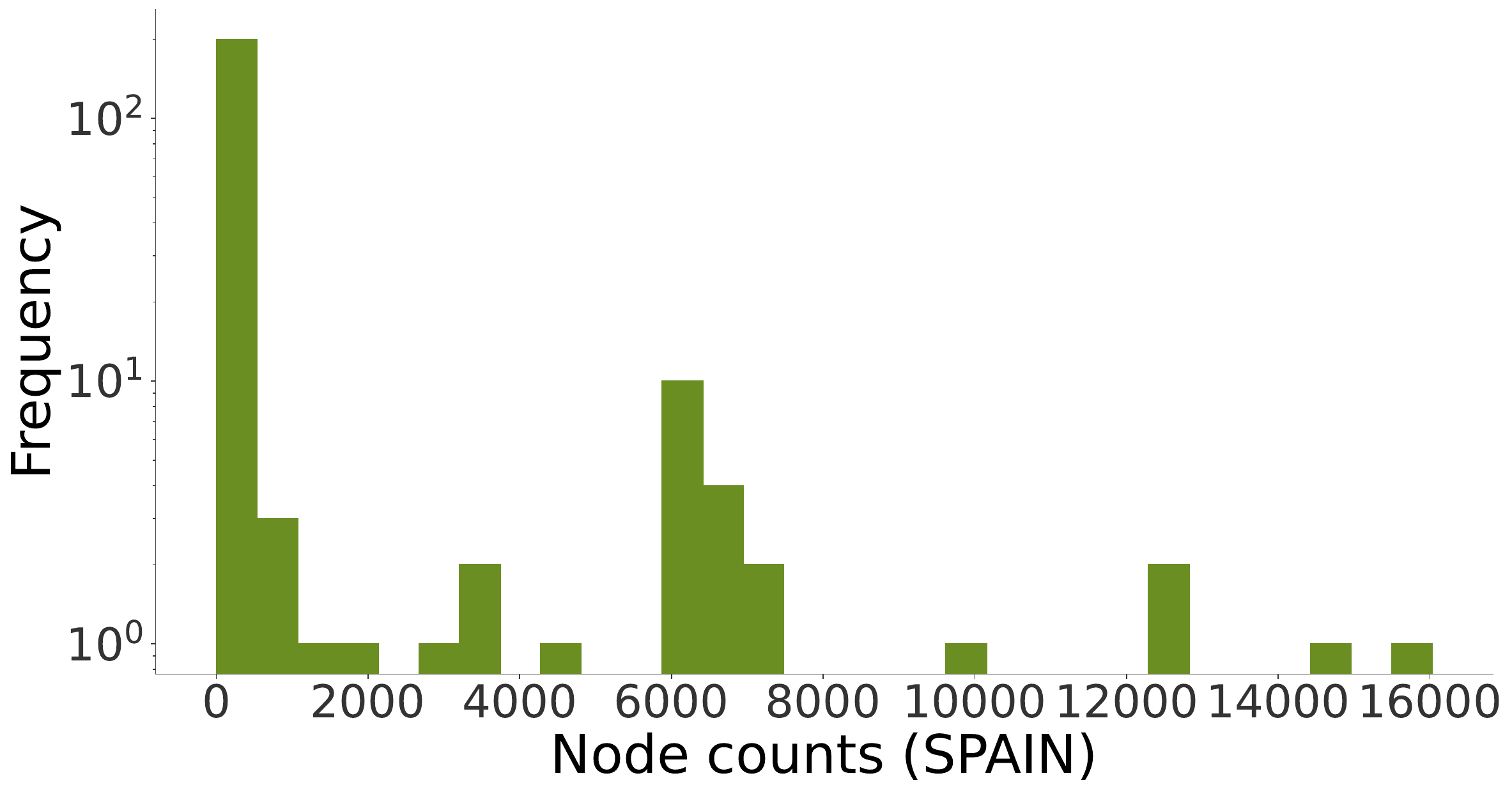}
\includegraphics[width=0.45\linewidth]{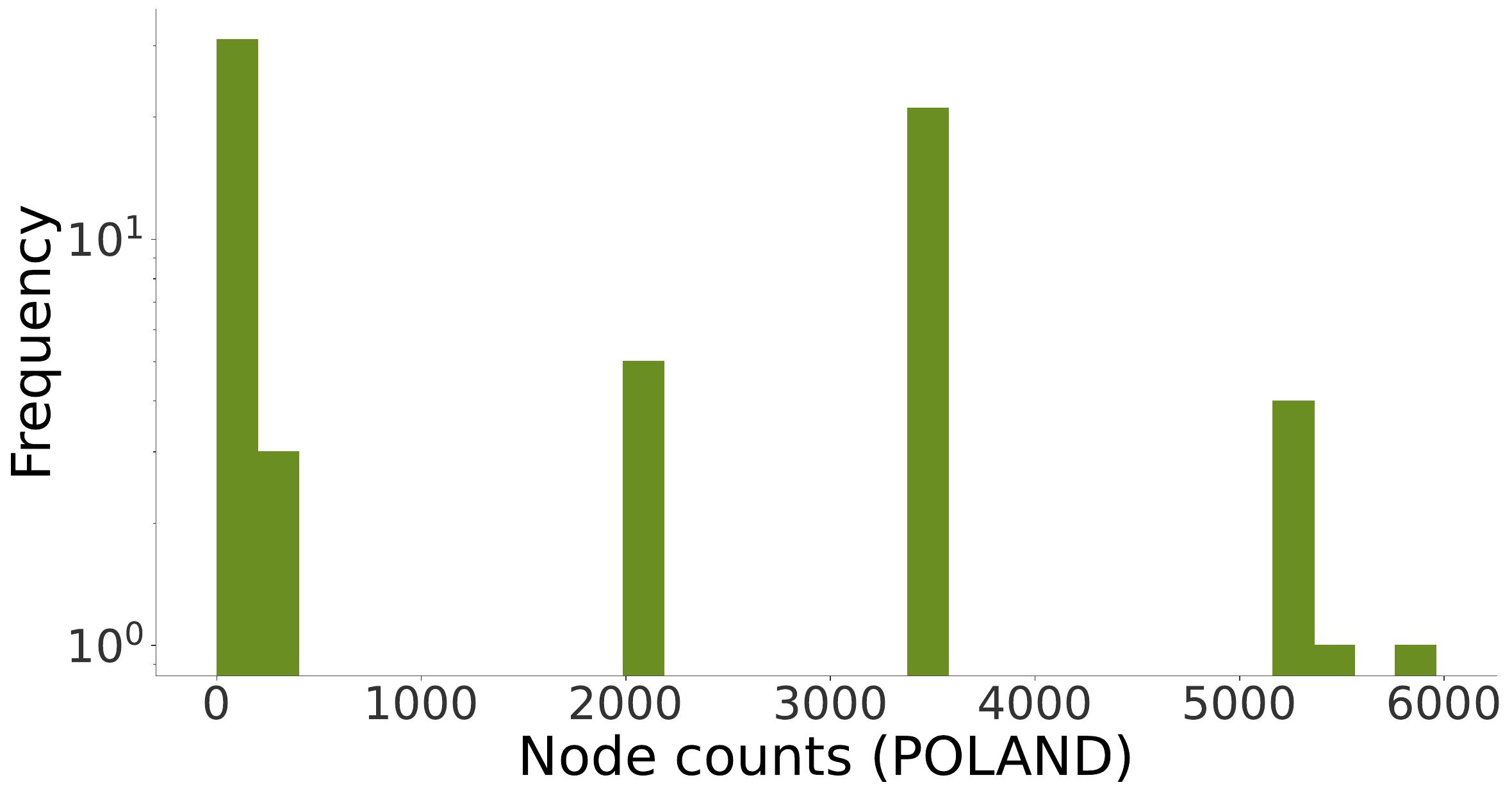}
\includegraphics[width=0.45\linewidth]{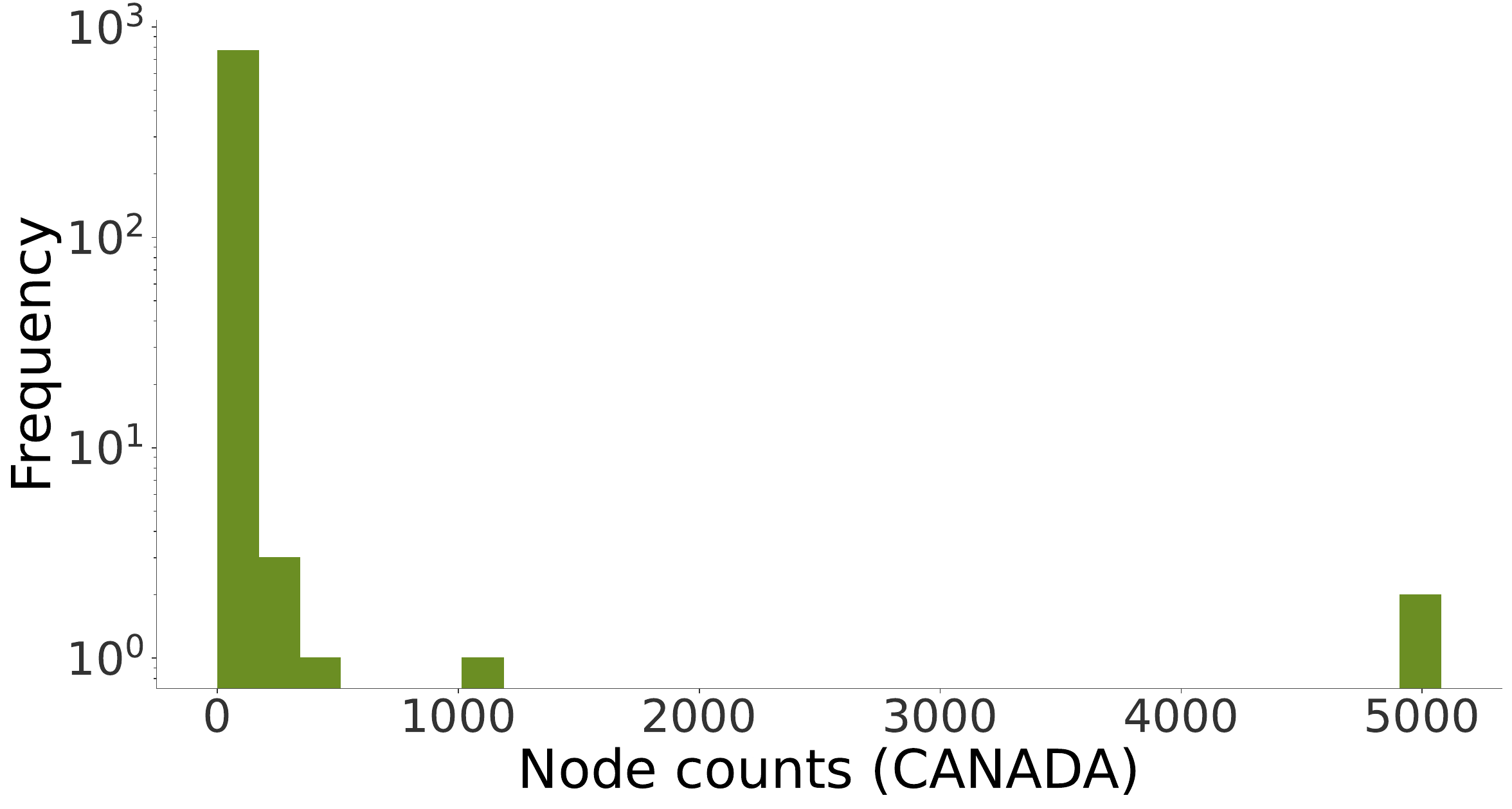}
\includegraphics[width=0.45\linewidth]{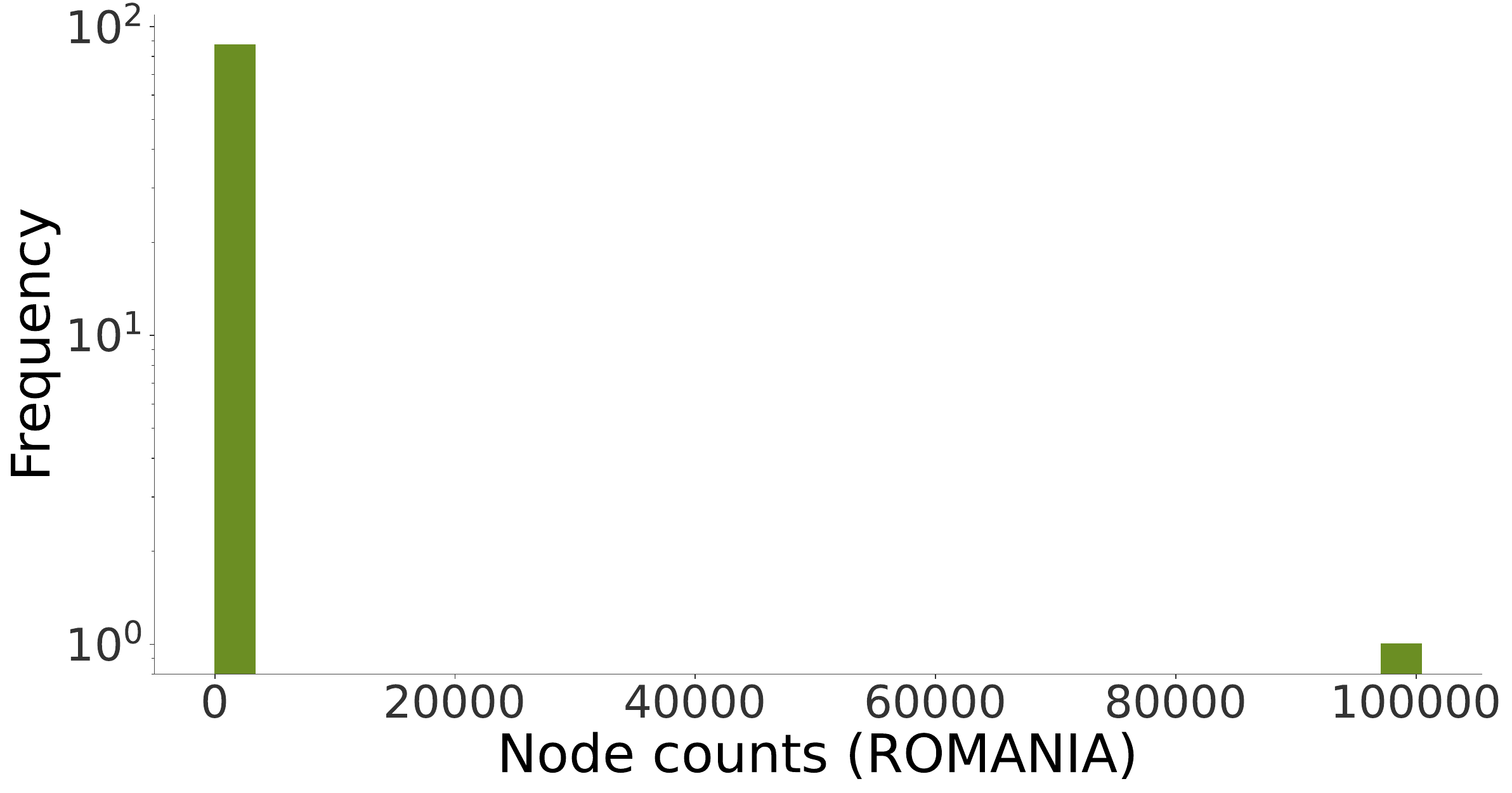}
\includegraphics[width=0.45\linewidth]{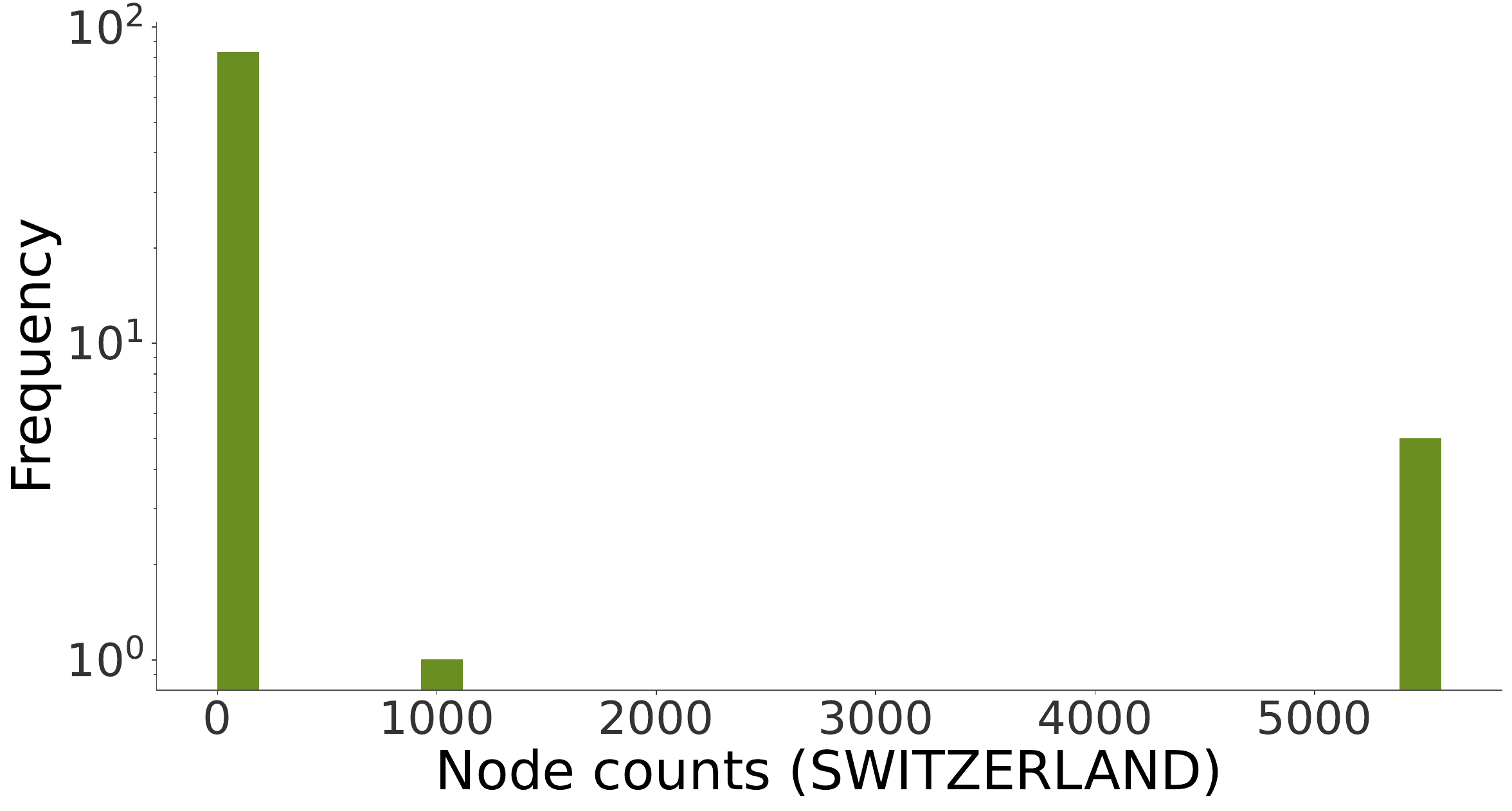}
\begin{figurenotes}
The figure shows the distribution of node counts by country, for all outlets, including those not specializing in political content. The node counts show the total number of owning entities in an outlet's ownership network, traversed link by link over stakes $>1\%$ in the entity directly below, so that multiplied-through stakes in the outlet itself can be smaller. This figure provides country-level detail for comparison with Figure~\ref{fig:app:node_counts_by_country}.
\end{figurenotes}
\end{figure}

\paragraph{Node counts by country and by traffic. } Figure \ref{fig:app:node_level_distributions} (left panel) shows the full distribution of node counts, aggregated for the ten countries examined. Ownership complexity varies substantially across countries (see Online Appendix Table~\ref{tab:ownership_transparency_by_country}). The United States and United Kingdom have the most tractable ownership structures, with 64\% and 58\% of outlets having fewer than five nodes, respectively. In contrast, Germany (21\%), Poland (24\%), Italy (26\%), and Spain (26\%) have the lowest share of tractable structures; roughly three-quarters of outlets have complex ownership networks. Strikingly, the pattern that larger outlets have harder-to-track ownership holds across nearly all countries. When weighted by traffic, the share of outlets with simple ownership declines substantially in some countries (e.g., from 55\% to 11\% in Romania and from 26\% to 7\% in Spain).\footnote{Figures~\ref{fig:app:node_counts_by_country} and \ref{fig:app:node_counts_by_country_nonpol} show the country-specific node distributions for hard news outlets and all outlets.} Consistent with the cross-country aggregate result, the \textit{market share} of hard-to-track outlets is higher than the share of hard-to-track outlets in most countries.

Similar to large outlets having higher ultimate owner counts, we also observe that the prevalence of hard-to-track ownership networks increases when we weight domains by their traffic: only 45\% of the market has networks of fewer than five nodes. Even when focusing on controlling or majority owners, almost half of the market still has hard-to-track ownership (see last column of Panel B in Table~\ref{tab:ownership_transparency}).

\begin{figure}[htbp]
\centering
\caption{Level Count Distribution, by Country}
\label{fig:app:level_counts_by_country}
\includegraphics[width=0.45\linewidth]{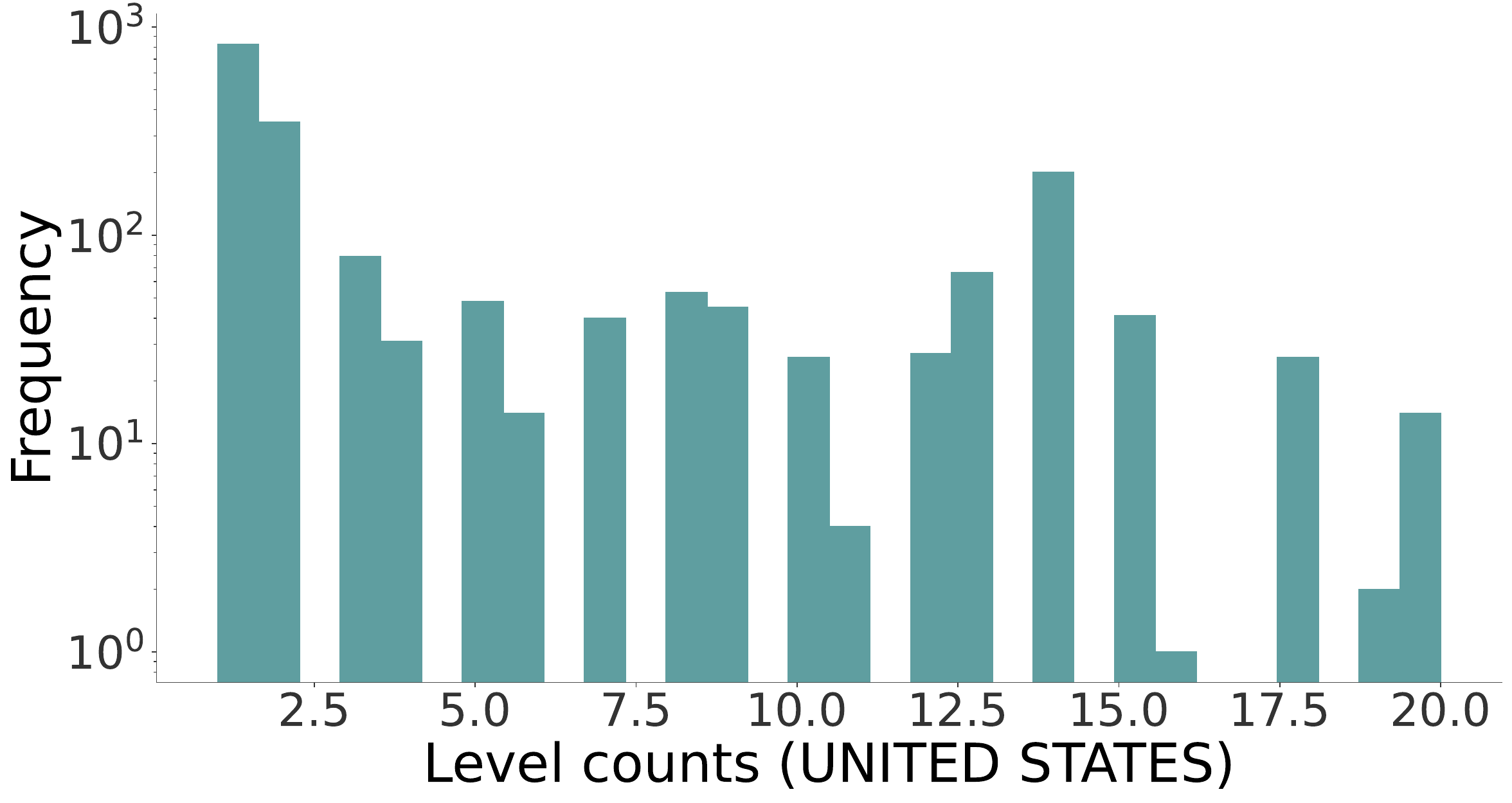}
\includegraphics[width=0.45\linewidth]{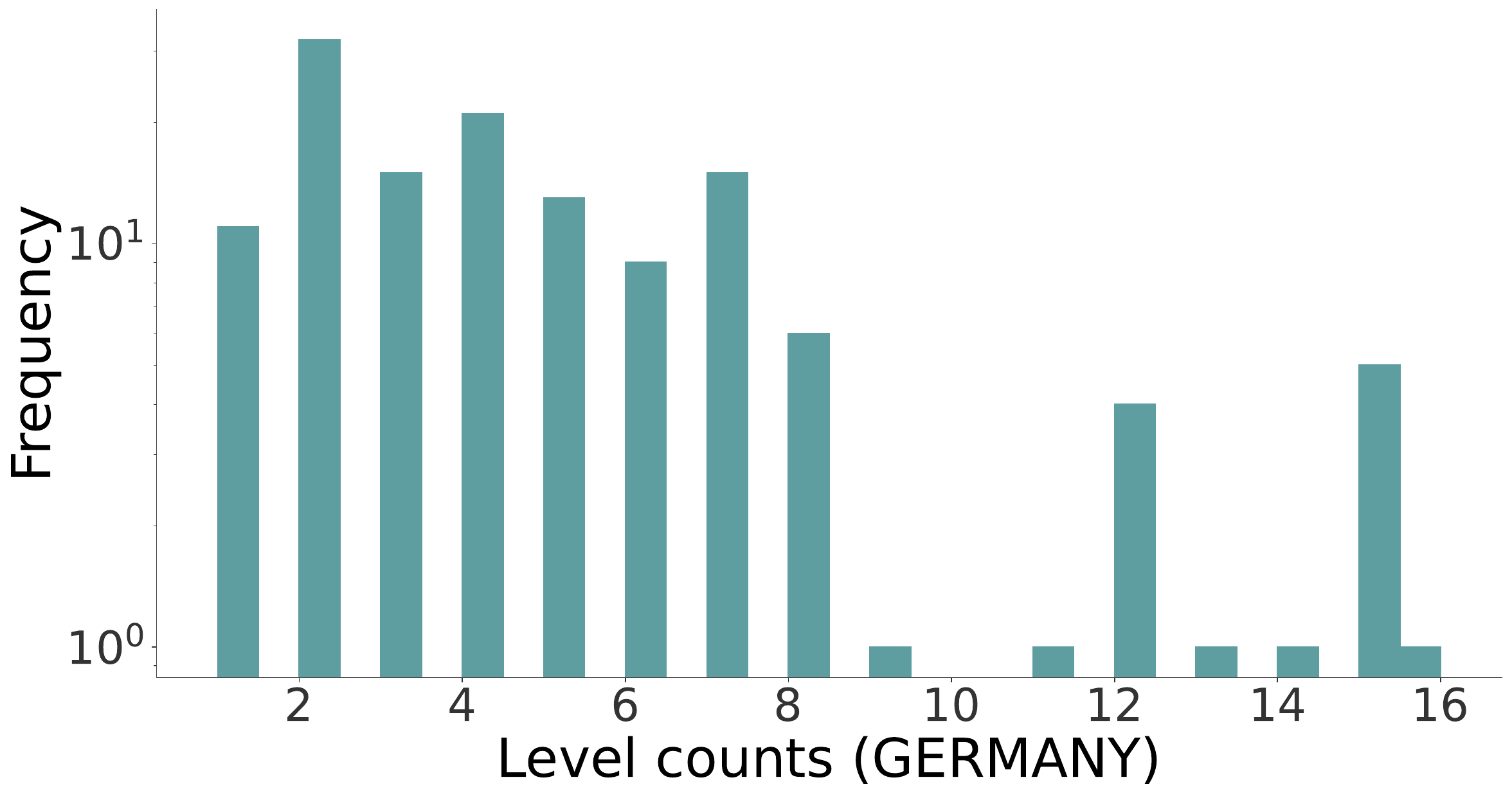}
\includegraphics[width=0.45\linewidth]{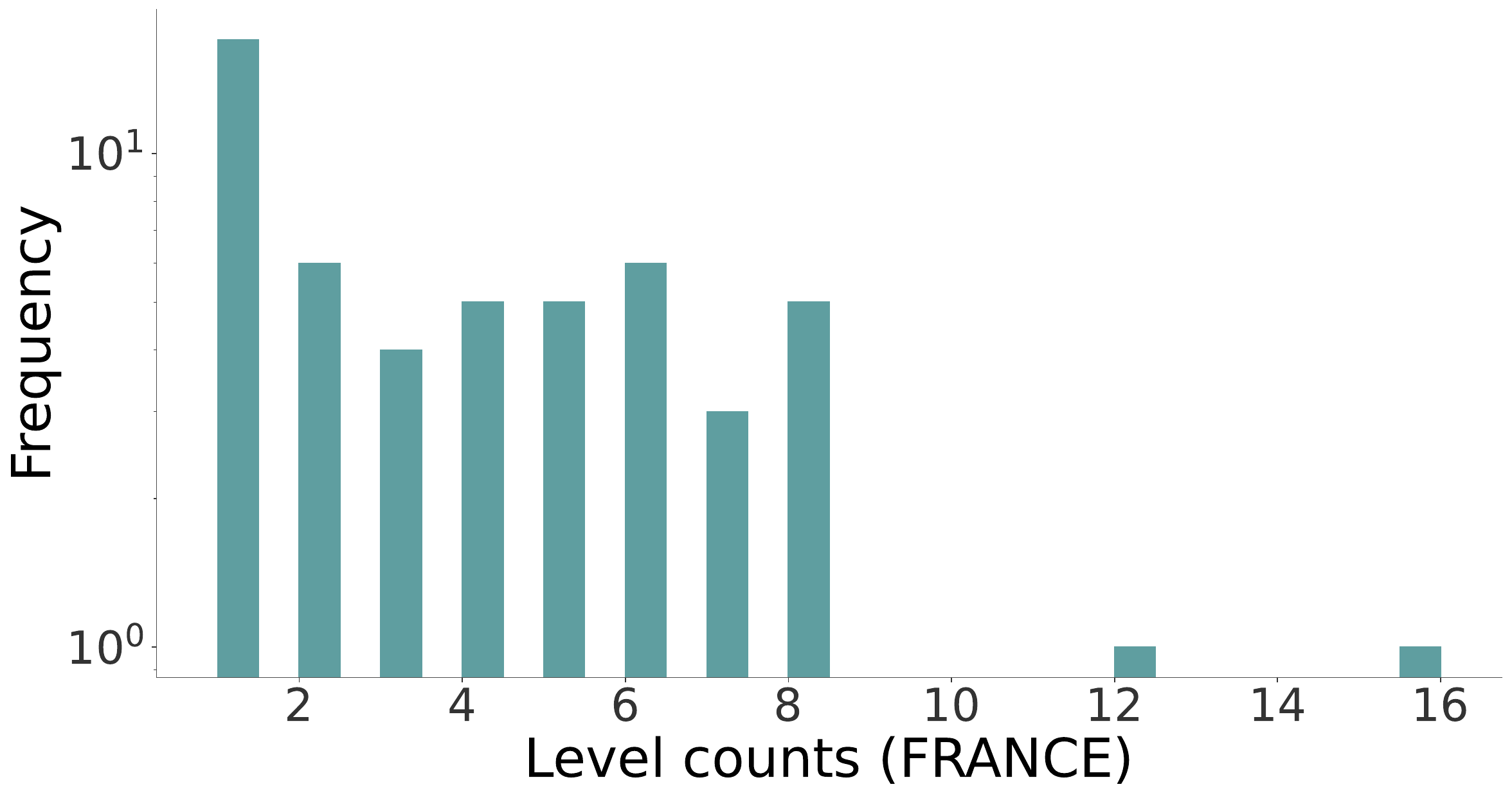}
\includegraphics[width=0.45\linewidth]{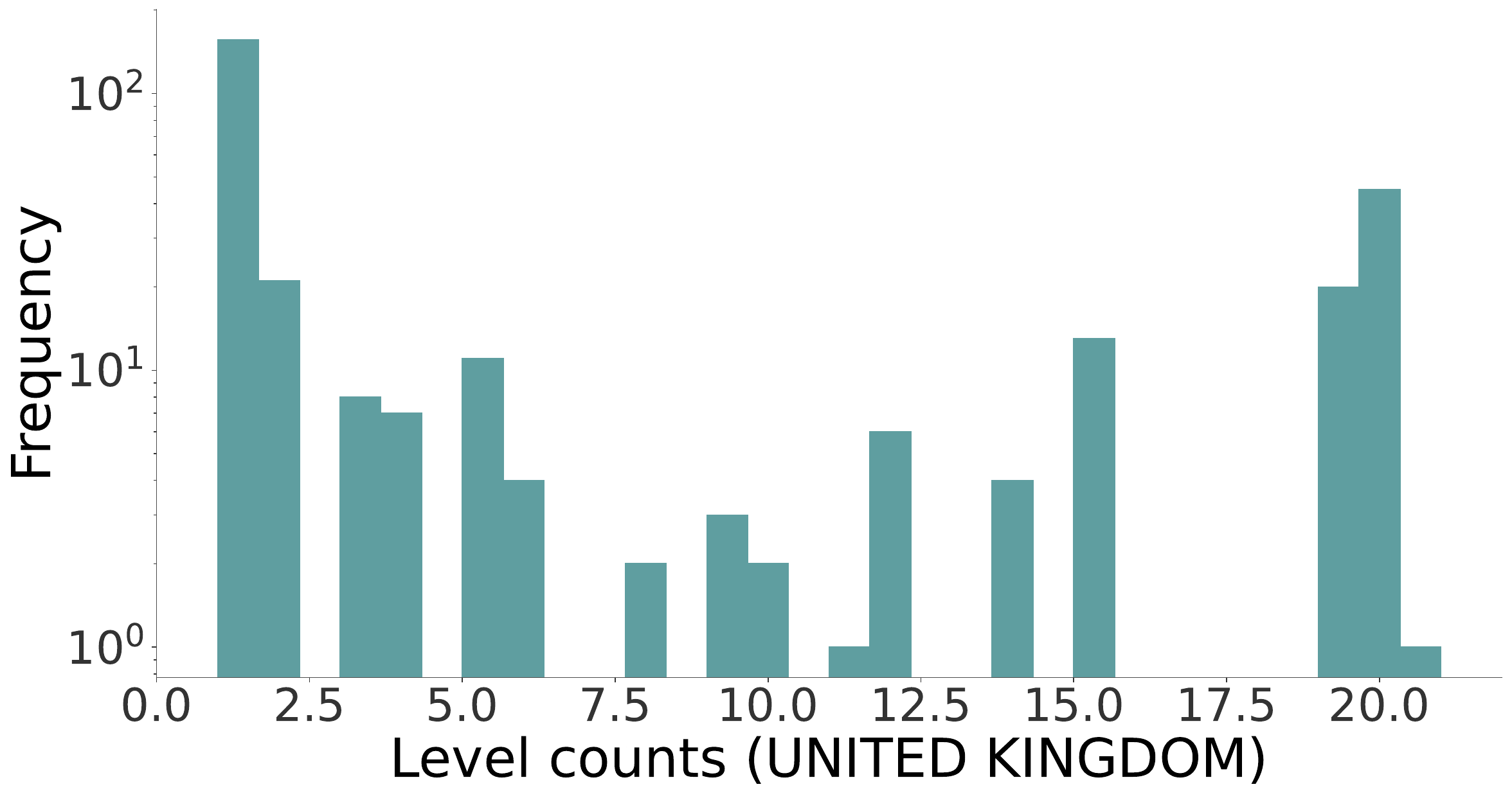}
\includegraphics[width=0.45\linewidth]{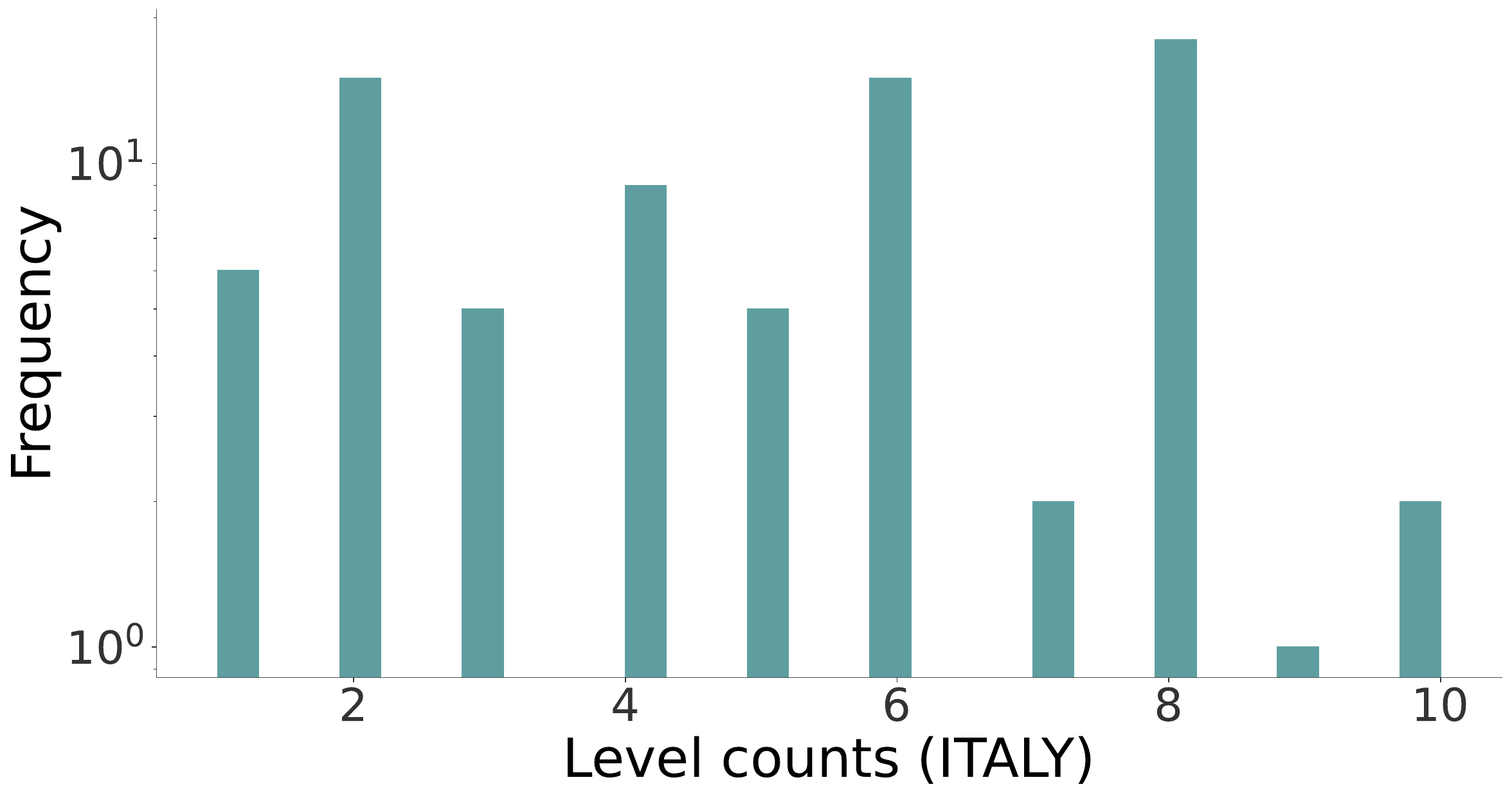}
\includegraphics[width=0.45\linewidth]{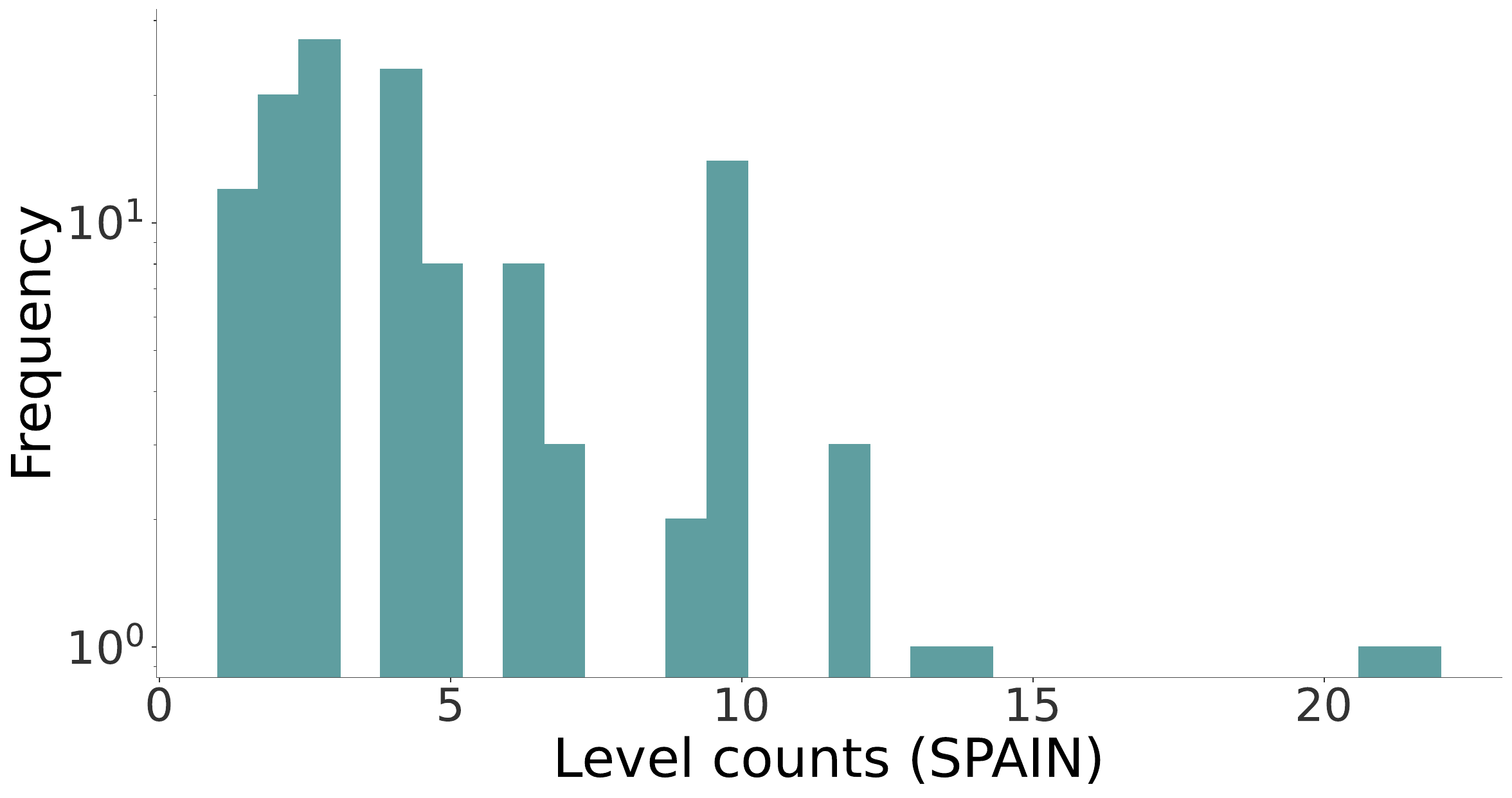}
\includegraphics[width=0.45\linewidth]{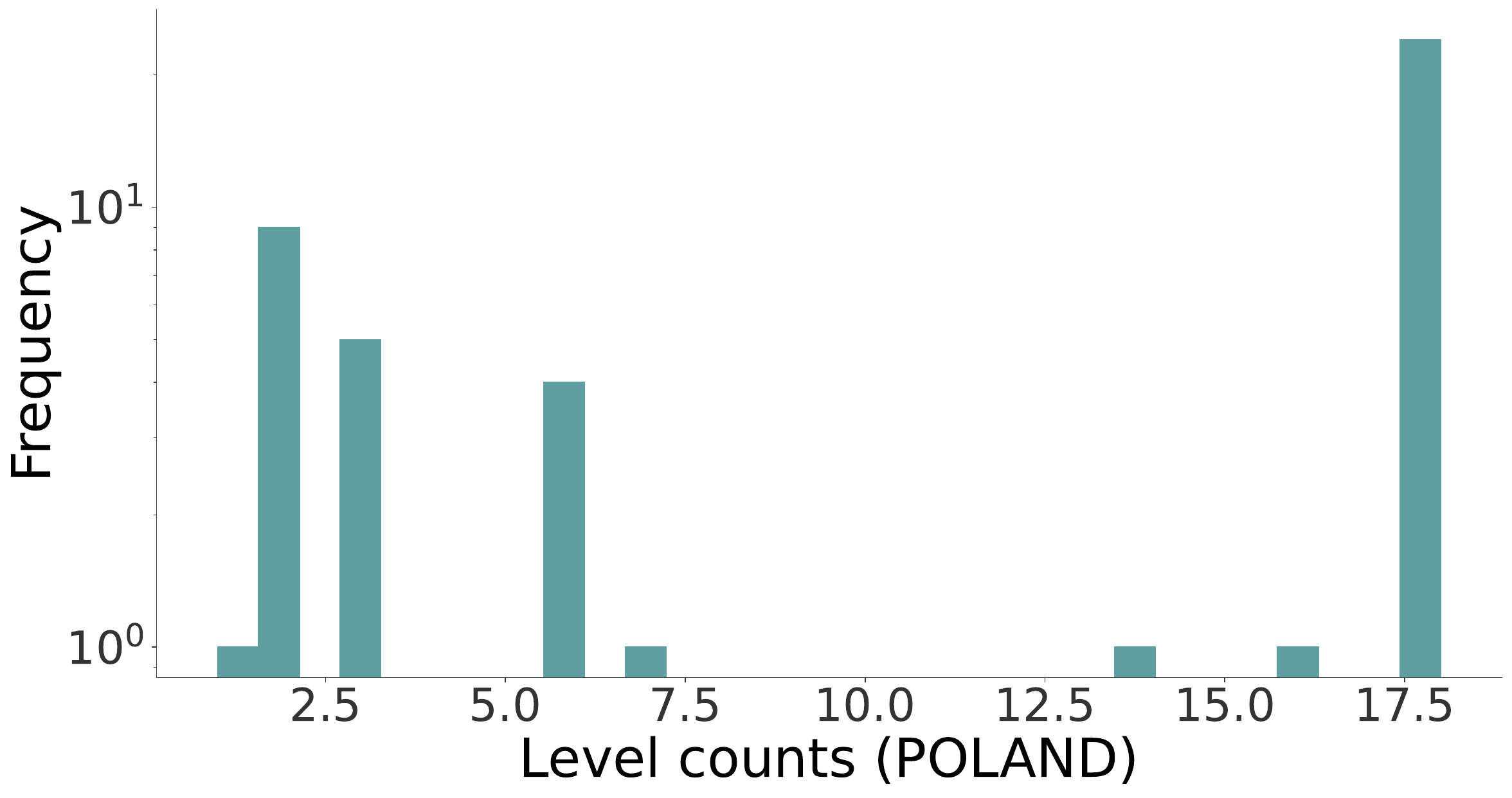}
\includegraphics[width=0.45\linewidth]{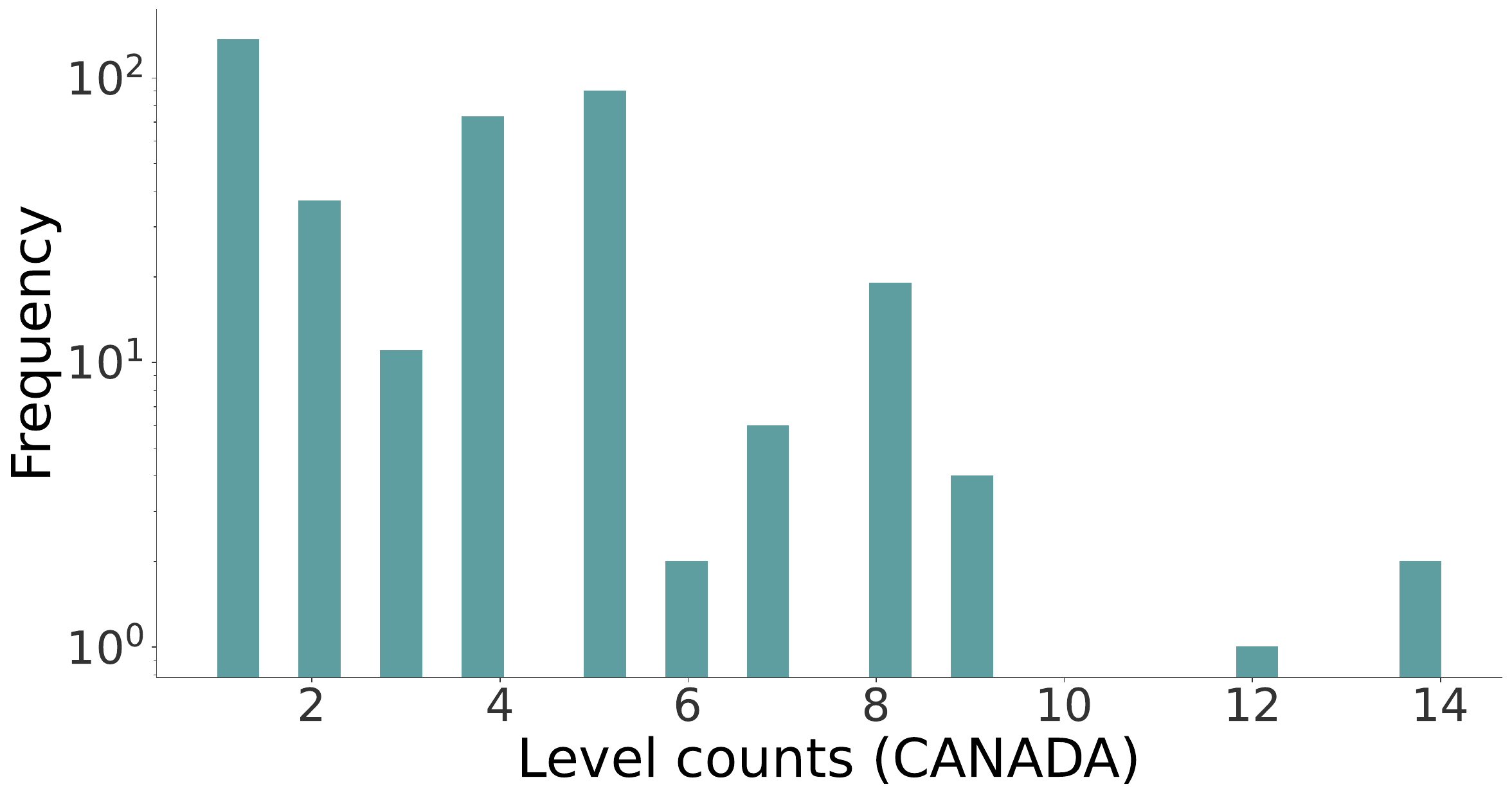}
\includegraphics[width=0.45\linewidth]{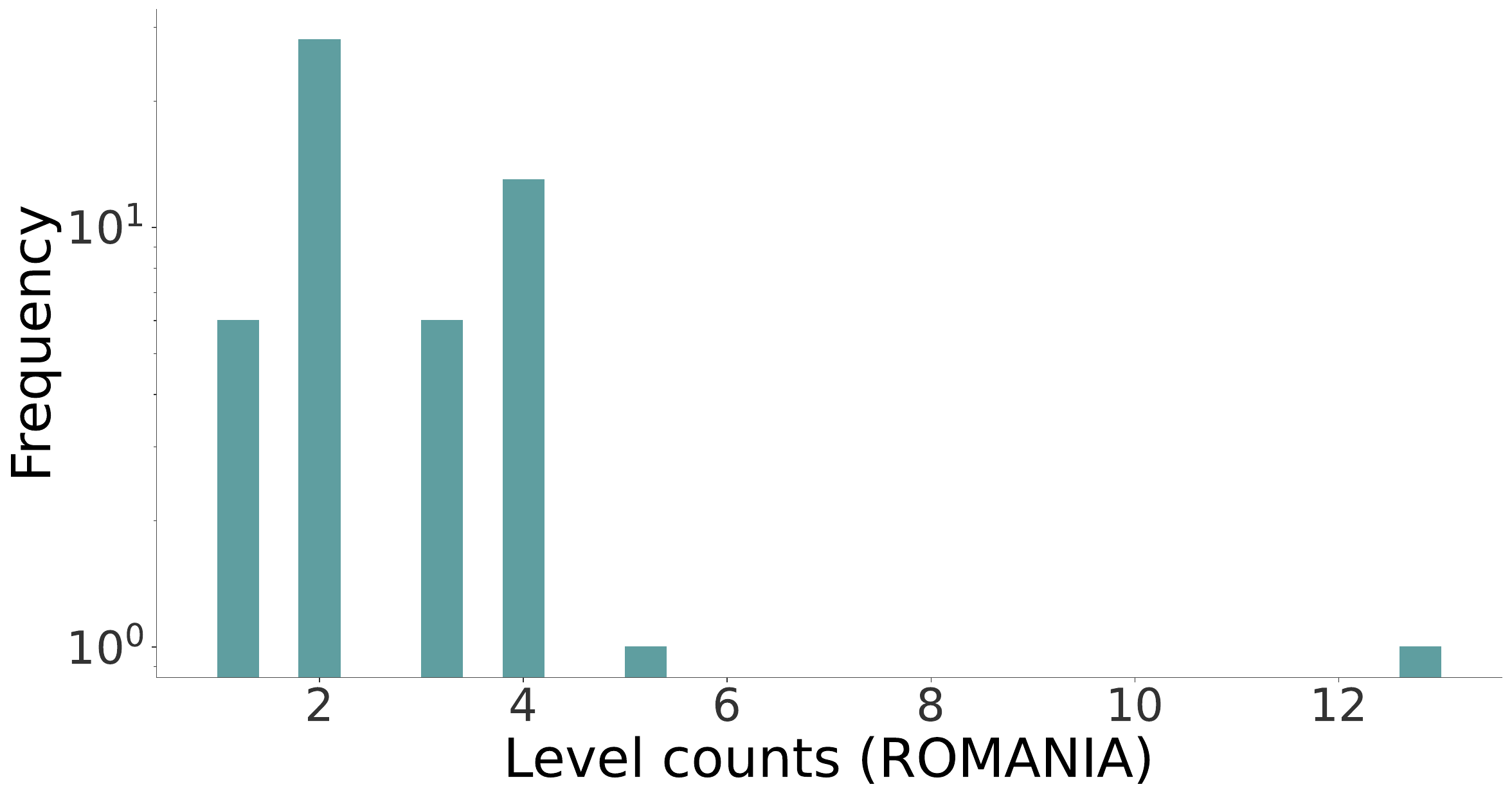}
\includegraphics[width=0.45\linewidth]{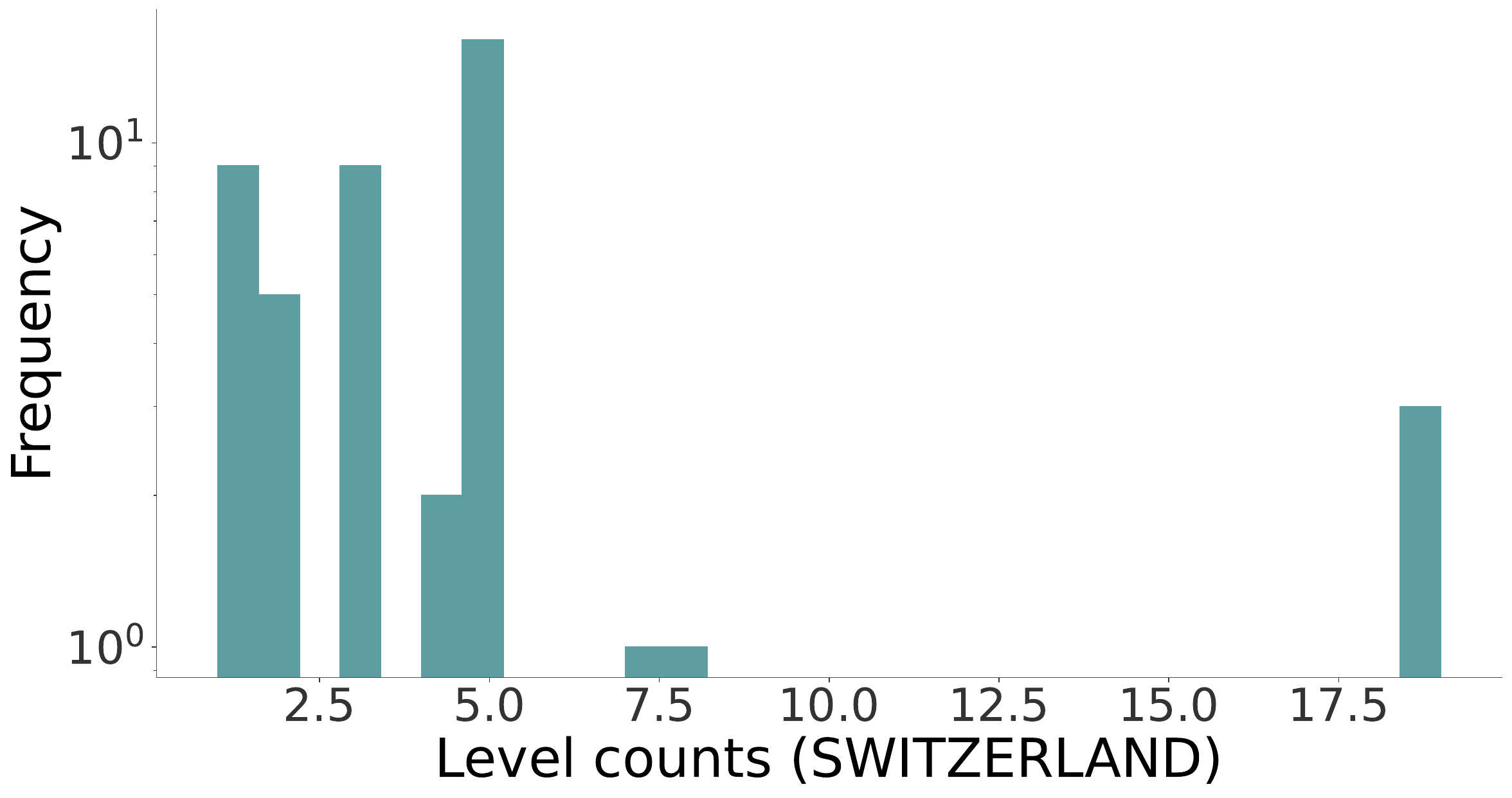}
\begin{figurenotes}
The figure shows the distribution of level counts by country, for all hard news outlets (more than 25\% political reporting). The level counts show the total number of levels in the ownership network, traversed link by link over stakes $>1\%$ in the entity directly below. This figure provides country-level detail for the aggregate distributions shown in Figure~\ref{fig:app:node_level_distributions}; see Section~\ref{sec:results} for discussion.
\end{figurenotes}
\end{figure}

\begin{figure}[htbp]
\centering
\caption{Level Count Distribution, by Country (All Outlets, Incl. Non-Political)}
\label{fig:app:level_counts_by_country_nonpol}
\includegraphics[width=0.45\linewidth]{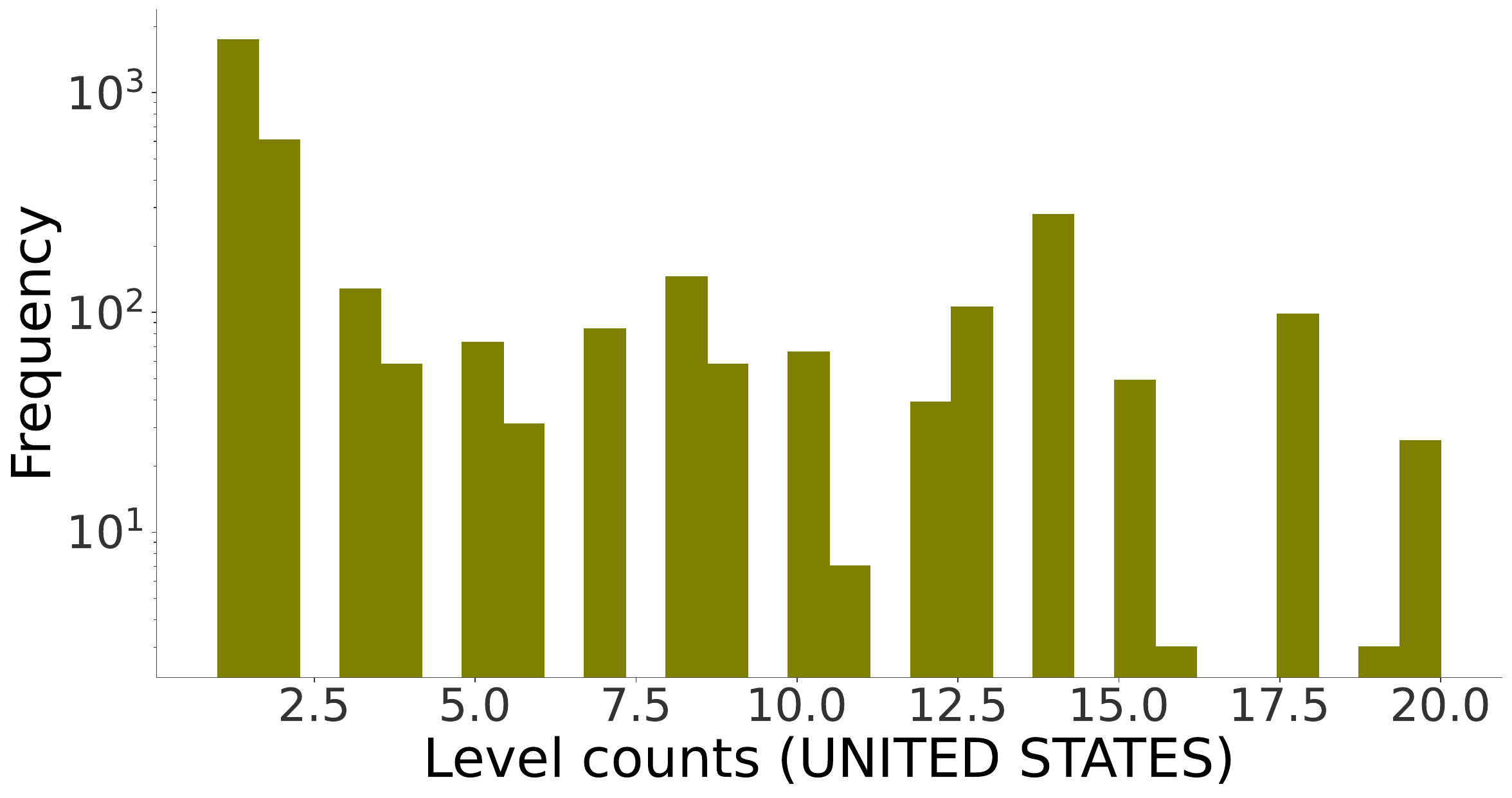}
\includegraphics[width=0.45\linewidth]{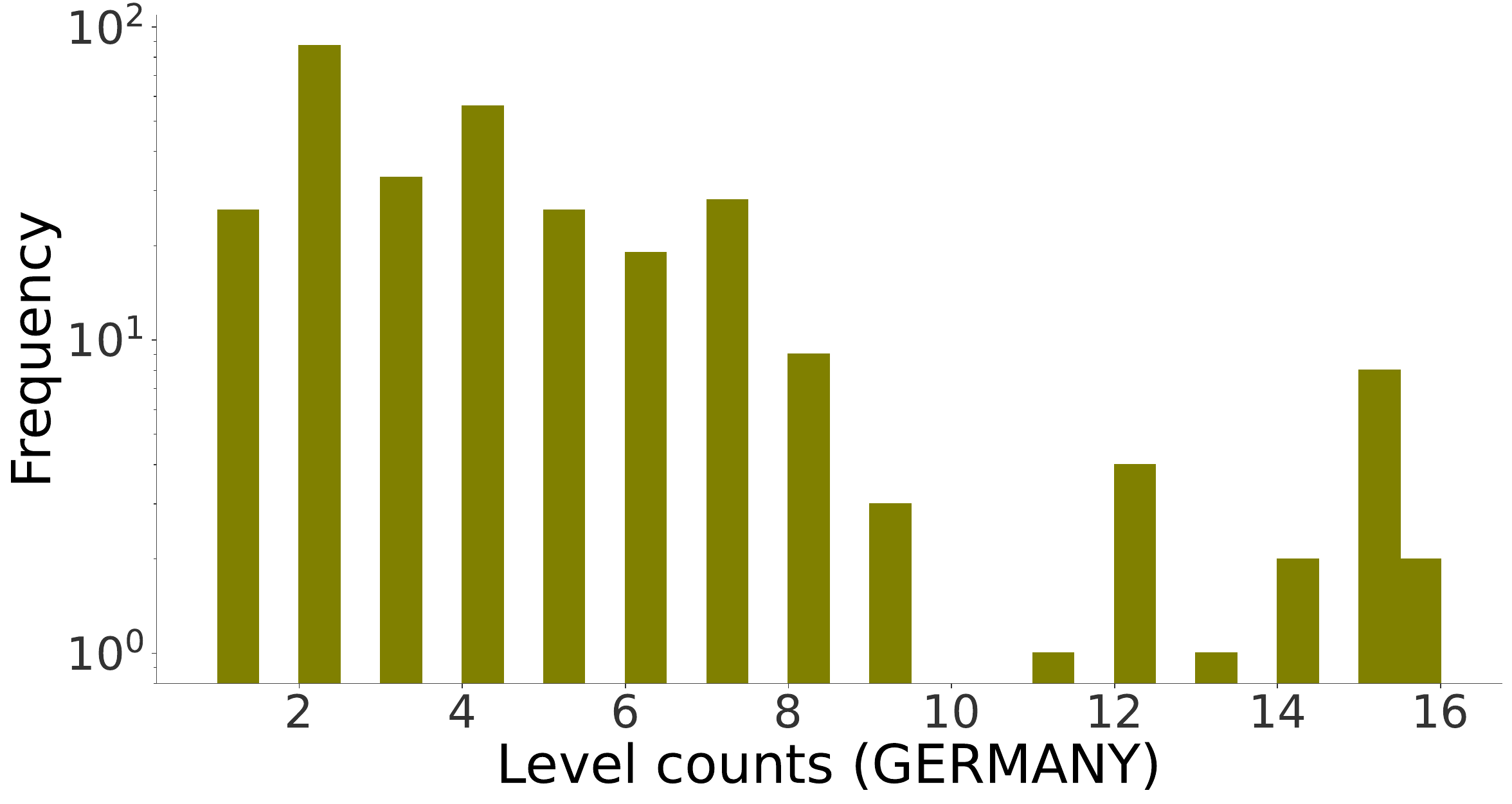}
\includegraphics[width=0.45\linewidth]{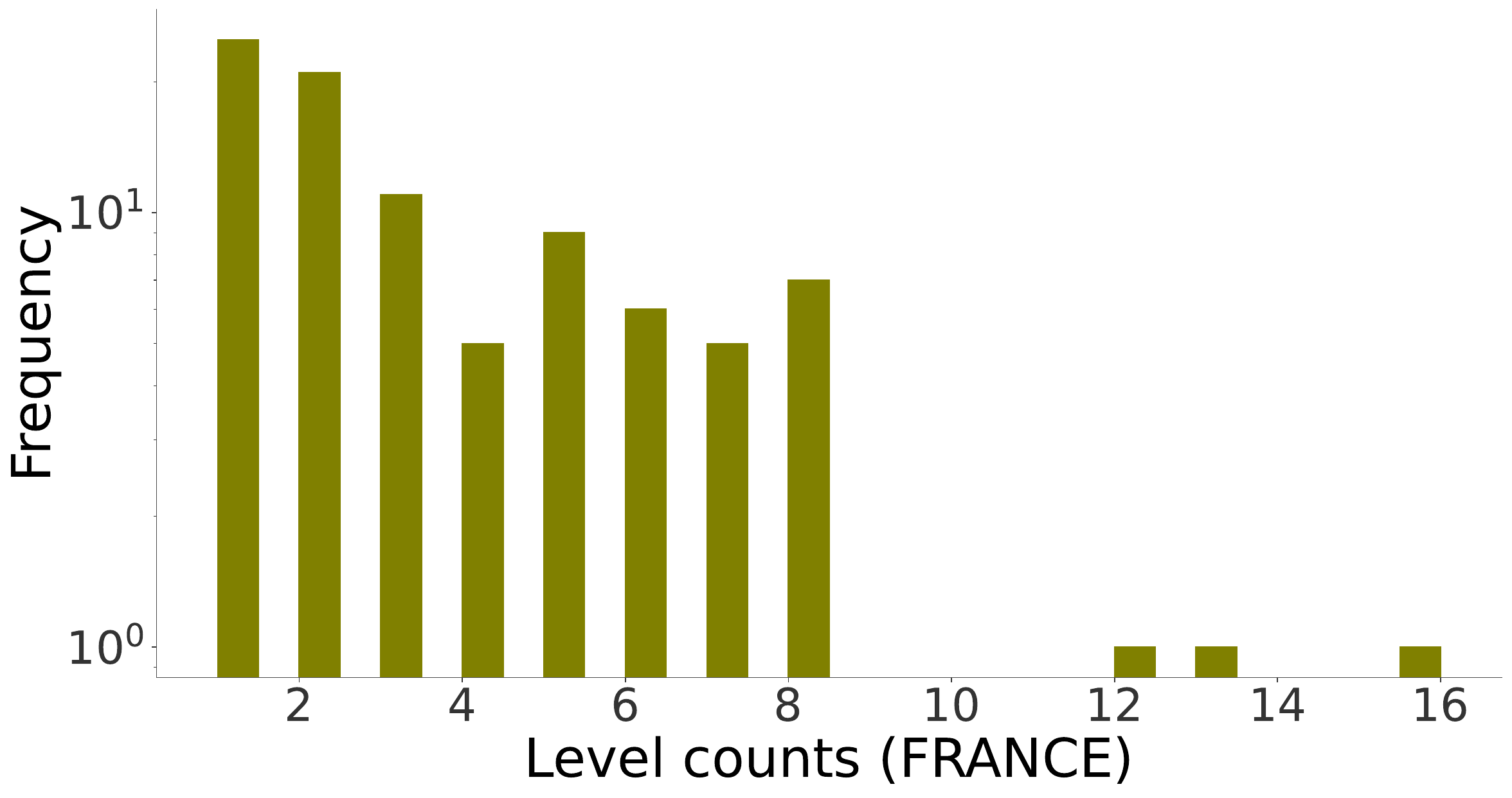}
\includegraphics[width=0.45\linewidth]{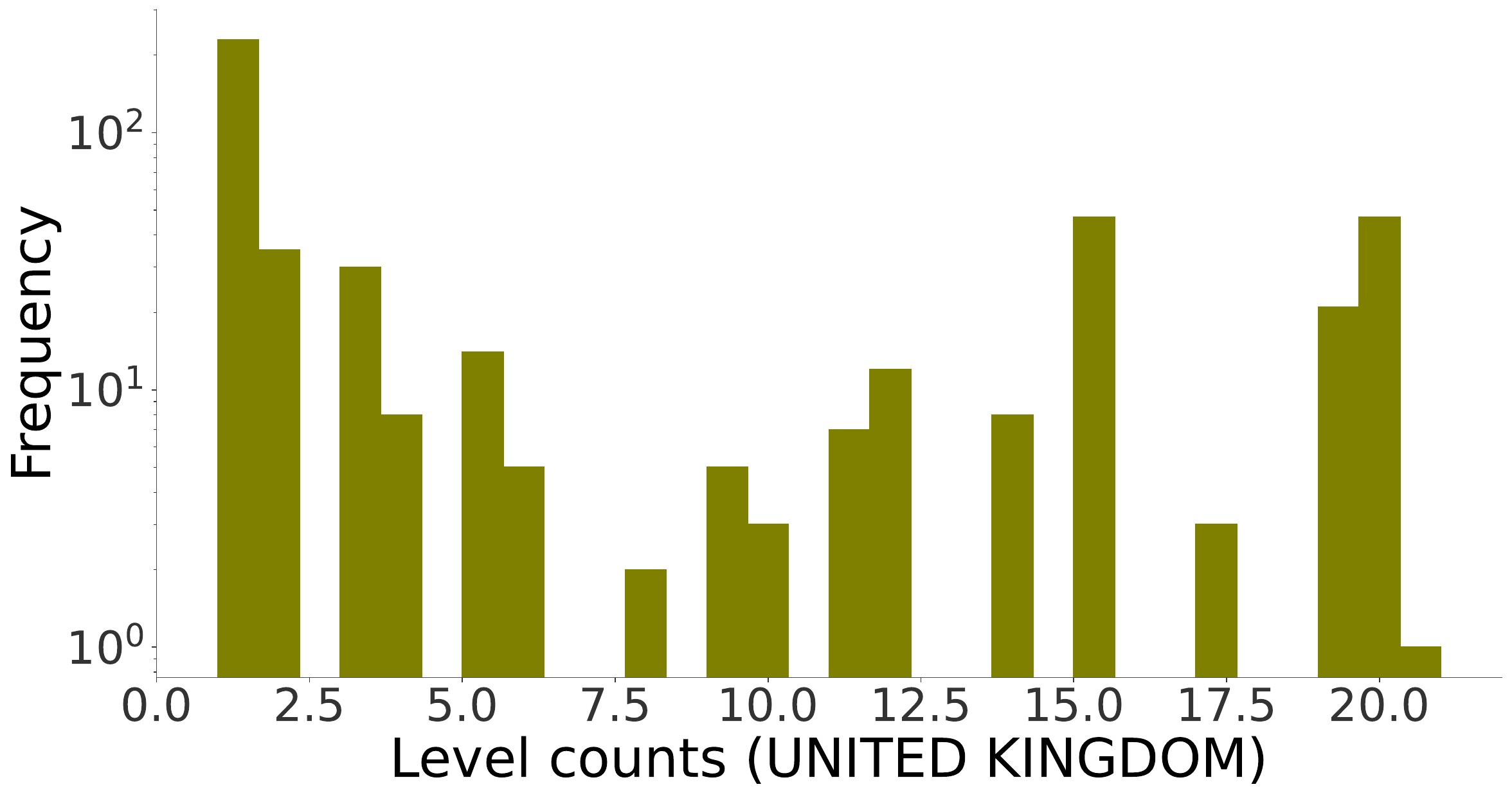}
\includegraphics[width=0.45\linewidth]{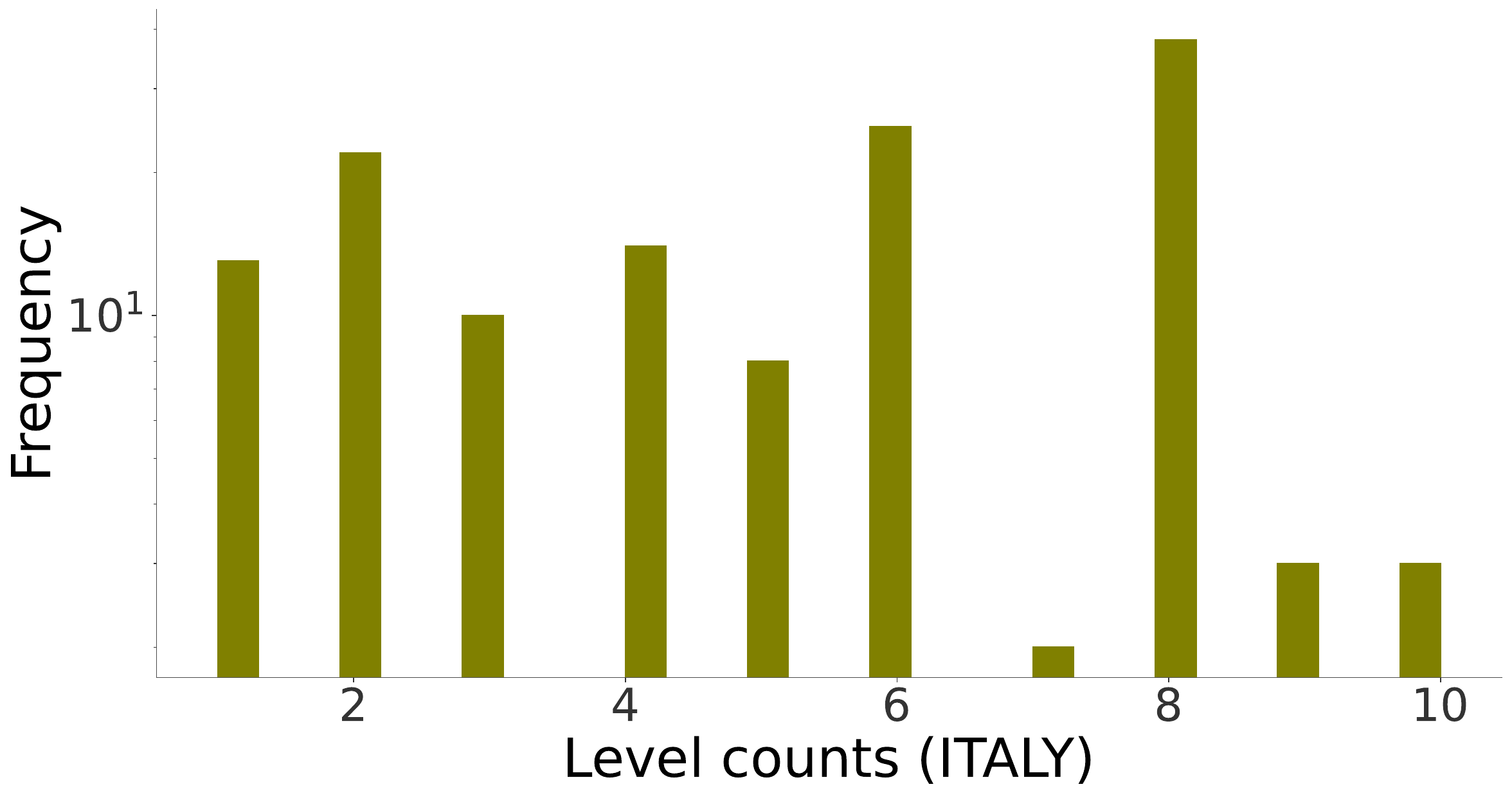}
\includegraphics[width=0.45\linewidth]{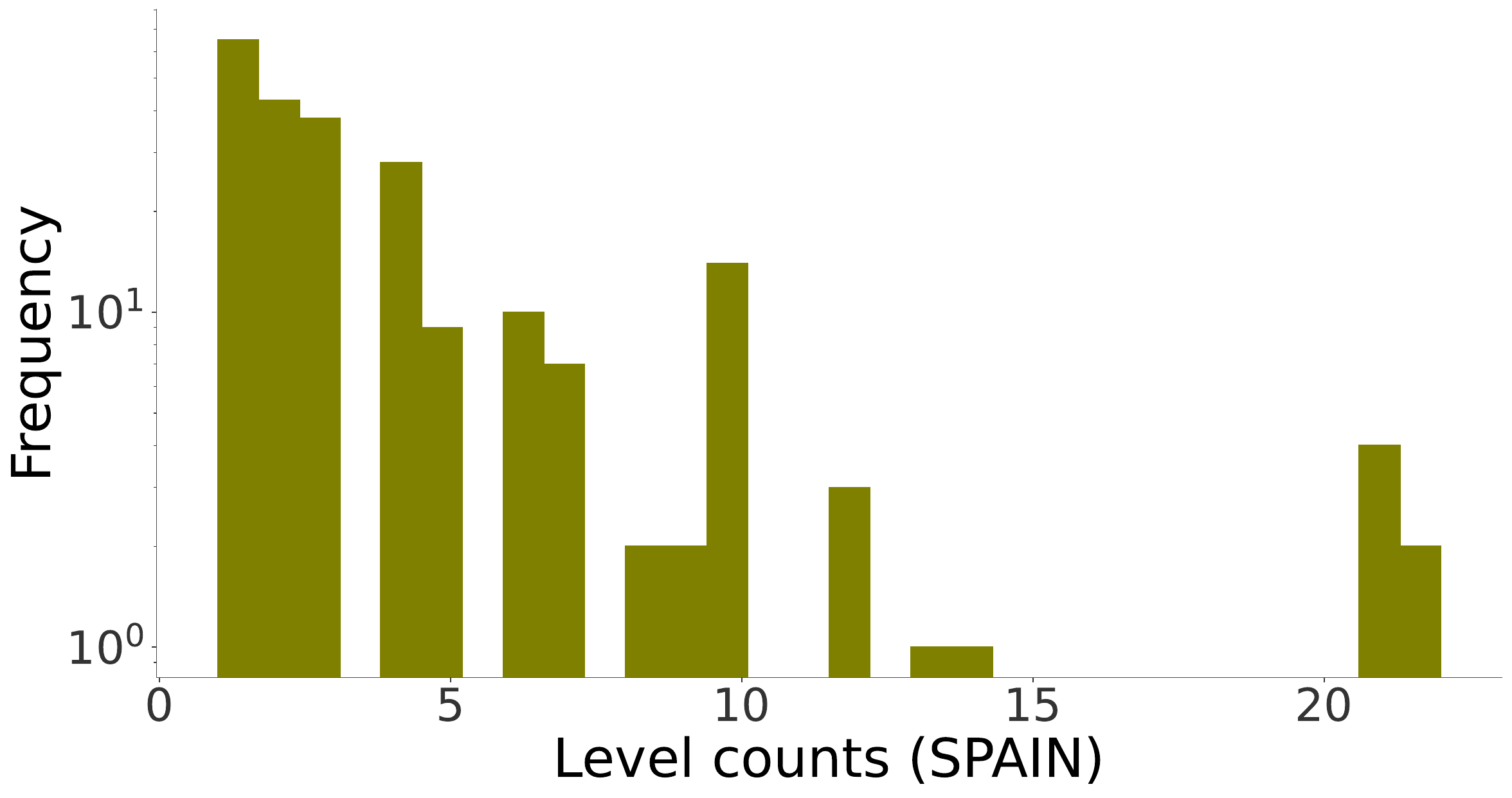}
\includegraphics[width=0.45\linewidth]{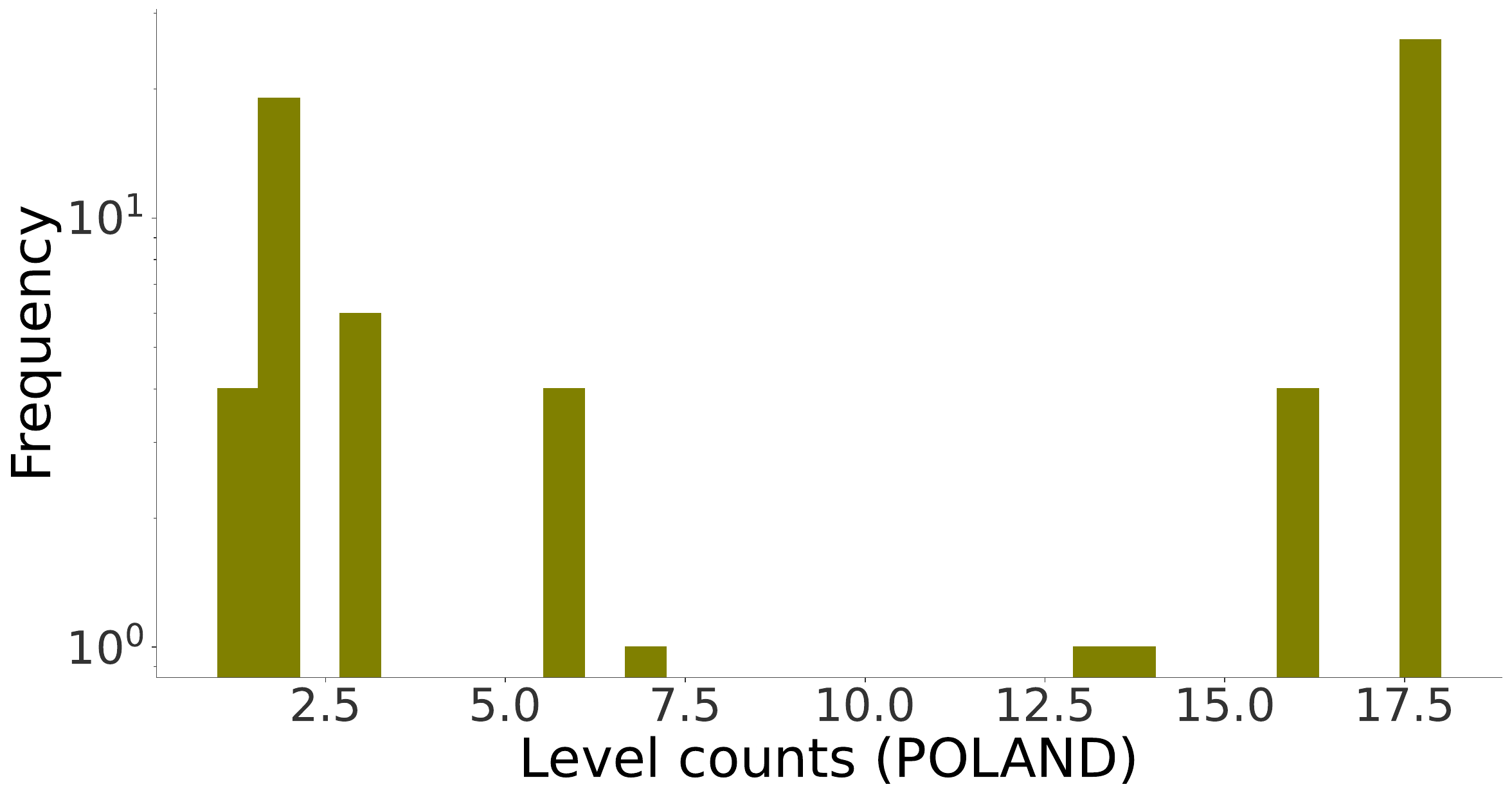}
\includegraphics[width=0.45\linewidth]{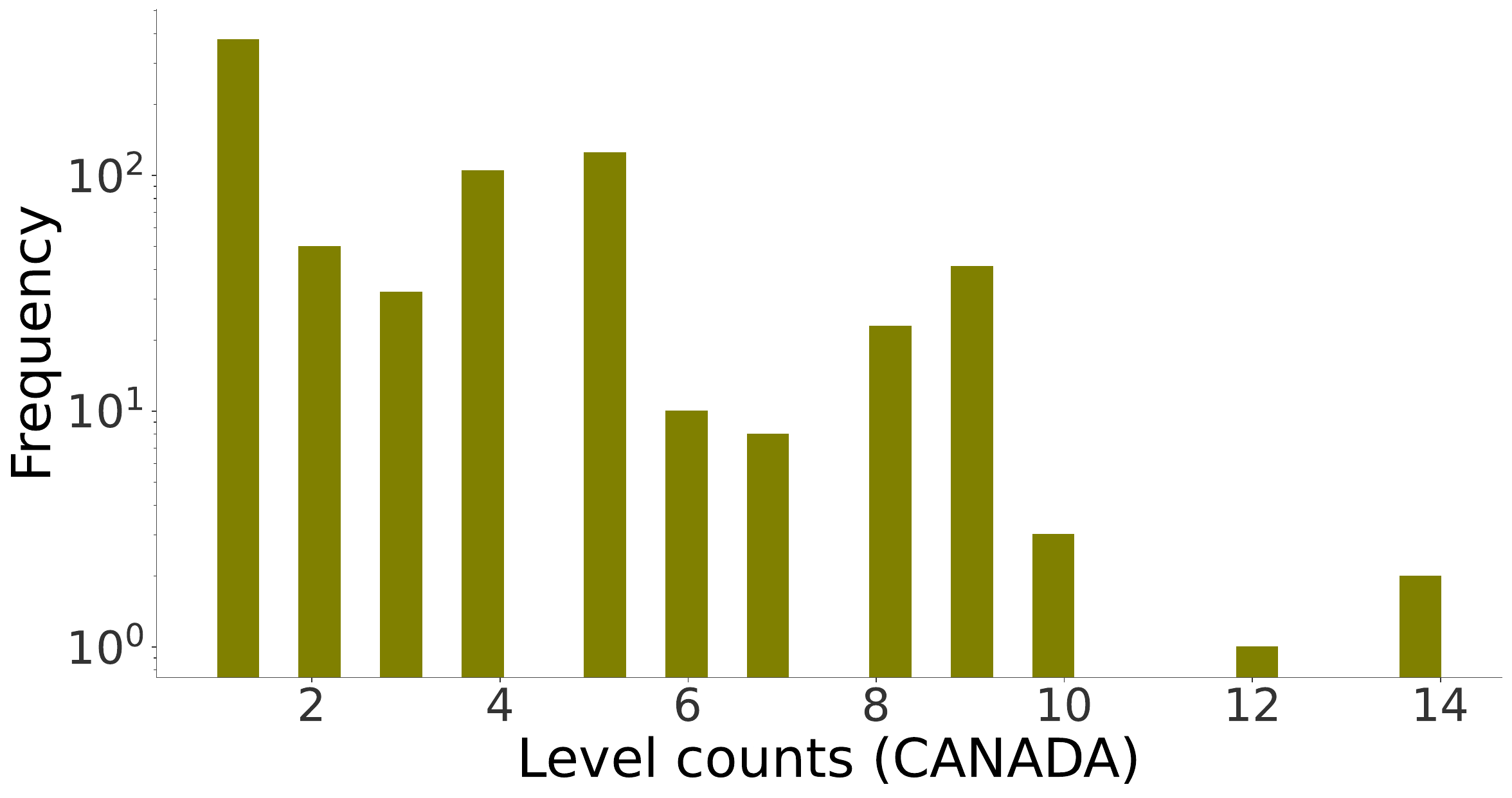}
\includegraphics[width=0.45\linewidth]{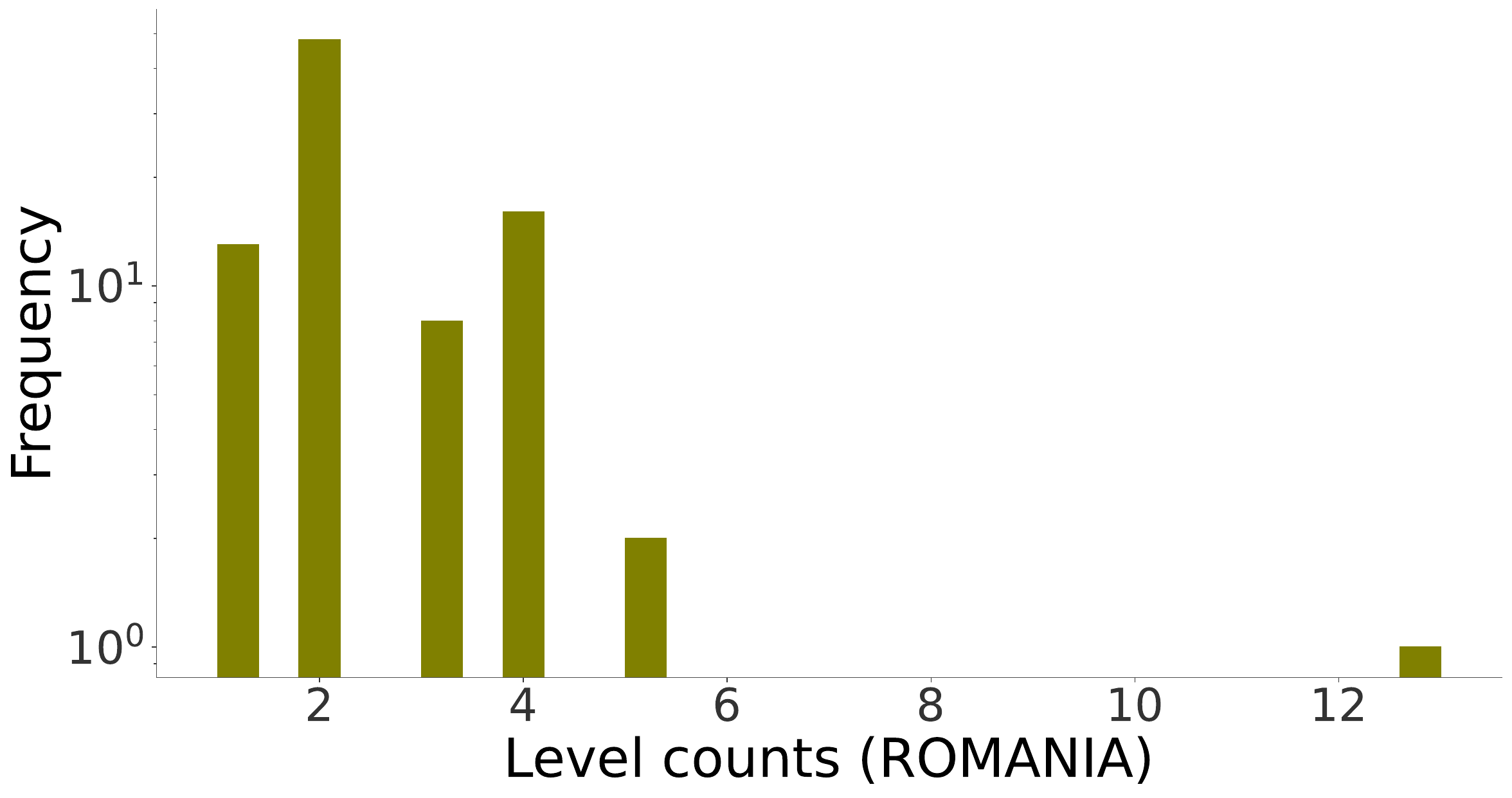}
\includegraphics[width=0.45\linewidth]{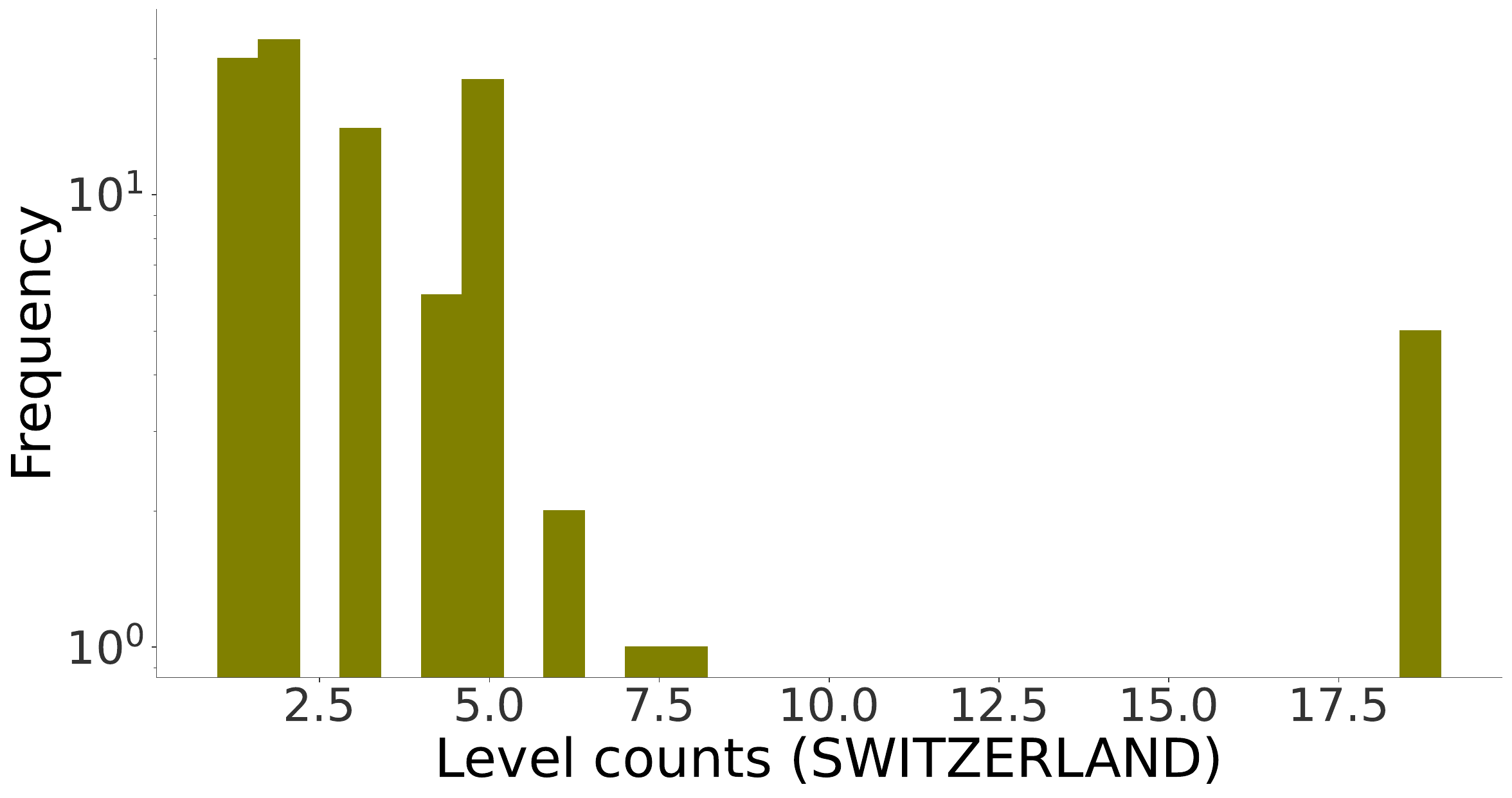}
\begin{figurenotes}
The figure shows the distribution of level counts by country, for all outlets, including those not specializing in political content. The level counts show the total number of levels in the ownership network, traversed link by link over stakes $>1\%$ in the entity directly below. This figure provides country-level detail for comparison with Figure~\ref{fig:app:level_counts_by_country}.
\end{figurenotes}
\end{figure}

\paragraph{Ownership levels. } An additional measure of ownership tractability is level counts. The right panel of Figure \ref{fig:app:node_level_distributions} shows that the number of levels ranges from 1 to 22. %\footnote{Given our coding procedure, ownership networks of outlets with more than one node mechanically come with more than one level. Recall that our registrant data is matched to one domain owner in Orbis (first level), which can then, in turn, be owned by $>1$ shareholders).}
Unsurprisingly, node counts and level are associated: The correlation is 72\% for outlets with fewer than 100 nodes and 39\% overall, suggesting level-node discrepancies are mostly driven by networks with already many nodes.\footnote{Figures~\ref{fig:app:level_counts_by_country} and \ref{fig:app:level_counts_by_country_nonpol} show the country-specific level distributions for hard news outlets and all outlets.}

\begin{figure}[htbp]
\centering
\caption{Shares of Controlled Firms} \label{fig:corr_ap_mw}
\includegraphics[width=0.79\textwidth]{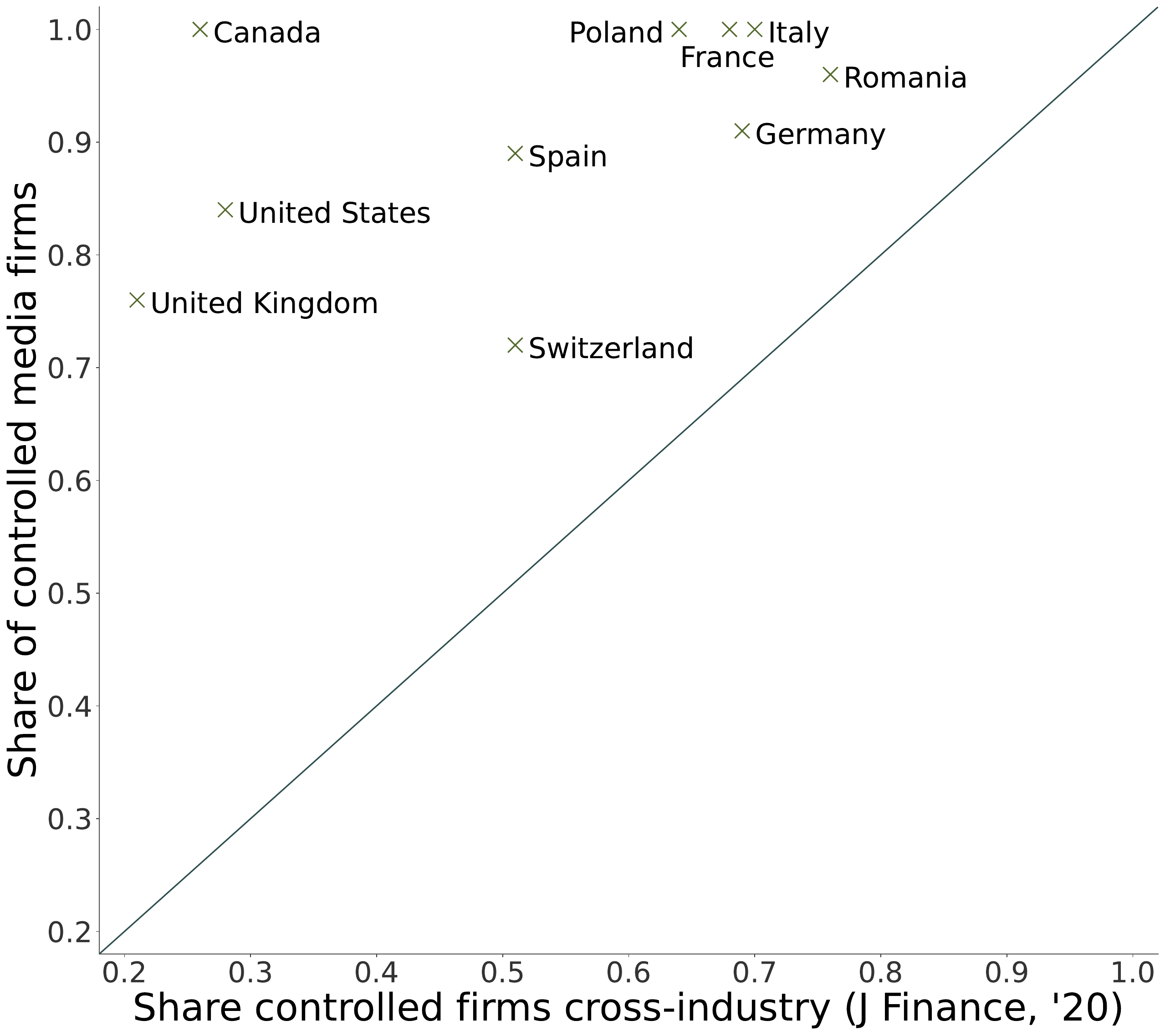}
\begin{figurenotes}
    Scatterplot of the share of ``controlled'' firms across sectors (firms where one owner holds 20\% of voting rights or more; horizontal axis) against the share of controlled firms in the media sector (vertical axis). The data on the cross-sector shares come from \citeauthor{aminadav2020corporate} (\citeyear{aminadav2020corporate}). The media sector shares are ours. Points represent shares at the country level. See Section~\ref{sec:discussion} for discussion.
\end{figurenotes}
%\begin{minipage}{0.7\textwidth}
\singlespacing
% \begin{center}
%{\footnotesize   \emph{Notes:} }
% \end{center}
%\end{minipage}
\end{figure}

\begin{table}[htbp]
\centering
\caption{Ownership Tractability by Country}
\label{tab:ownership_transparency_by_country}
\begin{threeparttable}
\begin{tabular}{lcccccc}
\toprule
 & \multicolumn{2}{c}{$<$5 Nodes} & \multicolumn{2}{c}{1 Owner} & \multicolumn{2}{c}{Direct} \\
\cmidrule(lr){2-3} \cmidrule(lr){4-5} \cmidrule(lr){6-7}
 & Share & Weighted & Share & Weighted & Share & Weighted \\
\midrule
United States & 64\% & 58\% & 64\% & 57\% & 45\% & 42\% \\
United Kingdom & 58\% & 36\% & 57\% & 34\% & 51\% & 28\% \\
Romania & 55\% & 11\% & 37\% & 9\% & 12\% & 2\% \\
France & 45\% & 47\% & 41\% & 45\% & 31\% & 11\% \\
Canada & 41\% & 53\% & 60\% & 55\% & 37\% & 46\% \\
Switzerland & 41\% & 26\% & 39\% & 25\% & 20\% & 23\% \\
Italy & 26\% & 8\% & 23\% & 12\% & 9\% & 1\% \\
Spain & 26\% & 7\% & 28\% & 4\% & 10\% & 1\% \\
Poland & 24\% & 5\% & 20\% & 4\% & 2\% & 0\% \\
Germany & 21\% & 18\% & 15\% & 10\% & 8\% & 9\% \\
\bottomrule
\end{tabular}
\begin{tablenotes}[flushleft]
\footnotesize
\item \textit{Notes}: This table reports ownership tractability metrics for hard news outlets (more than 25\% political reporting) by country. ``$<$5 Nodes'' shows the share of outlets with relatively simple ownership structures. ``1 Owner'' shows the share with a single ultimate owner. ``Direct'' shows outlets where the domain owner is itself the ultimate owner. The first column in each pair reports unweighted shares; the second column weights by web traffic (reach). Countries are ordered by share of outlets with $<$5 nodes (descending).
\end{tablenotes}
\end{threeparttable}
\end{table}

\begin{table}[htbp]
\centering
\caption{Ownership Tractability: Hard News vs.\ All Outlets}
\label{tab:ownership_transparency_all}
\begin{threeparttable}
\begin{tabular}{lcccc}
\toprule
 & \multicolumn{2}{c}{Hard News} & \multicolumn{2}{c}{All Outlets} \\
\cmidrule(lr){2-3} \cmidrule(lr){4-5}
 & Share & Weighted & Share & Weighted \\
\midrule
With $<$5 nodes & 55\% & 45\% & 59\% & 44\% \\
With 1 node & 39\% & 33\% & 43\% & 32\% \\
\bottomrule
\end{tabular}
\begin{tablenotes}[flushleft]
\footnotesize
\item \textit{Notes}: This table compares ownership network complexity between hard news outlets (more than 25\% political reporting) and all outlets in our sample. Ultimate owner data is only available for hard news outlets. The patterns are similar: hard news outlets have slightly more complex ownership structures (55\% vs.\ 59\% with $<$5 nodes), but when weighted by traffic, the shares are nearly identical (45\% vs.\ 44\%).
\end{tablenotes}
\end{threeparttable}
\end{table}

\clearpage

\subsection{Sector Shares}

This subsection reports the sector composition of the online news market. Table~\ref{tab:shares_sectors_political_grouped} contains the exact values underlying Figure~\ref{fig:sector_shift} in the main text, using the broad sector groupings defined there (the media group combines the Orbis categories ``Printing \& Publishing'' and ``Media \& Broadcasting''; the financial group combines ``Banking, Insurance \& Financial Services'' and ``Business Services'', which mostly comprise holding and asset-management companies). Tables~\ref{tab:shares_sectors_political} and \ref{tab:shares_sectors} report the disaggregated sector shares, without the broad groupings, for hard news outlets and for all outlets, respectively; similar patterns emerge in both.

\begin{table}[ht!]\centering \caption{Country-Level Ownership Shares by Sectors} \label{tab:shares_sectors_political_grouped}
\resizebox{0.8\textwidth}{!}{
\begin{threeparttable}
\small

\begin{tabular}{l*{3}{c}}
\toprule
Country & Sector & Share & Share \\
& & Owner Level 1 & Ult. Owners \\
& & Visitors & Visitors \\
\midrule
\textbf{U.S.} & Media Sector (broad) & 0.84 & 0.64 \\
 & Financial Sector (broad) & 0.07 & 0.27 \\
% \textbf{U.S.} & Communications & 0.04 & 0.05 \\
 & & & \\
\textbf{U.K.} & Media Sector (broad) & 0.98 & 0.09 \\
 & Financial Sector (broad) & 0.00 & 0.88 \\
& & & \\
\textbf{SWITZERLAND} & Media Sector (broad) & 0.90 & 0.98 \\
 & Financial Sector (broad) & 0.10 & 0.0 \\
 & & & \\
\textbf{SPAIN} & Media Sector (broad) & 0.82 & 0.14 \\
 & Financial Sector (broad) & 0.16 & 0.72 \\
% \textbf{SPAIN} & Public Administration & 0.00 & 0.09 \\
 & & & \\
\textbf{ROMANIA} & Media Sector (broad) & 0.79 & 0.3 \\
 & Financial Sector (broad) & 0.20 & 0.46 \\
% \textbf{ROMANIA} & Public Administration & 0.01 & 0.18 \\
& & & \\
\textbf{POLAND} & Media Sector (broad) & 0.98 & 0.32 \\
 & Financial Sector (broad) & 0.01 & 0.43 \\
% \textbf{POLAND} & Public Administration & 0.00 & 0.24 \\
 & & & \\
\textbf{ITALY} & Media Sector (broad) & 0.98 & 0.02 \\
 & Financial Sector (broad) & 0.02 & 0.89 \\
% \textbf{ITALY} & Public Administration & 0.00 & 0.07 \\
 & & & \\
\textbf{GERMANY} & Media Sector (broad) & 0.82 & 0.17 \\
 & Financial Sector (broad) & 0.08 & 0.71 \\
% \textbf{GERMANY} & Public Administration & 0.00 & 0.11 \\
 & & & \\
\textbf{FRANCE} & Media Sector (broad) & 0.98 & 0.43 \\
 & Financial Sector (broad) & 0.02 & 0.54 \\
 & & & \\
\textbf{CANADA} & Media Sector (broad) & 0.77 & 0.71 \\
 & Financial Sector (broad) & 0.18 & 0.27 \\
\bottomrule
\hline\hline
\end{tabular}
\begin{tablenotes}[flushleft]
\footnotesize{\item Notes: Ownership shares by sector based on unique visitors, based on 2020 ownership data and averaged 2018-2020 traffic data. The sample includes hard news outlets. Column 2 shows the sectors to which the owners belong according to the Orbis database. The Orbis categories ``Printing \& Publishing'' and ``Media \& Broadcasting'' are grouped into ``Media Sector (broad)''. Similarly, we group ``Banking, Insurance \& Financial Services'' and ``Business Services'' into ``Financial Sector (broad)''. Sectors with ownership shares $<$5\% both for immediate and ultimate owners are not shown. Column 3 shows the sector-specific ownership shares for domain owners (e.g., the immediate owner of the nytimes.com, according to the Orbis database, is the \textit{NEW YORK TIMES COMPANY}). In column 4, the shares are for ultimate owners.}
\end{tablenotes}
\end{threeparttable}
}
\end{table}

\begin{table}[tb]\centering \caption{Country-Level Ownership Shares by Sectors} \label{tab:shares_sectors_political}
\scalebox{0.60}[0.60]{
\begin{threeparttable}
\begin{tabular}{l*{5}{c}}
\toprule
Country & Sector &  Share &  Share & Share &  Share \\
& &  Owner &  Owner &  Ult. Owners &  Ult. Owners \\
& &  Level 1 &  Level 1 &  Ult. Owners &  Ult. Owners \\
& &  Page views &  Visitors &  Page views &.  Visitors \\
\midrule
\textbf{UNITED STATES} & Printing \& Publishing & 0.53 & 0.55 & 0.52 & 0.52 \\
 & Media \& Broadcasting & 0.31 & 0.28 &  searrow 0.1 &  searrow 0.09 \\
 & Business Services & 0.07 & 0.07 & 0.16 & 0.17 \\
 & Banking, Insurance \& Financial Services & 0.00 & 0.00 & 0.09 & 0.1 \\
   & Communications & 0.04 & 0.04 & 0.05 & 0.05 \\[1.8ex]
\textbf{UNITED KINGDOM} & Printing \& Publishing & 0.81 & 0.82 &  searrow 0.05 &  searrow 0.07 \\
 & Media \& Broadcasting & 0.15 & 0.14 &  searrow 0.0 &  searrow 0.0 \\
 & Business Services & 0.00 & 0.00 &  nearrow 0.65 &  nearrow 0.61 \\
 & Banking, Insurance \& Financial Services & 0.00 & 0.00 &  nearrow 0.26 &  nearrow 0.27 \\[1.8ex]
\textbf{SWITZERLAND} & Printing \& Publishing & 0.67 & 0.67 &  searrow 0.07 &  searrow 0.09 \\
 & Media \& Broadcasting & 0.25 & 0.23 &  nearrow 0.91 &  nearrow 0.89 \\
 & Business Services & 0.08 & 0.10 & 0.0 & 0.0 \\[1.8ex]
\textbf{SPAIN} & Printing \& Publishing & 0.70 & 0.71 &  searrow 0.11 &  searrow 0.14 \\
 & Business Services & 0.18 & 0.16 &  nearrow 0.45 &  nearrow 0.44 \\
 & Media \& Broadcasting & 0.08 & 0.09 & 0.0 & 0.0 \\
 & Banking, Insurance \& Financial Services & 0.00 & 0.00 &  nearrow 0.3 &  nearrow 0.28 \\
 & Public Administration, Education, Health Social Services & 0.00 & 0.00 & 0.09 & 0.09 \\[1.8ex]
\textbf{ROMANIA} & Printing \& Publishing & 0.50 & 0.47 & 0.4 &  searrow 0.3 \\
 & Media \& Broadcasting & 0.22 & 0.24 &  searrow 0.0 &  searrow 0.0 \\
 & Business Services & 0.19 & 0.20 &  searrow 0.08 &  searrow 0.09 \\
 & Information Services & 0.09 & 0.08 & 0.0 & 0.0 \\
 & Public Administration, Education, Health Social Services & 0.00 & 0.01 &  nearrow 0.13 &  nearrow 0.18 \\
 & Banking, Insurance \& Financial Services & 0.00 & 0.00 &  nearrow 0.33 &  nearrow 0.37 \\
   & Communications & 0.0 & 0.0 & 0.05 & 0.06 \\[1.8ex]
\textbf{POLAND} & Media \& Broadcasting & 0.67 & 0.65 &  searrow 0.0 &  searrow 0.0 \\
 & Printing \& Publishing & 0.32 & 0.33 & 0.29 & 0.32 \\
 & Business Services & 0.01 & 0.01 & 0.1 & 0.11 \\
 & Banking, Insurance \& Financial Services & 0.00 & 0.00 &  nearrow 0.34 &  nearrow 0.32 \\
 & Public Administration, Education, Health Social Services & 0.00 & 0.00 &  nearrow 0.27 &  nearrow 0.24 \\[1.8ex]
\textbf{ITALY} & Printing \& Publishing & 0.87 & 0.83 &  searrow 0.02 &  searrow 0.02 \\
 & Information Services & 0.09 & 0.12 & 0.0 &  searrow 0.0 \\
 & Business Services & 0.01 & 0.02 &  nearrow 0.79 &  nearrow 0.77 \\
 & Banking, Insurance \& Financial Services & 0.00 & 0.00 &  nearrow 0.12 &  nearrow 0.12 \\
 & Public Administration, Education, Health Social Services & 0.00 & 0.00 & 0.05 & 0.07 \\[1.8ex]
\textbf{GERMANY} & Printing \& Publishing & 0.44 & 0.48 &  searrow 0.03 &  searrow 0.05 \\
 & Media \& Broadcasting & 0.23 & 0.20 &  searrow 0.11 & 0.12 \\
 & Information Services & 0.19 & 0.14 &  searrow 0.0 &  searrow 0.0 \\
 & Business Services & 0.06 & 0.08 &  nearrow 0.66 &  nearrow 0.65 \\
 & Banking, Insurance \& Financial Services & 0.00 & 0.00 & 0.07 & 0.06 \\
 & Public Administration, Education, Health Social Services & 0.00 & 0.00 &  nearrow 0.12 &  nearrow 0.11 \\[1.8ex]
\textbf{FRANCE} & Printing \& Publishing & 0.95 & 0.94 &  searrow 0.49 &  searrow 0.43 \\
 & Business Services & 0.02 & 0.02 &  nearrow 0.42 &  nearrow 0.48 \\
 & Banking, Insurance \& Financial Services & 0.00 & 0.00 & 0.07 & 0.06 \\[1.8ex]
\textbf{CANADA} & Media \& Broadcasting & 0.42 & 0.47 & 0.42 & 0.48 \\
 & Printing \& Publishing & 0.28 & 0.30 & 0.2 & 0.23 \\
 & Business Services & 0.25 & 0.18 & 0.34 & 0.24 \\
\bottomrule
\hline\hline
\end{tabular}
\begin{tablenotes}[flushleft]
\footnotesize{\item Notes: Ownership shares by sector and country, based on page views (columns 3 and 5) and unique visitors (columns 4 and 6) -- considering all outlets reporting on politics in more than 25\% of their articles. Ownership data is from 2020, and traffic data are averaged for 2018-2020. Column 2 shows the sectors to which the owners belong according the Orbis database. Sectors with ownership shares below 5\% both for immediate and ultimate owners are not shown (for better readability). Columns 3 and 4 show the sector-specific ownership shares at the level of the immediate owner (e.g., the immediate owner of the nytimes.com, according to the Orbis database, is the \textit{NEW YORK TIMES COMPANY}). In column 7 and 8, the measures are calculated at the level of ultimate owners (e.g., there are around 3,000 ultimate owner entities for nytimes.com, among others \textit{CARLOS SLIM HELU Y FAMILIA}, \textit{JACKSON SQUARE LLC}, or \textit{VANGUARD GROUP INC}). Sectors where the share changes by more than 10 percentage points when moving from the immediate owner to the ultimate owner (i.e., moving from column 3 to 5 or from column 4 to 6) are highlighted with an arrow ($\nearrow$ for increases and $\searrow$ for decreases).}
% A table with all non-zero shares is shown in the Appendix.
\end{tablenotes}
\end{threeparttable}
}
\end{table}

\begin{table}[tb]\centering \caption{Country-Level Ownership Shares by Sectors (All Outlets, incl. Non-Political)} \label{tab:shares_sectors}
\scalebox{0.64}[0.64]{
\begin{threeparttable}
\begin{tabular}{l*{5}{c}}
\toprule
Country & Sector &  Share &  Share & Share &  Share \\
& &  Owner Level 1 &  Owner Level 1 &  Ult. Owners &  Ult. Owners \\
& &  Page views &  Visitors &  Page views &.  Visitors \\
\midrule
\textbf{UNITED STATES} & Printing \& Publishing & 0.48 & 0.51 & 0.48 & 0.48 \\
 & Media \& Broadcasting & 0.29 & 0.26 &  searrow 0.09 &  searrow 0.08 \\
 & Business Services & 0.10 & 0.08 & 0.17 & 0.17 \\
 & Communications & 0.06 & 0.06 & 0.05 & 0.05 \\
 & Information Services & 0.03 & 0.03 & 0.07 & 0.06 \\
 & Banking, Insurance \& Financial Services & 0.0 & 0.0 & 0.1 &  nearrow 0.11 \\[1.8ex]
\textbf{UNITED KINGDOM} & Printing \& Publishing & 0.77 & 0.79 &  searrow 0.06 &  searrow 0.08 \\
 & Media \& Broadcasting & 0.19 & 0.17 &  searrow 0.0 &  searrow 0.0 \\
 & Banking, Insurance \& Financial Services & 0.0 & 0.0 &  nearrow 0.27 &  nearrow 0.28 \\
 & Business Services & 0.0 & 0.0 &  nearrow 0.62 &  nearrow 0.58 \\[1.8ex]
\textbf{SWITZERLAND} & Printing \& Publishing & 0.65 & 0.66 &  searrow 0.09 &  searrow 0.1 \\
 & Media \& Broadcasting & 0.27 & 0.24 &  nearrow 0.89 &  nearrow 0.87 \\
 & Business Services & 0.07 & 0.09 & 0.0 & 0.0 \\[1.8ex]
\textbf{SPAIN} & Printing \& Publishing & 0.78 & 0.77 &  searrow 0.47 &  searrow 0.41 \\
 & Business Services & 0.12 & 0.12 & 0.22 &  nearrow 0.25 \\
 & Media \& Broadcasting & 0.06 & 0.07 & 0.0 & 0.0 \\
 & Public Administration, Education, Health Social Services & 0.01 & 0.01 & 0.06 & 0.07 \\
 & Banking, Insurance \& Financial Services & 0.0 & 0.0 &  nearrow 0.18 &  nearrow 0.2 \\[1.8ex]
\textbf{ROMANIA} & Printing \& Publishing & 0.46 & 0.48 & 0.49 & 0.41 \\
 & Computer Software & 0.17 & 0.10 &  searrow 0.0 & 0.0 \\
 & Media \& Broadcasting & 0.16 & 0.19 &  searrow 0.0 &  searrow 0.0 \\
 & Business Services & 0.14 & 0.16 & 0.06 & 0.07 \\
 & Information Services & 0.06 & 0.06 & 0.0 & 0.0 \\
 & Banking, Insurance \& Financial Services & 0.0 & 0.0 &  nearrow 0.24 &  nearrow 0.26 \\
 & Public Administration, Education, Health Social Services & 0.0 & 0.01 &  nearrow 0.16 &  nearrow 0.22 \\[1.8ex]
\textbf{POLAND} & Media \& Broadcasting & 0.68 & 0.65 &  searrow 0.09 &  searrow 0.14 \\
 & Printing \& Publishing & 0.31 & 0.34 &  searrow 0.19 &  searrow 0.2 \\
 & Business Services & 0.01 & 0.01 & 0.07 & 0.07 \\
 & Banking, Insurance \& Financial Services & 0.0 & 0.0 &  nearrow 0.4 &  nearrow 0.37 \\
 & Public Administration, Education, Health Social Services & 0.0 & 0.0 &  nearrow 0.25 &  nearrow 0.22 \\[1.8ex]
\textbf{ITALY} & Printing \& Publishing & 0.71 & 0.69 &  searrow 0.02 &  searrow 0.02 \\
 & Business Services & 0.11 & 0.10 &  nearrow 0.79 &  nearrow 0.77 \\
 & Media \& Broadcasting & 0.11 & 0.12 &  searrow 0.0 &  searrow 0.0 \\
 & Information Services & 0.07 & 0.09 & 0.0 & 0.0 \\
 & Banking, Insurance \& Financial Services & 0.0 & 0.0 &  nearrow 0.12 &  nearrow 0.12 \\
 & Public Administration, Education, Health Social Services & 0.0 & 0.0 & 0.05 & 0.06 \\[1.8ex]
\textbf{GERMANY} & Printing \& Publishing & 0.43 & 0.46 &  searrow 0.03 &  searrow 0.05 \\
 & Media \& Broadcasting & 0.24 & 0.21 &  searrow 0.13 & 0.13 \\
 & Information Services & 0.17 & 0.13 &  searrow 0.0 &  searrow 0.0 \\
 & Business Services & 0.07 & 0.08 &  nearrow 0.61 &  nearrow 0.6 \\
 & Public Administration, Education, Health Social Services & 0.01 & 0.01 &  nearrow 0.16 &  nearrow 0.16 \\
 & Banking, Insurance \& Financial Services & 0.0 & 0.0 & 0.06 & 0.06 \\[1.8ex]
\textbf{FRANCE} & Printing \& Publishing & 0.91 & 0.90 &  searrow 0.65 &  searrow 0.57 \\
 & Media \& Broadcasting & 0.08 & 0.08 & 0.0 & 0.0 \\
 & Business Services & 0.01 & 0.01 &  nearrow 0.26 &  nearrow 0.34 \\
 & Banking, Insurance \& Financial Services & 0.0 & 0.0 & 0.08 & 0.07 \\[1.8ex]
\textbf{CANADA} & Media \& Broadcasting & 0.43 & 0.47 & 0.46 & 0.5 \\
 & Business Services & 0.25 & 0.19 & 0.34 & 0.26 \\
 & Printing \& Publishing & 0.24 & 0.26 & 0.17 & 0.19 \\
 & Communications & 0.07 & 0.07 & 0.0 & 0.0 \\
\bottomrule
\hline\hline
\end{tabular}
\begin{tablenotes}[flushleft]
\footnotesize{\item Notes: Ownership shares by sector and country, based on page views (columns 3 and 5) and unique visitors (columns 4 and 6) -- considering all outlets, including those not specializing in political content. . Ownership data is from 2020, and traffic data are averaged for 2018-2020. Column 2 shows the sectors to which the owners belong according the Orbis database. Sectors with ownership shares below 5\% both for immediate and ultimate owners are not shown (for better readability). Columns 3 and 4 show the sector-specific ownership shares at the level of the immediate owner (e.g., the immediate owner of the nytimes.com, according to the Orbis database, is the \textit{NEW YORK TIMES COMPANY}). In column 7 and 8, the measures are calculated at the level of ultimate owners (e.g., there are around 3,000 ultimate owner entities for nytimes.com, among others \textit{CARLOS SLIM HELU Y FAMILIA}, \textit{JACKSON SQUARE LLC}, or \textit{VANGUARD GROUP INC}). Sectors where the share changes by more than 10 percentage points when moving from the immediate owner to the ultimate owner (i.e., moving from column 3 to 5 or from column 4 to 6) are highlighted with an arrow ($\nearrow$ for increases and $\searrow$ for decreases).}
% A table with all non-zero shares is shown in the Appendix.
\end{tablenotes}
\end{threeparttable}
}
\end{table}

\clearpage

\subsection{Market Concentration}

\begin{figure}[htbp]
\centering
\caption{Market Concentration ($C_{M}$) Based on Reach}
\label{fig:cum_shares}
\includegraphics[width=0.99\textwidth]{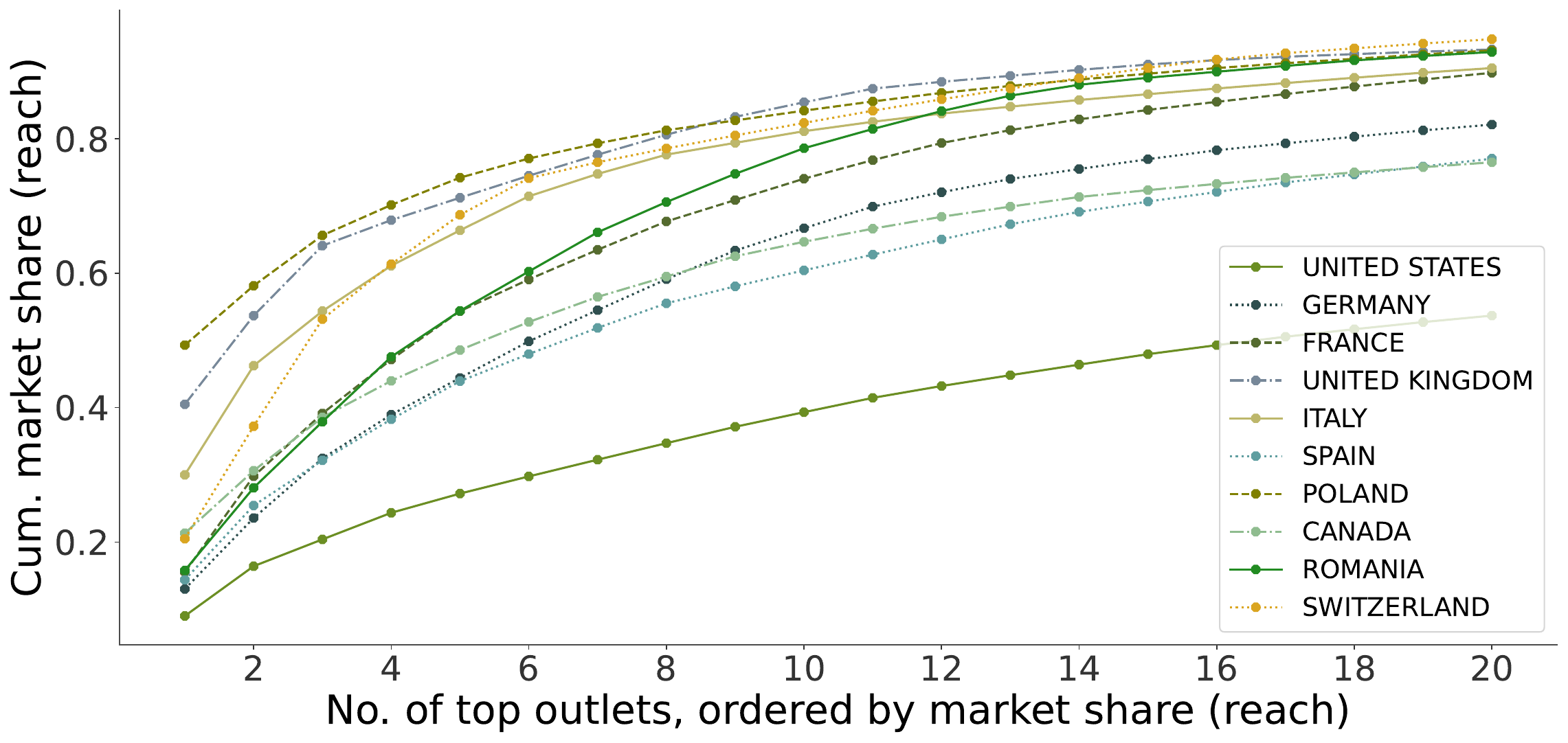}
\includegraphics[width=0.99\textwidth]{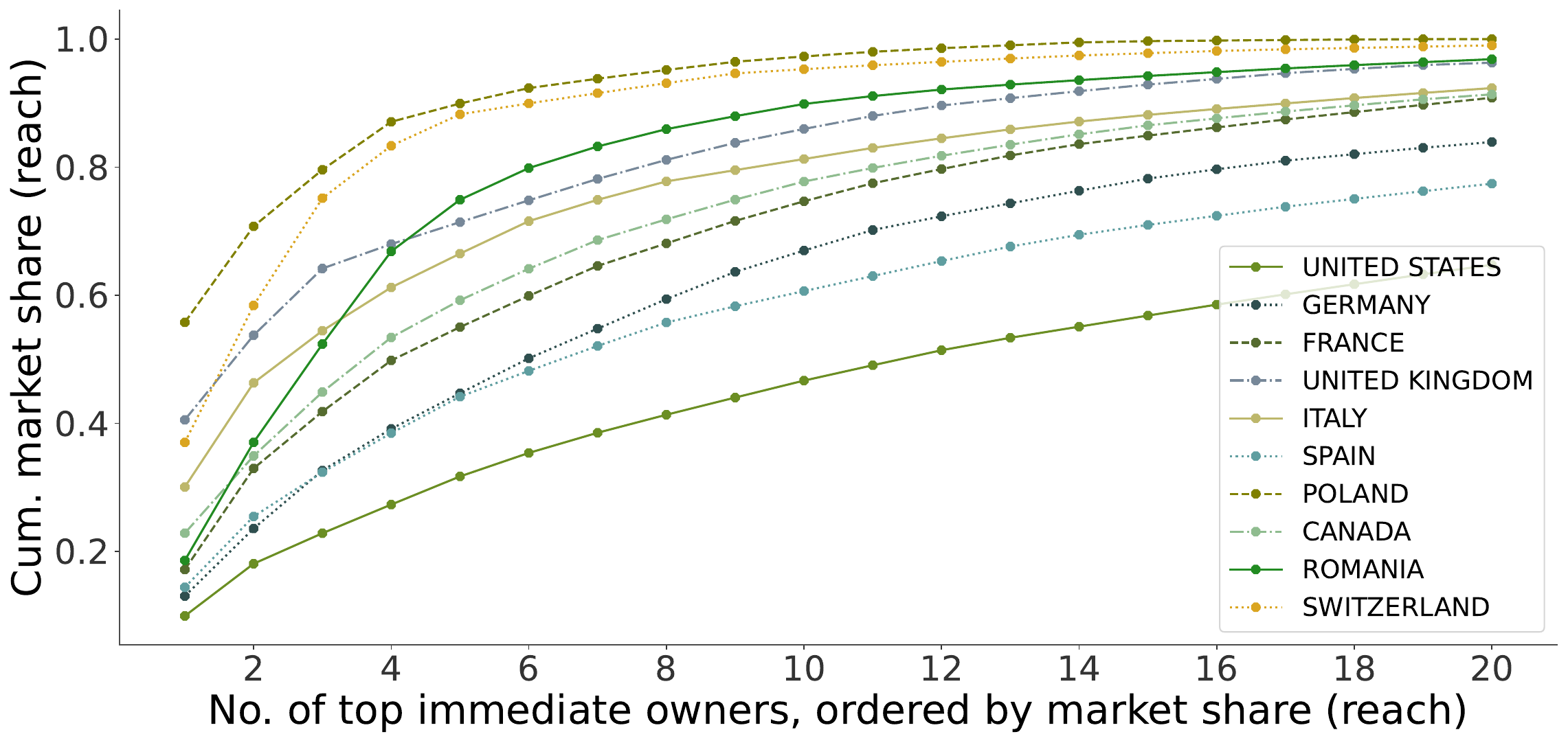}
\includegraphics[width=0.99\textwidth]{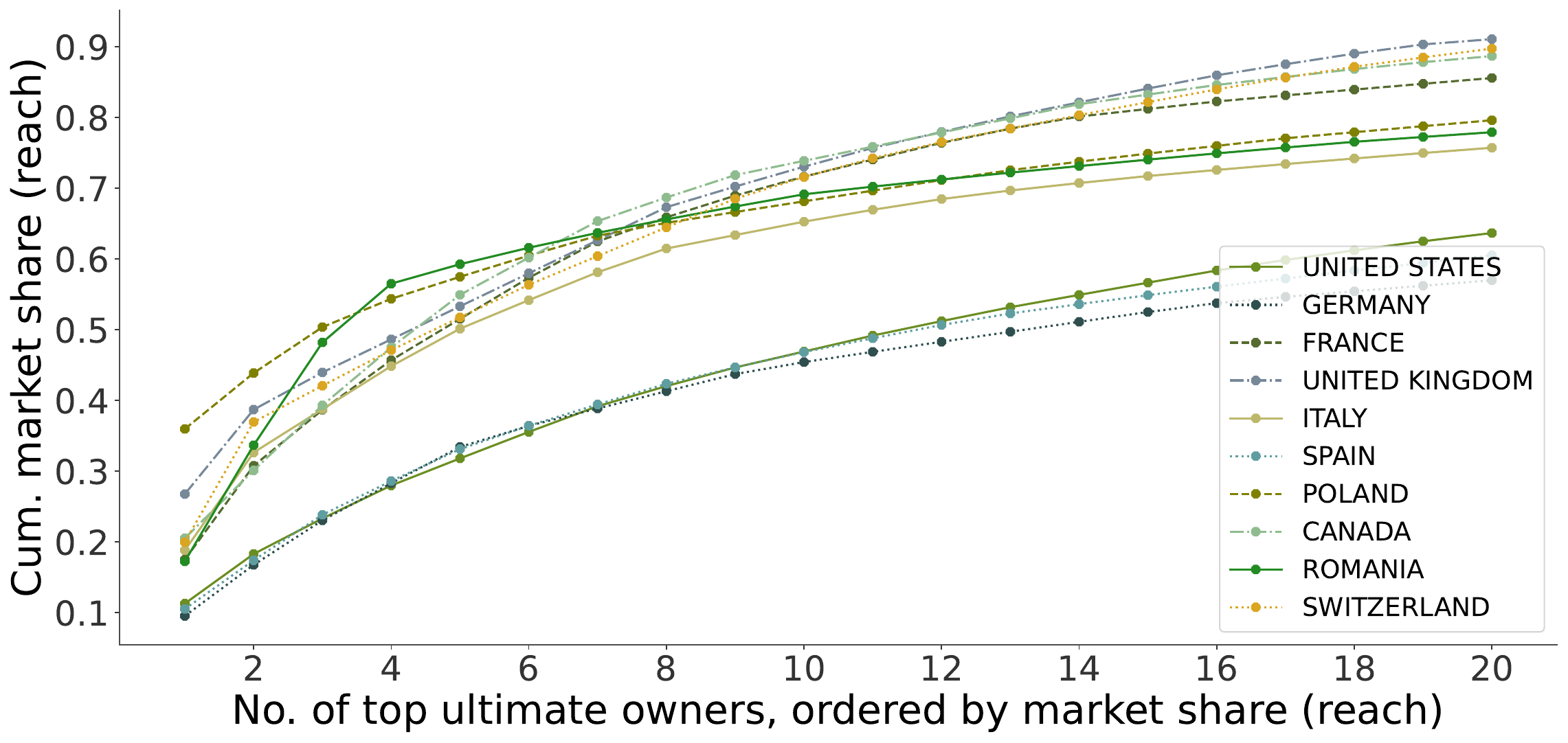}
\begin{figurenotes}
Cumulative market shares of the top $M$ entities by country, for all hard news outlets (more than 25\% political reporting). The $C_{M}$ (plotted on the $Y$-axis) is computed based on the top 1 to top 20 entities: domains, domain owners, and ultimate owners in the upper, middle, and lower panels, respectively. The market shares are based on individual visitors (i.e., reach). The number of entities $M$ (domains, domain owners, or ultimate owners, depending on the panel) is plotted on the $X$-axis.
\end{figurenotes}
\end{figure}

\begin{figure}[htbp]
\centering
\caption{Market Concentration ($C_{M}$) Based on Reach (All Outlets, Incl. Non-Political)}
\label{fig:cum_shares_nonpolitical}
\includegraphics[width=0.99\textwidth]{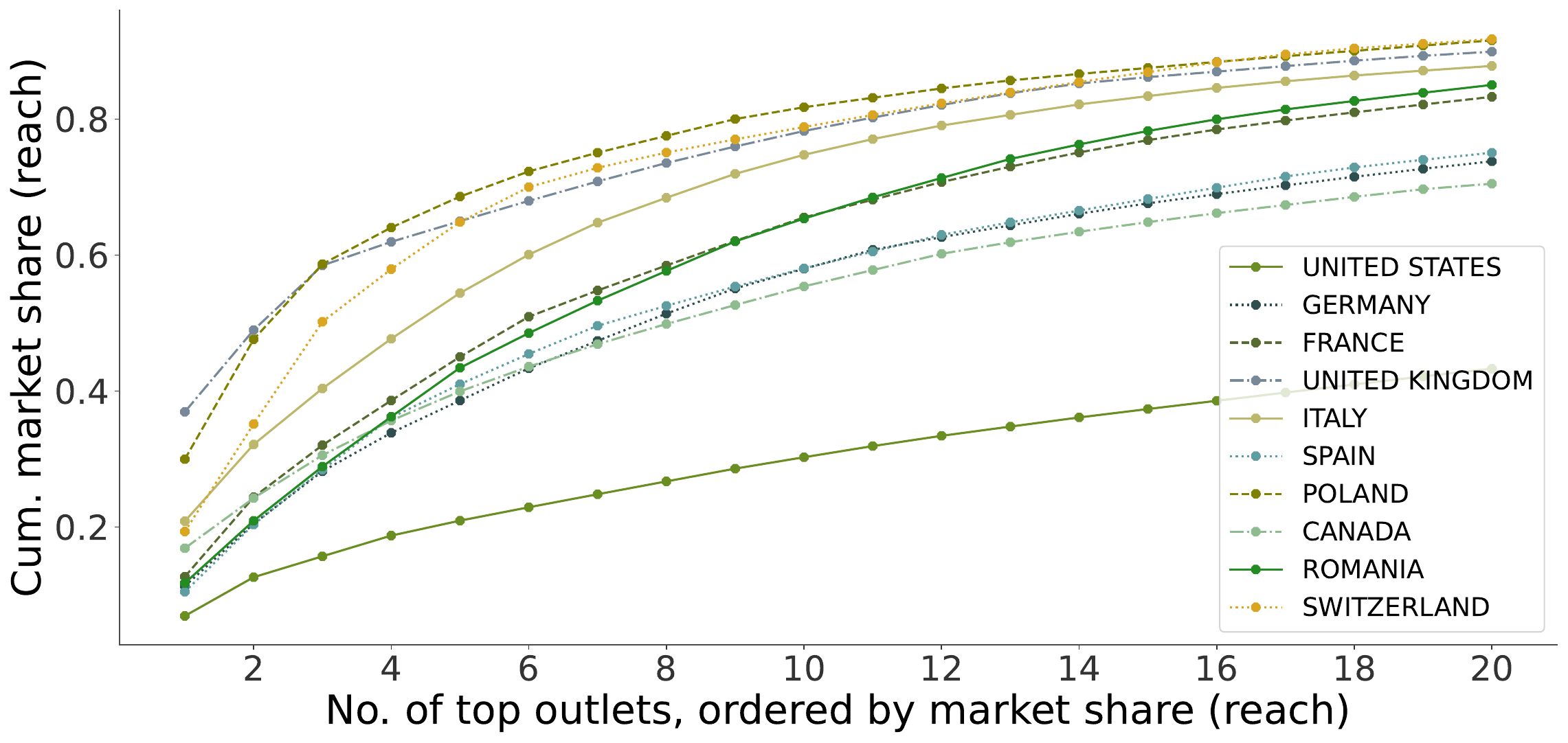}
\includegraphics[width=0.99\textwidth]{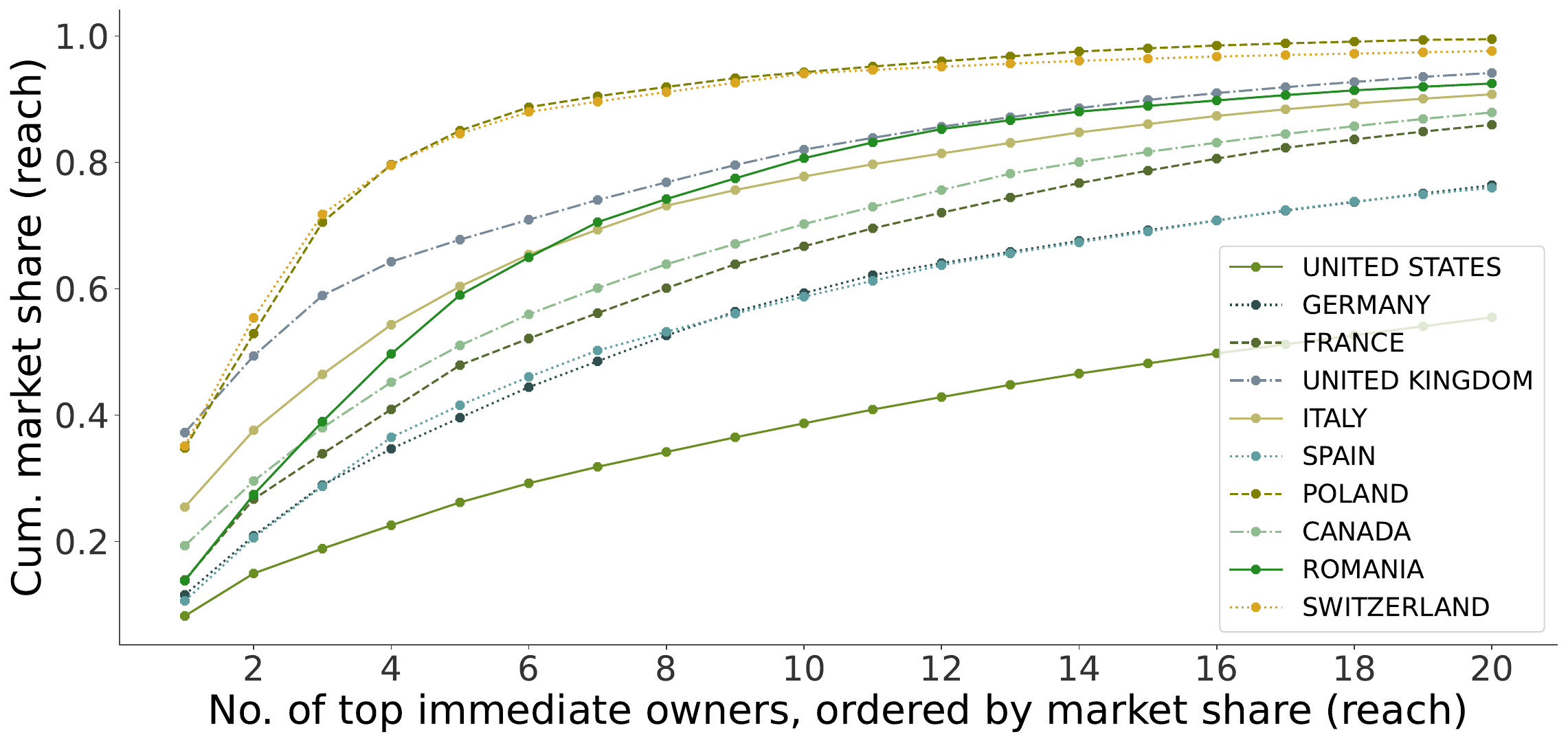}
\includegraphics[width=0.99\textwidth]{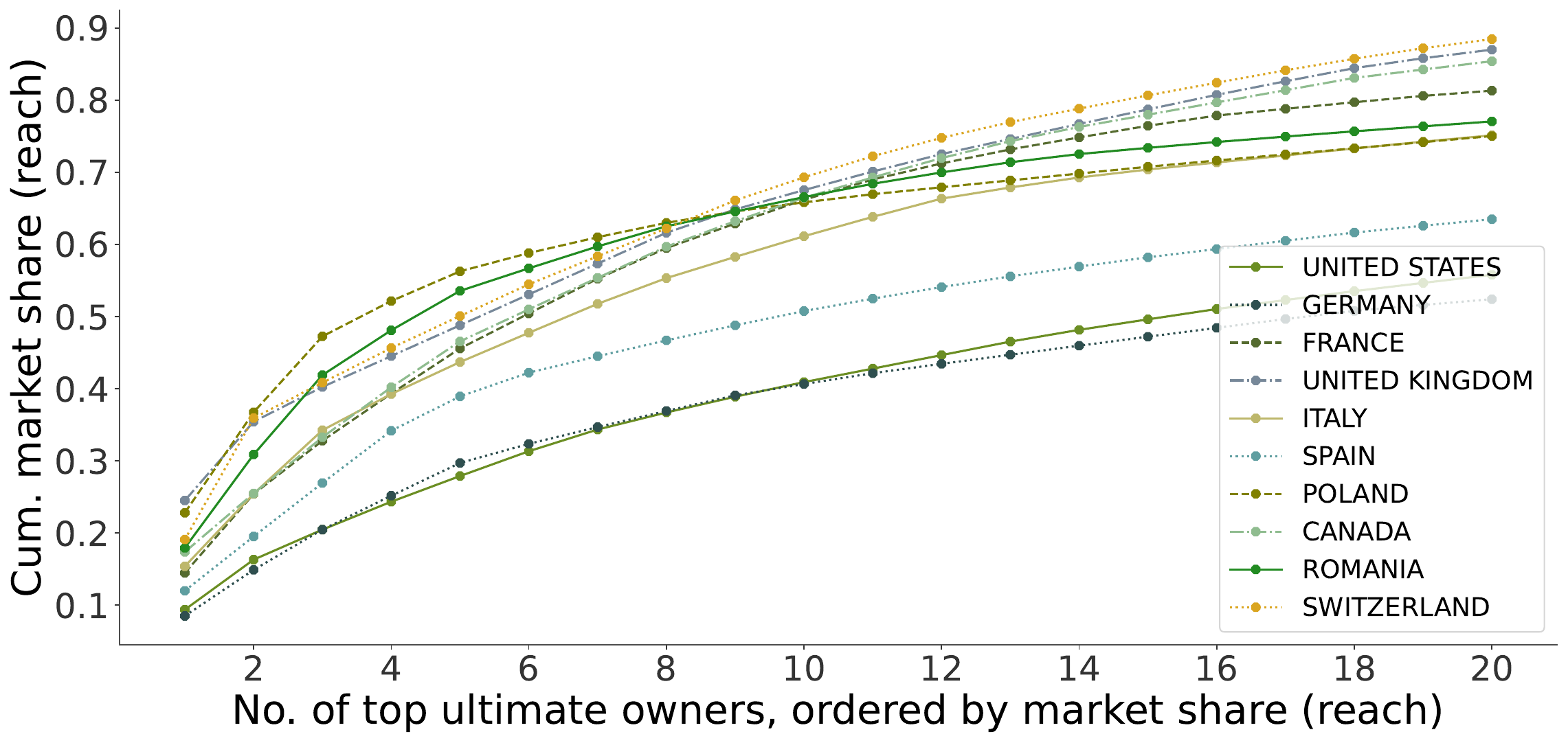}
\begin{figurenotes}
Cumulative market shares of the top $M$ entities by country, for all outlets, including those not specializing in political content. The $C_{M}$ (plotted on the $Y$-axis) is computed based on the top 1 to top 20 entities: domains, domain owners, and ultimate owners in the upper, middle, and lower panels, respectively. The market shares are based on individual visitors (i.e., reach). The number of entities $M$ (domains, domain owners, or ultimate owners, depending on the panel) is plotted on the $X$-axis.
\end{figurenotes}
\end{figure}

\begin{figure}[htb!]
\centering
\caption{Attention Shares and Traffic} \label{fig:attention_reach}
\includegraphics[width=0.99\linewidth]{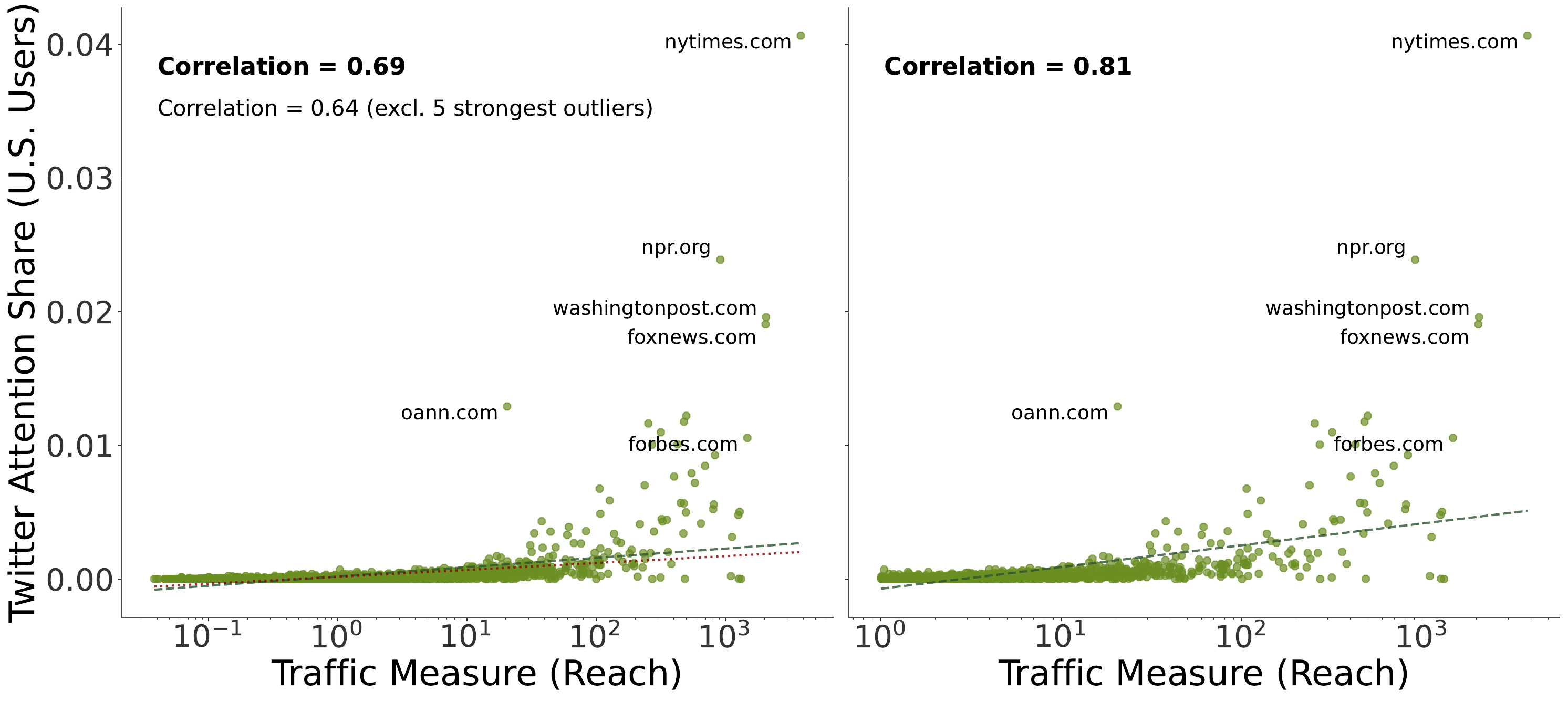}
\begin{figurenotes}
Scatterplot of the attention shares (in the spirit of \cite{Prat_2018}; here proxied by Twitter data) and outlet-level traffic. U.S. hard news outlets. Traffic data based on 2018--2020 reach (unique visitors per million). Attention shares are computed from a random sample of 6,475 U.S. Twitter users who follow at least one news outlet; we identify which news accounts each user follows and compute the share of attention each outlet receives, taking into account which other accounts the user also follows (October--December 2020). The left panel shows all outlets with two log-linear fits: the dashed green line includes all observations; the dotted red line excludes the five strongest outliers. The right panel restricts the sample to outlets with reach $\geq$ 1 and shows a single log-linear fit. Labeled outlets are the top 5 by reach, top 5 by attention share, and top 5 outliers from the regression line (deduplicated): nytimes.com, washingtonpost.com, foxnews.com, forbes.com, businessinsider.com, npr.org, and oann.com.
\end{figurenotes}
\end{figure}

This subsection complements the market concentration results of Section~\ref{sec:results:concentration} with a series of robustness checks, each presented in the same form as main-text Figure~\ref{fig:hhi_slope}, where every line traces a country's concentration across the news-site (domain), domain-owner, and ultimate-owner levels. We report the HHI based on page views instead of reach, based on the 2023 ownership data, and for all outlets including non-political ones, as well as the four-firm concentration ratio ($C_4$) as an alternative measure (with its own thresholds). Figures~\ref{fig:cum_shares} and~\ref{fig:cum_shares_nonpolitical} additionally report cumulative concentration ratios ($C_M$). Finally, Figure~\ref{fig:attention_reach} validates our reach-based market shares against attention shares in the spirit of \cite{Prat_2018}, as discussed in Section~\ref{sec:extensions}.

\begin{figure}[htbp]
\centering
\caption{HHI Measures by Country (Page Views Instead of Reach)}
\label{fig:hhi_political_views}
\includegraphics[width=0.85\textwidth]{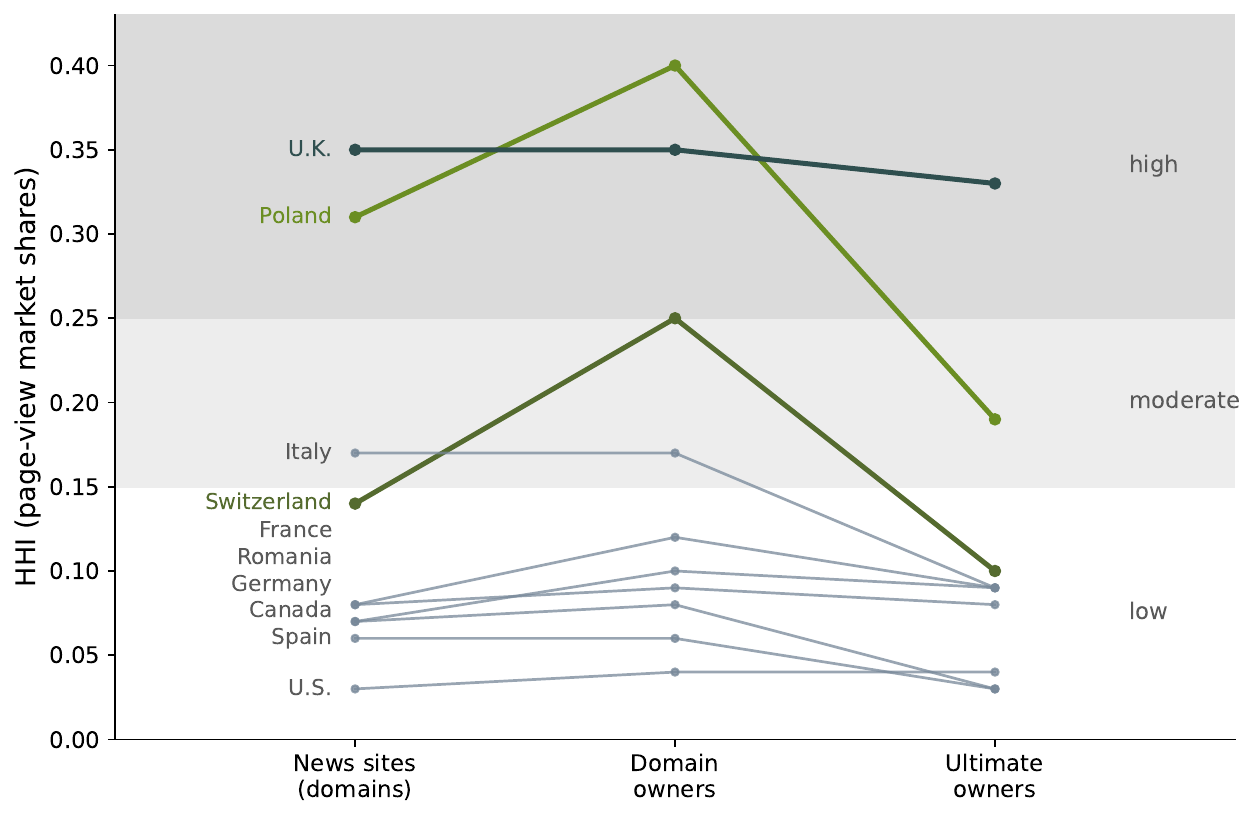}
\begin{figurenotes}
This figure replicates main-text Figure~\ref{fig:hhi_slope} using page views instead of reach as the traffic measure; each line traces a country's HHI across the news-site (domain), domain-owner, and ultimate-owner levels, and shaded bands mark the moderate (0.15--0.25) and high ($>$0.25) thresholds. The qualitative patterns are similar to those based on reach, though some countries show higher concentration with page views (e.g., the United Kingdom's domain-level HHI is 0.35 with page views versus 0.20 with reach).
\end{figurenotes}
\end{figure}

\begin{figure}[htbp]
\centering
\caption{HHI Measures by Country (2023 Ownership Data)}
\label{fig:hhi_2023_robustness}
\includegraphics[width=0.85\textwidth]{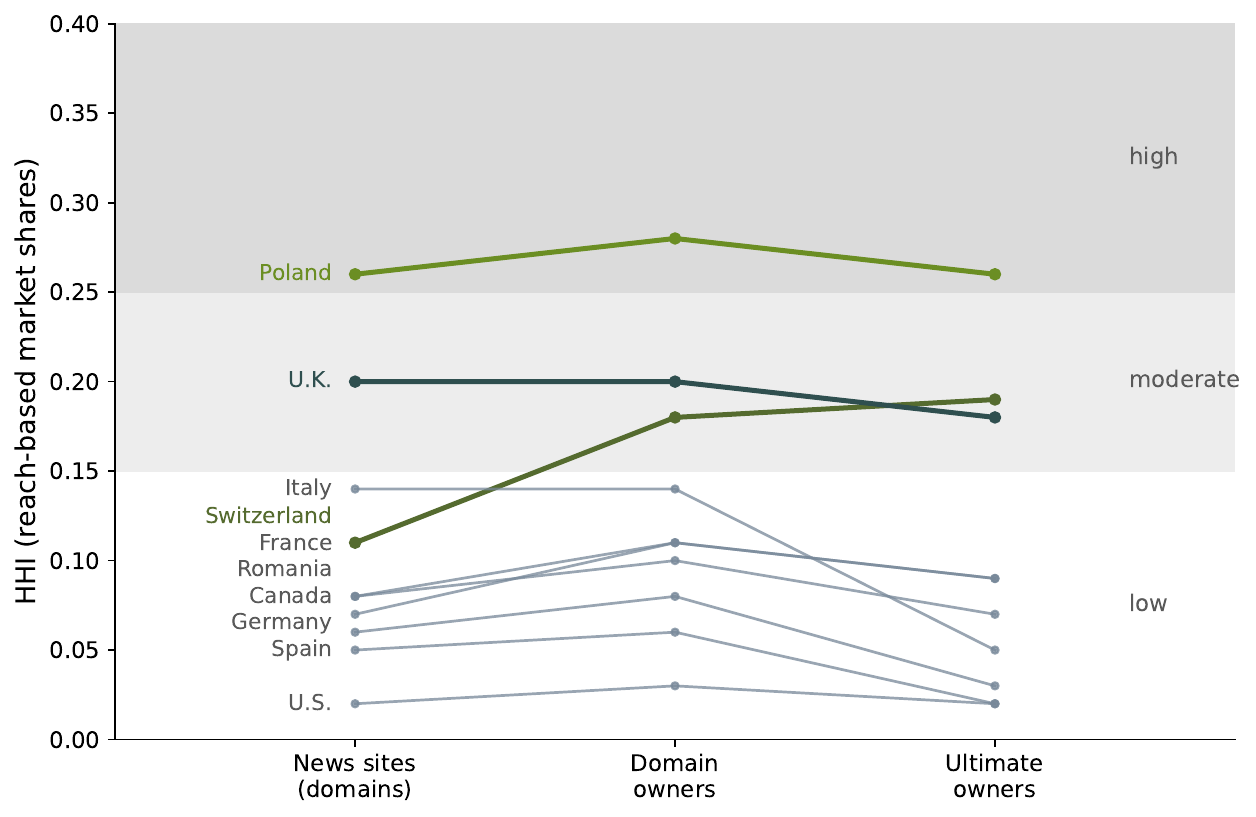}
\begin{figurenotes}
This figure replicates main-text Figure~\ref{fig:hhi_slope} using July 2023 ownership data instead of 2020 ownership data; each line traces a country's HHI across the news-site (domain), domain-owner, and ultimate-owner levels. The news-site (domain) level is unchanged, as the traffic data are held constant, while the domain-owner and ultimate-owner levels are based on July 2023 ownership snapshots.
\end{figurenotes}
\end{figure}

\begin{figure}[htbp]
\centering
\caption{HHI Measures by Country (All Outlets, Incl. Non-Political)}
\label{fig:hhi_all_outlets}
\includegraphics[width=0.85\textwidth]{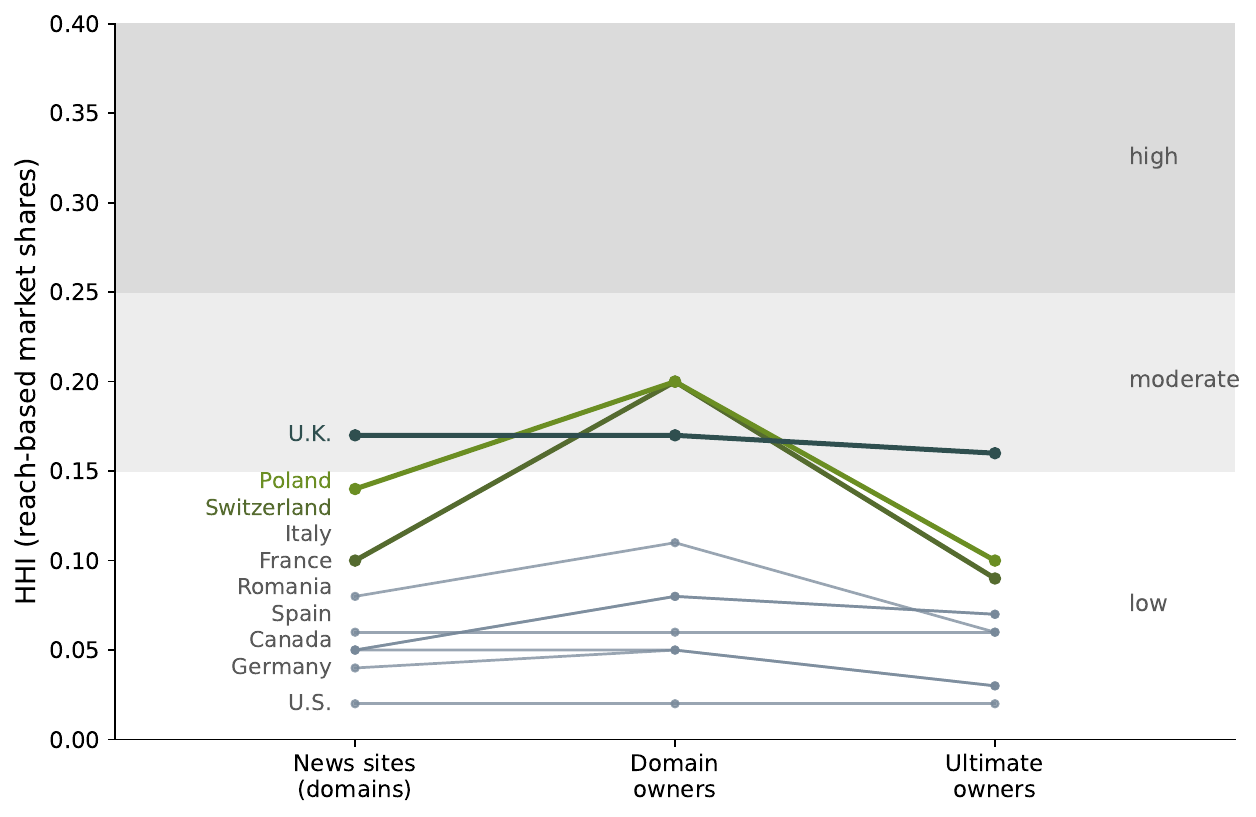}
\begin{figurenotes}
This figure replicates main-text Figure~\ref{fig:hhi_slope} using all media outlets rather than only those specializing in political content; each line traces a country's HHI across the news-site (domain), domain-owner, and ultimate-owner levels. Concentration levels are generally lower when including non-political outlets, as the sample size increases substantially.
\end{figurenotes}
\end{figure}

\begin{figure}[htbp]
\centering
\caption{$C_{4}$ Measures by Country for Domains, Domain Owners, and Ultimate Owners}
\label{fig:c4_domain_to_ultimate}
\includegraphics[width=0.85\textwidth]{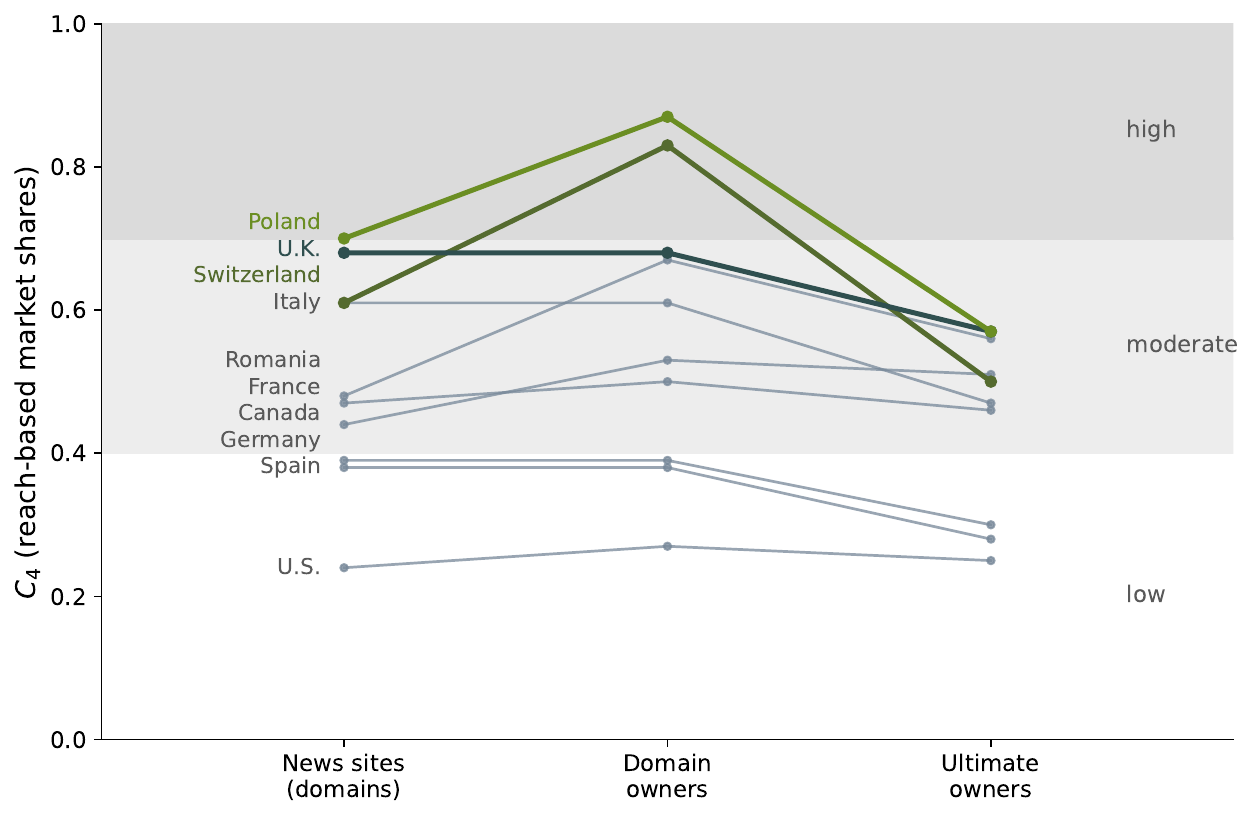}
\begin{figurenotes}
This figure replicates main-text Figure~\ref{fig:hhi_slope} using the four-firm concentration ratio ($C_4$, the combined market share of the top four entities, in terms of reach) instead of the HHI; each line traces a country's $C_4$ across the news-site (domain), domain-owner, and ultimate-owner levels. Shaded bands mark the standard $C_4$ thresholds for medium (40\%--70\%) and high ($>$70\%) concentration \citep{legalclarity2024cr4}. For example, Poland's $C_4$ is 70\% at the domain level, 87\% at the domain-owner level, and 57\% at the ultimate-owner level.
\end{figurenotes}
\end{figure}

\clearpage

\subsection{Concentration Measures: Summary Table}

\begin{table}[htbp]
\centering
\begin{threeparttable}
\caption{Market Concentration Measures by Country and Ownership Level}
\label{tab:concentration_all}
\scriptsize
\singlespacing
\setlength{\tabcolsep}{4pt}
\begin{tabular}{l r cc cc cc cc cc cc}
\toprule
 & & \multicolumn{4}{c}{Domain} & \multicolumn{4}{c}{Domain Owner} & \multicolumn{4}{c}{Ultimate Owner} \\
\cmidrule(lr){3-6} \cmidrule(lr){7-10} \cmidrule(lr){11-14}
 & & \multicolumn{2}{c}{HHI} & \multicolumn{2}{c}{$C_4$} & \multicolumn{2}{c}{HHI} & \multicolumn{2}{c}{$C_4$} & \multicolumn{2}{c}{HHI} & \multicolumn{2}{c}{$C_4$} \\
\cmidrule(lr){3-4} \cmidrule(lr){5-6} \cmidrule(lr){7-8} \cmidrule(lr){9-10} \cmidrule(lr){11-12} \cmidrule(lr){13-14}
Country & $N$ & Reach & Views & Reach & Views & Reach & Views & Reach & Views & Reach & Views & Reach & Views \\
\midrule
United States & 3213 & 0.02 & 0.03 & 24\% & 30\% & 0.03 & 0.04 & 27\% & 34\% & 0.03 & 0.04 & 25\% & 31\% \\
Canada & 461 & 0.07 & 0.07 & 44\% & 45\% & 0.10 & 0.10 & 53\% & 52\% & 0.09 & 0.09 & 51\% & 52\% \\
United Kingdom & 348 & 0.20 & 0.35 & 68\% & 81\% & 0.20 & 0.35 & 68\% & 81\% & 0.18 & 0.33 & 57\% & 70\% \\
Germany & 140 & 0.06 & 0.07 & 39\% & 46\% & 0.06 & 0.08 & 39\% & 46\% & 0.03 & 0.03 & 30\% & 29\% \\
Spain & 130 & 0.05 & 0.06 & 38\% & 41\% & 0.05 & 0.06 & 38\% & 42\% & 0.03 & 0.03 & 28\% & 28\% \\
Italy & 75 & 0.14 & 0.17 & 61\% & 65\% & 0.14 & 0.17 & 61\% & 65\% & 0.08 & 0.09 & 47\% & 49\% \\
Romania & 62 & 0.08 & 0.08 & 48\% & 47\% & 0.13 & 0.12 & 67\% & 65\% & 0.09 & 0.09 & 56\% & 56\% \\
France & 55 & 0.08 & 0.08 & 47\% & 52\% & 0.08 & 0.09 & 50\% & 52\% & 0.07 & 0.08 & 46\% & 48\% \\
Poland & 48 & 0.26 & 0.31 & 70\% & 74\% & 0.35 & 0.40 & 87\% & 92\% & 0.16 & 0.19 & 57\% & 63\% \\
Switzerland & 46 & 0.11 & 0.14 & 61\% & 67\% & 0.22 & 0.25 & 83\% & 87\% & 0.10 & 0.10 & 50\% & 52\% \\
\bottomrule
\end{tabular}
\begin{tablenotes}[flushleft]
\footnotesize{
\item \textit{Notes}: This table reports the Herfindahl-Hirschman Index (HHI) and four-firm concentration ratio ($C_4$) for hard news outlets (more than 25\% political reporting) across ten countries. Measures are computed at three ownership levels: domain (individual websites), domain owner (registrant), and ultimate owner (tracing ownership networks). Traffic is measured using reach (unique visitors) and page views. Ownership data is from July 2020. HHI values range from 0 to 1, with values below 0.15 indicating low concentration, 0.15--0.25 moderate, and above 0.25 high. $C_4$ values below 40\% indicate low concentration, 40--70\% medium, and above 70\% high. $N$ counts all sampled hard news outlets; each level is computed on the subsample with the required data (traffic data at the domain level, a 2020 registrant match at the domain-owner level, and a matched ownership network at the ultimate-owner level). Market shares are within-country traffic shares of the included outlets; at the owner levels, outlet traffic is assigned to owners -- split in proportion to ownership stakes at the ultimate-owner level -- and summed by owner before computing the measures.
}
\end{tablenotes}
\end{threeparttable}
\end{table}

\clearpage

\section{Robustness to Missing Traffic and Ownership Information} \label{app:sec:simulation}

This appendix section reports the simulation referenced in Section~\ref{sec:robustness_missing} of the main text: it assesses how missing outlets, missing traffic data, or missing ownership data would affect our market concentration estimates.

\begin{figure}[htbp]
\centering
    \caption{Simulation: Effect of outlets potentially missing in our data}
\label{fig:simulation}
\includegraphics[width=0.99\textwidth]{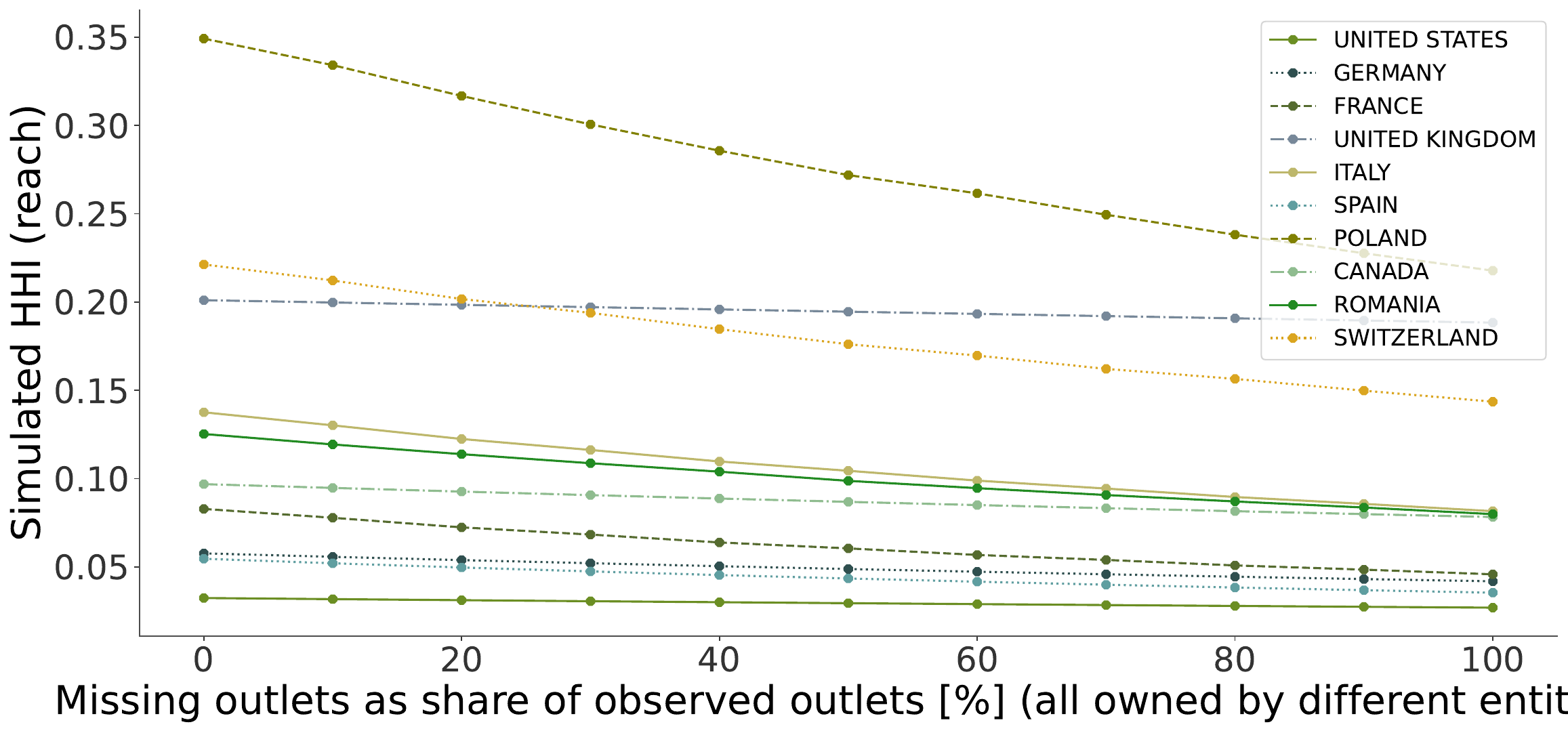}
\includegraphics[width=0.99\textwidth]{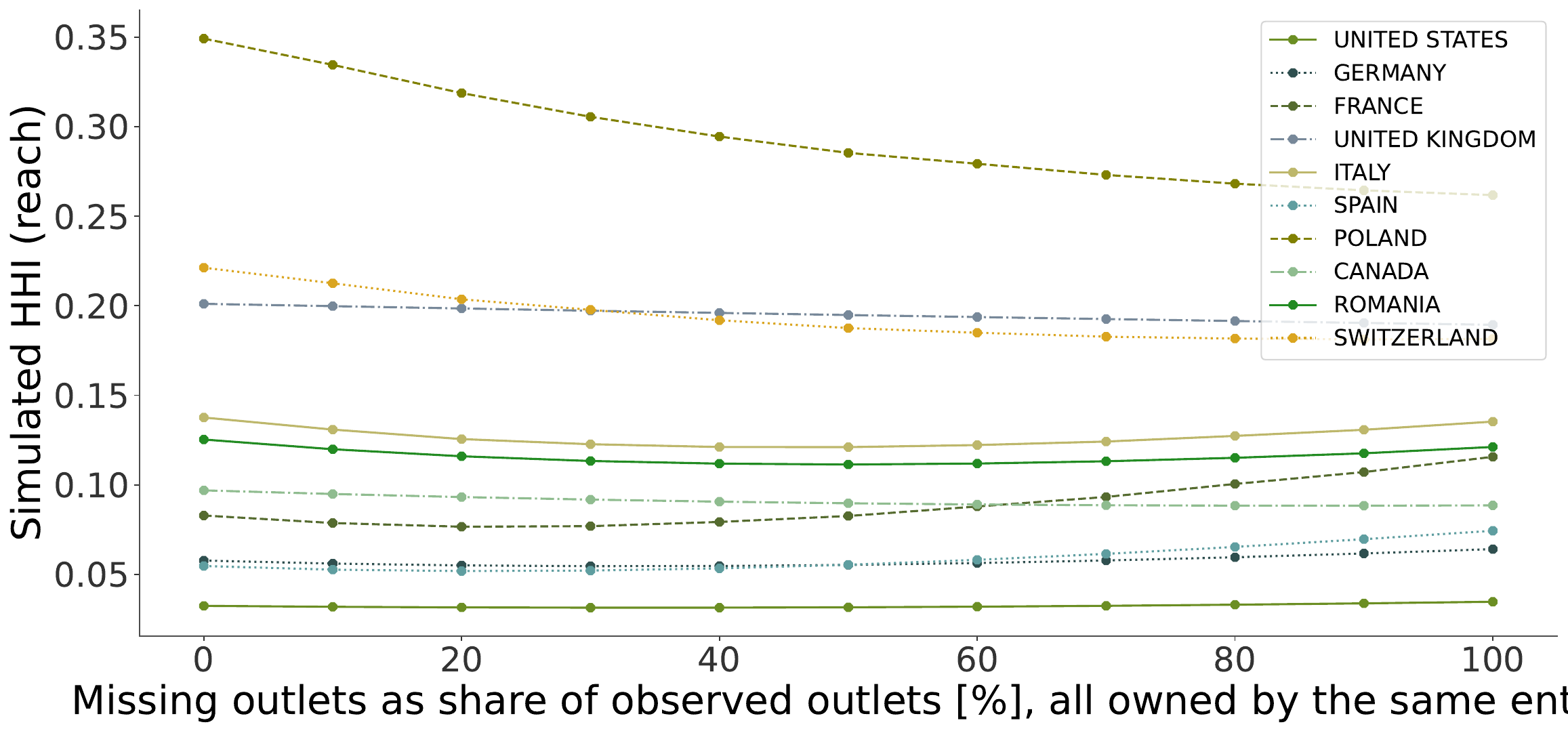}
\begin{figurenotes}
HHI values by country (vertical axis) under the assumption that a certain percentage of outlets, out of the currently observed outlets (e.g., 10\% of currently observed outlets) are missing. The percentage of hypothetically missing outlets is shown on the horizontal axis. Missing outlets are assumed to have a median reach (based on our actual country sample). In the upper panel, we simulate missing outlets that are all owned by different entities. In the lower panel, we simulate missing outlets that are all owned by the same owner.
\end{figurenotes}
\end{figure}

There are two potential ways in which data coverage issues could produce biased market concentration estimates in our approach. First, we could miss outlets entirely or, equivalently, not have traffic data for them (recall that all concentration measures require traffic data; hence, an outlet that does not enter our list and an outlet that enters our list but comes without traffic data, distort the concentration measure in the same way). Second, we could capture these outlets and their traffic but remain without ownership data. Given how we compile our sample, it is unlikely that we are missing a large outlet; we cross-check our outlet list against multiple open-source/academic datasets (ABYZ, AWIS, and Media Cloud). In particular, Amazon's AWIS (from which we obtain our traffic data) is a for-profit service used by marketing departments of large companies worldwide. It is thus unlikely that a site with high traffic that is not covered would remain unnoticed for long. Similarly, outlets listed in AWIS for which traffic data are missing are likely small: AWIS estimates a domain's web traffic from a large sample of individual internet users' browsing data. When a domain is visited very few times/by very few users, AWIS does not provide traffic data for this domain, since a reliable estimate of its overall traffic is infeasible. Therefore, we expect (if at all) potentially missed outlets to be small or of moderate size. Following this reasoning, we devise a simulation approach to assess the impact of missing outlets/data. Suppose we miss one or several hypothetical outlets in our data collection procedure. Our simulation procedure asks how the market concentration (specifically, the HHI) would change if we included such hypothetical outlets. Accordingly, we assume that the missing outlets would receive traffic equal to the median of all outlets for which we have complete data (both traffic and ownership). Then, we ask how the HHI value would change if, in our sample, $Z$ such outlets were missing. Overlooking $Z$ outlets in, say, Switzerland, where the total number of outlets is relatively low, would mean missing more of the market than, say, in the United States. Therefore, we scale $Z$ by the total number of observed outlets in the country. In other words, we evaluate how the Herfindahl index would change if we overlooked 10\%, 20\%, ..., up to 100\% of the current outlets (all assumed to have a circulation amounting to the current median). That is, for each country, our checks extend up to a doubling of the number of outlets. Figure \ref{fig:simulation} shows the simulation results. We distinguish between two corner cases: all overlooked outlets are owned by separate owners (upper panel), or they are all owned by the same owner (lower panel). The latter case also tests if it would be problematic if many of the outlets without ownership information belonged to the same conglomerate (recall that the outlets with missing ownership data tend to be small, see coverage statistics above).The simulation is based on the domain-owner-level HHI (reach-based, hard news outlets).
With different owners (upper panel), HHI values remain relatively stable across countries, provided that we do not miss more than 50\% of the currently observed market. In this (unlikely) scenario, even for countries that react most sensitively in our simulation, such as Switzerland and Poland, our HHI would not overstate concentration by more than 0.08. With the same owner (lower panel), the HHI values remain rather stable for all countries, even if we had overlooked as many median-sized outlets as the currently observed market contains (here, the missing outlets have opposing effects; they increase the market, but also become a large player as we move along the $x$-axis). Overall, assuming that our cross-checking renders it unlikely that large outlets are missing from our dataset, we conclude that our estimates are relatively robust to missing data.

\clearpage

\section{Additional Tables: Ownership and News Content}

This appendix section collects additional material for the ownership--content analyses of Sections~\ref{sec:coverage} and~\ref{sec:mechanisms}. It reports descriptive statistics of coverage similarity by ownership status and the path from the outlet universe to each estimation sample; the anatomy of the identifying ownership variation and pre-trends for switching pairs; intensive-margin robustness checks for the tone results (an intersection-normalized tone measure and diagnostics of the jointly covered event sets); the event-level tone regressions; the heterogeneity of the results by event theme (political versus non-political), including an event-level theme split of tone alignment; the U.S.-versus-non-U.S. comparison referenced in the Gentzkow--Shapiro discussion and the demand- versus supply-side benchmark; robustness checks based on alternative similarity measures and sample restrictions; splits by media type; and estimates based on within-country pairs only.

\clearpage
\subsection{Descriptive statistics and estimation samples}

\begin{table}[htbp]
\centering
\begin{threeparttable}
\singlespacing
\caption{From the Outlet Universe to the Estimation Samples} \label{tab:sample_overview}
\footnotesize
% Table created by code/tab_sample_overview.R
\begin{tabular}{@{\extracolsep{5pt}} lp{5.4cm}cc}
\\[-1.8ex]\hline
\hline \\[-1.8ex]
Sample & Key restriction & News sites & Pair-period obs. \\
\hline \\[-1.8ex]
News domain universe & ABYZ/AWIS/Media Cloud directories, ten countries & 11,967 & --- \\
Hard news outlets & political-content ratio $>$25\% & 4,578 & --- \\
Coverage panel (Tables 1, E3--E4) & GDELT-visible, ownership data, pairs in both periods & 2,019 & 4,074,342 \\
Tone panel (Table 2) & additionally: tone data and $\geq$1 jointly covered tone-scored event & 1,346 & 986,512 \\
Coverage robustness (Table E5) & additionally: complete country/state/media-type metadata & 1,284 & 1,647,372 \\
Tone robustness (Table E6) & tone panel restrictions plus complete metadata & 1,281 & 800,010 \\
\hline
\hline \\[-1.8ex]
\end{tabular}

\begin{tablenotes}[flushleft]
\footnotesize{
\item \textit{Notes}: This table traces the path from the initial news domain universe to each estimation sample of the content analysis. The first two rows describe the outlet universe from the sample construction of Section~\ref{sec:data}; the remaining rows describe the pair-period panels behind the content analyses, each restricted to pairs observed in both periods (balanced panels). The coverage panel requires GDELT-visible outlets with ownership data; the tone panel additionally requires tone data and at least one jointly covered tone-scored event per pair; the robustness samples additionally require complete country, state, and media-type metadata. Re-estimating the coverage specification of Table~\ref{tab:coverage_pairs} (column 2) on the tone panel yields 0.321 (dyadic SE 0.041), close to the baseline 0.330, indicating that sample selection across panels does not drive comparisons between the coverage and tone results.
}
\end{tablenotes}
\end{threeparttable}
\end{table}

\begin{table}[htbp]
\centering
\begin{threeparttable}
\singlespacing
\caption{Coverage Similarity by Ownership Status} \label{tab:dsc_by_ownership}
\footnotesize
% Table created by code/text_economic_significance_coverage.R
% Date and time: Fri, Jul 17, 2026 - 05:03:31 PM
\begin{tabular}{@{\extracolsep{5pt}} lccc}
\\[-1.8ex]\hline
\hline \\[-1.8ex]
 & Mean DSC & SD & N \\
\hline \\[-1.8ex]
All pairs & 0.037 & 0.088 & 4,074,342 \\
Same domain owner & 0.416 & 0.329 & 24,395 \\
Different domain owner & 0.035 & 0.079 & 4,049,947 \\
Same country, same domain owner & 0.416 & 0.329 & 24,383 \\
Same country, different domain owner & 0.055 & 0.108 & 1,600,493 \\
\hline \\[-1.8ex]
Cross-sectional gap (same $-$ different) & 0.381 &  &  \\
Within-dyad SD (all pairs) &  & 0.043 &  \\
Within-dyad SD (switcher pairs) &  & 0.245 &  \\
\hline
\hline \\[-1.8ex]
\end{tabular}

\begin{tablenotes}[flushleft]
\footnotesize{
\item \textit{Notes}: This table reports descriptive statistics of the Dice similarity coefficient (DSC) of event coverage by domain-ownership status, pooled over both sample periods (2019/2020 and 2022/2023). The sample is the balanced outlet-pair panel used in the regression analyses (pairs observed in both periods with complete coverage and ownership data). The cross-sectional gap is the difference between the mean DSC of same-domain-owner pairs and that of different-domain-owner pairs; Section~\ref{sec:results_coverage} compares the within-dyad estimate of Table~\ref{tab:coverage_pairs} to this gap. The within-dyad standard deviations are computed after removing pair means and describe the variation exploited by the fixed-effects estimator, reported for all pairs and for the subsample of pairs whose domain-ownership status changes between the two periods (``switcher pairs'').
}
\end{tablenotes}
\end{threeparttable}
\end{table}
\clearpage
\subsection{The identifying variation and pre-trends}

\begin{table}[htbp]
\centering
\begin{threeparttable}
\singlespacing
\caption{The Identifying Variation: Owner Groups Behind the Switching Pairs} \label{tab:switcher_anatomy}
\footnotesize
% Table created by code/tab_switcher_anatomy.R
\begin{tabular}{@{\extracolsep{5pt}} lccc}
\\[-1.8ex]\hline
\hline \\[-1.8ex]
 & Coverage (DSC) & Tone (cosine) & Switcher pairs \\
\hline \\[-1.8ex]
Baseline (all switchers) & 0.330$^{***}$ (0.031) & 0.342$^{***}$ (0.043) & 4,287 \\
Excl. largest owner group 1 (1,829 pairs) & 0.183$^{***}$ (0.027) & 0.222$^{***}$ (0.040) & 2,458 \\
Excl. largest owner group 2 (563 pairs) & 0.350$^{***}$ (0.033) & 0.375$^{***}$ (0.050) & 3,724 \\
Excl. largest owner group 3 (345 pairs) & 0.306$^{***}$ (0.033) & 0.301$^{***}$ (0.047) & 3,942 \\
Excl. largest owner group 4 (249 pairs) & 0.337$^{***}$ (0.032) & 0.345$^{***}$ (0.044) & 4,038 \\
Excl. largest owner group 5 (184 pairs) & 0.340$^{***}$ (0.032) & 0.352$^{***}$ (0.044) & 4,103 \\
\hline
\hline \\[-1.8ex]
\end{tabular}

\begin{tablenotes}[flushleft]
\footnotesize{
\item \textit{Notes}: Switching pairs are grouped by the common domain owner involved: pairs losing common ownership by their period-0 owner, pairs gaining it by their period-1 owner. The 4{,}287 switching pairs in the coverage estimation sample trace back to 92 distinct owner groups involving 326 outlets whose domain owner changes between the two waves; because pairs are combinations of outlets, large portfolios generate disproportionately many pairs (the largest group, a portfolio of 102 U.S.\ local news outlets, accounts for 43\% of switching pairs, and the five largest groups for 74\%). Each row re-estimates the main within-dyad specifications for coverage (Table~\ref{tab:coverage_pairs}, column 2) and tone (Table~\ref{tab:reporting_tone}, column 2) after excluding all switching pairs of the indicated owner group. Standard errors in parentheses are dyadic two-way clustered by domain (\texttt{domain\_a} and \texttt{domain\_b}), following \citet{cameron_miller_2015}. $^{*}$p$<$0.1; $^{**}$p$<$0.05; $^{***}$p$<$0.01.
}
\end{tablenotes}
\end{threeparttable}
\end{table}

\begin{table}[htbp]
\centering
\begin{threeparttable}
\singlespacing
\caption{Ownership-Event-Level Estimates and Inference} \label{tab:switcher_events}
\footnotesize
% Table created by code/tab_switcher_anatomy.R (ownership-event-level panel)
\begin{tabular}{@{\extracolsep{5pt}} lcc}
\\[-1.8ex]\hline
\hline \\[-1.8ex]
 & Coverage (DSC) & Tone (cosine) \\
\hline \\[-1.8ex]
Pair-weighted estimate & 0.330$^{***}$ & 0.342$^{***}$ \\
\quad (ownership-event-clustered SE) & (0.092) & (0.106) \\
\quad bootstrap 95\% CI & [0.088, 0.443] & [0.089, 0.516] \\
Event-weighted estimate & 0.061$^{***}$ & 0.072$^{*}$ \\
\quad (SE over events) & (0.022) & (0.040) \\
\quad bootstrap 95\% CI & [0.019, 0.107] & [-0.004, 0.152] \\
\hline \\[-1.8ex]
Mean per-event effect, events $>$100 pairs & 0.227 (7 events) & 0.534 (3 events) \\
Mean per-event effect, events $\leq$100 pairs & 0.047 (85 events) & 0.036 (39 events) \\
No. ownership events & 92 & 42 \\
No. switching pairs & 4,287 & 1,043 \\
\hline
\hline \\[-1.8ex]
\end{tabular}

\begin{tablenotes}[flushleft]
\footnotesize{
\item \textit{Notes}: This table re-expresses the within-dyad estimates at the level of ownership events. For each switching pair, the pair-level effect is the sign-adjusted change in the outcome net of the period trend estimated from non-switching pairs; per-event effects average these within the owner groups of Table~\ref{tab:switcher_anatomy}. The pair-weighted mean of the per-event effects reproduces the fixed-effects estimates of Tables~\ref{tab:coverage_pairs} and~\ref{tab:reporting_tone}; its standard error is clustered at the ownership-event level, and bootstrap confidence intervals resample events. The event-weighted mean gives each ownership event equal weight and is the relevant magnitude for a typical ownership change; the pair-weighted estimate is the relevant magnitude for market-level content homogenization, since larger portfolios affect more outlet pairs. The size split shows that content alignment is concentrated in large portfolio events. Per-event effects for small events are individually noisy (a single-pair event reflects one pair's change net of the common trend), so the size-split means, rather than medians of individual per-event effects, are the informative statistics. The tone column uses the Table~\ref{tab:reporting_tone} estimation sample. $^{*}$p$<$0.1; $^{**}$p$<$0.05; $^{***}$p$<$0.01.
}
\end{tablenotes}
\end{threeparttable}
\end{table}

\begin{figure}[htbp]
\centering
\caption{Quarterly Coverage-Similarity Trajectory by Ownership-change Group}
\label{fig:pretrend}
\includegraphics[width=0.99\textwidth]{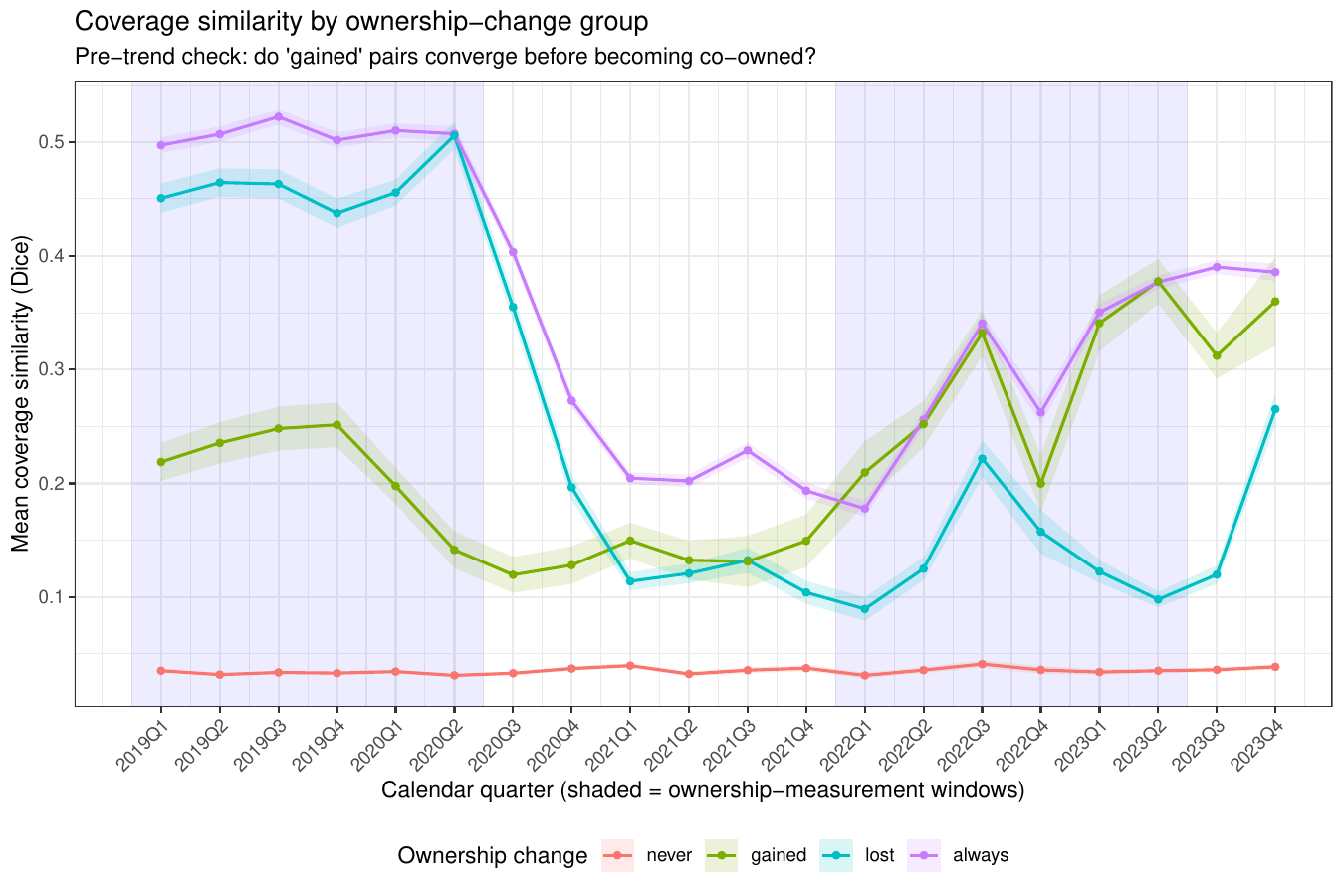}
\begin{figurenotes}
Groups: \emph{gained} = co-owned only in 2023 (FALSE$\to$TRUE); \emph{lost} = co-owned only in 2020 (TRUE$\to$FALSE); \emph{always} = both periods; \emph{never} = neither (reference). Lines are cross-pair mean Dice coefficients per quarter; ribbons are $\pm1.96\times$ SEM; shaded bands mark the two ownership-measurement windows (2019Q1--2020Q2 and 2022Q1--2023Q2), constructed around the July 2020 and July 2023 ownership snapshots. The differential within-period-0 trend for \emph{gained} pairs is $-0.015$ per quarter (SE $0.005$); Table~\ref{tab:pretrend_trend} reports the full differential-trend estimates. Standard errors are dyadic two-way clustered by domain.
\end{figurenotes}
\end{figure}

\begin{table}[htbp]
\centering
\begin{threeparttable}
\singlespacing
\caption{Pre-trend Test: Differential Coverage-Similarity Trends Within Period 0} \label{tab:pretrend_trend}
\footnotesize

% Table created by stargazer v.5.2.3 by Marek Hlavac, Social Policy Institute. E-mail: marek.hlavac at gmail.com
% Date and time: Tue, Jul 21, 2026 - 11:25:49 AM
\begin{tabular}{@{\extracolsep{5pt}}lc} 
\\[-1.8ex]\hline 
\hline \\[-1.8ex] 
 & \multicolumn{1}{c}{Dependent variable:} \\ 
\cline{2-2} 
\\[-1.8ex] & DSC (quarterly) \\ 
\hline \\[-1.8ex] 
 Quarterly trend (baseline: never co-owned) & $-$0.0004$^{**}$ \\ 
  & (0.0002) \\ 
  & \\ 
 Trend $\times$ gained co-ownership & $-$0.0154$^{***}$ \\ 
  & (0.0052) \\ 
  & \\ 
 Trend $\times$ lost co-ownership & 0.0058 \\ 
  & (0.0067) \\ 
  & \\ 
 Trend $\times$ always co-owned & 0.0001 \\ 
  & (0.0032) \\ 
  & \\ 
\hline \\[-1.8ex] 
Pair FE & Yes \\ 
Sample & 2019Q1--2020Q2 \\ 
Pairs: gained / lost / always / never & 853 / 3,432 / 10,052 / 7,991 \\ 
No. pairs & 22,328 \\ 
No. observations & 126,023 \\ 
Within R$^2$ & 0.004 \\ 
R$^{2}$ & 0.9268 \\ 
\hline 
\hline \\[-1.8ex] 
\end{tabular} 

\begin{tablenotes}[flushleft]
\footnotesize{
\item \textit{Notes}: This table reports the differential-trend regression underlying Figure~\ref{fig:pretrend}. The sample covers the quarters of the first ownership-measurement window (2019Q1--2020Q2), and the dependent variable is the quarterly Dice similarity coefficient of event coverage for each outlet pair. The estimation sample comprises all pairs that gain or lose common domain ownership between the two waves, all always-co-owned pairs, and a fixed random reference sample of 8{,}000 never-co-owned pairs (the reference group serves only to pin down the baseline quarterly trend); pairs enter with the quarters in which both outlets record covered events, yielding the number of pairs reported in the table. The linear quarterly trend is interacted with indicators for the ownership-change groups defined in Figure~\ref{fig:pretrend} (\emph{gained}, \emph{lost}, \emph{always}), with \emph{never} co-owned pairs as the reference group; pair fixed effects absorb level differences between groups. A positive trend differential for \emph{gained} pairs would indicate convergence in coverage predating common ownership (consistent with reverse causality); the estimated differential is negative. Transaction dates are unobserved and the switcher subsample is small, so this evidence is suggestive rather than a clean event study. Standard errors are shown in parentheses below the respective coefficient estimates and are dyadic two-way clustered by domain (\texttt{domain\_a} and \texttt{domain\_b}), following \citet{cameron_miller_2015}. The statistical significance of the coefficient estimates is indicated as follows: $^{*}$p$<$0.1; $^{**}$p$<$0.05; $^{***}$p$<$0.01.
}
\end{tablenotes}
\end{threeparttable}
\end{table}
\clearpage
\subsection{Intensive-margin robustness of the tone results}

\begin{table}[htbp]
\centering
\begin{threeparttable}
\singlespacing
\caption{Tone Similarity: Intersection-Normalized Measure} \label{tab:tone_intersection}
\footnotesize

% Table created by stargazer v.5.2.3 by Marek Hlavac, Social Policy Institute. E-mail: marek.hlavac at gmail.com
% Date and time: Tue, Jul 21, 2026 - 10:47:04 AM
\begin{tabular}{@{\extracolsep{5pt}}lccccccc} 
\\[-1.8ex]\hline 
\hline \\[-1.8ex] 
 & \multicolumn{7}{c}{Dependent variable:} \\ 
\cline{2-8} 
\\[-1.8ex] & \multicolumn{7}{c}{Intersection-normalized cosine tone similarity} \\ 
\\[-1.8ex] & (1) & (2) & (3) & (4) & (5) & (6) & (7)\\ 
\hline \\[-1.8ex] 
 Same domain owner & 0.1685$^{***}$ & 0.0888$^{***}$ &  &  &  &  &  \\ 
  & (0.0074) & (0.0268) &  &  &  &  &  \\ 
  & & & & & & & \\ 
 Ownership shared perc. &  &  & 0.0002$^{***}$ &  & $-$0.0006 &  &  \\ 
  &  &  & (0.0001) &  & (0.0005) &  &  \\ 
  & & & & & & & \\ 
 Ownership cosine simil. &  &  &  & 0.0148$^{*}$ &  &  &  \\ 
  &  &  &  & (0.0082) &  &  &  \\ 
  & & & & & & & \\ 
 Above 50 shared perc. &  &  &  &  &  & 0.0157$^{**}$ &  \\ 
  &  &  &  &  &  & (0.0078) &  \\ 
  & & & & & & & \\ 
 Own. sh. perc. Q1-2 &  &  &  &  &  &  & $-$0.0073 \\ 
  &  &  &  &  &  &  & (0.0177) \\ 
  & & & & & & & \\ 
 Own. sh. perc. Q2-3 &  &  &  &  &  &  & $-$0.0314$^{**}$ \\ 
  &  &  &  &  &  &  & (0.0125) \\ 
  & & & & & & & \\ 
 Own. sh. perc. Q3+ &  &  &  &  &  &  & $-$0.0346$^{*}$ \\ 
  &  &  &  &  &  &  & (0.0188) \\ 
  & & & & & & & \\ 
\hline \\[-1.8ex] 
Period FE & Yes & Yes & Yes & Yes & Yes & Yes & Yes \\ 
Dyad FE & No & Yes & Yes & Yes & Yes & Yes & Yes \\ 
Mean dep. var. & 0.786 & 0.786 & 0.784 & 0.784 & 0.906 & 0.906 & 0.906 \\ 
No. news sites & 1,346 & 1,346 & 1,346 & 1,346 & 870 & 870 & 870 \\ 
No. observations & 986,512 & 986,512 & 977,249 & 977,249 & 58,735 & 58,735 & 58,735 \\ 
\hline 
\hline \\[-1.8ex] 
\end{tabular} 

\begin{tablenotes}[flushleft]
\footnotesize{
\item \textit{Notes}: This table re-estimates all specifications of Table~\ref{tab:reporting_tone} with an intersection-normalized cosine tone similarity as the dependent variable: both outlets' tone vectors are restricted to the events covered by both before normalizing, so the measure isolates how similarly the two outlets report the events they both cover, independent of the extent of coverage overlap. The sample construction is identical to Table~\ref{tab:reporting_tone}, so columns are comparable one-for-one. The mean of the intersection-normalized measure is high because tone on jointly covered events is broadly aligned across all outlets; see Section~\ref{sec:results_tone} for the interpretation of the baseline and intersection-normalized estimates as brackets on the tone effect, and Tables~\ref{tab:intersection_descriptives} and~\ref{tab:intersection_ownership} for why the intersection-normalized estimate is conservative. \emph{Ownership shared perc.} is expressed in percentage points (0--100). Standard errors are shown in parentheses below the respective coefficient estimates and are dyadic two-way clustered by domain (\texttt{domain\_a} and \texttt{domain\_b}), following \citet{cameron_miller_2015}. The statistical significance of the coefficient estimates is indicated as follows: $^{*}$p$<$0.1; $^{**}$p$<$0.05; $^{***}$p$<$0.01.
}
\end{tablenotes}
\end{threeparttable}
\end{table}

\begin{table}[htbp]
\centering
\begin{threeparttable}
\singlespacing
\caption{The Jointly Covered Event Set, by Ownership Status} \label{tab:intersection_descriptives}
\footnotesize
% Table created by code/tab_intersection_diagnostics.R
\begin{tabular}{@{\extracolsep{5pt}} lcc}
\\[-1.8ex]\hline
\hline \\[-1.8ex]
 & Different domain owner & Same domain owner \\
\hline \\[-1.8ex]
Median no. jointly covered events & 11.0 & 253.0 \\
Mean overlap share & 0.068 & 0.497 \\
Mean salience, jointly covered events & 197.4 & 169.5 \\
Mean salience, own event sets & 170.6 & 167.3 \\
Median salience ratio (joint/own) & 1.119 & 1.009 \\
Pair-period observations & 2,662,679 & 21,103 \\
\hline
\hline \\[-1.8ex]
\end{tabular}

\begin{tablenotes}[flushleft]
\footnotesize{
\item \textit{Notes}: Descriptive statistics of each pair's jointly covered (tone-scored) event set, pooled over both periods, for pairs with at least one jointly covered event. An event's salience is the number of sample outlets covering it (within the tone-event universe); the salience ratio compares the mean salience of the pair's jointly covered events to the mean salience of the two outlets' own event sets. For pairs with different domain owners, the jointly covered set is small and selected toward the most widely covered events (salience ratio above one); for co-owned pairs it is large and close to representative of the outlets' own event sets. Intersection-conditional tone comparisons across ownership status therefore condition on systematically different event sets.
}
\end{tablenotes}
\end{threeparttable}
\end{table}

\begin{table}[htbp]
\centering
\begin{threeparttable}
\singlespacing
\caption{Common Ownership and the Jointly Covered Event Set} \label{tab:intersection_ownership}
\footnotesize

% Table created by stargazer v.5.2.3 by Marek Hlavac, Social Policy Institute. E-mail: marek.hlavac at gmail.com
% Date and time: Tue, Jul 21, 2026 - 10:47:49 AM
\begin{tabular}{@{\extracolsep{5pt}}lcccc} 
\\[-1.8ex]\hline 
\hline \\[-1.8ex] 
 & \multicolumn{4}{c}{Dependent variable: jointly covered event set of the pair} \\ 
\cline{2-5} 
\\[-1.8ex] & log(1+size) & Overlap share & Mean salience & Salience ratio \\ 
\\[-1.8ex] & (1) & (2) & (3) & (4)\\ 
\hline \\[-1.8ex] 
 Same domain owner & 1.7374$^{***}$ & 0.2491$^{***}$ & $-$16.5231$^{***}$ & $-$0.0708$^{***}$ \\ 
  & (0.2814) & (0.0305) & (3.9011) & (0.0160) \\ 
  & & & & \\ 
\hline \\[-1.8ex] 
Period FE & Yes & Yes & Yes & Yes \\ 
Dyad FE & Yes & Yes & Yes & Yes \\ 
Mean dep. var. & 3.215 & 0.085 & 200.3 & 1.16 \\ 
No. observations & 1,967,854 & 1,967,854 & 1,967,854 & 1,967,854 \\ 
\hline 
\hline \\[-1.8ex] 
\end{tabular} 

\begin{tablenotes}[flushleft]
\footnotesize{
\item \textit{Notes}: Within-dyad regressions of properties of the pair's jointly covered (tone-scored) event set on the same-domain-owner indicator, with period and dyad fixed effects, on the balanced tone-analysis sample. \emph{log(1+size)} is the log number of jointly covered events; \emph{overlap share} is the number of jointly covered events divided by the geometric mean of the two outlets' own event counts; \emph{mean salience} is the average number of sample outlets covering the jointly covered events; the \emph{salience ratio} relates this to the mean salience of the outlets' own event sets. Becoming co-owned expands the jointly covered set roughly six-fold and shifts its composition toward less widely covered events. The jointly covered set is thus itself an outcome of ownership, so intersection-normalized tone comparisons (Table~\ref{tab:tone_intersection}) condition on a post-treatment object and are likely conservative for the tone margin. Standard errors are shown in parentheses below the respective coefficient estimates and are dyadic two-way clustered by domain (\texttt{domain\_a} and \texttt{domain\_b}), following \citet{cameron_miller_2015}. $^{*}$p$<$0.1; $^{**}$p$<$0.05; $^{***}$p$<$0.01.
}
\end{tablenotes}
\end{threeparttable}
\end{table}
\clearpage
\subsection{Event-level tone regressions}

\begin{table}[htbp]
\centering
\begin{threeparttable}
\singlespacing
\caption{Analysis of Tone Using Quartile Bins} \label{tab:events_tone_bin}
\footnotesize

% Table created by stargazer v.5.2.3 by Marek Hlavac, Social Policy Institute. E-mail: marek.hlavac at gmail.com
% Date and time: Mon, Jun 02, 2025 - 12:14:24 PM
\begin{tabular}{@{\extracolsep{5pt}}lcccccc} 
\\[-1.8ex]\hline 
\hline \\[-1.8ex] 
 & \multicolumn{6}{c}{Dependent variable:} \\ 
\cline{2-7} 
\\[-1.8ex] & \multicolumn{6}{c}{Tone in which event was reported} \\ 
\\[-1.8ex] & (1) & (2) & (3) & (4) & (5) & (6)\\ 
\hline \\[-1.8ex] 
 Avg. tone same d-owner (Q2) & 0.246$^{***}$ &  & 0.194$^{***}$ & 0.111$^{***}$ & 0.234$^{***}$ &  \\ 
  & (0.012) &  & (0.012) & (0.018) & (0.015) &  \\ 
  & & & & & & \\ 
 Avg. tone same d-owner (Q3) & 0.475$^{***}$ &  & 0.366$^{***}$ & 0.302$^{***}$ & 0.371$^{***}$ &  \\ 
  & (0.020) &  & (0.020) & (0.031) & (0.024) &  \\ 
  & & & & & & \\ 
 Avg. tone same d-owner (Q4) & 0.795$^{***}$ &  & 0.529$^{***}$ & 0.445$^{***}$ & 0.638$^{***}$ &  \\ 
  & (0.035) &  & (0.037) & (0.060) & (0.043) &  \\ 
  & & & & & & \\ 
 Avg. tone same u-owner (Q2) &  & 0.229$^{***}$ & 0.086$^{***}$ & 0.144$^{***}$ & 0.144$^{***}$ &  \\ 
  &  & (0.013) & (0.013) & (0.021) & (0.016) &  \\ 
  & & & & & & \\ 
 Avg. tone same u-owner (Q3) &  & 0.419$^{***}$ & 0.145$^{***}$ & 0.152$^{***}$ & 0.285$^{***}$ &  \\ 
  &  & (0.020) & (0.020) & (0.033) & (0.025) &  \\ 
  & & & & & & \\ 
 Avg. tone same u-owner (Q4) &  & 0.769$^{***}$ & 0.348$^{***}$ & 0.460$^{***}$ & 0.629$^{***}$ &  \\ 
  &  & (0.036) & (0.037) & (0.063) & (0.043) &  \\ 
  & & & & & & \\ 
 Avg. tone same d-owner &  &  &  &  &  & 0.145$^{***}$ \\ 
  &  &  &  &  &  & (0.010) \\ 
  & & & & & & \\ 
 Avg. tone same u-owner &  &  &  &  &  & 0.058$^{***}$ \\ 
  &  &  &  &  &  & (0.011) \\ 
  & & & & & & \\ 
 Avg. tone same region & 0.041$^{***}$ & 0.051$^{***}$ & 0.042$^{***}$ & 0.029$^{***}$ & 0.042$^{***}$ & 0.033$^{***}$ \\ 
  & (0.005) & (0.005) & (0.005) & (0.005) & (0.006) & (0.005) \\ 
  & & & & & & \\ 
 Avg. tone same country & 0.345$^{***}$ & 0.342$^{***}$ & 0.330$^{***}$ & 0.423$^{***}$ & 0.310$^{***}$ & 0.245$^{***}$ \\ 
  & (0.018) & (0.018) & (0.018) & (0.024) & (0.021) & (0.019) \\ 
  & & & & & & \\ 
 Avg. tone same media type & 0.123$^{***}$ & 0.119$^{***}$ & 0.110$^{***}$ & 0.037 & 0.113$^{***}$ & 0.038$^{**}$ \\ 
  & (0.017) & (0.017) & (0.017) & (0.023) & (0.018) & (0.016) \\ 
  & & & & & & \\ 
\hline \\[-1.8ex] 
Website FE & Yes & Yes & Yes & Yes & Yes & Yes \\ 
Event FE & Yes & Yes & Yes & Yes & Yes & Yes \\ 
Mean dep. var. & -3.81 & -3.81 & -3.81 & -3.81 & -3.81 & -3.81 \\ 
Sample & All & All & All & Below Median & Above Median & All \\ 
Number of websites & 1,199 & 1,199 & 1,199 & 1,199 & 1,199 & 1,199 \\ 
Within R$^2$ & 0.057 & 0.054 & 0.059 & 0.047 & 0.078 & 0.072 \\ 
Total R$^2$ & 0.87 & 0.869 & 0.87 & 0.887 & 0.876 & 0.872 \\ 
Observations & 3,597,598 & 3,597,598 & 3,597,598 & 1,799,540 & 1,798,058 & 3,597,598 \\ 
\hline 
\hline \\[-1.8ex] 
\end{tabular} 

\begin{tablenotes}[flushleft]
\footnotesize{
\item \textit{Notes}: This table reports the results from estimating the relationship between the tone in which an event is reported by the focal outlet and the average upstream tone under common ownership. Columns 1–3 present specifications using the full sample, where the average upstream tone---the mean tone with which the other outlets under the same domain owner (d-owner) or ultimate owner (u-owner) report the same event---is divided into quartiles. The reference category for these quartile bins is the lowest quartile (Q1). Columns 4 and 5 split the sample at the median of the average ownership share of the outlet's largest ultimate owner and re-estimate the specification using quartile bins for both d-owner and u-owner upstream tone (quartiles recomputed within each subsample). Column 6 reports results from a specification using the continuous measures of average tone similarity rather than quartile bins. All regressions include website and event fixed effects, as indicated. The sample is restricted to observations for which the event was reported and to websites for which complete information is available. ``Avg. tone same d-owner'' refers to the mean tone of other outlets with the same domain owner reporting the same event $e$, ``Avg. tone same u-owner'' refers to the mean tone of other outlets with the same ultimate owner, ``Avg. tone same region'' and ``Avg. tone same country'' capture whether outlets in the same geographic region or country report the event with a similar tone, and ``Avg. tone same media type'' accounts for similarity in tone among outlets of the same media type. ``Below Median'' and ``Above Median'' refer to subsets of the data split at the median of the average ownership share of the outlet's largest ultimate owner. Standard errors are clustered by website and event.
}
\end{tablenotes}
\end{threeparttable}
\end{table}
\clearpage
\subsection{Heterogeneity by event theme}

\begin{table}[htbp]
\centering
\begin{threeparttable}
\singlespacing
\caption{Event-Level Tone Alignment by Event Theme} \label{tab:events_tone_bytheme}
\footnotesize

% Table created by stargazer v.5.2.3 by Marek Hlavac, Social Policy Institute. E-mail: marek.hlavac at gmail.com
% Date and time: Tue, Jul 21, 2026 - 10:50:45 AM
\begin{tabular}{@{\extracolsep{5pt}}lcc} 
\\[-1.8ex]\hline 
\hline \\[-1.8ex] 
 & \multicolumn{2}{c}{Dependent variable: focal outlet's tone for event $e$} \\ 
\cline{2-3} 
\\[-1.8ex] & \multicolumn{2}{c}{Tone} \\ 
 & Political events & Non-political events \\ 
\\[-1.8ex] & (1) & (2)\\ 
\hline \\[-1.8ex] 
 Avg. tone same d-owner & 0.1563$^{***}$ & 0.1306$^{***}$ \\ 
  & (0.0104) & (0.0102) \\ 
  & & \\ 
 Avg. tone same u-owner & 0.0512$^{***}$ & 0.0686$^{***}$ \\ 
  & (0.0117) & (0.0121) \\ 
  & & \\ 
 Avg. tone same state & 0.0304$^{***}$ & 0.0377$^{***}$ \\ 
  & (0.0055) & (0.0061) \\ 
  & & \\ 
 Avg. tone same country & 0.2474$^{***}$ & 0.2598$^{***}$ \\ 
  & (0.0218) & (0.0255) \\ 
  & & \\ 
 Avg. tone same media type & 0.0248 & 0.0614$^{***}$ \\ 
  & (0.0174) & (0.0236) \\ 
  & & \\ 
\hline \\[-1.8ex] 
Website FE & Yes & Yes \\ 
Event FE & Yes & Yes \\ 
Diff. political $-$ non-political (d-owner) & 0.0257$^{***}$ & (0.0076) \\ 
Diff. political $-$ non-political (u-owner) & -0.0175$^{**}$ & (0.0081) \\ 
\hline 
\hline \\[-1.8ex] 
\end{tabular} 

\begin{tablenotes}[flushleft]
\footnotesize{
\item \textit{Notes}: This table re-estimates the continuous event-level tone regression of Table~\ref{tab:events_tone_bin} (column 6) separately for political and non-political events (an event is political if a majority of its GDELT mentions carry the POLITICS theme tag, as in Table~\ref{tab:bytheme_interaction}). The dependent variable is the focal outlet's tone for event $e$; \emph{Avg.\ tone same d-owner} (\emph{u-owner}) is the mean tone of the other outlets under the same domain (ultimate) owner reporting the same event. Event fixed effects absorb event-specific tone, and the sample conditions only on the focal outlet's own coverage, so this specification isolates the intensive tone margin without the intersection-conditioning issues of the pair-level measures. Because the key regressors are leave-out means of the same outcome within the owner group, the coefficients measure within-event tone co-movement among co-owned outlets rather than a directional pass-through (the reflection problem of \citet{manski1993}; see Section~\ref{sec:mechanisms}); the theme differences remain interpretable under symmetric joint determination of co-owned outlets' tone across themes. The bottom rows report the political$-$non-political differences in the co-movement coefficients from a fully interacted stacked model with website-by-theme and event fixed effects (difference estimate with significance stars in the first column, standard error in the second). Standard errors are clustered two-way by website and event. $^{*}$p$<$0.1; $^{**}$p$<$0.05; $^{***}$p$<$0.01.
}
\end{tablenotes}
\end{threeparttable}
\end{table}

\begin{table}[htbp]
\centering
\begin{threeparttable}
\caption{Coverage Similarity by Event Theme (Political vs.\ Non-political)} \label{tab:cov_bytheme}
\footnotesize
\singlespacing

% Table created by stargazer v.5.2.3 by Marek Hlavac, Social Policy Institute. E-mail: marek.hlavac at gmail.com
% Date and time: Thu, May 28, 2026 - 02:23:34 PM
\begin{tabular}{@{\extracolsep{0.0pt}}lcccc} 
\\[-1.8ex]\hline 
\hline \\[-1.8ex] 
 & \multicolumn{4}{c}{Dependent variable: DSC (coverage), by event theme} \\ 
\cline{2-5} 
 & Political & Non-political & Political & Non-political \\ 
\\[-1.8ex] & (1) & (2) & (3) & (4)\\ 
\hline \\[-1.8ex] 
 Same domain owner & 0.327$^{***}$ & 0.262$^{***}$ &  &  \\ 
  & (0.034) & (0.037) &  &  \\ 
  & & & & \\ 
 Ownership shared perc. &  &  & 0.002$^{***}$ & $-$0.0003$^{**}$ \\ 
  &  &  & (0.0001) & (0.0001) \\ 
  & & & & \\ 
\hline \\[-1.8ex] 
Period FE & Yes & Yes & Yes & Yes \\ 
Dyad FE & Yes & Yes & Yes & Yes \\ 
Sample & All & All & Not same d. o. & Not same d. o. \\ 
Mean dep. var. & 0.04 & 0.041 & 0.038 & 0.039 \\ 
No. news sites & 1,885 & 1,829 & 1,885 & 1,829 \\ 
No. observations & 3,551,340 & 3,343,412 & 3,527,720 & 3,319,626 \\ 
Within R$^2$ & 0.024 & 0.015 & 0.042 & 0.001 \\ 
R$^{2}$ & 0.768 & 0.716 & 0.759 & 0.714 \\ 
\hline 
\hline \\[-1.8ex] 
\end{tabular} 

\begin{tablenotes}[flushleft]
\footnotesize{
\item \textit{Notes}: An event is \emph{political} if a majority ($>50\%$) of its GDELT mentions carry the POLITICS theme tag; coverage similarity is recomputed within each theme's event universe. The same-domain owner coefficient is 0.327 (political) vs.\ 0.262 (non-political). Columns 3--4 use the continuous shared-ownership measure. The main-text Table~\ref{tab:bytheme_interaction} formally tests the political$-$non-political difference. Standard errors are dyadic two-way clustered by domain (\texttt{domain\_a} and \texttt{domain\_b}), following \citet{cameron_miller_2015}. $^{*}$p$<$0.1; $^{**}$p$<$0.05; $^{***}$p$<$0.01.
}
\end{tablenotes}
\end{threeparttable}
\end{table}

\begin{table}[htbp]
\centering
\begin{threeparttable}
\caption{Tone Similarity by Event Theme (Political vs.\ Non-political)} \label{tab:tone_bytheme}
\footnotesize
\singlespacing

% Table created by stargazer v.5.2.3 by Marek Hlavac, Social Policy Institute. E-mail: marek.hlavac at gmail.com
% Date and time: Thu, May 28, 2026 - 02:24:41 PM
\begin{tabular}{@{\extracolsep{0.0pt}}lcccc} 
\\[-1.8ex]\hline 
\hline \\[-1.8ex] 
 & \multicolumn{4}{c}{Dependent variable: cosine-similarity(tone), by event theme} \\ 
\cline{2-5} 
 & Political & Non-political & Political & Non-political \\ 
\\[-1.8ex] & (1) & (2) & (3) & (4)\\ 
\hline \\[-1.8ex] 
 Same domain owner & 0.265$^{***}$ & 0.127$^{***}$ &  &  \\ 
  & (0.036) & (0.032) &  &  \\ 
  & & & & \\ 
 Ownership shared perc. &  &  & 0.002$^{***}$ & 0.0004$^{***}$ \\ 
  &  &  & (0.0001) & (0.0002) \\ 
  & & & & \\ 
\hline \\[-1.8ex] 
Period FE & Yes & Yes & Yes & Yes \\ 
Dyad FE & Yes & Yes & Yes & Yes \\ 
Sample & All & All & Not same d. o. & Not same d. o. \\ 
Mean dep. var. & 0.088 & 0.089 & 0.083 & 0.086 \\ 
No. news sites & 1,878 & 1,819 & 1,878 & 1,819 \\ 
No. observations & 1,634,182 & 1,286,824 & 1,617,642 & 1,275,750 \\ 
Within R$^2$ & 0.01 & 0.002 & 0.057 & 0.002 \\ 
R$^{2}$ & 0.757 & 0.711 & 0.748 & 0.692 \\ 
\hline 
\hline \\[-1.8ex] 
\end{tabular} 

\begin{tablenotes}[flushleft]
\footnotesize{
\item \textit{Notes}: As Table~\ref{tab:cov_bytheme} but with cosine tone similarity as the dependent variable. The same-domain-owner coefficient is 0.265 (political) vs.\ 0.127 (non-political), roughly a $2\times$ political concentration of the tone effect, much sharper than for coverage. Standard errors are dyadic two-way clustered by domain (\texttt{domain\_a} and \texttt{domain\_b}), following \citet{cameron_miller_2015}. $^{*}$p$<$0.1; $^{**}$p$<$0.05; $^{***}$p$<$0.01.
}
\end{tablenotes}
\end{threeparttable}
\end{table}
\clearpage
\subsection{Demand- versus supply-side and the Gentzkow--Shapiro contrast}

\begin{table}[htbp]
\centering
\begin{threeparttable}
\caption{Same-Domain-Owner Effect: U.S.\ vs.\ Non-U.S.\ Within-Country Dyads} \label{tab:gs_contrast}
\footnotesize
\singlespacing

% Table created by stargazer v.5.2.3 by Marek Hlavac, Social Policy Institute. E-mail: marek.hlavac at gmail.com
% Date and time: Tue, Jun 02, 2026 - 03:57:44 PM
\begin{tabular}{@{\extracolsep{0.0pt}}lcccc} 
\\[-1.8ex]\hline 
\hline \\[-1.8ex] 
 & \multicolumn{4}{c}{Dependent variable: pair-level content similarity} \\ 
\cline{2-5} 
 & \multicolumn{2}{c}{Coverage (DSC)} & \multicolumn{2}{c}{Tone (cosine)} \\ 
\\[-1.8ex] & (1) & (2) & (3) & (4)\\ 
\hline \\[-1.8ex] 
 Same domain owner & 0.325$^{***}$ & 0.325$^{***}$ & 0.267$^{***}$ & 0.250$^{***}$ \\ 
  & (0.036) & (0.051) & (0.038) & (0.070) \\ 
  & & & & \\ 
\hline \\[-1.8ex] 
Sample & US & Non-US & US & Non-US \\ 
Period FE & Yes & Yes & Yes & Yes \\ 
Dyad FE & Yes & Yes & Yes & Yes \\ 
Mean dep. var. & 0.055 & 0.129 & 0.098 & 0.191 \\ 
Switcher pairs & 3,509 & 772 & 1,674 & 463 \\ 
No. news-site pairs & 759,528 & 52,910 & 433,103 & 30,128 \\ 
No. observations & 1,519,056 & 105,820 & 866,206 & 60,256 \\ 
\hline 
\hline \\[-1.8ex] 
\end{tabular} 

\begin{tablenotes}[flushleft]
\footnotesize{
\item \textit{Notes}: Headline within-dyad specification (same domain owner, with period and dyad fixed effects) estimated separately on U.S.\ and pooled non-U.S.\ within-country dyads, for coverage (Dice) and tone (cosine) similarity. The U.S.\ same-owner effect is precise and comparable to the non-U.S.\ estimate, contrasting with the demand-driven, ownership-invariant slant documented for U.S.\ print newspapers by \citet{Gentzkow_Shapiro_2010}. Standard errors are dyadic two-way clustered by domain (\texttt{domain\_a} and \texttt{domain\_b}), following \citet{cameron_miller_2015}. $^{*}$p$<$0.1; $^{**}$p$<$0.05; $^{***}$p$<$0.01.
}
\end{tablenotes}
\end{threeparttable}
\end{table}

\begin{table}[htbp]
\centering
\begin{threeparttable}
\caption{Demand- vs.\ Supply-side Benchmark of Pair-level Content Similarity} \label{tab:demand_supply}
\footnotesize
\singlespacing

% Table created by stargazer v.5.2.3 by Marek Hlavac, Social Policy Institute. E-mail: marek.hlavac at gmail.com
% Date and time: Tue, Jun 02, 2026 - 03:58:41 PM
\begin{tabular}{@{\extracolsep{0.0pt}}lcccc} 
\\[-1.8ex]\hline 
\hline \\[-1.8ex] 
 & \multicolumn{4}{c}{Dependent variable: pair-level content similarity} \\ 
\cline{2-5} 
 & Coverage (DSC) & Tone (cosine) & Coverage (DSC) & Tone (cosine) \\ 
\\[-1.8ex] & (1) & (2) & (3) & (4)\\ 
\hline \\[-1.8ex] 
 Same domain owner & 0.407$^{***}$ & 0.405$^{***}$ & 0.403$^{***}$ & 0.401$^{***}$ \\ 
  & (0.015) & (0.016) & (0.015) & (0.016) \\ 
  & & & & \\ 
 Same country & 0.041$^{***}$ & 0.043$^{***}$ &  &  \\ 
  & (0.002) & (0.002) &  &  \\ 
  & & & & \\ 
 Same pub. state & 0.031$^{***}$ & 0.027$^{***}$ & 0.031$^{***}$ & 0.026$^{***}$ \\ 
  & (0.005) & (0.005) & (0.005) & (0.005) \\ 
  & & & & \\ 
 Same media type & 0.008$^{***}$ & 0.008$^{***}$ & 0.018$^{***}$ & 0.019$^{***}$ \\ 
  & (0.001) & (0.001) & (0.002) & (0.002) \\ 
  & & & & \\ 
 Period (after) & 0.015$^{***}$ & 0.018$^{***}$ & 0.012$^{***}$ & 0.015$^{***}$ \\ 
  & (0.002) & (0.002) & (0.003) & (0.003) \\ 
  & & & & \\ 
 Constant & 0.025$^{***}$ & 0.031$^{***}$ & 0.062$^{***}$ & 0.070$^{***}$ \\ 
  & (0.002) & (0.002) & (0.003) & (0.003) \\ 
  & & & & \\ 
\hline \\[-1.8ex] 
Sample & Full & Full & Within-cty. & Within-cty. \\ 
Dyad FE & No & No & No & No \\ 
No. observations & 1,097,811 & 1,097,811 & 562,245 & 562,245 \\ 
\hline 
\hline \\[-1.8ex] 
\end{tabular} 

\begin{tablenotes}[flushleft]
\footnotesize{
\item \textit{Notes}: Cross-sectional specification (no dyad fixed effects) regressing pair-level coverage (Dice) and tone (cosine) similarity on a same-domain-owner indicator alongside shared-market controls (same country, same state, same media type) and a period dummy. Columns 1--2 use all pairs with non-missing tone similarity, so that the coverage and tone benchmarks are estimated on identical samples; columns 3--4 restrict this sample to within-country dyads (where \emph{same country} is constant and is dropped). The same-domain-owner coefficient is roughly 10--50 times the size of any shared-market control. This is a \emph{descriptive} benchmark of relative magnitudes at the pair level; the better identification rests on the within-dyad specifications (Tables~\ref{tab:coverage_pairs}, \ref{tab:reporting_tone}), and it is distinct from the event-level tone-prediction regression in Table~\ref{tab:events_tone_bin}, which is in levels rather than pair similarity. Standard errors are dyadic two-way clustered by domain (\texttt{domain\_a} and \texttt{domain\_b}), following \citet{cameron_miller_2015}. $^{*}$p$<$0.1; $^{**}$p$<$0.05; $^{***}$p$<$0.01.
}
\end{tablenotes}
\end{threeparttable}
\end{table}
\clearpage
\subsection{Alternative specifications and sample restrictions}

\begin{table}[htbp]
\centering
\begin{threeparttable}
\caption{Alternative Similarity Measures and Ownership} \label{tab:coverage_alt_similarity_pairs}
\footnotesize
\singlespacing

% Table created by stargazer v.5.2.3 by Marek Hlavac, Social Policy Institute. E-mail: marek.hlavac at gmail.com
% Date and time: Mon, Jul 20, 2026 - 10:25:41 AM
\begin{tabular}{@{\extracolsep{5pt}}lcccccc} 
\\[-1.8ex]\hline 
\hline \\[-1.8ex] 
 & \multicolumn{6}{c}{Dependent variable:} \\ 
\cline{2-7} 
\\[-1.8ex] & \multicolumn{3}{c}{DSC} & DSC (weighted) & Jaccard & Cosine \\ 
\\[-1.8ex] & (1) & (2) & (3) & (4) & (5) & (6)\\ 
\hline \\[-1.8ex] 
 Same domain owner & 0.330$^{***}$ & 0.261$^{***}$ & 0.163$^{***}$ & 3.097$^{***}$ & 0.274$^{***}$ & 0.337$^{***}$ \\ 
  & (0.031) & (0.022) & (0.036) & (0.340) & (0.028) & (0.031) \\ 
  & & & & & & \\ 
\hline \\[-1.8ex] 
Period FE & Yes & Yes & Yes & Yes & Yes & Yes \\ 
Dyad FE & Yes & Yes & Yes & Yes & Yes & Yes \\ 
Sample & All dyads & Switchers & Top 1/3 domains & All dyads & All dyads & All dyads \\ 
Mean dep. var. & 0.037 & 0.273 & 0.143 & 0.29 & 0.022 & 0.048 \\ 
Mean ownership shared perc. & 1.656 & 61.806 & 3.276 & 1.656 & 1.656 & 1.656 \\ 
No. news sites & 2,019 & 779 & 678 & 2,019 & 2,019 & 2,019 \\ 
No. observations & 4,074,342 & 8,574 & 459,006 & 4,074,342 & 4,074,342 & 4,074,342 \\ 
Within R$^2$ & 0.03 & 0.265 & 0.003 & 0.036 & 0.04 & 0.03 \\ 
R$^{2}$ & 0.765 & 0.705 & 0.768 & 0.754 & 0.757 & 0.776 \\ 
\hline 
\hline \\[-1.8ex] 
\end{tabular} 

\begin{tablenotes}[flushleft]
\footnotesize{
\item \textit{Notes}: This table presents regressions relating different measures of coverage similarity between pairs of news outlets (``dyads'') to their ownership structure. The dependent variables capture how similar or overlapping the coverage of two outlets is, based on different similarity metrics: ``DSC'' (Dice similarity coefficient), ``DSC (weighted)'' (Dice similarity weighted by the logarithm of the sum of the outlets' domain sizes), ``Jaccard'' similarity, and ``Cosine'' similarity. All measures are constructed from binary event-coverage data, indicating whether each outlet covered a given event. The key explanatory variable is ``Same domain owner,'' an indicator for whether the two outlets share the same domain-level owner. Specifications are estimated on different subsamples: ``All dyads'' includes the full sample; ``Switchers'' restricts to outlet pairs that change their domain-level ownership status between periods; ``Top 1/3 domains'' focuses on outlet pairs drawn from the top third of domains by coverage size. All regressions include period and dyad fixed effects. Standard errors are shown in parentheses below the respective coefficient estimates and are dyadic two-way clustered by domain (\texttt{domain\_a} and \texttt{domain\_b}), following \citet{cameron_miller_2015}.
}
\end{tablenotes}
\end{threeparttable}
\end{table}

\begin{table}[htbp]
\centering
\begin{threeparttable}
\caption{Alternative Similarity Measures and Ownership (Continuous Share)} \label{tab:coverage_alt_similarity_pairs_II}
\footnotesize
\singlespacing

% Table created by stargazer v.5.2.3 by Marek Hlavac, Social Policy Institute. E-mail: marek.hlavac at gmail.com
% Date and time: Mon, Jul 20, 2026 - 10:26:48 AM
\begin{tabular}{@{\extracolsep{5pt}}lcccccc} 
\\[-1.8ex]\hline 
\hline \\[-1.8ex] 
 & \multicolumn{6}{c}{Dependent variable:} \\ 
\cline{2-7} 
\\[-1.8ex] & \multicolumn{3}{c}{DSC} & DSC (weighted) & Jaccard & Cosine \\ 
\\[-1.8ex] & (1) & (2) & (3) & (4) & (5) & (6)\\ 
\hline \\[-1.8ex] 
 Ownership shared percentage & 0.0016$^{***}$ & 0.0023$^{***}$ & 0.0015$^{***}$ & 0.0097$^{***}$ & 0.0012$^{***}$ & 0.0016$^{***}$ \\ 
  & (0.0001) & (0.0003) & (0.0001) & (0.0011) & (0.0001) & (0.0001) \\ 
  & & & & & & \\ 
\hline \\[-1.8ex] 
Period FE & Yes & Yes & Yes & Yes & Yes & Yes \\ 
Dyad FE & Yes & Yes & Yes & Yes & Yes & Yes \\ 
Sample & All dyads & Switchers & Top 1/3 domains & All dyads & All dyads & All dyads \\ 
Mean dep. var. & 0.037 & 0.273 & 0.143 & 0.29 & 0.022 & 0.048 \\ 
Mean ownership shared perc. & 1.656 & 61.806 & 3.276 & 1.656 & 1.656 & 1.656 \\ 
No. news sites & 2,019 & 779 & 678 & 2,019 & 2,019 & 2,019 \\ 
No. observations & 4,074,342 & 8,574 & 459,006 & 4,074,342 & 4,074,342 & 4,074,342 \\ 
Within R$^2$ & 0.046 & 0.17 & 0.032 & 0.023 & 0.049 & 0.043 \\ 
R$^{2}$ & 0.7688 & 0.6666 & 0.7754 & 0.7503 & 0.7597 & 0.7787 \\ 
\hline 
\hline \\[-1.8ex] 
\end{tabular} 

\begin{tablenotes}[flushleft]
\footnotesize{
\item \textit{Notes}: This table presents regressions that relate different measures of coverage similarity between pairs of news outlets (``dyads'') to the continuous share of ultimate ownership they have in common. The dependent variables are the same as in the previous table: ``DSC'' (Dice similarity coefficient), ``DSC (weighted)'' (Dice similarity weighted by the logarithm of the sum of the outlets' domain sizes), ``Jaccard'' similarity, and ``Cosine'' similarity. Each measure is based on binary event-coverage data, indicating whether each outlet covered a particular event. Here, the key explanatory variable is ``Ownership shared percentage,'' which captures the share of ultimate ownership that is held in common by the two outlets, measured in percentage points (0--100), as in the main-text tables. Different columns present results for distinct subsamples: ``All dyads'' includes the full sample; ``Switchers'' restricts to outlet pairs that change their domain-level ownership status between periods; and ``Top 1/3 domains'' focuses on the largest third of domains by coverage size. All regressions include period and dyad fixed effects. Standard errors are shown in parentheses below the respective coefficient estimates and are dyadic two-way clustered by domain (\texttt{domain\_a} and \texttt{domain\_b}), following \citet{cameron_miller_2015}
}
\end{tablenotes}
\end{threeparttable}
\end{table}

\begin{landscape}
\begin{table}[htbp]
\centering
\begin{threeparttable}
\caption{Robustness Checks: Coverage Similarity and Ownership} \label{tab:coverage_robustness}
\footnotesize
\singlespacing

% Table created by stargazer v.5.2.3 by Marek Hlavac, Social Policy Institute. E-mail: marek.hlavac at gmail.com
% Date and time: Tue, Jul 21, 2026 - 09:09:06 AM
\begin{tabular}{@{\extracolsep{5pt}}lcccccccccc} 
\\[-1.8ex]\hline 
\hline \\[-1.8ex] 
 & \multicolumn{10}{c}{Dependent variable:} \\ 
\cline{2-11} 
\\[-1.8ex] & \multicolumn{10}{c}{Dice-Similarity(coverage news site a, coverage news site b)} \\ 
\\[-1.8ex] & (1) & (2) & (3) & (4) & (5) & (6) & (7) & (8) & (9) & (10)\\ 
\hline \\[-1.8ex] 
 Same domain owner & 0.3631$^{***}$ & 0.3844$^{***}$ & 0.2920$^{***}$ & 0.4324$^{***}$ & 0.1517$^{***}$ &  &  &  &  &  \\ 
  & (0.0171) & (0.0363) & (0.0257) & (0.0378) & (0.0350) &  &  &  &  &  \\ 
  & & & & & & & & & & \\ 
 Ownership shared percentage &  &  &  &  &  & 0.0032$^{***}$ & 0.0017$^{***}$ & 0.0025$^{***}$ & 0.0017$^{***}$ & 0.0015$^{***}$ \\ 
  &  &  &  &  &  & (0.0001) & (0.0002) & (0.0004) & (0.0002) & (0.0002) \\ 
  & & & & & & & & & & \\ 
 Same country & 0.0339$^{***}$ &  &  &  &  & 0.0257$^{***}$ &  &  &  &  \\ 
  & (0.0021) &  &  &  &  & (0.0019) &  &  &  &  \\ 
  & & & & & & & & & & \\ 
 Same state & 0.0261$^{***}$ &  &  &  &  & 0.0295$^{***}$ &  &  &  &  \\ 
  & (0.0043) &  &  &  &  & (0.0048) &  &  &  &  \\ 
  & & & & & & & & & & \\ 
 Same media type & 0.0043$^{***}$ &  &  &  &  & $-$0.0005 &  &  &  &  \\ 
  & (0.0011) &  &  &  &  & (0.0010) &  &  &  &  \\ 
  & & & & & & & & & & \\ 
 Period (after) & $-$0.0078$^{***}$ &  &  &  &  & $-$0.0089$^{***}$ &  &  &  &  \\ 
  & (0.0013) &  &  &  &  & (0.0013) &  &  &  &  \\ 
  & & & & & & & & & & \\ 
 constant & 0.0233$^{***}$ &  &  &  &  & 0.0270$^{***}$ &  &  &  &  \\ 
  & (0.0014) &  &  &  &  & (0.0014) &  &  &  &  \\ 
  & & & & & & & & & & \\ 
\hline \\[-1.8ex] 
Period FE & No & Yes & Yes & Yes & Yes & No & Yes & Yes & Yes & Yes \\ 
Dyad FE & No & Yes & Yes & Yes & Yes & No & Yes & Yes & Yes & Yes \\ 
No "Same owner" variability & Yes & Yes & No & Yes & Yes & Yes & Yes & No & Yes & Yes \\ 
Variability Same$\rightarrow\textnormal{Not same}$ & Yes & Yes & Yes & Yes & No & Yes & Yes & Yes & Yes & No \\ 
Variability Not same$\rightarrow\textnormal{Same}$ & Yes & Yes & Yes & No & Yes & Yes & Yes & Yes & No & Yes \\ 
Mean dep. var. & 0.041 & 0.041 & 0.297 & 0.041 & 0.041 & 0.041 & 0.041 & 0.297 & 0.041 & 0.041 \\ 
No. news sites & 1,284 & 1,284 & 469 & 1,284 & 1,284 & 1,284 & 1,284 & 469 & 1,284 & 1,284 \\ 
No. observations & 1,647,372 & 1,647,372 & 4,276 & 1,646,642 & 1,643,826 & 1,647,372 & 1,647,372 & 4,276 & 1,646,642 & 1,643,826 \\ 
Within R$^2$ &  & 0.041 & 0.283 & 0.043 & 0.001 &  & 0.053 & 0.175 & 0.052 & 0.037 \\ 
R$^{2}$ & 0.1716 & 0.7560 & 0.7326 & 0.7561 & 0.7537 & 0.2488 & 0.7592 & 0.6921 & 0.7584 & 0.7625 \\ 
\hline 
\hline \\[-1.8ex] 
\end{tabular} 

\begin{tablenotes}[flushleft]
\footnotesize{
\item \textit{Notes}: This table presents additional regression specifications examining the robustness of the relationship between coverage similarity and media ownership. The dependent variable is the Dice similarity coefficient (``Dice-Similarity''), which measures the overlap in event coverage between two news outlets (a dyad). The first five columns use ``Same domain owner'' as the key explanatory variable, an indicator for whether both outlets in the pair share a common domain-level owner. Columns 6 to 10 instead use the ``Ownership shared percentage,'' a continuous measure of the share of ultimate ownership held in common by the two outlets, measured in percentage points (0--100), as in the main-text tables. All regressions are estimated at the outlet-pair (dyad) level over two periods, allowing for within-pair comparisons of how coverage similarity changes as ownership structures evolve.  The specifications vary in the inclusion of additional controls and fixed effects. Specifications (1) and (6) do not account for dyad FEs, but control for whether the outlets are based in the same country, the same state, or share the same media type. Those specifications also explicitly show the estimate for the period FE (in the form of a dummy for the post-change period; ``Period (after)''). The lines labeled ``No 'Same owner' variability,'' ``Variability Same$\rightarrow$Not same,'' and ``Variability Not same$\rightarrow$Same'' describe whether the sample includes pairs that change their domain-level ownership status between the first and second period. Because these specifications require the additional shared-market controls, the estimation sample is restricted to pairs with complete information on country, state, and media type, observed in both periods (a balanced panel, as in the main tables), which reduces the sample relative to Table~\ref{tab:coverage_pairs}. Standard errors are shown in parentheses below the respective coefficient estimates and are dyadic two-way clustered by domain (\texttt{domain\_a} and \texttt{domain\_b}), following \citet{cameron_miller_2015}
}
\end{tablenotes}
\end{threeparttable}
\end{table}
\end{landscape}

\begin{landscape}
\begin{table}[htbp]
\centering
\begin{threeparttable}
\caption{Robustness Checks: Tone Similarity and Ownership} \label{tab:tone_robustness}
\footnotesize
\singlespacing

% Table created by stargazer v.5.2.3 by Marek Hlavac, Social Policy Institute. E-mail: marek.hlavac at gmail.com
% Date and time: Tue, Jul 21, 2026 - 09:09:36 AM
\begin{tabular}{@{\extracolsep{5pt}}lcccccccccc} 
\\[-1.8ex]\hline 
\hline \\[-1.8ex] 
 & \multicolumn{10}{c}{Dependent variable:} \\ 
\cline{2-11} 
\\[-1.8ex] & \multicolumn{10}{c}{Cosine-Similarity(tone news site a, tone news site b)} \\ 
\\[-1.8ex] & (1) & (2) & (3) & (4) & (5) & (6) & (7) & (8) & (9) & (10)\\ 
\hline \\[-1.8ex] 
 Same domain owner & 0.4013$^{***}$ & 0.3446$^{***}$ & 0.2979$^{***}$ & 0.4213$^{***}$ & 0.1746$^{***}$ &  &  &  &  &  \\ 
  & (0.0151) & (0.0431) & (0.0321) & (0.0489) & (0.0429) &  &  &  &  &  \\ 
  & & & & & & & & & & \\ 
 Ownership shared percentage &  &  &  &  &  & 0.0035$^{***}$ & 0.0020$^{***}$ & 0.0030$^{***}$ & 0.0020$^{***}$ & 0.0019$^{***}$ \\ 
  &  &  &  &  &  & (0.0001) & (0.0002) & (0.0004) & (0.0002) & (0.0002) \\ 
  & & & & & & & & & & \\ 
 Same country & 0.0464$^{***}$ &  &  &  &  & 0.0330$^{***}$ &  &  &  &  \\ 
  & (0.0026) &  &  &  &  & (0.0025) &  &  &  &  \\ 
  & & & & & & & & & & \\ 
 Same state & 0.0298$^{***}$ &  &  &  &  & 0.0373$^{***}$ &  &  &  &  \\ 
  & (0.0050) &  &  &  &  & (0.0062) &  &  &  &  \\ 
  & & & & & & & & & & \\ 
 Same media type & 0.0121$^{***}$ &  &  &  &  & 0.0027$^{**}$ &  &  &  &  \\ 
  & (0.0017) &  &  &  &  & (0.0013) &  &  &  &  \\ 
  & & & & & & & & & & \\ 
 Period (after) & 0.0009 &  &  &  &  & $-$0.0018 &  &  &  &  \\ 
  & (0.0023) &  &  &  &  & (0.0023) &  &  &  &  \\ 
  & & & & & & & & & & \\ 
 constant & 0.0462$^{***}$ &  &  &  &  & 0.0537$^{***}$ &  &  &  &  \\ 
  & (0.0026) &  &  &  &  & (0.0024) &  &  &  &  \\ 
  & & & & & & & & & & \\ 
\hline \\[-1.8ex] 
Period FE & No & Yes & Yes & Yes & Yes & No & Yes & Yes & Yes & Yes \\ 
Dyad FE & No & Yes & Yes & Yes & Yes & No & Yes & Yes & Yes & Yes \\ 
No "Same owner" variability & Yes & Yes & No & Yes & Yes & Yes & Yes & No & Yes & Yes \\ 
Variability Same$\rightarrow\textnormal{Not same}$ & Yes & Yes & Yes & Yes & No & Yes & Yes & Yes & Yes & No \\ 
Variability Not same$\rightarrow\textnormal{Same}$ & Yes & Yes & Yes & No & Yes & Yes & Yes & Yes & No & Yes \\ 
Mean dep. var. & 0.085 & 0.085 & 0.367 & 0.379 & 0.338 & 0.085 & 0.085 & 0.367 & 0.379 & 0.338 \\ 
No. news sites & 1,281 & 1,281 & 395 & 345 & 87 & 1,281 & 1,281 & 395 & 345 & 87 \\ 
No. observations & 800,010 & 800,010 & 2,106 & 1,452 & 654 & 800,010 & 800,010 & 2,106 & 1,452 & 654 \\ 
Within R$^2$ &  & 0.019 & 0.363 & 0.02 & 0.002 &  & 0.065 & 0.27 & 0.064 & 0.058 \\ 
R$^{2}$ & 0.1796 & 0.7408 & 0.6834 & 0.7407 & 0.7396 & 0.2778 & 0.7530 & 0.6371 & 0.7522 & 0.7543 \\ 
\hline 
\hline \\[-1.8ex] 
\end{tabular} 

\begin{tablenotes}[flushleft]
\footnotesize{
\item \textit{Notes}: This table presents additional regression specifications examining the robustness of the relationship between tone similarity and media ownership. The dependent variable is the cosine similarity in tone between two news outlets (a dyad), computed from the outlets' event-level tone vectors as described in the main text. The first five columns use ``Same domain owner'' as the key explanatory variable, indicating whether both outlets in the pair share a common domain-level owner. Columns 6 to 10 use ``Ownership shared percentage,'' a continuous measure of the share of ultimate ownership held in common by the two outlets, measured in percentage points (0--100), as in the main-text tables. All regressions are estimated at the outlet-pair level over two periods, allowing for within-pair comparisons of how tone alignment changes as ownership structures evolve. The specifications vary in the inclusion of controls and fixed effects. Specifications (1) and (6) do not include dyad fixed effects but control for whether the outlets are in the same country, same state, or share the same media type; they also explicitly include a period dummy (``Period (after)'') for the post-change period. The lines labeled ``No `Same owner' variability,'' ``Variability Same$\rightarrow$Not same,'' and ``Variability Not same$\rightarrow$Same'' indicate whether the samples include pairs that change their domain-level ownership status between the two periods, facilitating identification from these transitions. Because these specifications require the additional shared-market controls, the estimation sample is restricted to pairs with complete information on country, state, and media type and non-missing tone similarity, observed in both periods (a balanced panel, as in the main tables), which reduces the sample relative to Table~\ref{tab:reporting_tone}. Standard errors are shown in parentheses below the respective coefficient estimates and are dyadic two-way clustered by domain (\texttt{domain\_a} and \texttt{domain\_b}), following \citet{cameron_miller_2015}
}
\end{tablenotes}
\end{threeparttable}
\end{table}
\end{landscape}
\clearpage
\subsection{Subsamples by media-type}

\begin{table}[htbp]
\centering
\begin{threeparttable}
\caption{Coverage Similarity and Ownership Across Media Types} \label{tab:coverage_similarity_media_types}
\footnotesize
\singlespacing

% Table created by stargazer v.5.2.3 by Marek Hlavac, Social Policy Institute. E-mail: marek.hlavac at gmail.com
% Date and time: Mon, Jul 20, 2026 - 10:27:48 AM
\begin{tabular}{@{\extracolsep{5pt}}lcccccc} 
\\[-1.8ex]\hline 
\hline \\[-1.8ex] 
 & \multicolumn{6}{c}{Dependent variable:} \\ 
\cline{2-7} 
\\[-1.8ex] & \multicolumn{6}{c}{DSC(events covered by news site a, events covered by news site b)} \\ 
\\[-1.8ex] & (1) & (2) & (3) & (4) & (5) & (6)\\ 
\hline \\[-1.8ex] 
 Same domain owner & 0.168$^{***}$ & 0.405$^{***}$ & 0.280$^{***}$ &  &  &  \\ 
  & (0.052) & (0.019) & (0.028) &  &  &  \\ 
  & & & & & & \\ 
 Ownership shared perc. &  &  &  & 0.005$^{***}$ & 0.003$^{***}$ & 0.003$^{***}$ \\ 
  &  &  &  & (0.002) & (0.0001) & (0.0002) \\ 
  & & & & & & \\ 
\hline \\[-1.8ex] 
Period FE & Yes & Yes & Yes & Yes & Yes & Yes \\ 
Dyad FE & No & No & No & No & No & No \\ 
Sample & Digital & Print & Broadcast & Digital & Print & Broadcast \\ 
Mean dep. var. & 0.022 & 0.04 & 0.078 & 0.018 & 0.035 & 0.073 \\ 
No. news sites & 35 & 957 & 292 & 35 & 957 & 292 \\ 
No. observations & 1,190 & 914,892 & 84,972 & 1164 & 901766 & 83092 \\ 
Within R$^2$ & 0.14 & 0.183 & 0.095 & 0.233 & 0.189 & 0.114 \\ 
R$^{2}$ & 0.141 & 0.184 & 0.102 & 0.233 & 0.190 & 0.119 \\ 
\hline 
\hline \\[-1.8ex] 
\end{tabular} 

\begin{tablenotes}[flushleft]
\footnotesize{
\item \textit{Notes}: This table presents regressions like the ones shown in the main text, but this time separately per media type. The dependent variable is the Dice similarity coefficient (``DSC''), which measures the overlap in event coverage between two outlets. Columns (1)--(3) use ``Same domain owner'' as the key explanatory variable, indicating whether both outlets share the same domain-level owner. Columns (4)--(6) replace this binary measure with ``Ownership shared perc.'', a continuous measure of the percentage of ultimate ownership held in common. Each column restricts the sample to a specific medium: digital-native outlets (``Digital''), print-native outlets (``Print''), or broadcast (TV/radio-native) outlets (``Broadcast''), examining either all pairs of outlets from that medium (columns 1--3) or the subset of pairs that never share a domain-level owner (columns 4--6). All specifications include period fixed effects but no dyad fixed effects; these media-type splits are therefore identified from cross-sectional variation across pairs, unlike the within-dyad specifications of the main tables. Standard errors are shown in parentheses below the respective coefficient estimates and are dyadic two-way clustered by domain (\texttt{domain\_a} and \texttt{domain\_b}), following \citet{cameron_miller_2015}. % By conditioning on media type, we assess whether ownership structures affect coverage similarity differently in different segments of the news market. All specifications include period fixed effects, and standard errors are clustered at the dyad level. The table also reports the mean of the dependent variable and the number of observations and news sites in each subset, along with the within R$^2$.
}
\end{tablenotes}
\end{threeparttable}
\end{table}
\clearpage
\subsection{Within-country robustness}

\begin{landscape}
\begin{table}[htbp]
\centering
\begin{threeparttable}
\caption{Coverage Similarity, Within-Country Dyads Only} \label{tab:cov_withincountry}
\footnotesize
\singlespacing

% Table created by stargazer v.5.2.3 by Marek Hlavac, Social Policy Institute. E-mail: marek.hlavac at gmail.com
% Date and time: Thu, May 28, 2026 - 02:14:58 PM
\begin{tabular}{@{\extracolsep{0.0pt}}lccccccc} 
\\[-1.8ex]\hline 
\hline \\[-1.8ex] 
 & \multicolumn{7}{c}{Dependent variable:} \\ 
\cline{2-8} 
\\[-1.8ex] & \multicolumn{7}{c}{DSC(events covered by news site a, events covered by news site b)} \\ 
\\[-1.8ex] & (1) & (2) & (3) & (4) & (5) & (6) & (7)\\ 
\hline \\[-1.8ex] 
 Same domain owner & 0.361$^{***}$ & 0.328$^{***}$ &  &  &  &  &  \\ 
  & (0.013) & (0.031) &  &  &  &  &  \\ 
  & & & & & & & \\ 
 Ownership shared perc. &  &  & 0.001$^{***}$ &  & 0.003$^{***}$ &  &  \\ 
  &  &  & (0.0001) &  & (0.0003) &  &  \\ 
  & & & & & & & \\ 
 Ownership cosine simil. &  &  &  & 0.132$^{***}$ &  &  &  \\ 
  &  &  &  & (0.012) &  &  &  \\ 
  & & & & & & & \\ 
 Above 50 shared perc. &  &  &  &  &  & 0.191$^{***}$ &  \\ 
  &  &  &  &  &  & (0.035) &  \\ 
  & & & & & & & \\ 
 Own. sh. perc. Q1-2 &  &  &  &  &  &  & 0.041$^{***}$ \\ 
  &  &  &  &  &  &  & (0.012) \\ 
  & & & & & & & \\ 
 Own. sh. perc. Q2-3 &  &  &  &  &  &  & 0.119$^{***}$ \\ 
  &  &  &  &  &  &  & (0.013) \\ 
  & & & & & & & \\ 
 Own. sh. perc. Q3+ &  &  &  &  &  &  & 0.258$^{***}$ \\ 
  &  &  &  &  &  &  & (0.029) \\ 
  & & & & & & & \\ 
\hline \\[-1.8ex] 
Period FE & Yes & Yes & Yes & Yes & Yes & Yes & Yes \\ 
Dyad FE & No & Yes & Yes & Yes & Yes & Yes & Yes \\ 
Sample & Within-cty. & Within-cty. & Within-cty., not same d. o. & Within-cty., not same d. o. & Within-cty., some sh. o. & Within-cty., some sh. o. & Within-cty., some sh. o. \\ 
Mean dep. var. & 0.06 & 0.06 & 0.055 & 0.055 & 0.132 & 0.132 & 0.132 \\ 
No. news sites & 2,019 & 2,019 & 2,019 & 2,019 & 721 & 721 & 721 \\ 
No. observations & 1,624,876 & 1,624,876 & 1,596,212 & 1,596,212 & 63,118 & 63,118 & 63,118 \\ 
Within R$^2$ & 0.129 & 0.037 & 0.046 & 0.042 & 0.067 & 0.029 & 0.094 \\ 
R$^{2}$ & 0.130 & 0.758 & 0.752 & 0.751 & 0.732 & 0.721 & 0.739 \\ 
\hline 
\hline \\[-1.8ex] 
\end{tabular} 

\begin{tablenotes}[flushleft]
\footnotesize{
\item \textit{Notes}: Replication of Table~\ref{tab:coverage_pairs} restricted to within-country dyads. The headline same-domain-owner coefficient moves from 0.330 to 0.328 (SE 0.031). Sub-sample labels carry the additional within-country restriction. Standard errors are dyadic two-way clustered by domain (\texttt{domain\_a} and \texttt{domain\_b}), following \citet{cameron_miller_2015}. $^{*}$p$<$0.1; $^{**}$p$<$0.05; $^{***}$p$<$0.01.
}
\end{tablenotes}
\end{threeparttable}
\end{table}
\end{landscape}

\begin{landscape}
\begin{table}[htbp]
\centering
\begin{threeparttable}
\caption{Tone Similarity, Within-Country Dyads Only} \label{tab:tone_withincountry}
\footnotesize
\singlespacing

% Table created by stargazer v.5.2.3 by Marek Hlavac, Social Policy Institute. E-mail: marek.hlavac at gmail.com
% Date and time: Thu, May 28, 2026 - 02:15:21 PM
\begin{tabular}{@{\extracolsep{0.0pt}}lccccccc} 
\\[-1.8ex]\hline 
\hline \\[-1.8ex] 
 & \multicolumn{7}{c}{Dependent variable:} \\ 
\cline{2-8} 
\\[-1.8ex] & \multicolumn{7}{c}{Cosine‐Similarity(tone news site a, tone news site b)} \\ 
\\[-1.8ex] & (1) & (2) & (3) & (4) & (5) & (6) & (7)\\ 
\hline \\[-1.8ex] 
 Same domain owner & 0.416$^{***}$ & 0.341$^{***}$ &  &  &  &  &  \\ 
  & (0.015) & (0.043) &  &  &  &  &  \\ 
  & & & & & & & \\ 
 Ownership shared perc. &  &  & 0.002$^{***}$ &  & 0.002$^{***}$ &  &  \\ 
  &  &  & (0.0002) &  & (0.001) &  &  \\ 
  & & & & & & & \\ 
 Ownership cosine simil. &  &  &  & 0.186$^{***}$ &  &  &  \\ 
  &  &  &  & (0.019) &  &  &  \\ 
  & & & & & & & \\ 
 Above 50 shared perc. &  &  &  &  &  & 0.033$^{**}$ &  \\ 
  &  &  &  &  &  & (0.016) &  \\ 
  & & & & & & & \\ 
 Own. sh. perc. Q1-2 &  &  &  &  &  &  & 0.036$^{*}$ \\ 
  &  &  &  &  &  &  & (0.020) \\ 
  & & & & & & & \\ 
 Own. sh. perc. Q2-3 &  &  &  &  &  &  & 0.135$^{***}$ \\ 
  &  &  &  &  &  &  & (0.020) \\ 
  & & & & & & & \\ 
 Own. sh. perc. Q3+ &  &  &  &  &  &  & 0.115$^{***}$ \\ 
  &  &  &  &  &  &  & (0.024) \\ 
  & & & & & & & \\ 
\hline \\[-1.8ex] 
Period FE & Yes & Yes & Yes & Yes & Yes & Yes & Yes \\ 
Dyad FE & No & Yes & Yes & Yes & Yes & Yes & Yes \\ 
Sample & Within-cty. & Within-cty. & Within-cty., not same d. o. & Within-cty., not same d. o. & Within-cty., some sh. o. & Within-cty., some sh. o. & Within-cty., some sh. o. \\ 
Mean dep. var. & 0.104 & 0.104 & 0.096 & 0.096 & 0.163 & 0.163 & 0.163 \\ 
No. news sites & 1,340 & 1,340 & 1,339 & 1,339 & 471 & 471 & 471 \\ 
No. observations & 495,842 & 495,842 & 485,538 & 485,538 & 23,280 & 23,280 & 23,280 \\ 
Within R$^2$ & 0.16 & 0.023 & 0.077 & 0.07 & 0.021 & 0.001 & 0.067 \\ 
R$^{2}$ & 0.160 & 0.734 & 0.715 & 0.713 & 0.640 & 0.633 & 0.657 \\ 
\hline 
\hline \\[-1.8ex] 
\end{tabular} 

\begin{tablenotes}[flushleft]
\footnotesize{
\item \textit{Notes}: As Table~\ref{tab:cov_withincountry} but with cosine tone similarity as the dependent variable; replication of Table~\ref{tab:reporting_tone}. The headline same-domain-owner coefficient moves from 0.342 to 0.341 (SE 0.043). Standard errors are dyadic two-way clustered by domain (\texttt{domain\_a} and \texttt{domain\_b}), following \citet{cameron_miller_2015}. $^{*}$p$<$0.1; $^{**}$p$<$0.05; $^{***}$p$<$0.01.
}
\end{tablenotes}
\end{threeparttable}
\end{table}
\end{landscape}

\end{document}